\documentclass[11pt,reprint,showpacs,showkeys,nofootinbib,ssfamily]{article}
\usepackage{axodraw2}
\usepackage{amsmath,fontenc,calrsfs}
\usepackage{amsfonts,amssymb,mathtools,slashed,empheq}
\usepackage{cite} 
\usepackage{hyperref} 
\hypersetup{backref, colorlinks=true  } 
\usepackage{multirow}
\usepackage{float}
\usepackage{appendix}
\usepackage{color}
\usepackage{url}
\usepackage{subfigure}
\usepackage{footnote}
\usepackage{authblk} 

\def\beq{\begin{equation}}
\def\eeq{\end{equation}}
\def\bea{\begin{eqnarray}}
\def\eea{\end{eqnarray}}
\def\nn{\nonumber}

\def\chic1{\chi_{c1}}

\newcommand{\hg}{\mathfrak{g}}
\newcommand{\ND}{${\rm ND}$}
\newcommand{\vsi}{\vec{\sigma}}
\newcommand{\mff}{\mathfrak{f}}
\newcommand{\hht}{\hat{t}}
\newcommand{\hhv}{\hat{v}}
\newcommand{\hft}{\mathfrak{t}}
\newcommand{\on}{\overline{n}}
\newcommand{\ep}{\epsilon}
\newcommand{\ve}{\varepsilon}
\newcommand{\vep}{\varepsilon}

\def\Xint#1{\mathchoice
   {\XXint\displaystyle\textstyle{#1}}%
   {\XXint\textstyle\scriptstyle{#1}}%
   {\XXint\scriptstyle\scriptscriptstyle{#1}}%
   {\XXint\scriptscriptstyle\scriptscriptstyle{#1}}%
   \!\int}
\def\XXint#1#2#3{{\setbox0=\hbox{$#1{#2#3}{\int}$}
    \vcenter{\hbox{$#2#3$}}\kern-.5\wd0}}

\def\dashint{\Xint-}

\newcommand*\widefbox[1]{\fbox{\hspace{.20em}#1\hspace{.0em}}}

\def\Xint#1{\mathchoice
   {\XXint\displaystyle\textstyle{#1}}%
   {\XXint\textstyle\scriptstyle{#1}}%
   {\XXint\scriptstyle\scriptscriptstyle{#1}}%
   {\XXint\scriptscriptstyle\scriptscriptstyle{#1}}%
   \!\int}

\newcommand{\vk}{\mathbf{k}}

\newcommand{\vp}{\mathbf{p}}
\newcommand{\vq}{\mathbf{q}}

\newcommand{\vr}{\mathbf{r}}

\newcommand{\tv}{\widetilde{v}}

\newcommand{\Ima}{{\rm Im}\,}
\newcommand{\Rea}{{\rm Re}}
\newcommand{\sgn}{{\rm sgn}}

\usepackage{hyperref}
\hypersetup{%
  colorlinks = true,
  linkcolor  = black
}

\title{The exact discontinuity of a partial wave along the left-hand cut and the exact $N/D$ method 
in non-relativistic scattering} 
\author[]{J.~A. Oller}
\affil[]{\it Departamento de F\'{\i}sica, Universidad de Murcia, E-30071 Murcia, Spain}
\author[]{D.~R.~Entem}
\affil[]{\it Grupo de F\'{\i}sica Nuclear and IUFFyM, Universidad de Salamanca, E-37008 Salamanca, Spain}

\begin{document}
\maketitle
\begin{abstract}
We first deduce the analytical continuation  in the complex planes of the initial and final three-momenta 
of the Lippmann-Schwinger equation in coupled or uncoupled partial-wave amplitudes. 
This  result allows us to deduce a master equation whose solution is the exact discontinuity
of the on-shell partial-wave amplitudes along the left-hand cut. 
This equation is always a linear non-singular integral equation whose solution is fixed 
 exclusively by the knowledge of the potential, applicable to either regular or singular potentials. 
 The capability of calculating exactly this discontinuity allows one to 
 settle  the exact $N/D $ method in two-body non-relativistic scattering for coupled and uncoupled waves.
  We exemplify this new advance in scattering theory by explicitly checking 
the agreement between the Lippmann-Schwinger equation with the corresponding solutions
of the exact $N/D$ method for some examples that involve regular and singular potentials, 
either attractive or repulsive. 
\end{abstract}

\newpage
\tableofcontents
\newpage

\section{Introduction}
\label{ref.180608.1}
\def\theequation{\arabic{section}.\arabic{equation}}
\setcounter{equation}{0}

 The interest for studying collisions between two bodies
in quantum mechanics with an eye on phenomenology leads oneself to the problem of 
scattering with singular potentials rather sooner than later. 
This name stems from the fact that a singular potential behaves too pathologically when the separation distance $r$ 
between the two particles  tends to zero, such that it overcomes de centrifugal barrier. 
 This fact might invalidate  standard results and prevent  the application of ordinary techniques 
for the two-body scattering problem in quantum mechanics with regular potentials 
\cite{Gottfried.170929.2,Faddeev.170929.1}. 
 For example, the terms in the Born series to study scattering are divergent  above some order in the iteration of the potential 
if the latter diverges faster than $1/r^{2+\vep}$ for $r\to 0$ and $\vep>0$. 
 The case $V(r)\to C_2/r^2$ for $r\to0$  is of marginal type  with its regular or singular character 
 depending on the value of $C_2$. These two results are analyzed with further detail in Sec.~\ref{sec:190116.1}.  
 
The problem of collision theory is intimately linked with spectroscopy.
Optical spectroscopy has its origin in 1672, when Newton observed that sunlight could
be separated in colors when passing through a prism.
  In 1802 Wollaston observed numerous dark lines in Newton's spectrum
of sunlight.
  Subsequently since 1814 von Fraunhofer studied and classified in detail this type of lines
by mounting a prism in front of a theodolite with another farther prism in 
 which the light originally impinged. Today his method of
 characterization of the most prominent spectral lines by letters, such as the
 {\rm D} absorption lines of Sodium, is still in use. 
It was also noticed by Fraunhofer that light from different sources can have very different spectra (for example
 the light of the star Sirius compared with that of the Sun).
 Around 45 years later Kirchoff and Bunsen explained the origin of Fraunhofer lines
 in a series of classic articles published in 1859 and 1860. 
They determined that the dark lines in the spectrum were due to absorption by chemical elements
present in the stellar medium, which give rise to  characteristic emission lines with the 
same frequency when heated.

In typical collision experiments we have a background of scattered particles on top of which
 structures that resemble ``lines'' of absorption/emission appear at characteristic energies, similarly
to these pioneering studies of optical spectroscopy. The results of these experiences are used
to learn and delve into the interactions that govern the collisions of the particles and
that give account of the experimental facts.

\subsection{Classical analysis of a collision process}
\label{sec.180429.1}

  The basis for the analysis in classical mechanics of a scattering process is direct for the case
of regular potentials, while requires  extra considerations for a singular potential. 
However, the result in both cases is quite intuitive, and worth keeping in mind 
 for later more abstract  developments.

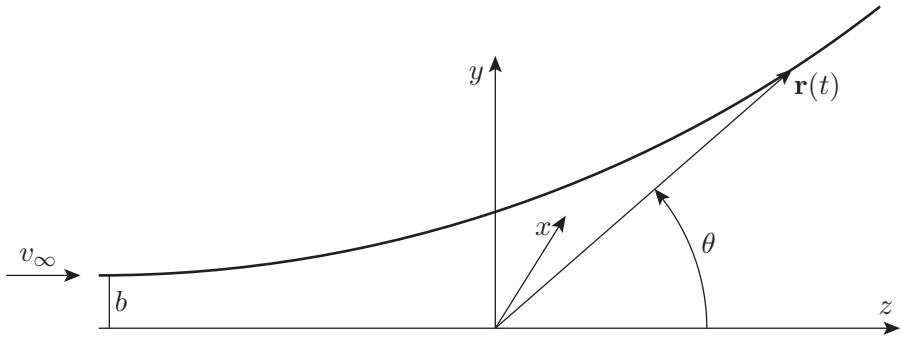
\begin{figure}
\begin{center}
\begin{picture}(300,100)(0,0)
\SetColor{Black}
\LongArrow(0,0)(300,0)
\Text(295,10)[lt]{$z$}
\LongArrow(150,0)(150,100)
\Text(140,99)[lt]{$y$}
\LongArrow(150,0)(175,40)
\Text(165,40)[lt]{$x$}
\SetWidth{1}
\CArc(0,500)(480,270,308)
\SetWidth{0.5}
\Line(4,0)(4,20)
\Text(6,14)[lt]{$b$}
\LongArrow(150,0)(260,96)
\Text(263,97)[lt]{${\bf r}(t)$}
\LongArrowArc(150,0)(80,0,39)
\Text(228,36)[lt]{$\theta$}
\LongArrow(-35,20)(-10,20)
\Text(-30,30)[lt]{$v_{\infty}$}
\end{picture}
\caption{{\small Schematic diagram of a scattering process for a central potential $V(r)$. 
\label{fig.180429.1}}}
\end{center}
\end{figure}

Let us describe the collision process in the center of mass system of the two particles with reduced mass $\mu$, 
 which is represented in Fig.~\ref{fig.180429.1}. 
The projectile and target interact by means of a potential $V(r)$ that is supposed to vanish for $r \to \infty $ 
and  be spherically symmetric (which is enough for our illustrative purposes at this point). 
The energy $E$ of the two particles at infinity is purely kinematical and this fixes the incident relative 
velocity  $v_\infty=\sqrt{2\mu E}$~.
 Because of the conservation of the orbital angular momentum the collision plane is fixed and we take it as 
 the $zy$ plane, with the polar angle $\theta$ such that 
 $y=r\sin\theta$ and $z=r\cos\theta$. The impact parameter $b$ is the minimum distance between the particles in case of 
 absence of interaction. In terms of it the orbital angular momentum is $L_x=b \mu v_\infty $.
 As the particles move closer, the  the polar angle $\theta(t)$
 varies with time according to the potential, being $\theta(t=-\infty)=\pi $ and $\theta(t=+\infty)=\digamma (b)$, 
 the  scattering angle. 
 
 The fundamental equations that govern the collision process at the classical level are the conservation
 of the orbital angular momentum, $L_x=-\mu r^2 \dot{\theta}$, and of the energy, 
 $E=\frac{\mu \dot{r}^2}{2}+\frac{L^2}{2\mu r^2}+V(r)$. For the configuration in 
 Fig.~\ref{fig.180429.1} the conservation of $L_x$ implies that $\dot{\theta} \leq 0 $, so
 that the polar angle, which starts at $\pi$, decreases monotonically with time 
 and if there were relative rotation between the particles this would be clockwise.
  
  Now, let us move to the main point of this presentation of scattering in classical mechanics. 
An important concept for our following discussions is the classical turning point, $r_c$, which is the 
smallest distance between the target and projectile. We also take the origin of time, $t=0$, when $r=r_c$. 
If this point is finite, $r_c\neq 0$, then it is also a turning point 
so that $\dot r=0$ there, which implies the equation $E=\frac{L^2}{2\mu r_c^2}+V(r_c)$ 
with the constraint $E>0$. It is easy to convince oneself that $r_c$ is always finite 
 for a regular as well as for a repulsive singular potential. 
 However, in the case of an attractive singular potential $r_c=0$ 
 for high enough energy, because then any potential barrier has finite height 
 as the potential overcomes the centrifugal potential for $r\to 0$. 
In this case the  two particles crash into  $r=0$, where the velocity becomes infinity. 
One has there a singular point which prevents the continuation in time ($t>0$) of the collision process.  
Indeed, to proceed forward one is required to take further assumptions on the way the particles 
eventually start separating each other by reversing the sign of the radial velocity, $\dot r$. 
If we take a positive infinitesimal time $t=0^+$ with an infinitesimal separation between the particles $r=\varepsilon$,   
one can use the conservation of orbital angular momentum and energy to fix $\dot\theta$ and $\dot r$, 
but there is no way to fix the polar angle. Thus, it may have an arbitrary value for the onset of the 
separation process, a value that can be taken as a function of $E$ and orbital angular momentum, $\theta_\ell(E)$.

Thus, in the case of falling into $r=0$ for an attractive singular potential 
we can still describe classically the next phase of separation between the particles but at the expenses of
 requiring a new function of the
 energy and orbital angular momentum that is not encoded neither in the initial state nor in the knowledge
 of the potential for $r\neq 0$. In this sense there is a loss of information because of having reached 
 the singular point at $r=0$, which should be counterbalanced by some kind of microscopic degrees of freedom.
 As a result, the point $r=0$ would represent a region of short distances whose dynamics
 is unknown to us but whose effects can be classically parameterized through the knowledge of the
 potential at finite (or relatively large) distances and the a priori unknown function $\theta_\ell(E)$.

In summary, for the singularly attractive case we can have
classically two spiral trajectories, one in-going and another out-going, which are not part of
the same hyperbolic orbit that would connect them uniquely. Additional information is required
 beyond the knowledge of the potential and the initial conditions of problem to
  characterize the out-going branch 
 once the singular point $r=0$ is reached. Of course, there is also the physical possibility that there is no 
 out-going trajectory and that the target and projectile fuse together in a system with some other dynamics.

\subsection{Quantum mechanical considerations}
\label{sec.180429.2}

It is well-known that the understanding of the atomic spectra played an important role in the (early) development 
 of quantum mechanics. It is also true, as it is our main point to show in the following exposition, 
that very soon this phenomenological interest drove to the necessity 
to perform calculations involving singular potentials. 

 The energy levels of a Hydrogen atom for a pure Coulomb force are given by the 
 Bohr formula in non-relativistic quantum mechanics,
\begin{align}
\label{180426.1}
E_n&=-\frac{\mu \alpha^2}{2n^2}~,
\end{align}
where $\mu=m_e/(1+m_e/M)$ is the reduced mass of the electron and nucleus, whose individual masses are  
$m_e$ y $M$, in order, $\alpha$ is the fine structure and $n$ is the principal quantum number \cite{Gottfried.170929.2}. 
 From this equation the frequency $T_n$ of a photon emitted in the transition between $n\to n_0$, with  
 $n>n_0$, is
\begin{align}
\label{180426.2}
T_n&=\frac{E_{n}-E_{n_0}}{2\pi}=R\left(\frac{1}{n^2}-\frac{1}{n_0^2}\right)~.
\end{align}
This equation was obtained phenomenologically by Rydberg in 1889 and in his honor the constant
$R=m_e\alpha^2/4\pi $ is called the Rydberg constant.\footnote{We are neglecting here the tiny difference between 
the electron mass and the reduced mass $\mu$ of the electron-nucleon system.}
The first formula for the series of spectral lines, encompassing the visible ones,
was discovered a few years earlier by Balmer (1885) with $n_0=2$,
 which in turn allowed him to predict new lines.
 In addition Rydberg also  obtained an empirical formula for the spectral lines of the
  alkali atoms,\footnote{Alkali atoms  have $ Z-1 $ electrons forming
  closed layers corresponding to the noble gas that immediately precedes the element and the last
  electron on a $ns$ level. It is $Li$, $Na$, $ K $, etc.}
\begin{align}
\label{180426.3}
T_n&=T_\infty-\frac{R}{n-\upsilon}~,
\end{align}
with $T_\infty$ the series limit and $\upsilon$ is known as the quantum defect. 
 Subsequent works showed that better fits to experiment could be obtained by allowing  $\upsilon$ to depend
 slowly on $T_n$. The Eq.~\eqref{180426.3} suggests that
the alkali atoms have energy levels that go like 
\begin{align}
\label{180426.4}
E_n&=-\frac{\mu \alpha^2}{2(n-\upsilon)^2}~.
\end{align}
The presence of the quantum defect is the result of the variation of the potential
 with respect to the pure Coulomb one 
 due to the electrons in the innermost layers. It follows then that  $\upsilon$ 
 is in general a function of $E$ and $\ell$, denoted by $\upsilon_{\ell}(E_n)$, 
that decreases as $\ell$ increases because 
 the atomic wave function of the most exterior electronic state is less sensitive 
 to the inner structure of the electron core.
 The function $\upsilon_{\ell}(E_n)$ is fitted for a given $\ell$ with a low-degree polynomial in $E$, 
and it summarizes concisely important information both for spectroscopy and scattering.

 The explanation of quantum defects is an old subject in physics. 
 Sommerfeld \cite{sommerfeld.180501.1} in the years 1916 and 1920 published 
 some works based on the old quantum mechanics where he 
showed that the quantum defect is related to the precession of the elliptical orbit
of the external electron because the potential is not a pure $1/r$ function. 
 When the orbit penetrates in the inner region 
 the speed of the electron increases because the Coulomb attraction of the nucleus is greater as the shielding 
 of the internal electronic cloud diminishes, causing the precession of the orbit. 
 For states with greater orbital angular momentum, the penetration into the internal zone is lower and the
 quantum defect $\upsilon_{\ell}(E_n)$ is smaller. To improve the agreement with experience in the case of
large $\ell$ one should take into account the polarization phenomena that originate an extra term in the
 potential of the form $\beta/r^4 $, as first noted by Bohr (1923) \cite{bohr.180501.1}.
  A perturbative analysis was carried out by Born and Heisenberg (1926) \cite{born.180501.1} 
  within the old quantum mechanics, and by Waller (1926) \cite{waller.180501.1} using standard quantum mechanics.
  Potentials that depends on $r$ as $1/r^4$ over long distances play an important role in 
 electron-atom and ion-atom interactions.
  This is an example that clearly shows the appearance of a singular 
 potential (since it diverges as $1/r^4 $ for $r \to 0$) that is important at the phenomenological level.
 Of course, there are many other familiar examples of important singular potentials like the 
 van der Waals potential,
\begin{align}
\label{180429.1}
V(r)&=-\frac{C_6}{r^6}~,~C_6>0~.
\end{align}
  This potential is crucial in molecular physics to characterize 
 the long-distance interaction between neutral atoms in a molecule. 
At smaller distances the intermolecular potential becomes repulsive, e.g. because of the 
Pauli repulsion, a phenomenological fact that is reflected in the well-known Leonard-Jones potential.

Historically the quantum theory that accounted for the quantum defects of the energy levels 
 of a valence electron due to the distortion of the Coulomb interaction by the electrons in the
internal layers is called Quantum Defect Theory (QDT),  
and its basic formulation is due to Hartree in 1928 \cite{hartree.180428.1} and
 subsequently by others, remarkably by Seaton \cite{seaton.180427.1,seaton.180427.2}.
 A phenomenological scale $r_0$ is introduced so that the potential is decomposed into a short-distance potential
  for $r<r_0$, and another for long distances and applied wherever $r>r_0$.
 The potential is not explicitly available at short distances since the dynamics in the internal region
 is much more complicated. The potential at long distances
 in the works of Hartree and Seaton is a Coulomb potential $-Ze/r$, with $ Z = Z_0-N$, being
 $+Z_0 e$ the core charge and $N$ the number of electrons in the internal region that shield the 
 nuclear charge.  The two linearly independent electronic wave functions are solved 
 algebraically for $r>r_0$, the so called Coulomb functions.
  The relative coefficient in their linear superposition is
   fixed by matching with a slowly energy-dependent parameterization 
  of the logarithmic derivative at $r=r_0$,  
controlled by the small dimensionless parameters $r_0 k\ll 1$ ($k=\sqrt{2 \mu |E|}$). 
 The explicit evaluation of the long-distance part of the electronic wave function allows to take into 
 account the much more important energy dependence that is controlled by a longer distance scale $\beta_n$ 
 associated with the long-range potential. So for a potential $C_n r^{-n}$ the new scale of distances $\beta_n$
 is determined by dimensional analysis from the Schr\"odinger equation and is given by the expression 
 \begin{align}
 \label{180514.1}
\beta_n&= (2 \mu |C_n|/\ell(\ell+1))^{1/(n-2)}~.
 \end{align}
In this way the energy dependence for $k r_0 \ll 1$ is controlled by the dimensionless variable
 $\beta_n k$, being much more relevant than the weak dependence on energy generated
 from the logarithmic derivative at short distances $\widetilde{\lambda} (E)$. 
 The latter is related with the short-distance $K$ matrix which is an analytical function of 
 the energy \cite{gao.180610.1}. As first noticed by Hartree \cite{hartree.180428.1},  
the fact that the short-range potential is much larger than the energy in the region $r\leq r_0$ 
 implies the slow dependence in $E$ of $\lambda(E)$.  
  In this way, it is enough to parameterize the latter with a low-degree polynomial in energy whose coefficients 
can be fixed from the knowledge of the spectrum of a few bound states close 
to the threshold of the system. Then it is possible to predict other bound states, 
resonances or the scattering between these systems.

Along the time this formulation has been extended to long-range potentials of 
the type $-1/r^\alpha $ with $\alpha \leq 2$ \cite{greene.180427.1,greene.180427.2,greene.180427.3},
 and then for $\alpha>2$, thus entering the region of singular potentials.
 For the long-range region of the potential, algebraic solutions have been obtained
 for singular potentials of the type $1/r^n $ with $n=4$ \cite{vogt.180503.1,gao.180428.1}, 
 $n=3$ \cite{gao.180428.2} and 6 \cite {gao.180428.3}. 
 The potential $1/r^3$ generates the long distance part
 for the dipole-dipole interaction as well as the tensor interaction by the exchange of a pion
in nuclear physics;
the interaction between a neutral atom and a charged atomic system is governed by an interaction
of the type $1/r^4$, and the $1/r^6$ van der Waals potential describe
the interaction at relatively large distances between neutral atoms,
 cf. the equation \eqref{180429.1}.
 Recently Bao {\it et al.} \cite{gao.180428.4} solved analytically
  the potential $V(r)=-\left(C_1/r + C_4/r^4 \right)$ and applied it within the multiscale QDT.
 This is the first time in which there is an
  analytical solution for a long distance potential that involves two scales.
  This new development allowed them to characterize the dependence on the principal quantum number $n$
 of the quantum defects for states more tightly bound than the so-called Rydberg states (which
 must be sufficiently excited).
 
Now, let us discuss typical solutions of the Schr\"odinger equation with a singular potential 
for all $r>0$. 
The results that follow are quite plausible if one keeps in mind those discussed above 
in classical mechanics. 
 So, for a repulsive singular  for $r\to 0$ we have a turning point at $r=r_c$, 
with an infinite potential barrier, so that 
 there is only one independent solution   that vanishes 
at $r=0$ and the solution for the reduced wave function is fixed. 
In this way, the classical in- and out-going trajectories form the rays of a quantum wave which is 
uniquely defined by the initial conditions and the potential $V(r)$.
 Therefore, the treatment of  repulsive singular potentials within the Schr\"odinger equation is 
 analogous to the case of a regular potential. 

 For an  attractive singular potential and sufficiently high angular momentum we still have a  
 well-defined turning point at some $r_c>0$. The situation, however, is not analogous to ordinary potential 
 scattering because the potential barrier is finite, as the attractive singular potential overcomes 
the centrifugal one in in the limit $r\to 0$, and again we have an inner zone where the energy 
 is greater than the potential. This implies the existence of two linearly independent solutions 
 for the Schr\"odinger equation in that region. 
 On the other hand, when the angular momentum is small enough the decrease of the
 strength of the centrifugal potential causes that there is no turning point and that classically the two
 particles clash at the origin. As we discussed along the classical treatment, because of the singularity at $ r = 0 $ 
 we have two separate spiral trajectories for $r\to 0$, one in-going and another out-going, which are not part of a
 continuously connected hyperbolic trajectory that relates them uniquely. 
 Even  if we impose classically the conservation of energy and angular momentum
and we know the radial position, it would not be possible to fix the value of the polar angle 
 when leaving $r=0$. This has its imprint in the quantum case since
the classical trajectories are the rays of the quantum wave front, which will thus be clearly undefined
in the area around the singularity at $r=0$. That is, the solution of the Schr\"odinger equation is not unique 
 in terms of the initial incident conditions (energy and angular momentum) 
 together with the usual boundary condition of vanishing the radial wave function at $r=0$.
  This ambiguity was associated in the classical case with the need to provide the leaving value of the polar angle,
which we denoted by $\theta_\ell(E)$. At the quantum level, the value of this angle manifests in the need
to specify the relative phase between the in-going and out-going waves, which we denominate by $\varphi_\ell(E) $.

It is also a well-known fact \cite{case.180502.1} that the semiclassical approximation for the region 
$r\to 0$ is well suited. One way to see it is to realize that the quantum fluctuations 
of a wave with a typical size $h$ imply a typical momentum uncertainty of around $1/h$,
 which in turn induces a contribution to the kinetic energy that diverges like $1/h^2$ 
for $h\to 0$. However, this behavior is overtaken by the singular attractive potential in the same limit which plays 
an increasingly dominant role for $r\to 0$. Of course, one can also invoke the fulfilling for $r\to 0$ of the 
standard condition for the WKB method, $|d\lambda(r)/dr|\ll1$ with $\lambda(r)$ the de Broglie wave length. 
 Simple classical considerations can be applied to derive in a quite intuitively way 
the most remarkable features of the wave function for $r\to 0$.
 Let us suppose the attractive case first, with a potential that in the limit $r\to 0$ 
 diverges as $-1/r^n$, $n>2$, and so the relative three-momentum $p(r)=\sqrt {2\mu(E-V (r))}$ tends to $r^{-n/2}$
 for $r \to 0 $. Since the probability of finding a particle at a certain point of its orbit is
inversely proportional to its speed, we have that the probability of finding the particles in the region of short
distances decrease as $1/r^{-n/2}=r^ {n/2} $. The square root of the previous probability gives us
the magnitude of the reduced wave function, which therefore decreases as $r^{n/4} $ for $r \to 0$.
This solution is oscillating with a wave number that coincides with $p(r)$ and that, therefore, diverges as
$r^{-n/2} $. Thus,  for $r\to 0$ we have a wave oscillating increasingly fast in $r$ (with a phase  
that goes like $r^{-n/2+1}$) and having a vanishing amplitude as $r^{n/4}$.
 To treat the repulsive singular potentials, just transform $p(r)\to i |p(r)|$
and instead of two  oscillating solutions we have a unique wave function that is finite for 
$r\to 0$, that behaves as $r^{n / 4} \exp(-B r^{-n/2+1})$,
 with $B$ a constant.
 It is also clear that as the potential is more singular at short distances
  they generate wave functions increasingly suppressed
 in the limit $r \to 0$.

A related aspect of the previous discussion has to do with the characteristic fact of
 quantization of the binding energies in quantum mechanics, which is a property in all studies of 
 spectroscopy. However, we have concluded above that for an attractive singular potential  
the relative phase $\varphi_\ell(E)$ is a continuous function of $E$, as if the spectrum were always 
continuum and unbound. This conclusion was already obtained by Plesset in 
1932 \cite{plesset.180501.1} in the study of solutions to the Dirac equation 
 with singular potentials expressed as polynomials in  $1/r$. 
 The quantization of the binding energies calculated employing an attractive singular potential 
 in the Schr\"odinger equation (as well as in relativistic equations, like the Dirac one)
follows by imposing the vanishing of the wave function for $r\to \infty$ together with 
the extra condition of orthogonality between the eigenstates with different energies 
as introduced by  Case \cite{case.180502.1}. Even more, because $|V(r)|>> E$ for $r\to 0$ 
the orthogonality condition requires that the phase is the same for all the eigenstates of energy, 
so that only one experimental input, e.g. a binding energy, is necessary to fix the whole 
spectrum, which typically still remains unbounded from below \cite{frank.180502.1}, cf. 
Sec.~\ref{sec:190116.1}.

In the previous treatment of the Schr\"odinger equation with an attractive singular potential
one insisted of keeping unitarity, which requires an in-going and an out-going wave in the collision process 
through the origin. This was also the case analyzed classically, where we considered having these two branches 
in the trajectory of the particles. There we also mentioned the possibility about 
 having the fusion between the particles so that there exists only an incoming  
branch. This is precisely the point of view developed by Vogt and Wannier \cite{vogt.180503.1}, who  
considered the solution of the Schr\"odinger with only an ingoing wave 
and calculated the cross section of ion absorption by atoms. 
One is taking the center of an attractive singular potential as a sink and, of course, 
there is no conservation of probability and unitarity is violated. 
 It should be noted that recently using  cold atoms
\cite{denschlag.180503.1} it was possible to study experimentally an
 attractive interaction of the type $V(r)=-C_2/r^2$ by charging a thin wired. 
 The coefficient $C_2\geq 0$ 
can be controlled by the voltage applied  to the wire. 
 The experiment \cite{denschlag.180503.1} actually observed an 
acceleration in the rate of loss of atoms in the optical trap and
other characteristics of the interaction $-1/r^2$,
 as the linear increase of the cross section with the linear density charge (voltage) in the wire.
Thus, the absorption of atoms in the singularity of $V(r)=-C_2/r^2$ was experimentally established as a fact.

This result shows  that to impose or not unitarity in the scattering solution is ultimately
 a physical criterion that will have to be justified according to the nature of the phenomenon that is being studied.
For example, for the interaction between neutral atoms governed at long distances by the interaction of van der
Waals forces one would expect the solution to comprise  in- and out-going waves  
because at short distances we have the Pauli repulsion between the electronic orbitals. 
Similarly, for nuclear interactions one would expect that
this  is also the case given the repulsion of nucleon centers (repulsion between constituent quarks at a 
deeper level).
A comprehensive review of the techniques
and applications of singular potentials up to the year 1970 is given in  Ref.~\cite{frank.180502.1}.

The imposition of the orthogonality condition to solve the Schr\"odinger equation 
in the presence of singular potentials  to determine the eigenfunctions of 
the continuum spectrum with positive $E>0 $ seems to be rather recent. 
 Up to the best of our knowledge, it has not occurred until the works of Arriola and Pav\'on \cite{arriola.180502.1}, where 
the authors  apply this technique to study the nucleon-nucleon ($NN$) interactions in vacuum and 
also derive its generalization to coupled waves.

The idea of ​​Case to impose the aforementioned orthogonality between eigenfunctions
with different energy values ​​has been  focused later on within the problem of the 
self-adjoint extension of an hermitian operator, in this case the Hamiltonian with a
singular potential \cite{meetz.180503.1}.
 Of course, the resulting scattering process is unitary.
 The basic idea resides in the so-called deficiency indices of the  Hamiltonian.\footnote {In the discussion that follows
to simplify matters we do not take into account some peculiarity of the potential $1/r^2 $. A deeper discussion 
on this potential can be found in Sec.~\ref{sec:190116.1} and in Refs.~\cite{landau.180720.1,gopa.180502.1}.} 
 If the Hamiltonian is
essentially self-adjoint (as it happens for singular repulsive potentials) 
 the solution is analogous to the regular case.
However, for the attractive singular case one has to impose an additional boundary condition to
characterize the domain of wave functions on which $ H $ is self adjoint, hence the need for 
an additional parameter. The latter defines the particular self-adjoint extension that has been made
 \cite{meetz.180503.1}.  Its implementation for the study of dispersion problems is again very late and
it seems that it appears for the first time in the study of Gopalakrishnan \cite{gopa.180502.1}.

\subsection{Renormalization techniques, taking $r_0\to 0^+$}
\label{sec.180911.1}

In more recent times, the problem of finding  (approximate) solutions to the Schr\"odinger equation 
 with singular potentials  has also begun to be addressed 
from the point of view of effective field theories and the renormalization group \cite{kolck.180504.1}. 
Techniques of quantum field theory were previously used for the study of the Dirac delta-type (zero-range) potentials in
 Refs.~\cite{zeldovich.180504.1,berezin.180504.1, tarrach.180504.1}.
 
 At the intuitive level, renormalization can be understood as the technique to  integrate out
  degrees of freedom associated with distances much smaller than those of interest,
   in favor of effective or more ``explicit'' degrees of freedom.
 Given the fundamental Heisenberg uncertainty relation for position-momentum, if we denominate by
 $ d $ the upper limit of the typical distances associated with the short-distance degrees of freedom,
 there is at the same time an associated uncertainty in their three-momentum $ p $ that obeys the relationship
$ d \, p \simeq 1$.
 So the degrees of freedom associated with much smaller distances
 that those of the study would be associated to much larger linear momenta (and typically also to higher
 energies). In other examples, it is not the spatial scale that is much smaller, but the temporal scale, that is,
we have degrees of freedom that oscillate with characteristic frequencies much higher than those of the problem
 under study, which dynamics evolves much more slowly. From the energy-time uncertainty relation (or from the 
 relation between frequency and energy)
it is deduced  that these fast movements are associated with greater energies than  
the typical ones in the problem.

 A rather intuitive but important example of renormalization associated with integrating out energetic 
 degrees of freedom, corresponding to short time scales, is the approximation 
 of Born-Oppenheimer for the study of the equilibrium positions of ions in molecules, 
 so central in quantum Chemistry. 
 Here we have the heavy nuclei with a size much smaller than the typical distances
on which the electrons are distributed in the atoms and  in the molecule.
 The electrons, given their lightness, move much more quickly than
the nuclei, which barely move around certain equilibrium positions.
In this way, for calculating the electronic movement is sufficient to consider the nuclei 
fixed at certain positions. 
 Once this is solved, we can calculate the equilibrium positions
of the nuclei from the average energy of the electronic system as a function of the positions of the nuclei.
As a result, the Coulomb repulsive interaction between the nuclei is modified (or ``dressed'')
by the presence of electronic degrees of freedom much more rapidly oscillating. It then reads
\begin{align}
\label{180504.1}
V_{ef}&=\sum_{i=1}^N\sum_{j>i}^N \frac{Z_i Z_j e^2}{|\vr_i-\vr_j|}+{\cal{E}}(\vr_i)~,
\end{align}
where ${\cal E}(\vr_i)$ is the mean energy of the electrons over its fast movement taking into account their 
mutual repulsion and attraction with the positively charged ions. In the previous equation the 
 vectors $\vr_i$ correspond to the position of the ions and the subscript {\it ef} refers to effective.

 This example illustrates the important point  that
 renormalization typically implies a change in the interaction between the systems whose dynamics is 
 considered explicitly.
Thus, a new term $ {\cal{E}}(\vr_i)$ is added to the purely Coulomb repulsion 
between the ions (which would be their interaction
without further degrees of freedom). This extra term can be in fact a complicated function
 of the positions of the ions $\vr_i$.
This last point is relevant because it indicates that in full non-perturbative calculations
 one should expect in general complicated  dependencies on the explicit degrees of freedom
for the extra contribution required by renormalizing the effective potential.
In quantum field theory, arguments usually based on symmetry 
and dimensional analysis are often employed to deduce the resulting functional dependence
 of the modification to the ``bare'' interaction between explicit degrees of freedom  in the
variables of the problem. In this way, if the renormalized effective interaction can be deduced
 successfully  the theory is said  renormalizable. 
On the contrary, if this modification is more complicated and cannot be given in closed form
 we consider the theory as a non-renormalizable one.
 In fact, it is still nowadays a serious problem to make sense of a non-renormalizable theory within 
 non-perturbative calculations.

In connection with this discussion, there has always been an interest in the study of singular potentials
  given that non-renormalizable quantum field theories are expected to give rise 
  to such potentials \cite{frank.180502.1}.
So if these are attractive, they necessarily require to fix at least 
an extra parameter beyond the knowledge of the potential,
 as we have discussed before. For illustrative purposes, let us perform  simple manipulations 
 in relation with the well-known non-renormalizable character of the General Theory of Relativity.
 To see the origin of the difficulties let us take the Newton law for the interaction between two 
 particles of energies $E_1$ and $E_2$ at positions $\vr_1$ and $\vr_2$.  Considering too the 
 equivalence between mass and energy, we would write for the potential
\begin{align}
\label{180504.2}
V(r)&\approx-\frac{G E_1 E_2}{r}~,~r=|\vr_1-\vr_2|~.
\end{align}
Next, we apply once more the position-momentum  Heisenberg uncertainty relation 
 for a system with characteristic distance $r\to 0$, which implies that the radial momentum 
 $p$ diverges  as $1/r$. The relativistic relation between energy and momentum then becomes 
 simply $E=p$, that when substituted in Eq.~\eqref{180504.2} gives the potential
\begin{align}
\label{180504.2b}
V(r)&\xrightarrow[r\to 0]{}-\frac{G}{r^3}~,
\end{align}
which is a singular one.

An important formal point of similarity between solving scattering with a singular potential
 in the Schr\"odinger equation  and a non-renormalizable  quantum field theory is the presence in both cases
of divergent series. 
In this way, the study of singular potentials can be taken as a testing ground 
 for methods to resum series in quantum field theory. 
This program indeed deserved particular attention during the sixties 
 as a trial method to give meaning to non-renormalizable quantum field theories 
 prior to the current one of weak interactions.

 For example, for a general singular potential  no term in the Born series exists 
 above some order, cf. Sec.~\ref{sec:190116.1}, while for the 
 repulsive singular case one can solve  the  $T$ matrix that does not involve any additional free parameter 
 from the Schr\"odinger equation (or the equivalent Lippmann-Schwinger (LS)  equation). 
 Therefore, it is possible to check whether the latter is obtained after applying some resummation method 
to  the diverging Born series. As in non-renormalizable quantum field theories these divergences in the Born 
series cannot be compensated with just the terms in the original Lagrangian.  
 This program was a point of great importance for the development  of the method of {\it peratization}
 proposed in the reference \cite{feinberg.180505.1} in connection with weak interactions.
 The method consists of using the perturbation theory series in powers of the coupling constant
 and identify the most singular terms  in each order of perturbation theory  employing a cutoff regularization.
 Then, one resums the series of the most leading divergences and check whether the result remains finite when
 the cutoff is sent to {\it infinity}. 
 Terms that are less singular than the ones  already resummed (or the first set)  in the perturbation series 
are then considered similarly.
 This procedure of isolating the singular terms by their degree of divergence when the cutoff is sent to 
 infinity and resum them is repeated until at the end there are only finite terms in the perturbative series.
 Despite some hopeful initial results,  it was soon found that this method did not always give acceptable results
 for singular repulsive potentials. So in the reference \cite{aly.180505.1} a logarithmic singular potential 
 of the type $ V(r)=\frac{g (\ln r)^2}{r^4}~,~g>0~,$ was discussed. 
 The authors were able to resum the perturbative series of leading divergences in the cutoff but when taking 
 its infinite limit the resummed series was not finite, but it diverged. 
 This result is then a counter-example of the peratization method and drove the authors to warn
against the use of this method which in their words should be employed 
with ``extreme caution". It is then not a method that could be applied generally, though in some cases it could 
 deliver correct results \cite{khuri.180505.1}. 
 
With the techniques of the renormalization group the aim is to solve scattering by singular potentials
  but now taking the limit $r_0 \to 0^+$, with $r_0$ the separation distance between the short- and long-range 
 potentials as introduced in QDT. 
 We discussed above that the influence of short-distance physics in QDT is parameterized by the function
 $\widetilde{\lambda}(E)$,  which is typically written as a low-degree polynomial
in energy (whose coefficients are fitted for example to the positions of few bound states
determined by experience or by first-principle calculations).
  However, in QDT the distance $r_0$ is kept fixed as a one more phenomenological parameter 
 to reproduce data the best as possible. 
 The presence of $r_0$ induces an ambiguity in QDT because, at the end,  
 it is an auxiliary scale and one should probe that its effects 
are reabsorbed in the short-distance function $\widetilde{\lambda}(E)$ (a point that is not explicitly treated 
 in QDT \cite{seaton.180427.2,gao.180428.4}). 
 The techniques of effective quantum field theories and renormalization group precisely aim to fill this  gap. 
They pursue to take the limit $r_0 \to 0^+$, and relegate the introduction  of $r_0$ as just an intermediate 
step without physical implications. 
 Some specific techniques used in the literature for $NN$ interactions along these lines are discussed 
 in the next section.

\subsection{Nuclear physics}
\label{sec.180910.1}

The $NN$ scattering is the building block for nuclear physics and it is a long-standing problem that 
has been treated along the years within different approaches. 
The interactions between two nucleons constitute typically a more 
complicated problem than the examples already treated from atomic and molecular physics, 
 mainly because there is no such a clear separation between long and short distances as in the 
 latter case. 
There are also several scales that are involved simultaneously. 
On the one hand, there is  the pion mass ($m_\pi=138$~MeV) which settles the natural scale for 
the three-momentum of the nucleons. 
 However, their typical energy is much smaller because the nucleon mass is almost an 
 order of magnitude larger than the pion mass (around $-16$ MeV is the binding energy per nucleon in 
  symmetric nuclear matter or the first term in the semi-empirical mass formula). 
  Additionally, one also has the 
 $S$-wave scattering lengths which are much larger in absolute size
 than $m_\pi^{-1}$. 
 On the other hand, due to the fact that 
 the pion weak-decay constant $f_\pi$ ($f_\pi=92.4$~MeV) is of similar size as $m_\pi$  
and the axial coupling of the nucleon $g_A$ is around 1 ($g_A=1.26$),
 the  pion-nucleon coupling is  of order 1 [$(g_A m_\pi/f_\pi)^2=3.88$ to be compared 
 with $e^2= 0.092$], clearly signalling towards a non-perturbative physics.  

Nonetheless, for distances large than  1.5--2~fm \cite{arriola.180502.1} one can identify a long-distance potential 
between the two nucleons given by the exchange of one pion (OPE). This potential can be expressed by the 
equation \cite{kolck.180505.1}
\begin{align}
\label{180505.2}
V(r)&=C_\pi \mathbf{i}_1\cdot \mathbf{i}_2 
\left[T(r)S_{12}+Y(r)\mathbf{s}_1\cdot\mathbf{s}_2\right]~,\\
Y(r)&=\frac{e^{-m_\pi r}}{r}~,\nn\\
T(r)&=\frac{e^{-m_\pi r}}{r}\left[1+\frac{3}{m_\pi r}+\frac{3}{(m_\pi r)^2}\right]~.\nn
\end{align} 
 We identify here the spins (isospins) of the nucleons $\mathbf{s}_j$ ($\mathbf{i}_j$), 
 the coupling constant $C_\pi=( g_A m_\pi)^2/(3\pi f_\pi^2)$ and the tensor operator 
 $S_{12}=3(\mathbf{s}_1\cdot \hat{\mathbf{r}})(\mathbf{s}_2\cdot \hat{\mathbf{r}})
-\mathbf{s}_1\cdot\mathbf{s}_2$. 
 The function $Y(r)$ in Eq.~\eqref{180505.2}, the only contribution for singlet waves,
  is a regular potential of Yukawa type.  
  However, the function $T(r)$ also contributes  in the triplet waves and 
  it gives rise to a  singular potential that diverges as $1/r^3$ for $r\to 0$. 
At sufficiently short distances the $NN$ interactions should be repulsive 
because of the Pauli repulsion due to the overlapping of the orbitals 
of the fermionic constituent quarks.
This basic  observation about the expected behavior of the $NN$ interactions at 
 short distances implies that when dealing with the singular potentials 
 involved in the low-energy description of $NN$ scattering 
 one  must impose boundary conditions that guarantee unitarity. 
 In addition, this is also an experimental fact.

The interest in the application of effective field theories for the study of the $NN$ 
interactions was triggered by the seminal articles of Weinberg \cite{weinberg.180505.1,weinberg.180505.2}.
 It is argued in these papers that the low-energy nuclear interaction can be described by a potential
$ V (r) $ that can be calculated in a systematic and controlled way in the chiral 
 effective field theory of QCD at low energies ($\chi$PT).
Each new chiral order in the perturbative series adds an extra derivative on the potential with respect to $ r $, 
which would then increase its singular character as the order of the calculation of the potential increases. 
This fact has been an important complication for not having brought to a fully 
satisfactory state the idea of ​​applying
 $\chi$PT  to the calculation of the low-energy $NN$ scattering amplitude.

 Usually in canonical effective field theories  the interactions are supposed to be weak enough 
 so as to justify the application of perturbation theory.
 The non-perturbative character of the $ NN $ interaction is based on two facts.
One of them is a quantum effect of kinematical origin within the typical scales of the problem.
 The typical separation distance that two nucleons can travel when propagating as
virtual particles is $l_{NN} \sim 1/ E_{\text{kinetic}} \sim m/p^2 \sim (m/m_\pi) b_\pi $, with
$ b_\pi $ the maximum range of the $ NN $ interaction that is given by the Compton wavelength of the pion, 
$ b_\pi \sim m_\pi^{- 1}$. 
As $ m/m_\pi $ is large  there is more than enough distance for having several repetitive collisions  
 when the two nucleons separate.
The same conclusion follows if we consider that the velocity of a real 
nucleon with typical three-momentum $m_\pi$ is $\sim m_\pi/m$. 
This is so small  that the time to go through 
the range of strong interactions is just
$\sim m/ m_\pi^2$, which is much larger than $b_\pi$.

Nonetheless, if the strength of the interaction between the two nucleons 
were small enough then the problem  would become perturbative. 
 This is not the case because the potential between two nucleons due to the exchange of one pion 
goes like $g_A^2 m_\pi^3/f_\pi^2$. 
Once this factor is multiplied by $l_{NN}$ we get the dimensionless number
\begin{align}
\label{180507.2}
\frac{l_{NN}}{16\pi}\frac{g_A^2 m_\pi^3}{f_\pi^2}=\frac{m g_A^2}{16\pi f_\pi^2} m_\pi~,
\end{align}
which is around  0.5, and this indicates that one has to treat 
the $NN$ interaction non-perturbatively. 
In the previous expression we have included the phase-space factor 
 $1/16\pi$, that counts the propagation of the two nucleons in all directions.
We also distinguish in Eq.~\eqref{180507.2} the scale \cite{oller.180722.1,birse.180506.1}
\begin{align}
\label{180613.1}
\Lambda_{NN}= \frac{16\pi f_\pi^2}{m g_A^2}\simeq 2 m_\pi~,
\end{align}
 having a relatively small size (despite this scale is not proportional to $m_\pi$), 
 which is behind the non-perturbative nature of the $NN$ interactions,

The waves with higher orbital angular momentum are more perturbative 
as $\ell$ increases because then the centrifugal barrier 
screens a broader internal region where the nuclear forces are stronger.
The condition for the minimum value of $\ell$ for which the $NN$ interactions 
 tend to become  perturbative  is the solution of the equation 
 $\ell(\ell+1)/mr^2-|V_\pi(r)|=0$ for $r\simeq 1/m_\pi$. 
 This exercise shows that this happens for $\ell$ around a few units. 
 At the phenomenological level the $D$ waves are those that typically establish the transition 
 towards to a more perturbative $NN$ dynamics \cite{oller.180722.2,carbonell.180920.1}.

Another point to take into account and that complicates the  $NN$ interactions, 
 adding a new scale to low energies, is the unnaturally large size of the $NN$ $S$ waves ($a_s$), 
being much larger than the Compton wavelength of the pion, $|a_s|\gg b_\pi $. 
For example, we have $a_s \simeq -25$~fm for the partial wave $^1 S_0$. 
 This makes the dimensionless number in Eq.~\eqref{180507.2} be even greater 
 for this other distance scale by a factor $|a_s|m_\pi \gg 1 $. 
 Therefore, for small  three-momenta compared to $m_\pi$, where the effective range expansion applies,
 we have that the $ NN $ interactions  must be iterated in the $ S $ waves.

We now discuss some important types of approaches to  $NN$ scattering when 
 the long-range potential is calculated within $\chi$PT. 
In one of them the  singularity at the origin is avoided by introducing a separation distance
 $r_0$ above which the chiral potential $V(r)$ is used.
 Below this distance the potential is modified by hand and becomes well behaved, for example
historically much use has been made of choice $V (r) = V(r_0) \theta(r_0-r)$.
 However, in this field  it is more frequent to consider the momentum representation than the
 configuration one  and so, typically, a sharp or smooth cutoff 
 that cuts the high three-momenta above a scale $\Lambda \sim 1/r_0$ is employed.
 This working scheme was indeed the one discussed in the original articles by Weinberg and 
 developed firstly in Refs.~\cite{ordo.180611.1}.
However, it has been criticized for having explicit dependence on  $r_0$ or $\Lambda $,
that finally acts as a fit parameter of the theory. 
 Its dependence cannot be 
reabsorbed in the counterterms that characterize the contact terms in the potential $V (r)$ calculated 
 in $\chi$PT up to some 
order in the standard chiral power counting \cite{kaplan.180611.1,arriola.180502.1,kolck.180505.1,soto.180921.1}. 
Nonetheless, there have been interesting attempts to justify this remnant dependence on $r_0$ 
\cite{epelbaum.180623.1} based on Lepage's ideas \cite{lepage.180623.1} on cutoff renormalization.
 Relevant review articles within this method are the references \cite{machleidt.180505.1,hammer.180505.1}.
 Despite the previous criticism it is quite true that this procedure is the one that, by far, has reached
 the highest numerical precision when reproducing the experimental data. 
 Remarkably Ref.~\cite{entem.180621.1} fitted  the 2016 database of $ pp $ and $ np $  
 with a $\chi^2/$datum = 1.15.
 This reference employs a chiral potential calculated to order $p^5$ (N$^4$LO) containing
  28 low-energy constants that adjust to about 4900 experimental points of $pp$ and $np$.
 Similar results regarding the $\chi^2/$datum and the number of free parameters
 have been obtained by the reference \cite{epelbaum.180621.1}, that adjusts
 the self-consistent database of Granada-2013 using a regularization method with
 cutoff introduced differently from \cite{entem.180621.1}.
 Without adjusting directly to the $NN$ database we have Ref.~\cite{epelbaum.180621.2} that
employs a chiral potential calculated at N$^4$LO.
 Currently, the dominant contributions to the exchange of two and three pions have already been calculated
up to sixth order in $\chi$PT and applied to the study of the upper partial waves with 
$\ell \geq 4 $ \cite{kaiser.180621.1}. 
  An excellent agreement with the data on the $NN$ phase shifts is obtained without including any free parameter.

Another method consists of solving the Schr\"odinger equation with the potential $ V (r) $
for all $ r> 0 $,  following the method
of Case \cite{case.180502.1} but extended to positive energies.
Here we highlight the works of the Granada group by Arriola {\it et al.}
which solve the problem of scattering by imposing, 
together with the condition of finitude of the wave function at the origin, 
the orthogonality of the eigenfunctions with different energies \cite{arriola.180502.1}.
For the decoupled case, a constant remains per partial wave
that must be fixed for a singular attractive potential, while the dispersion is
completely set for a regular potential as well as for a singular repulsive one.
Here we always refer to the attractive/repulsive character of the potential in the limit $ r \to 0 $.
For the coupled case, as originally concluded in the works of
Arriola and Pav\'on \cite{arriola.180502.1},
 one has to diagonalize first the potential in the limit
$r\to 0 $, having as many free constants  as attractive (negative) eigenvalues.
However, this method is not always phenomenologically successful, as it turns out 
from the study of the $NN$ interactions where there is not always a good agreement with 
the experimental phase shifts in the range of expected three-momentum 
\cite{arriola.180502.1,entem.180512.1,zeoli.180513.1}. 
To our opinion the main reasons for the rigidity of this approach are: 
i) The number of free parameters is fixed by the attractive or repulsive nature of the singular potential
 for $r \to 0 $, although originally it was intended to be a long-range potential. 
ii) No  extra considerations based on the need to have
an equally accurate representation of the interaction for both
 long and short distances emerge. This is surprising because the short-range physics 
should improve order by order as the long-range potential does. Intuitively this would 
drive towards an increasing number of free parameters corresponding to a more 
 accurate representation of the short-range physics.

The third working scheme is a method that mixes perturbative and non-perturbative techniques.
The basic idea is to solve the LS equation  non-perturbatively  when the potential
is given to dominant order, and  proceed perturbatively when higher orders are added to the potential. 
At the same time, new terms are provided to the potential to guarantee 
that the problem is renormalized, first non-perturbatively to dominant order
and then perturbatively to subdominant orders.
There is typically a promotion of higher-order counterterms with respect to 
the purely dimensional power counting, necessary to guarantee the renormalicity of the non-perturbative 
calculation with  dominant $V(r)$ \cite{kolck.180505.1,birse.180506.1}.
From here it is determined the flow of renormalization with the cutoff of the counterterms in the potential,
 for the purpose of determining additional local terms to be included in $V(r)$.
 The latter must also reabsorb the divergences that arise when solving the LS equation
 in a perturbative way when subdominant contributions (or higher orders) in the calculation of $V(r)$ 
are considered.

There is a whole series of works along these lines. 
Within this theoretical scheme one has the so-called KSW approximation  \cite{kaplan.180611.1},
that treats pions in a perturbative way, which is not realistic as demonstrated later \cite{stewart.180612.1}. 
This is understood by virtue of the smallness of the scale $\Lambda_{NN}$, 
  introduced above in Eq.~\eqref{180613.1}, despite not involving
 any  power of $m_\pi$ \cite{oller.180722.1,birse.180506.1}.
Later works include the OPE potential as part of the potential that
 has to be iterated to all orders, with the corresponding promotion to lower orders
of terms that in the standard chiral counting would be subleading, since they are associated
to waves with $\ell \geq 1$ \cite{kolck.180505.1}. 
Also this reference advocates to treat perturbatively the contributions of higher orders in the potential.
  The references \cite{pavon.180622.1} (in coordinate space) and \cite{yang.180622.1} (in momentum space) 
develop and implement at the quantitative level in the base of distorted waves 
the perturbative procedure of entering higher orders 
of the potential (they employ the chiral potential up to N$^2$LO), as introduced in Ref.~\cite{kolck.180622.1} 
with an illustrative example.  
 It is intended to find the necessary counterterms order by order that absorb the cutoff dependence 
of the partial waves  in the process of taking the limit $\Lambda \to \infty $.
 Another set up is that of the works of Birse {\it et al.} that use the Wilsonian analysis of the
renormalization group to determine how the counterterms scale with the cutoff around a fixed infrared point,
 from which their chiral order is determined \cite{birse.180506.1,birse.180622.1}.
Requiring that the off-shell amplitude   be independent of the cutoff
  the following equation for the flow of the potential with the
cutoff (sharp cutoff) \cite{birse.180622.1} results:
\begin{align}
\label{180622.1}
\frac{\partial V_\Lambda(p',p;E)}{\partial \Lambda}&=
\frac{m}{2\pi^2}V_\Lambda(p',\Lambda;E)\frac{\Lambda^2}{\Lambda^2-p^2}V_\Lambda(\Lambda,p;E)
\end{align}
Its complete study is of course very complex.
In practice it has been carried out in a systematic way but not exempt 
of certain assumptions, such as the functional form of the infrared fixed point when it is not
 the trivial one. One  controversial point is that while
 Ref.~\cite{kolck.180505.1} finds  a limit cycle type dependence of the contract terms 
at leading order for attractive singular potentials,
the renormalization group analysis of the references \cite{birse.180506.1,birse.180622.1}
  is based on the existence of a fixed infrared point.
 The schemes within this third block still seem to be distant
to achieve its full phenomenological potential.

Despite the great efforts  dedicated to the study of $NN$ interactions within the paradigm of 
effective field theories from the papers \cite{weinberg.180505.1,weinberg.180505.2}, 
continuation of many other previous studies making use of techniques based on dispersion relations and
quantum field theory  \cite{machleidt.180612.1,brown.180612.1},\footnote{Other methods used 
in connection with the calculation of the $NN$ long-range potential in $\chi$PT are 
 the ``subtractive regularization", which is rather a technical way
  to send the cutoff to infinity, and the $V_{\rm low-k}$ approach that allows one to  use a 
small three-momentum cutoff and, up to now, is rather a phenomenological approach \cite {bogner.180514.1}.} 
 none of the three schemes presented is entirely satisfactory.  
The main reason lies in the difficulty to establish a priori
 the correct way of modifying the potential used in the Schr\"odinger or LS equations  
 due to the short-range degrees of freedom. 
 Within the jargon of studies based on the methods of 
effective field theories we refer to the appropriate structure of the potential so as to end up 
with a properly renormalized scattering amplitude when the separation 
distance $r_0 \to 0 $ (or when the three-momentum cutoff $\Lambda \to \infty$). 
At the same time one also demands that it can capture the short-distance physics in a phenomenological successful way. 
This is not at all a simple problem within non-perturbative calculations, 
where many times arguments  based on symmetries and dimensional analysis, which
 work well for perturbative calculations,  are invoked though it is not clear whether
 they are appropriate for the non-perturbative case.
  Bear in mind that in the latter case the situation is much more restrictive, since
in the perturbative calculations new terms arise in the Lagrangian as the order of calculation increases
in the perturbative series. Strictly speaking, therefore,
 we would have infinite counterterms for a given partial wave if  calculated to all 
orders in the perturbative expansion for a singular potential.
In contrast, in a non-perturbative calculation the number of free parameters  is finite and restricted.
 In addition, it is also known that not every regularization method that is suitable, and even optimal, in 
 a perturbative calculation, such as dimensional regularization, can be applied in the non-perturbative case.
In this sense, the references \cite{cohen.180506.1,oller.180506.1} offer examples of non-perturbative calculations
which are independent of cutoff and the way it is introduced in the limit $\Lambda \to \infty $,
 while they give nonsense results if calculated with dimensional regularization.

Another possibility to obtain regularization independent results
is to use the $N/D$ method \cite{chew.180624.1,oller.211116.5}. 
The method exploits the unitarity 
and analytical properties of partial-wave amplitudes (PWAs) and gives
rise to a linear integral equation, from which the scattering amplitude can be calculated. 
The input for the $N/D$ integral equation
is the discontinuity of the partial-wave projected $T$ matrix along
the left-hand cut (LHC), that we denote by $\Delta(p^2)$. This discontinuity stems
from the explicit degrees of freedom included in the theory. 
 In this way, an advantage of the method is that the counterterms of
the effective field theory, which correspond to zero-range interactions, 
do not give any contribution to this discontinuity. 
Nonetheless, their physical effects can be accounted for by including appropriate  
subtraction constants in the $N/D$ equations. 
The traditional shortcut of the $N/D$ method  is that for a
given potential the discontinuity $\Delta(p^2)$ for a general partial wave
is not known a priori, and the approximation typically made is
to calculate it perturbatively. This approach has been pursued by
Oller {\it et al.} using $\chi$PT up to N$^2$LO
and reproducing low-energy $N N$ phase shifts with good precision
\cite{oller.180724.1,oller.180722.2,albaladejo.180724.1}.\footnote{For a somewhat 
different implementation of the $N/D$ method in the same problem see Ref.~\cite{lutz.180921.1}.}
 A crucial and novel point in our work here is to derive an equation that allows one 
to calculate the exact discontinuity of a PWA along the LHC for both 
regular and singular potentials. 
When supplemented in the $N/D$ equations we then have the exact $N/D$ method whose solutions will be 
discussed in detail below along the manuscript. Some first results were already reported in 
Ref.~\cite{entem.170930.1}.

\subsection{Criterion for singular potentials}
\label{sec:190116.1}

We now consider some remarks concerning the singular potentials,  
which ultimately will allow us to define a criterion for their characterization.

 1.- Even for an attractive regular potential in $S$ wave we would expect to have the collapse of the two particles at 
$r=0$ because of the absence of the centrifugal barrier.  
However, quantum mechanically this does not occur because of the Heisenberg uncertainty principle. 
To show it let us consider a bound system with a characteristic size $a$. 
 As $a\to 0$ the uncertainty in the momentum, as required by the Heisenberg uncertainty principle, grows as $\Delta p\simeq 1/a$. 
Therefore the energy goes like 
\begin{align}
E=T+V\simeq \frac{1}{2 \mu a^2}+V~.
\label{170116.12}
\end{align}
For a regular potential, diverging in absolute value less strongly than $1/r^2$ for $r\to 0$, 
the kinetic term dominates and there is a distance $a_0\neq 0$ for which the energy is a minimum. 
Thus, the quantum mechanics effects prevent matter from collapsing and stabilize it 
for a regular potential thanks to the emergence of a  ``quantum-potential barrier''. 
 The latter is however overcome by a singular potential, which is defined as any potential 
that diverges  in absolute value  stronger  than $1/r^2$ for $r\to 0$. 
In such a case there is no a fundamental state of minimum energy \cite{plesset.180501.1} and the system 
would ultimately collapse by decaying to consecutively lower-energy states.   
Furthermore, the Hamiltonian for $r\to 0$ is increasingly dominated by the potential, as in the classical case, 
and we have the validity of the semiclassical methods.

2.- Let us now show that for a singular potential the terms of  the Born series diverge above some order. 
This is clear for a singular potential with $|V(r)|$ diverging faster than $1/r^3$ for $r\to 0$, for which 
 even the Fourier transform does not exist.

 Consider the iteration of the potential by the propagator in quantum mechanics,  
 which  reads in configuration space 
\begin{align}
G_k(r)&=\int\frac{d^3 q}{(2\pi)^3}\frac{1}{q^2-k^2-i\vep}=-\frac{e^{ik r}}{4\pi r}~.
\label{180713}
\end{align}
Increasing by an extra order the Born series implies to add another contribution with a one more  $GV$ factor, 
which in more detail is given by
\begin{align}
\label{180714}
\int d^3r G_k(|\vr'-\vr|)V(\vr)~.
\end{align}
If the potential for $r\to 0$  diverges more strongly than $1/r^\gamma$ with $\gamma>2$, then when
 $\vr'$ and $\vr$ tend to zero simultaneously the previous integral has a nominal diverging power of 
${r}^{3-\gamma-1}$ ($3-\gamma-1<0$). 
Therefore, the integrand in the Fourier transform of the $n$-times iterated potential  
 has, in the limit in which all the integration vectors $\vr_i\to 0$,
 a lower bound in absolute value that scales as $r\to 0$ to a power  of $n(3-\gamma-1)+3-\gamma$. 
As a result this power  becomes negative for $n>(3-\gamma)/(3-\gamma-1)$  
 and the corresponding term in the Born series diverges.\footnote{As a clarification remark, notice 
that there are $3(n+1)$ integrations in the coordinates, ${-n}$ propagators (each of them diverging as $1/r$)
 and $n+1$ diverging factors from the singular potentials bounded from below by  $1/r^{\gamma}$, $\gamma>2$.} 

3.- Let us take a potential of the form $V(r)=\alpha/2\mu r^2$. The  Schr\"odinger equation 
for the reduced wave function $u(r)=r\psi(r)$ is 
\begin{align}
-\frac{d^2 u(r)}{dr^2}+\frac{\alpha}{r^2}u(r)+\frac{\ell(\ell+1)}{r^2}u(r)=k^2 u(r)
\label{170116.4}
\end{align}

We try a solution in power series  of $r$ in the form
\begin{align}
u(r)=\sum_{n=0}^\infty a_n r^{\nu+n}~.
\label{170116.15}
\end{align}
Substituting it into Eq.~\eqref{170116.4} one obtains the indicial equation for $n\geq 0$
\begin{align}
a_{n+2} \left\{- (n+2+\nu)(n+1+\nu)+\alpha +\ell(\ell+1)\right\}-a_{n}k^2 =0~.
\label{170116.16}
\end{align}
The coefficient of the first term proportional to $a_0$ implies the equation
\begin{align}
\nu(\nu-1)-\bar{\alpha}&=0~,
\label{170116.20}
\end{align}
with $\bar{\alpha}=\alpha+\ell(\ell+1)$. While the coefficient of the next term proportional to $a_1$ gives rise to 
the equation  
\begin{align}
\nu(\nu+1)-\bar{\alpha}&=0~,
\label{170713.1}
\end{align}
The solutions to Eq.~\eqref{170116.20} are
\begin{align}
\nu=\frac{1}{2}\pm\frac{1}{2}\sqrt{1+4\bar{\alpha}}~,
\label{170116.21}
\end{align}
 and the solutions of Eq.~\eqref{170713.1} are
\begin{align}
\nu_1=-\frac{1}{2}\pm\frac{1}{2}\sqrt{1+4\bar{\alpha}}~.
\label{170116.21b}
\end{align}
We do not consider any longer the solutions starting with $a_1$ because they reduce to 
the ones starting with $a_0$ since $\nu_1+1=\nu$.

For the classification that follows it is worth keeping in mind 
that classically there is a turning point $r_c$ for $\bar{\alpha}>0$, being given by  $r_c=\sqrt{\bar{\alpha}/k^2}$. 
We distinguish several cases:
\begin{enumerate}
\item Repulsive effective potential, $\bar{\alpha}>0$
  There is only one acceptable solution behaving as
\begin{align}
u(r)\xrightarrow[r\to 0]{}r^{\frac{1}{2}+\frac{1}{2}\sqrt{1+4\bar{\alpha}}}~, 
\label{170116.22}
\end{align}
because the other solution for $\nu$ in Eq.~\eqref{170116.21} gives rise to a divergent wave function $u(r)$ 
in the limit $r\to 0$.

\item Zero effective potential, $\bar{\alpha}=0$. In this case  Eq.~\eqref{170116.20} reduces to 
\begin{align}
\nu(\nu-1)&=0\to \nu=0~,~1.
\label{170116.24}
\end{align}
 The series expansion of $u(r)$ in powers of $r$ provides two linearly independent solutions, 
one of them tends to a constant and the other vanishes as $r$ for $r\to 0$. 
 The subdominant terms are obtained by applying the recurrence relation of Eq.~\eqref{170116.16},  which now 
 simplifies to
\begin{align}
a_{n+2}(n+2+\nu)(n+1+\nu)+a_n k^2 &=0~,
\label{170116.25}
\end{align}
and one independent solution involves only even powers and the other only odd powers of $r$. 

Since the free case for an $S$-wave corresponds to this situation as well [the solution is proportional
to the spherical Bessel function of the fist kind $j_0(k r)$], it is clear that the only acceptable 
solution is the linearly independent one vanishing at the origin as $r$.

\item  Weakly attractive effective potential, $-1/4<\bar{\alpha}<0$. %
 The reduced wave function $u(r)$ has the limit behavior 
\begin{align}
u(r)\xrightarrow[r\to 0]{}r^{\frac{1}{2}\pm\frac{1}{2}\sqrt{1+4\bar{\alpha}}}~.
\label{170116.22a}
\end{align}
Both linearly independent solutions are real and vanish at the origin. However, the 
one that is connected smoothly by a continuous change in $\bar{\alpha}$ from positive to negative 
values going though $\bar{\alpha}=0$  is the solution with the larger exponent. 
Therefore, there is only one acceptable solution that in the limit $r\to 0$ tends to
\begin{align}
u(r)\xrightarrow[r\to 0]{}r^{\frac{1}{2}+\frac{1}{2}\sqrt{1+4\bar{\alpha}}}~.
\label{170116.22c}
\end{align}
For other arguments to reach the same conclusion
 the reader is referred to Refs.~\cite{landau.180720.1,gopa.180502.1}.

\item Transition attractive potential, $\bar{\alpha}=-1/4$. 
The  Eq.~\eqref{170116.20} becomes 
\begin{align}
(\nu-\frac{1}{2})^2 =0~,
\label{210116.1}
\end{align}
and  $\nu=1/2$ is a double root and only one linearly independent solution can be found as a power expansion 
of $r$ as in Eq.~\eqref{170116.15}. 
The other independent solution is of the form 
\begin{align}
u_2(r)&=\sum_{n=0}^\infty \left\{c_{n}r^{\nu+n}\log\beta r+c'_{n}r^{\nu+n}\right\}~.
\label{210116.2}
\end{align}
Substituting this expression in the original differential equation one has the condition for $\nu$
\begin{align}
(\nu-\frac{1}{2})^2 c_0&=0~,\nn\\
(\nu-\frac{1}{2})\left[ 2 c_0+ (\nu-\frac{1}{2})c'_0\right]&=0~.
\label{210116.3}
\end{align}
The first of these equations imply that $\nu=1/2$.
 One can also deduce the following recurrence relations, 
\begin{align}
\label{210116.4}
-(n+2)^2 c_{n+2}&=k^2c_n~,\\
\label{210116.5}
2c_{n+2}+(n+2)c'_{n+2}&=\frac{k^2}{n+2}c'_n~.
\end{align}
The former stems from the coefficient 
multiplying the log term and the latter  from the rest of terms without logarithm.
 The condition $c'_0=0$ can be imposed because any shift in $c'_0$ is reabsorbed 
in the linear superposition with the other linearly independent solution $u_1(r)$.\footnote{The recurrence 
 relation involving $c'_{n+2}$ and $c'_n$ of Eq.~\eqref{210116.5}, with $c'_0\neq 0$, would give rise 
 to a contribution to the $c'_n$ coefficients proportional to $c'_0$ that is precisely the same one already considered 
 in $u_1(r)$.}  Then, Eq.~\eqref{210116.5} allows one to fix all the $c'_n$
 in terms of the $c_n$, which are determined from Eq.~\eqref{210116.4} in terms of the free parameter $c_0$. 
The parameter $\beta$ in Eq.~\eqref{210116.2} has dimensions of inverse of distance, and it is required so as to 
have a dimensionless argument of the $\log$. 
 A change in its value is reabsorbed also in the linear superposition with $u_1(r)$. For this case with $\nu=1/2$ the indicial equation 
of Eq.~\eqref{170116.16} and Eq.~\eqref{210116.4} are the same.

We cannot use the same argument as in case 3 to single out one solution, 
based on the smooth continuation in $\bar{\alpha}$, 
 because for $\bar{\alpha}=-1/4$ the square root $\sqrt{1+4\bar{\alpha}}$ is singular. 
 Nonetheless, we can still apply the same argument as Ref.~\cite{landau.180720.1} applies for the 
 case $\bar{\alpha}>-1/4$. The point is to require the continuity of the logarithmic derivative of the 
reduced wave function $u(r)$ at $r_0$ in the limit $r_0\to 0$ after freezing the potential to its value at 
$r=r_0$ for $r<r_0$. In this region we have the Schr\"odinger equation
\begin{align}
&\frac{d^2 u(r)}{dr^2}+\frac{1}{4 r_0^2}u(r)=-k^2 u(r)~,
\label{180720.1}
\end{align}
whose solution for $r_0\to 0$ is $u(r)=C \sin r/2r_0$. For $r>r_0$ the wave function is written as 
the superposition $u(r)=A\sqrt{r}+B\sqrt{r}\log r$. Requiring the continuity of the logarithmic
 derivative at $r_0\to 0$ we have the equation
 \begin{align}
 \frac{1+x\log r_0+2 x}{1+x \log r_0}=\cot 1/2~.
\label{180720.2} 
\end{align}
 with $x=B/A$. Solving for Eq.~\eqref{180720.2} we find that $x$ vanishes as $-1/\log r_0$ for $r_0\to 0$. 
 Thus, we are also in this case left with a unique acceptable solution of the type $u_1(r)=\sum_{n=0} c_n r^{n+1/2}$.
 For another argument to reach the same conclusion the reader is referred to Ref.~\cite{gopa.180502.1}.

\item Strong attractive effective  potential, $\bar{\alpha}<-1/4$. 
 There are two linearly independent complex solutions behaving as
\begin{align}
u(r)\xrightarrow[r\to 0]{}r^{\frac{1}{2}\pm i\frac{1}{2}\sqrt{-1-4\bar{\alpha}}}~,
\label{170116.23}
\end{align}
and the two solutions vanish at the origin. To guarantee that the flux across a small surface of radius 
$r_0\to 0$ is zero, the superposition of the 
two linearly-independent solutions in Eq.~\eqref{170116.23} 
 should give rise to a real solution of the form 
\begin{align}
\label{170116.24b}
u(r)=C \sqrt{r} \cos\left( \sqrt{-1-4\bar{\alpha}}\log \beta r + B\right)~.
\end{align} 
Again $\beta$ has dimensions of length$^{-1}$ and its precise value has no physical implications since 
a change in it can be reabsorbed in the real phase $B$. 
  This is an example in a simple context of dimensional transmutation.
  To show indeed that the physical solution is  
real, modulo a global phase, one can also apply the same argument as in case 4 based on the continuity of the 
logarithmic derivative at $r_0\to 0$ after freezing the potential at $r=r_0$. This point can be 
 further consulted  in Ref.~\cite{landau.180720.1}.
One extra condition beyond the knowledge of the potential is then necessary to fix the solution in this case.  
 E.g. the phase $B$ can be adjusted so as to reproduce the binding energy of a bound state 
\cite{case.180502.1} or  some scattering data  \cite{arriola.180502.1}.

\end{enumerate}

Thus, a potential $V(r)$ is called singular if $|V(r)|$ diverges in the limit $r\to 0$ more strongly than 
$\alpha/r^\gamma$ with $\gamma>2$, or as $\alpha/r^2$ when  $\alpha+\ell(\ell+1)< -1/4$. 
In the opposite case the potential is characterized as regular. 
In the case of coupled channels one first diagonalizes the potential $V(r)$ in the limit 
$r\to 0$ and apply the previous criterion to qualify the eigenchannels as having regular or singular 
interactions \cite{arriola.180502.1}.

\section{Unitarity and partial-wave expansion}
\label{sec.180803}
\def\theequation{\arabic{section}.\arabic{equation}}
\setcounter{equation}{0}   

In the following, a potential indicated with a lowercase $v$ has an extra minus sign, so that  $v=-V$. 
Indeed, for the rest of this manuscript we are going to use mostly the sign convention of lowercase $v$ for the potential.  

Let us split the full Hamiltonian $H$ in the free one $H_0$ plus the potential $v$, $H=H_0-v$. 
We denote by $r_0(z)$ and $r(z)$ the resolvents of $H_0$ and $H$, in order, 
\begin{align}
\label{180803.1}
r_0(z)&=\left(H_0-z\right)^{-1}~,\nn\\
r(z)&=\left(H-z\right)^{-1}~,
\end{align}
with $\Ima z\neq 0$. 
From the equality 
\begin{align}
\label{180921.1}
r(z)=(H_0-z-v)^{-1}=((I-v r_0)(H_0-z))^{-1}=r_0(I-v r_0)^{-1}~,
\end{align}
one can obtain the following equations for $r(z)$
\begin{align}
\label{180803.2}
r(z)&=r_0(z)+r_0(z)v r(z)\\
\label{180803.3}
&=r_0(z)+r(z) v r_0(z)~. 
\end{align}
For instance, take the following steps
$r(z)=r_0(I-vr_0)^{-1}=r_0(I-v r_0+vr_0)(I-vr_0)^{-1}=r_0+r_0 v r(z)$. For the other equation 
 multiply Eq.~\eqref{180921.1} by $I-vr_0$ to the right.

The relation between the $T$ matrix $t(z)$ and the resolvent $r(z)$ is, by definition, 
\begin{align}
\label{180803.4}
t(z)r_0(z)&=v r(z)~,
\end{align}
such that from Eq.~\eqref{180803.2}
\begin{align}
\label{180803.5}
r(z)&=r_0+r_0(z) t(z) r_0(z)~.
\end{align}
By comparing this result with Eq.~\eqref{180803.3} we also obtain that 
\begin{align}
\label{180803.6}
r_0(z) t (z)&=r(z) v~.
\end{align}
The LS equation results by employing  Eq.~\eqref{180803.2} for $r(z)$, and its relation with $t(z)$, 
Eq.~\eqref{180803.4}, such that
\begin{align}
\label{180803.7}
v r(z)=v(r_0(z)+r_0(z)v r(z))=v(r_0(z)+r_0(z)t(z) r_0(z))=t(z) r_0(z)~.
\end{align}
Multiplying by $r_0(z)^{-1}$ to the right of the last equality we obtain the LS equation
\begin{align}
\label{180803.8}
t(z)=v+v r_0(z) t(z)~.
\end{align}
Had we used instead Eqs.~\eqref{180803.3} and \eqref{180803.6} we would have obtained this other form of the 
LS equation 
\begin{align}
\label{180803.8b}
t(z)=v+t(z) r_0(z) v~.
\end{align}
An interesting property of $t(z)$ is that 
\begin{align}
\label{180803.9}
t(z)^\dagger=t(z^*)
\end{align}
as follows from the fact that $r(z)^\dagger=r(z^*)$, as it is clear from its definition. This property will be used 
below in regarding the unitarity properties of the $T$ matrix.\footnote{The potential is also required to fulfill this property, 
$v^\dagger(z)=v(z^*)$.}

We can also obtain a relationship between two resolvent operators evaluated at different values of $z$, which is known as the 
Hilbert identity, and that it is very useful also to establish the unitarity properties of the $T$ matrix. 
The Hilbert identity reads
\begin{align}
\label{180803.10}
r(z_1)-r(z_2)=(z_1-z_2)r(z_1)r(z_2)~.
\end{align}
Its demonstration is direct.  One takes the difference
\begin{align}
\label{180803.11}
r(z_2)^{-1}-r(z_1)^{-1}&=z_1-z_2
\end{align}
and multiply it to the left by $r(z_1)$ and to the right by $r(z_2)$. 

We can write the Hilbert identity in terms of the $T$ matrix as 
\begin{align}
\label{180803.12}
t(z_1)-t(z_2)&=(z_1-z_2)t(z_1)r_0(z_1)r_0(z_2)t(z_2)~,
\end{align}
as follows by multiplying Eq.~\eqref{180803.10} to the left and right by $v$, so that
\begin{align}
\label{180803.13}
v r(z_1)v-vr(z_2)v=t(z_1)-t(z_2)=(z_1-z_2)vr(z_1)r(z_2)v=(z_1-z_2)t(z_1)r_0(z_1)r_0(z_2)t(z_2)~.
\end{align}

In the following we are going to be mainly interested in the partial-wave projected LS equation.  For that 
we consider two-body states projected in a given partial wave, $|k,\ell S,J\mu \rangle$, where $k$ refers to the modulus 
of the three-momentum, $\ell$ is the orbital angular momentum, $S$ the total spin of the 
system, $J$ the total angular momentum and $\mu$ its third component. 
To shorten the notation, we will typically  denote the discrete indices  globally as $\lambda$, and then 
 write $|k,\lambda\rangle$ for the same state. 
 This compact notation is also valuable because it readily shows that 
 the results can be applied to other choice of partial-wave projection, e.g. in the helicity basis.
  The partial-wave states are normalized such that 
\begin{align}
\label{180804.1}
\langle k\lambda|k',\lambda'\rangle&=2\pi^2\frac{\delta(k-k')}{k^2}\delta_{\lambda\lambda'}~.
\end{align}
For the case of particles without spin we have
\begin{align}
\label{180804.2}
|k,\ell \mu\rangle&=\frac{1}{\sqrt{4\pi}}\int d\hat{\vk}Y_{\ell \mu}(\hat{\vk})^*|\vk\rangle~,
\end{align}
with $|\vk\rangle$ the free-state with three-momentum $\vk$ and normalized such that $\langle \vk|\vk'\rangle=(2\pi)^3\delta(\vk-\vk')$.
 In the case of particles with spins $s_1$ and $s_2$ we have to combine them properly according to the Clebsch-Gordan series,
 \begin{align}
\label{180804.3}
|k,\ell S,J\mu\rangle&=\frac{1}{\sqrt{4\pi}}\int d\hat{\vk}
\sum_{m_1,m_2}C(s_1s_2S|m_1m_2M)C(\ell S J|\mu-M\, M\mu)Y_{\ell(\mu-M)}(\hat{\vk})^*
|\vk,s_1s_2,m_1 m_2\rangle~,
 \end{align}
 where  the $m_i$ refer to third components of the spins $s_i$ 
and $M=m_1+m_2$ is the third component of the total spin $S$. 
A Clebsch-Gordan coefficient for $\mathbf{j}_1+\mathbf{j}_2=\mathbf{j}_3$ is indicated by $(j_1 j_2 j_3|m_1 m_2 m_3)$. 

The matrix elements of the $T$ matrix between partial-wave states constitute the partial-wave projected $T$ matrix, whose matrix 
elements are indicated schematically by $t_{ij}(k,k';z)$,
\begin{align}
\label{180804.4}
t_{ij}(k,k';z)&=\langle k,\lambda_i|t(z)|k',\lambda_j\rangle~.
\end{align}
They are typically referred as the PWAs.

The LS equation in partial waves is obtained by taking the matrix element between partial-wave states of Eqs.~\eqref{180803.8} and 
\eqref{180803.8b}. 
\begin{align}
\label{180804.5}
t_{ij}(k,k';z)&=v_{ij}(k,k')+\frac{\mu}{\pi^2}\sum_{n} \int_0^\infty \frac{dq q^2}{q^2-2\mu z}v_{in}(k,q)t_{nj}(q,k';z)\nn\\
&=v_{ij}(k,k')+\frac{\mu}{\pi^2}\sum_{n} \int_0^\infty \frac{dq q^2}{q^2-2\mu z}t_{in}(k,q;z)v_{nj}(q,k')~.
\end{align}
In this equation, $v_{ij}(k,k')$ is the partial-wave projected potential given by 
\begin{align}
\label{180804.6}
v_{ij}(k,k')=\langle k,\lambda_i|v|k',\lambda_j\rangle~.
\end{align}

For the Hilbert identity in partial waves we have from Eq.~\eqref{180803.12}
\begin{align}
\label{180804.7}
t_{ij}(k,k';z_1)-t_{ij}(k,k';z_2)&=(z_1-z_2)\frac{2\mu^2}{\pi^2}\sum_n\int_0^\infty \frac{dq q^2}{(q^2-2\mu z_1)(q^2-2\mu z_2)}
t_{in}(k,q;z_1)t_{nj}(q,k';z_2)~.
\end{align}
Next, we are going to use the property that follows from Eq.~\eqref{180803.9},
\begin{align}
\label{180804.8}
t_{ij}(k,k';z^*)&=t_{ji}(k',k;z)^*~,
\end{align}
and take $z_2=z_1^*=z^*$ in Eq.~\eqref{180804.7}. It results
\begin{align}
\label{180804.9}
t_{ij}(k,k';z)-t_{ji}(k',k;z)^*&=2i \Ima z\,
\frac{2\mu^2}{\pi^2}\sum_n\int_0^\infty \frac{dq q^2}{(q^2-2\mu z)(q^2-2\mu z^*)}
t_{in}(k,q;z)t_{nj}(q,k';z)^*~.
\end{align}
Assuming time-reversal invariance the PWAs are symmetric such that\footnote{To show it take directly 
the matrix element of the $T$ matrix between partial-wave states as follows from the expansion of the latter ones according 
to Eq.~\eqref{180804.3}. The resulting expression involves a  double angular integration and sums over the 
Clebsch-Gordan coefficients, $\sum_{m_i,m'_i} \frac{1}{4\pi}\int d\hat{\vk}d\hat{\vk}' C(s_1 s_2 S|m_1 m_2 M)C(s_2 s_2 S'|m'_1 m'_2 M') 
C(\ell S J|\mu-M\, M \mu) C(\ell' S' J|\mu-M'\, M' \mu)
Y_{\ell(\mu-M')}(\hat{\vk}')^*Y_{\ell(\mu-M)}(\hat{\vk})\langle\vk,s_1m_1 s_2m_2|t(z)|\vk',s_1m'_1 s_2m'_2\rangle$. 
 We perform next the time-reversal transformation ($\vartheta$)  \cite{Gottfried.170929.2}, 
 $\vartheta|\vk,s_1 m_1 s_2 m_2\rangle=i^{2(m_1+m_2)}|-\!\vk,s_1 -\!m_1 s_2 -\!m_2\rangle$ and 
$\vartheta t(z) \vartheta^{-1}=t(z^*)=t^\dagger(z)$. Related to the changes of sign in the 
 third component of the spins and in $\vk$ we use the following properties 
 of the spherical harmonics, $Y_{\ell m}(\hat{\vk})^*=(-1)^{\ell+m} Y_{\ell -m}(-\hat{\vk})$, and Clebsch-Gordan coefficients, 
 $C(j_1 j_2 j_3|-\!m_1  -\!m_2 -\!m_3)=(-1)^{j_1+j_2-j_3}C(j_1 j_2 j_3|m_1  m_2 m_3)$. 
 Therefore, the previous expression, after 
 the $\vartheta$  transformation on the scattering amplitude, can be easily shown to be
  $\sum_{m_i,m'_i}\frac{1}{4\pi}\int d\hat{\vk}d\hat{\vk}'
 C(s_1 s_2 S|-m_1 -m_2 -M)C(s_2 s_2 S'|-m'_1 -m'_2 -M') (-1)^{s_1+s_2-S}(-1)^{s_1+s_2-S'} 
C(\ell S J|-\mu+M\, -M -\mu) C(\ell' S' J|-\mu+M'\, -M' -\mu) (-1)^{\ell+S-J}(-1)^{\ell'+S'-J}
Y_{\ell(-\mu+M')}(-\hat{\vk}')Y_{\ell(-\mu+M)}(-\hat{\vk})^* (-1)^{\ell+\ell'}(-1)^{M'-M} i^{2(m_1+m_2-m'_1-m'_2)}
\langle-\vk,s_1-m'_1 s_2-m'_2|t(z)|-\vk',s_1-m_1 s_2-m_2\rangle$. 
 Thus, all the phases cancel between each other and 
 by taking also into account that because of rotational symmetry  the PWA is independent of $\mu$ (Wigner-Eckart theorem), 
 it is clear the symmetric character of the PWAs. Shorter expressions for the partial-wave projection involving only 
 one sum over spherical harmonics can be derived from the one given here, see e.g. Refs.~\cite{dilege.180815.1,oller.180722.1}.
 Of course, this is also valid for the PWAs in the helicity basis 
because it is just a change of basis with real coefficients, see chapter 3, $\S$5 of \cite{martin.180804.1}. 
 A direct demonstration of this property in the helicity basis can be found in chapter 5, $\S$3.3 of 
 Ref.~\cite{martin.180804.1}. \label{foot.180607.1}}
\begin{align}
\label{180804.10}
t_{ij}(k,k';z)&=t_{ji}(k',k;z)~.
\end{align} 
Inserting this result into Eq.~\eqref{180804.9}, the latter simplifies as
\begin{align}
\label{180804.11}
\Ima t_{ij}(k,k';z)&=\Ima z\,\frac{2\mu^2}{\pi^2}\sum_n\int_0^\infty \frac{dq q^2}{(q^2-2\mu z)(q^2-2\mu z^*)}
t_{in}(k,q;z)t_{jn}(k',q;z)^* \nn \\
&=\Ima z\,\frac{2\mu^2}{\pi^2}\sum_n\int_0^\infty \frac{dq q^2}{(q^2-2\mu \Rea z)^2+(2\mu\Ima z)^2}
t_{in}(k,q;z)t_{jn}(k',q;z)^*
\end{align}
We now take the limit $\Ima z\to 0^+$ and use the result 
\begin{align}
\label{180804.12}
\lim_{\Ima z\to 0^+}\frac{2\mu\Ima z}{(q^2-2\mu\Rea z)^2+(2\mu\Ima z)^2}=
\pi
\delta(q^2-2\mu \Rea z)~.
\end{align}
We indicate by $\kappa$ the on-shell momentum, 
\begin{align}
\label{180804.13}
\kappa&=\sqrt{2\mu \Rea z}~,
\end{align}
 and then Eq.~\eqref{180804.11} reads, with $z=E+i\vep$ and $\vep\to 0^+$,
\begin{align}
\label{180804.14}
\Ima t_{ij}(k,k';z)&=\theta(E)\frac{\mu \kappa}{2\pi}\sum_n t_{in}(k,\kappa;z)t_{jn}(k',\kappa;z)^*~,
\end{align}
which is the off-shell unitarity relation. 
This imaginary part is the reason of the presence of the right-hand cut (RHC) of 
unitarity cut in the PWAs for positive real values of the energy (or physical ones).

Two important particular cases of Eq.~\eqref{180804.14} are the half-off-shell unitarity relation
by taking $E={k'}^2/2\mu$ (so that $\kappa=k'$). It then reads
\begin{align}
\label{180804.15}
\Ima t_{ij}(k,\kappa;E+i\vep)&=\frac{\mu \kappa}{2\pi}\theta(E)\sum_n 
t_{in}(k,\kappa;E+i\vep)t_{jn}(\kappa,\kappa;E+i\vep)^*~.
\end{align}
This is a unitarity relation of the same type as the one for form factors \cite{oller.180804.1,martin.180804.1}. 
We also have here another version of the Watson final-state theorem because Eq.~\eqref{180804.15}
 implies that along the RHC the phase of the half-off-shell PWA is the same as the phase of the on-shell PWA for 
 the uncoupled case.
  
 The on-shell unitarity relation, or simply unitarity, stems by taking the extra requirement $k'=k$  in the half-off-shell case. 
 Then Eq.~\eqref{180804.15} becomes
\begin{align}
\label{180804.16}
\Ima t_{ij}(\kappa,\kappa;E+i\vep)&=\theta(E)\frac{\mu \kappa}{2\pi}\sum_n 
t_{in}(\kappa,\kappa;E+i\vep)t_{jn}(\kappa,\kappa;E+i\vep)^*~.
\end{align}

In the present work we are going to make extensive use of the on-shell unitarity relation. 
For this case case we can also introduce the partial-wave decomposition of the $S$ matrix, given by 
 its matrix elements between states $|k,\lambda\rangle$, 
\begin{align}
S_{ij}(E)&=\delta_{ij}+i\frac{\mu \kappa}{\pi}t_{ij}(\kappa,\kappa;E+i\vep)~.
\label{130116.1} 
\end{align}
This is a unitary operator for $E\geq 0$ because it satisfies 
\begin{align}
S(E) S(E)^\dagger&=S(E)^\dagger S(E)=I~,~E\geq 0~,
\label{130116.3}
\end{align}
as an immediate consequence of on-shell unitarity in partial waves, Eq.~\eqref{180804.16}.

\section{Analytical properties of the potential}
\label{sec:dc}
\def\theequation{\arabic{section}.\arabic{equation}}
\setcounter{equation}{0}

We assume rotational invariance so that a local potential can be decomposed as a sum of terms, 
such that every one of them is a function of $r$ times a polynomial operator which is
 made out of the products of the spins of the particles with themselves and with $\vr$.  
To simplify the writing, a general factor function of $r$ in the previous decomposition 
is denoted as potential and represented by $v(r)$. 
Later one we will distinguish between different $v(r)$'s for the particular case analyzed 
of $NN$ interactions. 
Additionally, we assume firstly that the potential is local and afterwards we will discuss how the 
main results of our work can be straightforwardly generalized to a broad type of non-local 
potentials. Furthermore, since in the considerations of this section and the next one the 
coupling between different PWAs is not relevant (and could be easily accounted for), we omit 
 partial-wave subscripts to simplify the notation. 

Let us consider a local potential $v(r)$ and write it in terms of its 
Fourier transform $\tilde{v}(q)$,\footnote{To avoid confusion we use 
different symbols for the potential in configuration and momentum spaces 
meanwhile they are simultaneously used. Later we will just 
employ the potential in momentum space and we will drop the tilde on $\tv(q)$.} with 
$q$ the modulus of the momentum transfer $\vq$,  
\begin{align}
v(r)&=\int \frac{d^3 q}{(2\pi)^3}e^{i\vq \vr}\tv(q)=\int_0^\infty \frac{dq q^2}{(2\pi)^3}\int d\Omega_q e^{iq r \cos\theta}\tv(q)
=\frac{1}{(2\pi)^2}\int_0^\infty \frac{dq\, q}{i r}(e^{i q r}-e^{-i q r})\tv(q)\nn\\
&=\frac{1}{2\pi^2 r}\Ima \int_0^\infty dq\,q e^{iq r} 
\tv(q)~,
\label{140116.1}
\end{align}
where we have also taken into account that $\tv(q)$ is a real function.

\begin{figure}[ht]
\begin{center}
\includegraphics[width=.2\textwidth]{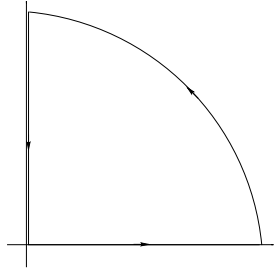}
\end{center}
\caption{{\small Wick rotation applied to the calculation of the inverse Fourier transform of Eq.~\eqref{140116.1}.}
\label{fig:140116.1}}
\end{figure} 
Next, we perform the Wick rotation in the variable $q$, as shown in Fig.~\ref{fig:140116.1}, 
and the integration variable $y$ along the imaginary axis is introduced as $q=i y$. 
 We apply the Cauchy integration theorem to the closed integration contour  
given by the infinity semicircle, the positive real semi-axis and the positive imaginary semi-axis slightly displaced 
to the right by an amount $\ve\to 0^+$, as indicated in Fig.~\ref{fig:140116.1}. This displacement is 
needed so as to avoid the left-hand cut in the potential, as it is clear a few lines below.
 It results  
\begin{align}
\oint dq\,q e^{iq r} \tv(q)=0~.
\label{140116.2}
\end{align}
 Note that  the integration contour is closed with positive imaginary part,  
 so that there is an exponential dumping along the semicircle at infinity because 
 of $e^{i q r}=e^{-r \Ima q} e^{ir\Rea q}$ as $r>0$. We can then rewrite Eq.~\eqref{140116.1} as 
\begin{align}
v(r)&=-\frac{1}{2\pi^2 r}\Ima \int_0^\infty dy\, e^{-yr} y \tv(iy+\ve)
=-\frac{1}{2\pi^2 r}\int_0^\infty dy\, e^{-yr} y\Ima \tv(iy+\ve)~.
\label{140116.3}
\end{align}

This expression shows that  the finite range part of $v(r)$ can be interpreted 
as a superposition of Yukawa potentials. 
The function $-\Ima \tv(i\mu+\ve)$ is called the spectral function, $\eta(\mu^2)$,
\begin{align}
\eta(\mu^2)=-\Ima \tv(i\mu+\ve)~.
\label{140116.4}
\end{align}
In terms of it Eq.~\eqref{140116.3} becomes
\begin{align}
v(r)&=\frac{1}{(2\pi)^2}\int_0^\infty d\mu^2 \eta(\mu^2)\frac{e^{-\mu r}}{r}~.
\label{140116.5}
\end{align}
The previous equation is also suitable to express $\tv(q)$ in terms of a dispersion relation (DR). 
Taking its Fourier transform  
\begin{align}
\tv(q)&=\int d^3 r e^{-i\vq\vr}v(r)=
\frac{1}{(2\pi)^2}\int_0^\infty d\mu^2\,\eta(\mu^2)
\int d^3 r e^{-i\vq\vr}\frac{e^{-\mu r}}{r}~,
\label{140116.6}
\end{align}
and keeping in mind that the Fourier transform of a Yukawa potential is
\begin{align}
\int d^3 r e^{-i\vq\vr}\frac{e^{-\mu r}}{r}&=\frac{4\pi}{q^2+\mu^2}~,
\label{140116.7}
\end{align}
the Eq.~\eqref{140116.6} becomes
\begin{align}
\tv(q)&=\frac{1}{\pi}\int_0^\infty d\mu^2\frac{\eta(\mu^2)}{\mu^2+q^2}~.
\label{140116.8}
\end{align}
This expression is an unsubtracted dispersion relation (DR) for the potential $\tv(q)$ in momentum space. 
It explicitly shows that 
$\tv(q)$ only depends on $q^2$. We can also directly check from Eq.~\eqref{140116.8} that 
\begin{align}
\Ima \tv(i\mu\pm \ve)=\mp \eta(\mu^2)~.
\label{140116.9}
\end{align}
 To show it let us substitute $q=i\mu\pm \ve$, $\mu>0$, in Eq.~\eqref{140116.8} with the 
 integration variable primed,
\begin{align}
\tv(i\mu\pm \ve)&=\frac{1}{\pi}\int_0^\infty d{\mu'}^2\frac{\eta({\mu'}^2)}{{\mu'}^2-\mu^2\pm i\ve}
=\frac{1}{\pi}\dashint_0^\infty d{\mu'}^2\frac{\eta({\mu'}^2)}{{\mu'}^2-\mu^2}
\mp i\int_0^\infty d{\mu'}^2\eta({\mu'}^2)\delta({\mu'}^2-\mu^2)\nn\\
&=\frac{1}{\pi}\dashint_0^\infty d{\mu'}^2\frac{\eta({\mu'}^2)}{{\mu'}^2-\mu^2}
\mp i \eta(\mu^2)~.
\label{140116.10}
\end{align}
Where we have made use of the well-known result
\begin{align}
\int dx\frac{f(x)}{x-i\ve}=\dashint dx\frac{f(x)}{x}+i\pi\delta(x)~,
\end{align}
with the dash in the integration symbol meaning the principal value of the integral.

An important consequence of Eq.~\eqref{140116.8} is that $\tv(q)$ is a function in the complex $q^2$ plane that only has 
a LHC, along which it is necessary that $q^2<0$. 
 For the exchange of massive force carriers there is a lower threshold $m_\pi^2$
 below which the spectral function is zero, so that the LHC extends for $q^2\in]-\infty,-m_\pi^2]$. 
Another important property that follows from Eq.~\eqref{140116.8} is that $\tv(q^2)$ satisfies the 
Schwarz reflection principle
\begin{align}
\tv(z)=\tv(z^*)^*~.
\label{140116.11b}
\end{align}
Any analytic function that is real along an open interval of the real axis fulfills Eq.~\eqref{140116.11b} 
in its domain of analyticity. Of course, the potential 
is real for $q^2>0$, which is the physical region.

\begin{figure}[ht]
\begin{center}
\includegraphics[width=.3\textwidth]{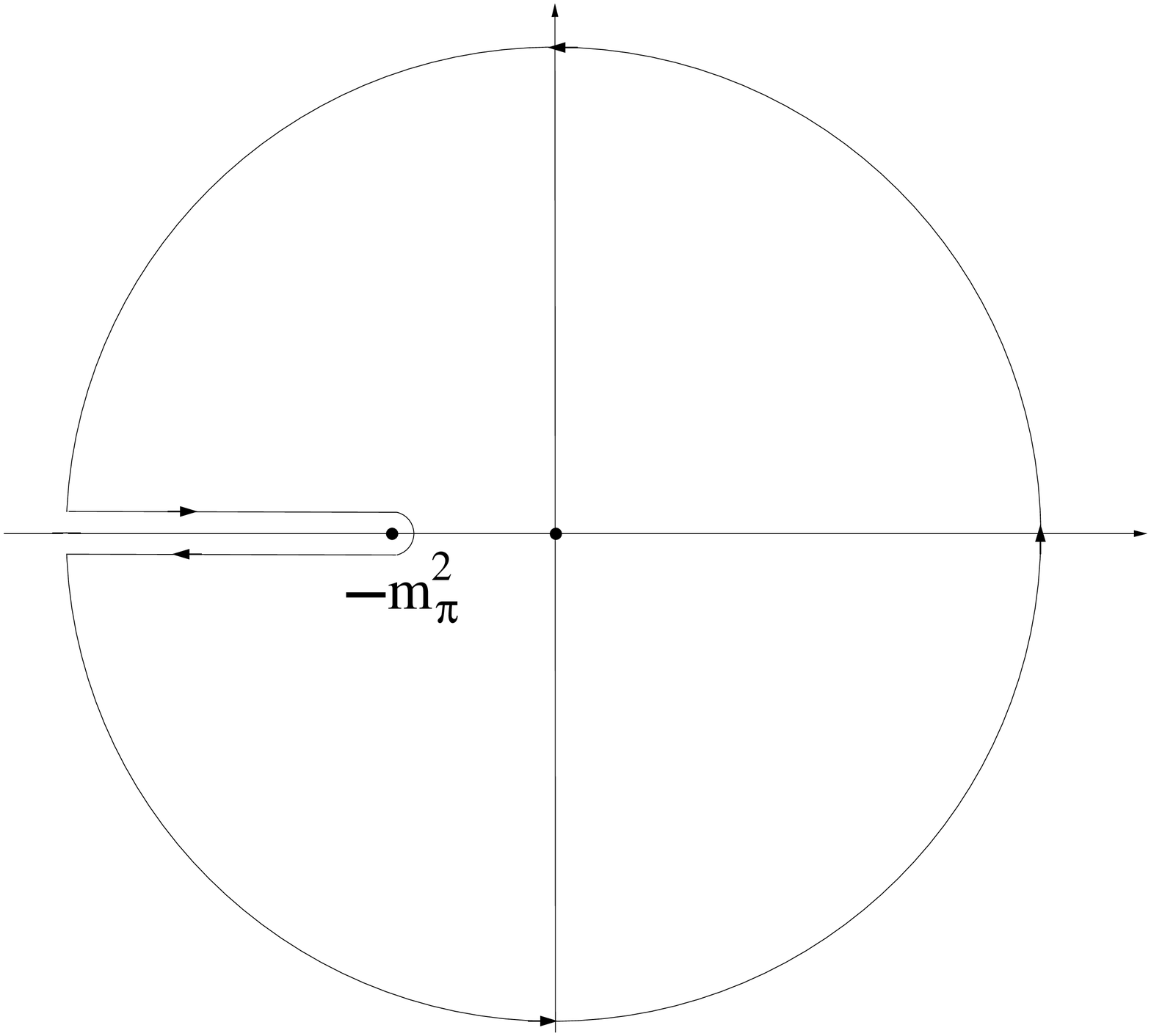}
\end{center}
\caption{{\small Contour of integration ${\cal C}$ used in the complex integration of Eq.~\eqref{140116.12} to derive the DR for the 
potential $\tv(q^2)$.
\label{fig:140116.2}
}}
\end{figure} 

If $\tv(q^2)$ does not vanish for $q^2\to \infty$ one needs to take more subtractions in the DR of Eq.~\eqref{140116.8}. 
This can be accomplished by considering the application of the
 Cauchy integration theorem to the integral of the function $\tv(z)/(z^n (z-q^2))$ along 
the closed integration contour ${\cal C}$ in the complex $q^2$ plane shown in Fig.~\ref{fig:140116.2}, which 
is a circle at infinity that engulfs the LHC. The application of the Cauchy theorem  gives
\begin{align}
\label{140116.12}
\oint dz\frac{\tv(z)}{z^n(z-q^2)}&=2\pi i \frac{\tv(q^2)}{(q^2)^n}+2\pi i \frac{1}{(n-1)!}
\frac{d^{n-1}}{dz^{n-1}}\frac{\tv(z)}{z-q^2}\Bigg|_{z=0}~,
\end{align}
while the direct integration along the integration contour is
\begin{align}
\label{150116.1}
\oint dz\frac{\tv(z)}{z^n(z-q^2)}&=\int_{-\infty}^{-m_\pi^2}dk^2\frac{\tv(k^2+i\ve)-\tv(k^2-i\ve)}{(k^2)^n(k^2-q^2)}
=2i\int_{-\infty}^{-m_\pi^2}dk^2 \frac{\Ima \tv(k^2+i\ve)}{(k^2)^n(k^2-q^2)}~.
\end{align}
 Equating Eqs.~\eqref{140116.12} and \eqref{150116.1} one has 
\begin{align}
\tv(q^2)&=-\frac{(q^2)^n}{(n-1)!}\frac{d^{n-1}}{dz^{n-1}}\frac{\tv(z)}{z-q^2}\Bigg|_{z=0}
+\frac{(q^2)^n}{\pi}\int_{-\infty}^{-m_\pi^2}dk^2\frac{\Ima  \tv(k^2+i\ve)}{(k^2)^n(k^2-q^2)}~.
\label{150116.2}
\end{align}
 The first term on the right-hand side (rhs) of the previous equation is a polynomial in $q^2$. 
Evaluating explicitly the derivative at $z=0$ by applying the binomial theorem,
\begin{align}
\frac{d^m}{dz^m}\frac{\tv(z)}{z-q^2}&=\sum_{j=0}^{m}
\left(
\begin{matrix}
m\\
j
\end{matrix}
\right)
\tv^{(m-j)}(z)\frac{d^j}{dz^j}\frac{1}{z-q^2}\nn\\
&=\sum_{j=0}^{m}
\left(
\begin{matrix}
m\\
j
\end{matrix}
\right)
\tv^{(m-j)}(z)\frac{(-1)^j j!}{(z-q^2)^{j+1}}~,
\label{150116.3}
\end{align}
one obtains
\begin{align}
-\frac{(q^2)^n}{(n-1)!}\frac{d^{n-1}}{dz^{n-1}}\frac{\tv(z)}{z-q^2}\Bigg|_{z=0}&=
-\frac{(q^2)^{n}}{(n-1)!}\sum_{j=0}^{n-1}
\left(
\begin{matrix}
n-1\\
j
\end{matrix}
\right)
\tv^{(n-1-j)}(0)\frac{(-1)^j j!}{(-q^2)^{j+1}}\nn\\
&=\sum_{j=0}^{n-1}\frac{1}{(n-1-j)!}\tv^{(n-1-j)}(0)(q^2)^{n-1-j}~.
\label{150116.4}
\end{align}
Introducing the new index for the sum in the previous equation
\begin{align}
p=n-1-j~,
\label{150116.5}
\end{align}
we finally rewrite Eq.~\eqref{150116.2} as
\begin{align}
\tv(q^2)&=\sum_{p=0}^{n-1}\frac{\tv^{(p)}(0)}{p!}(q^2)^p
+\frac{(q^2)^n}{\pi}\int_{-\infty}^{-m_\pi^2}dk^2\frac{\Ima  \tv(k^2+i\ve)}{(k^2)^n(k^2-q^2)}~.
\label{150116.6}
\end{align}

We now perform the inverse Fourier transform of the previous expression to calculate $v(r)$ in configuration space. 
\begin{align}
v(r)&=\int  \frac{d^3q}{(2\pi)^3}e^{i\vq \vr} \tv(q^2)
=\int  \frac{d^3q}{(2\pi)^3}e^{i\vq \vr}\left\{
\sum_{p=0}^{n-1}\frac{\tv^{(p)}(0)}{p!}(q^2)^p
+\frac{(q^2)^n}{\pi}\int_{-\infty}^{-m_\pi^2}dk^2\frac{\Ima  \tv(k^2+i\ve)}{(k^2)^n(k^2-q^2)}
\right\}\nn\\
&=\sum_{p=0}^{n-1}\frac{\tv^{(p)}(0)}{p!}(-\nabla^2)^p\int  \frac{d^3q}{(2\pi)^3}e^{i\vq\vr}
+\frac{(-\nabla^2)^n}{\pi}
\int_{-\infty}^{-m_\pi^2}dk^2\frac{\Ima\tv(k^2+i\ve)}{(k^2)^n}\int\frac{d^3q}{(2\pi)^3}\frac{e^{i\vq\vr}}{k^2-q^2}~.
\label{150116.7}
\end{align}
The last two integrations in $\vq$ in the last equation are  
\begin{align}
\int d^3q \frac{e^{i\vq\vr}}{(2\pi)^3}&=\delta(\vr)~,\nn\\
\int\frac{d^3q}{(2\pi)^3}\frac{e^{i\vq\vr}}{k^2-q^2}&=-\frac{e^{-\sqrt{-k^2}r}}{4\pi r}~.
\label{150116.8}
\end{align}
A change of the integration variable in the dispersive integral, $k^2=-\mu^2$, 
brings Eqs.~\eqref{150116.7} into the form
\begin{align}
v(r)&=\sum_{p=0}^{n-1}\frac{\tv^{(p)}(0)}{p!}(-\nabla^2)^p\delta(\vr)+\int_{m_\pi^2}^\infty d\mu^2\frac{\eta(\mu^2)}{(\mu^2)^n}
(\nabla^2)^n\frac{e^{-\mu r}}{4\pi^2 r}~.
\label{150116.9}
\end{align}
From this expression it is clear that the subtractive polynomial in the DR of Eq.~\eqref{150116.6} 
in momentum space becomes a polynomial of contact terms comprising a Dirac-delta function 
and its higher order derivatives in configuration space.
 We can simplify the dispersive integration in Eq.~\eqref{150116.9} by noticing that
\begin{align}
(\nabla^2)^n\frac{e^{-\mu r}}{r}=(\mu^2)^n\frac{e^{-\mu r}}{r}-4\pi\sum_{m=0}^{n-1}(\mu^2)^{n-1-m}(\nabla^2)^m\delta(\vr)~.
\label{150116.10}
\end{align}
Employing this result in Eq.~\eqref{150116.9} it becomes 
\begin{align}
v(r)&=\sum_{p=0}^{n-1}\frac{\tv^{(p)}(0)}{p!}(-\nabla^2)^p\delta(\vr)
-\frac{1}{\pi}\sum_{p=0}^{n-1}(\nabla^2)^p\delta(\vr)\int_{m_\pi^2}^\infty d\mu^2\frac{\eta(\mu^2)}{(\mu^2)^{p+1}}
+\int_{m_\pi^2}^\infty d\mu^2 \eta(\mu^2)\frac{e^{-\mu r}}{4\pi^2 r}~.
\label{150116.11}
\end{align}
Depending on the degree of divergence of the spectral function $\eta(\mu^2)$ in the limit $\mu^2\to\infty$ some (or all or any) of the 
integrations in the second sum of the rhs of the previous equations will actually diverge. 
But all these terms  can be reabsorbed in the coefficients of the first one 
 (this is called a renormalization procedure), 
 so that we can simply write the previous equation in terms of a 
 few coefficients, $\widetilde{\omega}^{(p)}(0)$, as
\begin{align}
v(r)&=\sum_{p=0}^{n-1}\frac{\widetilde{\omega}^{(p)}(0)}{p!}(-\nabla^2)^p\delta(\vr)
+\int_{m_\pi^2}^\infty d\mu^2 \eta(\mu^2)\frac{e^{-\mu r}}{4\pi^2 r}~.
\label{150116.12}
\end{align}
This expression implies that any potential can be written as the superposition of Yukawa potentials (weighted in terms of the 
spectral function) plus a zero-range part which is a polynomial in $\delta(\vr)$ and its higher order derivatives (according to the 
action of the Laplacian operator). 
 Giving sense to this polynomial is not a trivial matter and one needs to develop new techniques 
which are the main aim of our work.

The process of acting with $\nabla^2$ inside the integrand in Eq.~\eqref{150116.9} is equivalent to manipulate the factor $(q^2)^n$ 
in the DR in momentum space, Eq.~\eqref{150116.6}, in order to reduce the number of subtractions:
\begin{align}
&\frac{(q^2)^n}{\pi}\int_{-\infty}^{-m_\pi^2}dk^2\frac{\Ima  \tv(k^2+i\ve)}{(k^2)^n(k^2-q^2)}
=\frac{(q^2)^{n-1}}{\pi}\int_{-\infty}^{-m_\pi^2}dk^2\frac{(q^2-k^2+k^2)\Ima  \tv(k^2+i\ve)}{(k^2)^n(k^2-q^2)}\nn\\
&=-\frac{(q^2)^{n-1}}{\pi}\int_{-\infty}^{-m_\pi^2}dk^2\frac{\Ima  \tv(k^2+i\ve)}{(k^2)^n}
+\frac{(q^2)^{n-1}}{\pi}\int_{-\infty}^{-m_\pi^2}dk^2\frac{\Ima  \tv(k^2+i\ve)}{(k^2)^{n-1}(k^2-q^2)}
\label{160116.1}
\end{align}
We could iterate further this process for the last integral in the previous equation 
 while the resulting integrals are convergent.  
The coefficient in Eq.~\eqref{150116.6} that multiplies $(q^2)^{n-1}$ is 
\begin{align}
\frac{\tv^{n-1}(0)}{(n-1)!}
\label{160116.2}
\end{align}
that cancels with the first term on the rhs of the last line of Eq.~\eqref{160116.1}, which is an explicit expression for $v^{(n-1)}(0)$ 
that results from a lower than $n$-times subtracted DR, e.g. by taking $n-1$ derivatives in Eq.~\eqref{140116.8}. 
Namely,
\begin{align}
\tv^{(p)}(0)&=\frac{(-1)^p p!}{\pi}\int_{m_\pi^2}^\infty d\mu^2\frac{\eta(\mu^2)}{(\mu^2)^{p+1}}~,
\label{160116.3}
\end{align}
an expression that is valid for an $m$-times subtracted DR for $\tv(q^2)$ as along as $m<p$.

The degree of divergence of $\eta(\mu^2)$ does not only determine a lower bound on
 the number of subtractions needed to be taken in the 
DR of $\tv(q^2)$ but also contributes to the possible divergence of $v(r)$ for $r\to 0$ in configuration space. 
To see it let us assume that $\eta(\mu^2)$ scales stronger than $\mu^n$, with $n>-1$, in the limit $\mu\to \infty$. 
 Next, insert this asymptotic behavior for $\mu>\Lambda>>m_\pi$ in the integration of Eq.~\eqref{150116.12} (that gives 
$v(r)$ for $r\neq 0$). We have the following contribution to $v(r)$, we first assume that $n\geq 0$, 
\begin{align}
\frac{1}{2\pi^2 r}\int_{\Lambda}^\infty d\mu\, \mu^{n+1}e^{-\mu r}=\frac{(-1)^{n+1}}{2\pi^2 r}\frac{\partial^{n+1}}{\partial r^{n+1}}\int_\Lambda^\infty 
d\mu\, e^{-\mu r}\xrightarrow[r\to 0]{}\frac{(n+1)!}{2\pi^2 r^{n+3}}~,
\label{160116.4}
\end{align}
which gives rise to  a singular potential. 
For $n=-1$ we can make use of the result that 
$\int_{\Lambda}^\infty d\mu \mu^\ve e^{-\mu r}=r^{-1-\ve}\Gamma(1+\ve,r \Lambda)$, 
with $\ve\to 0^+$, to conclude that in the limit $r\to 0$ we have the contribution from the asymptotic region of the 
spectral function,
 \begin{align}
 \label{180827.1}
\frac{1}{2\pi^2 r^{2+\ve}}~,
 \end{align}
which is again of the singular-potential type.

\subsection{Calculation of the spectral function}
\label{sec:170116.2}

 In quantum field theory 
 one usually calculates directly the potential in momentum space, from which one can calculate by 
proper analytical extrapolation its spectral function by the knowledge of the imaginary part of $\tv(q^2)$ for negative values 
of $q^2$. 
However, in other cases the potential in configuration space $v(r)$ is the one known. Let us determine 
how to calculate then $\eta(\mu^2)$ in a way suited as well for singular potentials.

We first multiply by the factor $e^{-\mu r}$ so that there is no problem 
with the convergence of the Fourier transform integral of the potential in the limit $r\to \infty$, 
 with the understanding that at the end of the calculations one takes the limit $\mu\to 0^+$. 
In this way all the potentials have by construction a finite range and are suitable  
for a treatment within $S$-matrix theory.
\begin{align}
\tv(q)&=\int d^3r\,e^{-i\vq\vr}v(r) e^{-\mu r}
=\frac{4\pi}{q}\Ima \int dr \,r e^{(iq-\mu) r} v(r)~. 
\label{160116.5}
\end{align}
The previous Fourier transform does not exist for singular potentials at the origin, diverging faster or equal than 
$1/r^3$. For such cases we can apply the following iteration rule if the potential behaves as
\begin{align}
v(r)\xrightarrow[r\to 0]{}\alpha r^{-n}~,~n\geq 3~,
\label{160116.6}
\end{align}
with $\alpha$ a real constant (the constant is real if the potential is Hermitian, as we assume). 
The method could not be applied if $\alpha$ were actually a function of $r$ that cannot be expanded 
in power series of $r$ for $r\to 0$. 

Returning to the case of Eq.~\eqref{160116.6} with a constant $\alpha$ we can conveniently rewrite Eq.~\eqref{160116.5} as
\begin{align}
\tv(q)=\frac{4\pi}{q}\Ima\int_0^\infty dr\, e^{(i q-\mu)r}r\left\{v(r)-\frac{\alpha}{r^n}\right\}
+\frac{4\pi \alpha}{q}\Ima \int_0^\infty dr\,\frac{e^{(iq-\mu)r}}{r^{n-1}}~.
\label{160116.7}
\end{align}
The point is that the degree of divergence of $v(r)-\alpha/r^n$ is reduced compared with $v(r)$ 
in the limit $r\to 0$. Let us now discuss the last integral in the 
previous equation, keeping in mind that i) we are interested in calculating the spectral function of $\tv(q)$ and ii) we proceed by 
analytical continuation in $q$ starting in the region of positive real values. Performing an integration by parts and keeping only those 
terms relevant  for the purpose i) when $\mu\to 0^+$, one has
\begin{align}
\int_0^\infty dr\,e^{(iq-\mu)r}r^{-n+1}&=
 \frac{e^{(iq-\mu)r}r^{-n+2}}{-n+2}\Bigg|_0^\infty
+\frac{ i q}{n-2}\int_0^\infty dr\,e^{(iq-\mu)r}r^{-n+2}~.
\label{160116.8}
\end{align}
 The first term on the rhs of the last line is zero in the limit $r\to\infty$, while in the limit $r\to 0$ is divergent 
 but independent of $q$, so that it does not contribute to i).  
 Thus, we are only left with the last  term in the previous equation and 
Eq.~\eqref{160116.7} becomes
\begin{align}
\tv(q)\longrightarrow & \frac{4\pi}{q}\Ima \int_0^\infty dr\,e^{(iq-\mu)r}r\left\{v(r)-\frac{\alpha}{r^n}\right\}
+\frac{4\pi\alpha}{n-2}\Ima i \int_0^\infty dr\,e^{(iq-\mu)r}r^{-n+2}~.
\label{160116.9}
\end{align} 
The process could be further iterated for the first integral if $v(r)-\frac{\alpha}{r^n}$ diverges equal or faster than $1/r^3$ 
at short distances,  and for the second one if $n\geq 4$.

  Let us now work out explicitly the case of a potential
\begin{align}
v(r)=\frac{\alpha}{r^n}~,
\label{160116.10}
\end{align}
with integer $n\geq 3$. Performing the iteration process of Eq.~\eqref{160116.9} $n-2$ times one ends with
\begin{align}
\tv(q)\rightarrow \frac{4\pi\alpha q^{n-3}}{(n-2)!} \Ima \int_0^\infty dr\, \frac{ i^{n-2}e^{(iq-\mu)r}}{r}~.
\label{160116.11}
\end{align}
We distinguish between even and odd $n$ because of the factor $i^{n-2}$:

\noindent
Even-$n$:
\begin{align}
\tv(q)\xrightarrow[]{\text{even}~n}& \frac{4\pi \alpha(-1)^{\frac{n-2}{2}} q^{n-3}}{(n-2)!}\int_0^\infty dr
 \frac{\sin(qr)e^{-\mu r}}{r}\nn\\
&=\frac{4\pi\alpha (-1)^{\frac{n-2}{2}} q^{n-3}}{(n-2)!}\arctan\frac{q}{\mu}\nn\\
&=\frac{2\pi i \alpha (-1)^{\frac{n}{2}} q^{n-3}}{(n-2)!}\left\{\log(1+i\frac{q}{\mu})-\log(1-i\frac{q}{\mu})\right\}~.
\label{160116.12}
\end{align}
Here we have expressed
\begin{align}
\arctan x &=\frac{1}{2i}\left\{\log(1+i x)-\log(1-i x)\right\}~,
\label{160116.13}
\end{align}
and the log defined with the cut to the left.

\noindent
Odd-$n$:
\begin{align}
\tv(q)\xrightarrow[]{\text{odd}~n}& -\frac{4\pi\alpha i^{n-1}q^{n-3}}{(n-2)!}\Ima\int_0^\infty dr\,i \frac{e^{(iq-\mu)r}}{r}     \nn\\
&\to -\frac{4\pi\alpha (-1)^\frac{n-1}{2}q^{n-3}}{(n-2)!}\int_0^\infty dr\,(\cos qr-1)\frac{e^{-\mu r}}{r}\nn\\
&\to \frac{2\pi\alpha (-1)^\frac{n-1}{2}q^{n-3}}{(n-2)!}\log(1+\frac{q^2}{\mu^2})\nn\\
&= \frac{2\pi\alpha (-1)^\frac{n-1}{2}q^{n-3}}{(n-2)!}\left\{\log(1+i\frac{q}{\mu})+\log(1-i\frac{q}{\mu})\right\}~.
\label{160116.14}
\end{align}
From the first to the second line of the previous equation we have neglected a (divergent) term which is just an 
 integer power of $q$ and, therefore, it does not contribute to the spectral function of $\tv(q)$. 

When considering complex values of $q$ we take as final expressions those in the last lines of 
Eqs.~\eqref{160116.12} and \eqref{160116.14}. 
The reason is because they have the simplest analytical extrapolation with the branch points located at
\begin{align}
q=\pm i \mu~,
\label{160116.15}
\end{align}
and cuts extending for negative values of the arguments of the $\log$'s,
\begin{align}
q=\pm i\sqrt{\mu^2+x^2}~~,~~ x\in \mathbb{R}~,
\label{160116.16}
\end{align}
for $\log(1\pm i q/\mu)$, in order.

 \begin{figure}[ht]
\begin{center}
\includegraphics[width=.3\textwidth]{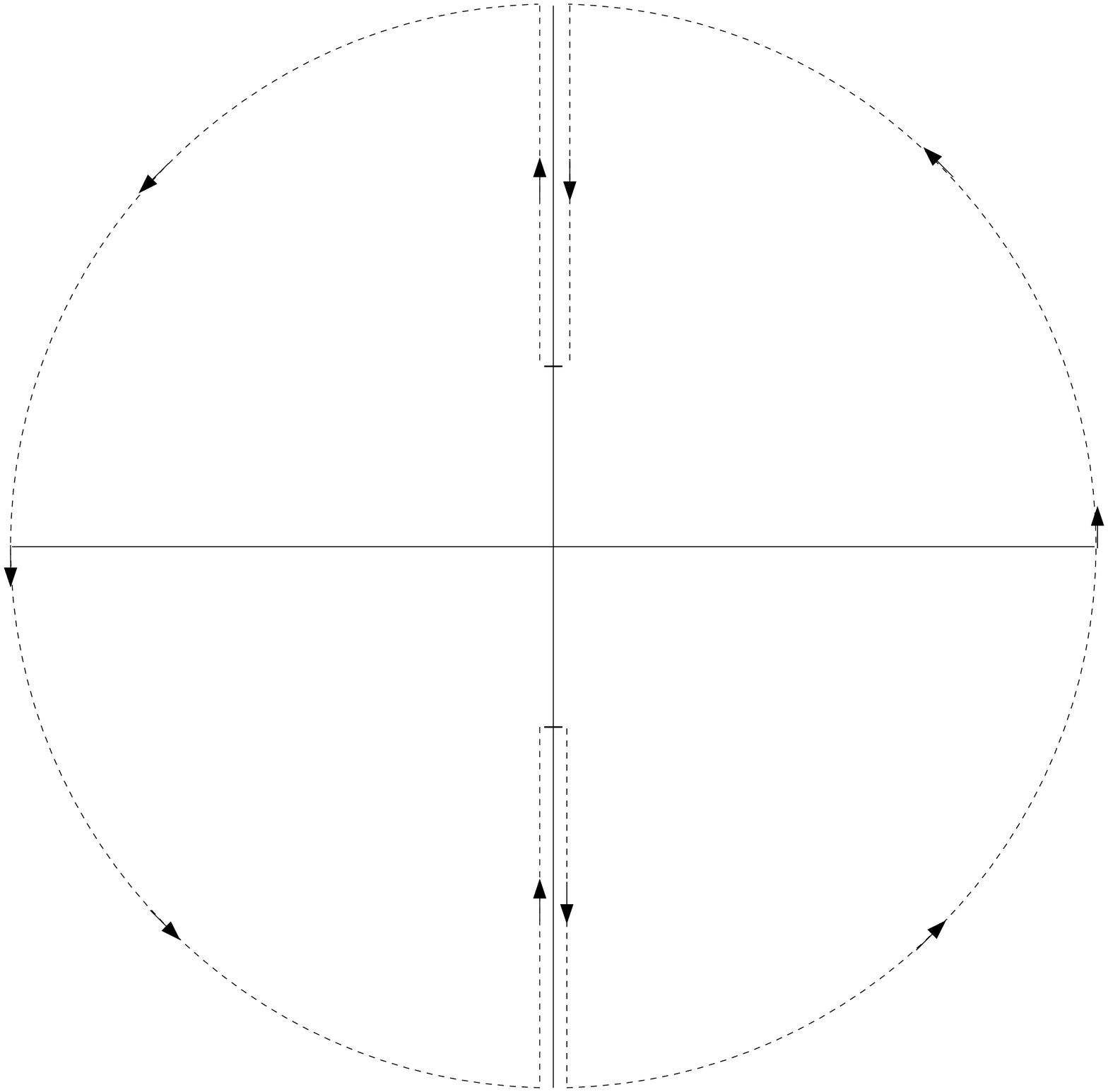}
\end{center}
\caption{{\small Contour of integration ${\cal C}$ used in the DR of $\tv_2(q)$,  Eq.~\eqref{170116.2}.
\label{fig:170116.1}}}
\end{figure} 

$\bullet$  As an example to illustrate the calculation of a spectral function and 
the implementation of a concrete DR let us take the potential $v_2(r)$, 
corresponding to $n=2$ in Eq.~\eqref{160116.10} and  $\alpha=1$. 
 In this case the Fourier transform corresponds exactly with the rhs of Eq.~\eqref{160116.12} and gives
\begin{align}
\tv_2(q)&=\frac{4\pi}{q}\arctan\frac{q}{\mu}=-\frac{2\pi i}{q}\left\{\log(1+i\frac{q}{\mu})-\log(1-i\frac{q}{\mu}) \right\}~.
\label{170116.1}
\end{align}
The previous expression has the cuts given in Eq.~\eqref{160116.16} and in order to write down a DR 
for $\tv_2(q)$ we consider the integration 
contour plotted in Fig.~\ref{fig:170116.1}. The function $\tv_2(q)$ vanishes for $q\to \infty$ and then it is not necessary to take 
any subtraction to remove the integration along the circle at infinity. We then have
\begin{align}
\label{170116.2}
\tv_2(q)&=\frac{1}{2\pi i}\oint_{{\cal C}} dz\frac{\tv_2(z)}{z-q}~,\\
\tv_2(q)&=\frac{1}{2\pi i}\int_{\infty}^{\mu} d(-i\nu)\frac{\tv_2(-i\nu-\ve)-\tv_2(-i\nu+\ve)}{-i\nu-q}
+\frac{1}{2\pi i}\int_\infty^\mu d(i\nu) \frac{\tv_2(i\nu+\ve)-\tv_2(i\nu-\ve)}{i\nu-q}\nn\\
&=\frac{1}{2\pi}\int_\mu^\infty d\nu\frac{\tv_2(-i\nu+\ve)-\tv_2(-i\nu-\ve)}{i\nu+q}
+\frac{1}{2\pi}\int_\mu^\infty  d\nu \frac{\tv_2(i\nu-\ve)-\tv_2(i\nu+\ve)}{i\nu-q}~.\nn
\end{align}
From Eq.~\eqref{170116.1} we directly obtain
\begin{align}
\label{170116.4b}
\tv_2(i\nu+\ve)-\tv_2(i\nu-\ve)&=-\frac{2\pi}{\nu}\left\{
\log(1-\frac{\nu}{\mu}+i\ve)-\log(1-\frac{\nu}{\mu}-i\ve)\right\}
=-\frac{4\pi^2 i}{\nu}\theta(\nu-\mu)~,\\
\label{170116.5}
\tv_2(-i\nu+\ve)-\tv_2(-i\nu-\ve)&=\frac{4\pi^2i}{\nu}\theta(\nu-\mu)~.
\end{align}
This is substituted in Eq.~\eqref{170116.2} with the result
\begin{align}
\tv_2(q)&=2\pi i \int_{\mu}^\infty \frac{d\nu}{\nu}\left\{\frac{1}{i\nu-q}+\frac{1}{i\nu+q}\right\}
=4\pi\int_\mu^\infty \frac{d\nu}{\nu^2+q^2}=\frac{4\pi}{q}\arctan\frac{q}{\mu}~,
\label{170116.6}
\end{align}
the same result as in Eq.~\eqref{170116.1}. It is also clear from Eq.~\eqref{170116.5} that 
the spectral function for this case is $\eta(\nu^2)=2\pi^2 \theta(\nu-\mu)/\nu$.

$\bullet$ As a second example, we calculate the spectral function of the repulsive short-range potential 
for the ${\rm He}$--${\rm He}$ interaction as derived in Ref.~\cite{tang.180728.1}. This potential stems from the 
exchange of a pair of electrons between two multielectronic systems. 
It is given by
\begin{align}
v_s(r)&=\alpha e^{-\beta r}r^{\frac{7}{2\beta}-1}~,
\label{170116.7}
\end{align}
with $\alpha$ and $\beta>0$.
A direct calculation gives for its Fourier transform
\begin{align}
\tv_s(q)&=\frac{2\pi\alpha }{i q}\Gamma(1+\frac{7}{2\beta})
\left\{(-i q+\beta)^{-1-\frac{7}{2\beta}}-(iq+\beta)^{-1-\frac{7}{2\beta}}\right\}~,
\label{170116.8}
\end{align}
which is an even function of $q$. 
Now, we can write in complex variable 
\begin{align}
(\pm iq +\beta)^{-1-\frac{7}{2\beta}}=e^{-(1+\frac{7}{2\beta})\log(\pm i q+\beta)}~,
\label{170116.7b}
\end{align}
which shows that this function has cuts of the same type as those already discussed for Eq.~\eqref{160116.16} 
\begin{align}
q=\pm i \sqrt{\beta^2+x^2}~~,~~x\in \mathbb{R}~.
\label{170116.8a}
\end{align}
To simplify the writing we introduce the notation
\begin{align}
\gamma&=1+\frac{7}{2\beta}~.
\label{170116.9}
\end{align}
The discontinuity of $\tv_2(q)$ across the pure imaginary axis and the spectral function are, in order,
\begin{align}
\tv_s(i\nu+\ve)-\tv_2(i\nu-\ve)&=
\frac{2\pi\alpha}{\nu}\Gamma(\gamma)(e^{-\gamma\log(-\nu+\beta+i\ve)}-e^{-\gamma\log(-\nu+\beta-i\ve)})\\
&=\frac{2\pi\alpha}{\nu}\Gamma(\gamma)e^{-\gamma\log(\nu-\beta)}(e^{-i\gamma \pi}-e^{i\gamma \pi})
\theta(\nu-\beta)\nn\\
&=-\frac{4\pi i \alpha}{\nu}\Gamma(\gamma)(\nu-\beta)^{-\gamma}\sin(\gamma\pi)\theta(\nu-\beta)~,\nn\\
\label{170116.10}
\eta_s(\nu^2)&=\frac{2\pi\alpha}{\nu}\Gamma(\gamma)(\nu-\beta)^{-\gamma}\sin(\gamma\pi)\theta(\nu-\beta)~.
\end{align}

\subsection{Partial-wave projected potential}
\label{sec:170116.1}

 In this section we discuss the analytical properties of the potential projected in a given partial 
wave as a function of two complex arguments, $p_1$, $p_2$ that correspond to the 
out- and in-going three-momenta, in order.  The projected 
potential as a function of one of the two complex variables 
 is an analytical function in the corresponding cut complex plane, 
with the position of the cuts depending on the other 
argument. The cuts that appear in the potential are called dynamical cuts (DCs). 
The same cuts  also appear 
in the on-shell and half-off-shell PWAs, with no generation of extra DCs, 
 as discussed in Sec.~\ref{sec:aphfs}. 
 For definiteness we generically refer as pions to the force-carrier particles.

 The results that follow concerning the analytical properties of a partial-wave 
 projected potential are explicitly obtained for the case of a Yukawa potential, 
 \begin{align}
 \label{180728.3}
 v(\vp_1,\vp_2)&=\frac{g^2}{\vq^2+m_\pi^2}~,
 \end{align}
 where $\vq=\vp_1-\vp_2$ is the momentum transfer and the range of the potential is $1/m_\pi$. 
 Nonetheless, as we discuss at the end of this section, the derivations that follow 
 can be straightforwardly generalized to any local potential
  by making use of its spectral decomposition
\begin{align}
v(\vp_1,\vp_2)&=\int_{m_\pi}^\infty d\mu \frac{2\mu\eta(\mu^2)}{\vq^2+\mu^2}~,
\label{eq.sdv}
\end{align}
that could also include subtractions, cf. Eq.~\eqref{150116.6}.

Since the analytical properties of the potential are encoded 
in the kinematical configurations that put one-shell the  pions exchanged (in other terms, when  
the denominator in Eq.~\eqref{eq.sdv} vanishes),  our results 
could be applied to non-local potentials too that fulfill Eq.~\eqref{eq.sdv} but with a more involved 
spectral function of the type  $\eta(\mu^2,\vp_1,\vp_2)$,  being an entire function
 of $\vp_1$ and $\vp_2$. Because of rotational invariant the dependence on these arguments enter as 
 $p_1^2$, $p_2^2$ or $\vp_1\cdot \vp_2$, with $p_i=|\mathbf{p}_i|$. 
  This is the case for potentials calculated in chiral effective field theory for the 
  $NN$ interactions, once relativistic interactions to the potential are calculated perturbatively in powers of 
 the inverse of the nucleon mass. 
  By the same token,   although we consider explicitly in the following the simpler $S$-wave projection, which is denoted as 
$v(p_1,p_2)$,  our results regarding the location of the branch points and cuts 
are also valid for any other partial wave.

We take the $S$-wave projection of the potential in Eq.~\eqref{180728.3}, 
\begin{align}
v(p_1,p_2)&=g^2\int_{-1}^{+1}\left(p_1^2+p_2^2-2p_1 p_2 z+m_\pi^2\right)^{-1}dz~,
\label{swpv}
\end{align}
where  $g^2$ is  a coupling constant and $z=\cos \theta$, with $\theta$ the relative angle. 
The denominator vanishes for 
\begin{align}
p_2&=p_1 z \pm i \sqrt{p_1^2(1-z^2)+m_\pi^2}~,~z\in [-1,+1]~,
\label{3.4tex}
\end{align}
which gives rise to a cut of logarithmic nature in the $p_2$ variable as a function of $p_1$, with the 
branch points obtained by substituting in the previous equation $z= \pm 1$. 
This is the origin of the DCs referred above. 
We might also characterize them because they owe their origin by putting on-shell the (particles) pions exchanged as 
force carriers. 

\begin{figure}
\begin{center}
\begin{tabular}{cc}
\includegraphics[width=.4\textwidth]{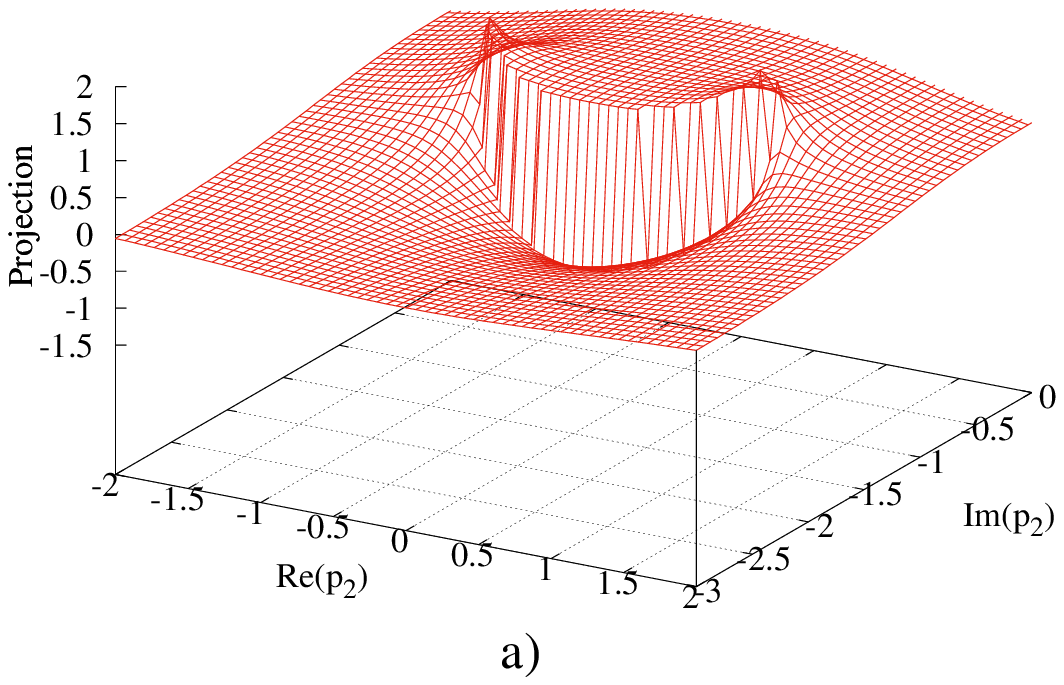} & 
\includegraphics[width=.4\textwidth]{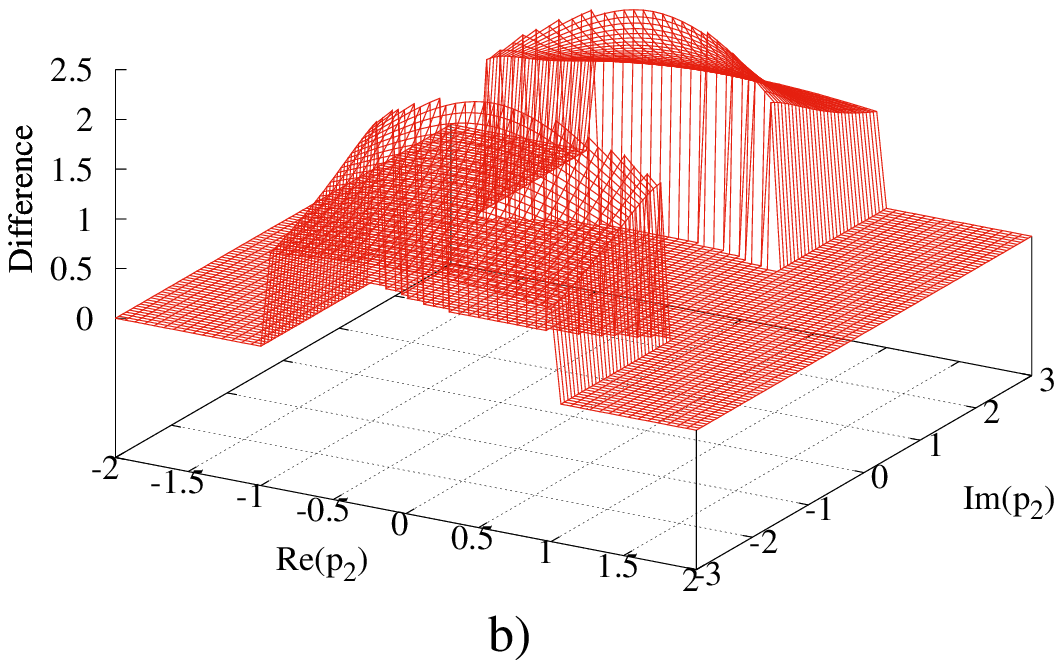}  
\end{tabular}
\end{center}
\caption[pilf]{{\small \protect {We plot in the left panel the real part of 
the $S$-wave projection of Eq.~\eqref{swpvs}. One branch of the cut in 
Eq.~\eqref{3.4tex} can be clearly seen. For the right panel we have 
the real part of the difference between the expressions in Eq.~\eqref{swpvm} and 
\eqref{swpvs}. 
One can clearly see the two strips in which the two forms of the $S$-wave  projected potential differ. 
For these plots we have taken $m_\pi=1$, $g=1$ and $p_1=1$.}
\label{fig:vcut}
}}
\end{figure}

 Indeed, Eq.~\eqref{swpv} can
 be integrated algebraically, although for complex $p_1$ and $p_2$ one has to be careful because 
the imaginary part of the denominator can change sign when varying $z$ in the interval of integration. 
The right process is to extract first $2 p_1 p_2$ as a common 
factor in the denominator and then perform the integration in $z$,
\begin{align}
v(p_1,p_2)&=-\frac{g}{2 p_1 p_2}\int_{-1}^{+1}\frac{dz}{z-(p_1^2+p_2^2+m_\pi^2)/(2p_1p_2)}\nn\\
&=-\frac{g}{2 p_1 p_2} \big\{ \log\big[1-\frac{p_1^2+p_2^2+m_\pi^2}{2p_1p_2}\big]
- \log\big[-1-\frac{p_1^2+p_2^2+m_\pi^2}{2p_1p_2}\big] \big\}\nn\\
&=\frac{g}{2 p_1 p_2}  \log\big[\frac{(p_1+p_2)^2+m_\pi^2}{(p_1-p_2)^2+m_\pi^2}\big]~.
\label{swpvs}
\end{align}
Notice that since $z$ is real the imaginary part of the denominator in the integral of the 
 equation above does not change sign and we do not cross the logarithmic cut in 
the result. We can combine together the two logs in the second line in 
 just one because their arguments have the same imaginary part. 
 In the complex $p_2$ plane the cut in Eq.~\eqref{3.4tex}  is a curved one
   parameterized by $z\in[-1,1]$ and its lower branch is represented 
in the left panel of Fig.~\ref{fig:vcut} in units for which  $m_\pi=1$, $g=1$ and $p_1=1$, 
 where we show the real part of Eq.~\eqref{swpvs} (that perfectly agrees with the direct 
 numerical evaluation of Eq.~\eqref{swpv}).

Next, had we considered $n$ pion exchanges this would have implied that the momentum transfer squared 
should be equal to $-(n m_\pi)^2$, giving rise to the curved cut in the complex $p_2$ plane:
\begin{align}
p_2&=p_1 z\pm i \sqrt{p_1^2(1-z^2)+n^2 m_\pi^2}~,~z\in[-1,+1]~.
\label{finitecuts}
\end{align}  
 For instance, when iterating the LS equation the value of $n$ rises with the order of the iteration 
 of the potential and this originates an 
infinite sequence of finite curved cuts. This certainly would make difficult the calculation of the 
discontinuity across the DCs.


We can however proceed differently with the analytical extrapolation of $v(p_1,p_2)$ so that 
the location of the cuts allows a much simpler manipulation. The point is that for real and positive 
$p_i$ we can rewrite the last line of Eq.~\eqref{swpvs} by expressing the logarithm of the quotient as the 
difference of the logarithms of the numerator and denominator as
\begin{align}
\label{swpvm}
v(p_1,p_2)&=
\frac{g}{2 p_1 p_2}  \left[
\log\left((p_1+p_2)^2+m_\pi^2\right)-
\log\left((p_1-p_2)^2+m_\pi^2\right)
\right ]~.
\end{align}
For complex values of $p_i$ it is not true that in general Eqs.~\eqref{swpvs} and 
\eqref{swpvm} agree, because we cannot  always 
combine the difference of the two logs in Eq.~\eqref{swpvm} as the log of the ratio of the arguments,
 because the latter could have imaginary parts of different sign. 

 However, we argue that we can legitimately use Eq.~\eqref{swpvm}, 
instead of Eq.~\eqref{swpvs}, for the analytical extrapolation into complex values of the arguments of the partial-wave potential. 
The argument is the following. When solving the LS equation in the physical case the same result
 is obtained by employing Eqs.~\eqref{swpvs} and 
\eqref{swpvm}, as they both agree on the physical axis, and also around it. Then, to go into the 
complex plane of the moduli of the momenta we can proceed by analytical continuation
of  the projected potential as given in Eq.~\eqref{swpvm},  
because it has simpler and more convenient analytical properties. 
This clearly exemplifies that although the physical limit is the same 
one can consider different analytical extrapolations in the complex plane. 

The cuts in the complex $p_2$ plane
 stemming from Eq.~\eqref{swpvm} are
\begin{align}
p_2&=p_1\pm i\sqrt{m_\pi^2+x^2}~,\nn\\
p_2&=-p_1\pm i\sqrt{m_\pi^2+x^2}~,~x\in \mathbb{R}~.
\label{vcuts}
\end{align}
 These cuts are vertical lines  that extend up to $\pm i \infty$ 
starting from the branch points that are given by the previous equation with $x=0$. 
Therefore, these cuts are much simpler to characterize and handle
 than the curved ones of Eq.~\eqref{finitecuts}. E.g. they keep their form if we 
 substituted $m_\pi^2$ by $(n m_\pi)^2$.

Now, let us take under consideration the function defined 
by the difference between the potential functions of 
Eqs.~\eqref{swpvm} and \eqref{swpvs}. 
This function vanishes along the real axis of the complex $p_2$ plane, 
and then it should be zero in all its analytical domain containing the real axis. 
However, this analytical domain is not the whole complex $p_2$ plane because 
it is cut in such a way that there are two strips, delimited by the curved cuts 
in Eq.~\eqref{swpvs} and the vertical ones of Eq.~\eqref{swpvm}, 
so that they are isolated from the real axis. This is explicitly shown 
in the right panel of Fig.~\ref{fig:vcut} for the same values of the parameters 
and $p_1$ as in the left panel. 
 In these strips the total partial wave that would result by solving the LS equation would be different 
regarding which form for the projected potential, 
either Eq.~\eqref{swpvm} or Eq.~\eqref{swpvs}, is used. 

We do not need to discuss the appropriate form for the analytical continuation 
of  the partial wave projection of an 
amplitude involving $\sin \theta$ because the latter will necessarily appear raised to an even power when 
taking into account also the angular dependence from the spherical harmonics. The point is that 
such dependence on powers of $\sin \theta$ always stems from powers of 
$q_x\pm i q_y=|q|\sin\theta e^{\pm i \phi}$, with 
$q_x=|q|\sin\theta \cos \phi$, $q_y=|q|\sin\theta \sin\phi$ 
 and $\phi$ is the azimuthal angle. 
 Now,  the complex dependence on the azimuthal angle stemming 
from powers of $\sin\theta e^{\pm i\phi}$ is compensated 
by the one of the same type originating from the spherical harmonics 
(the partial-wave projection of a potential is real for real three-momenta), so that one always ends with 
even powers of $\sin\theta$ in the formula for a partial-wave projection. 
For  explicit examples the reader can consult chapter 5, $\S$6.3 of Ref.~\cite{martin.180804.1} for 
the partial-wave analysis of $\pi N$ scattering or 
Ref.~\cite{kaiser.080516.1} for that of $NN$ scattering. 

 A partial-wave projection with higher powers of $\cos\theta$ in the numerator 
  can be treated by iteration so that it can be reduced to a form corresponding 
 to Eq.~\eqref{swpv}. For instance,
\begin{align}
\int_{-1}^{+1} \frac{z^2 dz}{z-\xi}&=\int_{-1}^{+1} z dz+\xi\int_{-1}^{+1}\frac{z dz}{z-\xi}
=\int_{-1}^{+1}dz z+\xi \int_{-1}^{+1} dz+\xi^2\int_{-1}^{+1}\frac{dz}{z-\xi}~,
\label{210615.6}
\end{align}
etc. Here we have introduced the symbol $\xi$ which means
\begin{align}
\label{180819.2}
\xi&=\frac{p_1^2+p_2^2+m_\pi^2}{2p_1p_2}~.
\end{align}

Now, for the spectral function representation in Eq.~\eqref{eq.sdv} 
we have for its partial-wave projection the same cuts as in Eq.~\eqref{3.4tex}. 
The lower limit of the integration 
in the spectral decomposition gives rise to the branch point singularities and 
the continuous rise in $\mu$ just provides new contributions to the discontinuity along 
the same vertical infinite cuts, as it can be easily worked out. 
For instance, its $S$-wave projection is
\begin{align}
\label{180923.1}
v(p_1,p_2)&=\frac{1}{2p_1p_2}\int_{m_\pi^2}^\infty d\mu^2 \eta(\mu^2)
\left[
\log\left((p_1+p_2)^2+\mu^2\right)-
\log\left((p_1-p_2)^2+\mu^2\right)
\right ]~,
\end{align}
 and the discontinuity between $p_2=-p_1\pm \vep+i\sqrt{m_\pi^2+x^2}$, stemming entirely from the 
 first log, is
 \begin{align}
\frac{\pi}{p_1p_2}\int_{m_\pi^2}^{m_\pi^2+x^2} d\mu^2 \eta(\mu^2)~.
 \end{align}
 Notice the upper limit in the last integration because it is required that 
  $(p_1+p_2)^2+\mu^2=-(m_\pi^2+x^2)+\mu^2<0$.

\section{DCs in the PWAs}
\label{sec:aphfs}
\def\theequation{\arabic{section}.\arabic{equation}}
\setcounter{equation}{0}   

 We want to study the half-off-shell $T$-matrix $t(k,k';{k'}^2/m)$  for complex arguments.  
The starting point for discussing the analytical extrapolation of 
the LS equation in partial waves, is taking  real and positive values for $k$ and $k'$. 
 The LS in partial waves for a complex energy $z$ reads, cf. Eq.~\eqref{180804.5}, 
\begin{align}
\label{180805.1}
t(k,k';z)&=v(k,k')+
\frac{m}{2\pi^2}
\int_0^\infty \frac{dp_1 p_1^2}{p_1^2-m z} v(k,p_1) t(p_1,k';z)~,
\end{align}
and the integration contour ${\cal C}$ consists of the positive real semi axis, that we can take as its original extension. 
Of course, we should give a vanishing imaginary 
part to $k'$ when $z={k'}^2/m$,  so as to make meaningful the integral because of the RHC, cf. Eq.~\eqref{180804.14}. 
In the following,  we equate 
the reduced mass $\mu$ to $m/2$.\footnote{This change is motivated because 
we will exemplify our theory below with examples corresponding to $NN$ scattering in partial waves, 
with $m$ the nucleon mass.} 
Nonetheless, we can undo this step by 
substituting  $m$ in the general expression by $2\mu$.

 \subsection{Integration contour for the LS equation in partial waves}
 \label{sec.180805.1}
 
 Now, let us consider complex values for $k$ and $k'$ and proceed with the analytical extrapolation
 of the LS equation in those variables by deforming the original integration contour, such that,
 in general terms, it reads for the half-off-shell $T$ matrix 
\begin{align}
t(k,k';\frac{{k'}^2}{m})&=v(k,k')+
\frac{m}{2\pi^2}
\int_{{\cal C}} \frac{dp_1 p_1^2}{p_1^2-{k'}^2} v(k,p_1) t(p_1,k';\frac{{k'}^2}{m})~.
\label{200315.1}
\end{align}
Let us discuss how to fix the integration contour ${\cal C}$. 
 This problem is by itself of interest, e.g. in order to determine the residues of 
 on-shell PWAs at resonance, virtual and bound-state poles, which determine the couplings of these states 
to the different waves. It is also a key point for our later developments, in particular,  
this will allow us to calculate the discontinuity of an on-shell PWA along the LHC. 
When considering a half-off-shell PWA we will often suppress the energy argument because it is 
 then assumed that the energy is given in terms of the second three-momentum argument, 
 so that $t(k,k')\equiv t(k,k';{k'}^2/m)$. 

Taking into account that the cuts of $v(k,p_1)$ are located at 
\begin{align}
p_1&=+k\pm i\sqrt{m_\pi^2+x^2}~,\nn\\
p_1&=-k\pm i \sqrt{m_\pi^2+x^2}~,
\label{200315.2}
\end{align}
according to Eq.~\eqref{vcuts}, it follows then that 
for complex $k$ with $|\Ima k|<m_\pi$ the cuts in Eq.~\eqref{200315.2} 
do not cross the original integration contour and hence do not induce any deformation in the contour.  
Contrarily, for $|\Ima k|>m_\pi$ one needs to deform the integration contour by circumventing 
the vertical line from $| \Rea k|$ up to $| \Rea k|\pm i(|\Ima k|-m_\pi)$, namely,
\begin{align}
\label{261215.1}
p_1=|\Rea k|\pm i\lambda (|\Ima k|-m_\pi)~,~|\Ima k|>m_\pi~,\lambda\in [0,1]~.
\end{align}
 The  sign $\pm$ in the previous equation is determined  according to  the possibilities:
\begin{align}
|\Ima k|&>m_\pi\nn\\
\Rea k&>0~,~  \pm = +\hbox{sign}(\Ima k)\nn\\  
\Rea k&<0~,~  \pm = -\hbox{sign}(\Ima k) 
\label{210315.1} 
\end{align}

\begin{figure}
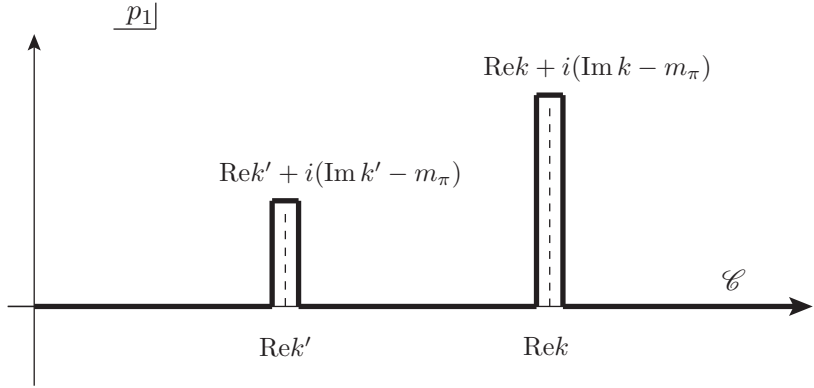

\begin{center}
\begin{axopicture}(300,150)(0,-20)
\Line(40,105)(56,105)
\Line(56,105)(56,115)
\Text(45,110)[l]{$p_1$}
\Text(270,10)[l]{${\cal C}$}
\Line[dash](105,0)(105,35)
\Line[dash](205,0)(205,75)
\Text(95,-15)[l]{{\small $\Rea k'$}}
\Text(195,-15)[l]{{\small $\Rea k$}}
\Text(80,50)[l]{{\small $\Rea k'+i(\Ima k'-m_\pi)$}}
\Text(180,90)[l]{{\small $\Rea k+i(\Ima k-m_\pi)$}}
\Line(0,0)(300,0)
\LongArrow(10,-30)(10,100)
\Line(0,0)(10,0)
\SetWidth{2}
\Line(10,0)(100,0)
\Line(100,0)(100,40)
\Line(100,40)(110,40)
\Line(110,40)(110,0)
\Line(110,0)(200,0)
\Line(200,0)(200,80)
\Line(200,80)(210,80)
\Line(210,80)(210,0)
\Line[arrow,arrowpos=1](210,0)(300,0)
\end{axopicture}\\
\end{center}
\caption{{\small Deformed integration contour for the LS equation in the complex $p_1$ plane. 
For the figure it is assumed that $k$ and $k'$ belong to the first quadrant of their corresponding complex planes 
and that their imaginary parts are larger than $m_\pi$.}}
\label{fig:210315.1}
\end{figure}

Regarding the deformation in the contour induced by $t(p_1,k')$ in Eq.~\eqref{200315.1} 
we proceed by iterating the LS equation in the Neumann series,
\begin{align}
t(k,k')&=v(k,k')+\frac{m}{2\pi^2}\int_{{\cal C}_1}\frac{dp_1 p_1^2}{p_1^2-{k'}^2} 
v(k,p_1)v(p_1,k')\nn\\
&+\left(\frac{m}{2\pi^2}\right)^2\int_{{\cal C}_2}\frac{dp_1 p_1^2}{p_1^2-{k'}^2} v(k,p_1)
\int_{{\cal C}_3}\frac{dp_2 p_2^2}{p_2^2-{k'}^2}v(p_1,p_2)v(p_2,k')\nn\\
&+\left(\frac{m}{2\pi^2}\right)^3\int_{{\cal C}_4}\frac{dp_1 p_1^2}{p_1^2-{k'}^2}
v(k,p_1)\int_{{\cal C}_5}\frac{dp_2 p_2^2}{p_2^2-{k'}^2}v(p_1,p_2)
\int_{{\cal C}_6}\frac{dp_3 p_3^2}{p_3^2-{k'}^2}v(p_2,p_3)v(p_3,k')
+\ldots 
\label{210315.2}
\end{align}
The induced deformations in the integration contour can be easily realized by reading 
the terms in the previous equation from left to right (the order in which the integration limits are 
fixed for every integral in a multiple integral). In this way, we have for 
$(n+1)m_\pi>|\Ima k|> m_\pi$ 
 a set of vertical segments in the 
deformed contours for each $p_i$, $n\geq i \geq 1$, of decreasing extent $|\Ima p_{i-1}|-m_\pi$ 
(according to the rules Eq.~\eqref{210315.1} applied to $p_i$)
 until $(n+1)m_\pi>|\Ima k|$. 
 In addition the rightmost $v$ factor in Eq.~\eqref{210315.2},
 $v(p_m,k')$ for every $m$-times iterated contribution, induces for $|\Ima k'|>m_\pi$ 
an additional vertical segment  in the last integration variable $p_m$, 
defined by $|\Rea k'|\pm i\lambda (|\Ima k'|-m_\pi)$, $\lambda\in [0,1]$,  
with the signs according to  Eq.~\eqref{210315.1} applied to $k'$.
 It is important to stress that the number of added vertical contours is finite for a given 
 $|\Ima k|<\infty$. Therefore, 
 all the finite set of vertical additions in the original contour
 due to $v(k,p_1) t(p_1,k';{k'}^2/m)$ coalesces in one 
at $|\Rea k|\pm i(|\Ima k|-m_\pi)$ and another at $|\Rea k'|\pm i(|\Ima k'|-m_\pi)$,
according to the rules  in Eq.~\eqref{210315.1} applied to $k$ and $k'$. 
The effect in the DCs from the possible vanishing of the energy denominator, $1/(p_m^2-{k'}^2)$, 
is already taking 
into account by the added vertical segments already introduced in ${\cal C}$ (see case ii) in the next section). 
This discussion fixes the contour ${\cal C}$ of integration in Eq.~\eqref{200315.1}. 
 For illustration we plot it in Fig.~\ref{fig:210315.1} for $k$ and $k'$ in the first quadrant 
 of their respective complex planes.

 \subsection{Analytical properties of the half-off-shell PWA}
 \label{sec.180805.2}

 In virtue of Eq.~\eqref{vcuts} the independent term in Eq.~\eqref{200315.1}, $v(k,k')$, establishes the cuts  in the 
 complex $k$ plane at 
\begin{align}
k&=(\pm) k'\pm i\sqrt{m_\pi^2+x^2}~,
\label{210315.3}
\end{align}
where the $\pm$ sign between brackets is uncorrelated with the other $\pm$ symbol.
 
Let us show now that the half-off-shell PWA $t(k,k';{k'}^2/m)$  is a regular function 
in the cut complex $k$ plane with cuts located according to Eq.~\eqref{210315.3}. 
 No more cuts than those in Eq.~\eqref{210315.3} arise by iterating the LS equation,  Eq.~\eqref{210315.2}. 
We could have the following four cases:

\begin{figure}[ht]
\begin{center}
\begin{tabular}{ll}
\begin{axopicture}(200,100)(20,0)
\LongArrow(10,0)(10,100) 
\LongArrow(0,40)(200,40) 
\Line[dash](90,60)(90,100) 
\Line[dash](90,30)(90,0) 
\Line[dash](105,40)(105,70) 
\Text(180,50)[l]{{\small ${\cal C}$}} 
\GCirc(90,45){1}{.1} 
\Text(88,52)[l]{{\tiny $k'$}} 
\Text(60,60)[l]{{\tiny $k'+im_\pi$}} 
\Text(60,30)[l]{{\tiny $k'-im_\pi$}} 
\Text(110,30)[l]{{\tiny $\Rea k$}} 
\Text(105,80)[l]{{\tiny $k-im_\pi$}}
\Line[dash](105,0)(105,40) 
\SetWidth{2}
\Line(10,40)(100,40) 
\Line(100,40)(100,75)
\Line(100,75)(110,75)
\Line(110,75)(110,40)
\Line[arrow,arrowpos=1](110,40)(200,40)
\Text(100,-10)[l]{{\scriptsize a)}}
\end{axopicture}
&
\begin{axopicture}(200,100)(-20,0)
\LongArrow(10,0)(10,100) 
\LongArrow(0,40)(200,40) 
\Line[dash](90,45)(90,100) 
\Line[dash](90,15)(90,0) 
\Line[dash](105,40)(105,70) 
\Text(180,50)[l]{{\small ${\cal C}$}} 
\GCirc(90,30){1}{.1} 
\Text(88,35)[l]{{\tiny $k'$}} 
\Text(60,50)[l]{{\tiny $k'+im_\pi$}} 
\Text(60,15)[l]{{\tiny $k'-im_\pi$}} 
\Text(110,30)[l]{{\tiny $\Rea k$}} 
\Text(105,80)[l]{{\tiny $k-im_\pi$}}
\Line[dash](105,0)(105,40) 
\SetWidth{2}
\Line(10,40)(100,40) 
\Line(100,40)(100,75)
\Line(100,75)(110,75)
\Line(110,75)(110,40)
\Line[arrow,arrowpos=1](110,40)(200,40)
\Text(100,-10)[l]{{\scriptsize b)}}
\end{axopicture}
\\
\\
\\
\begin{axopicture}(200,100)(20,0)
\LongArrow(10,0)(10,100) 
\LongArrow(0,40)(200,40) 
\Line[dash](90,60)(90,100) 
\Line[dash](90,30)(90,0) 
\Line[dash](105,15)(105,100) 
\Text(180,50)[l]{{\small ${\cal C}$}} 
\GCirc(90,45){1}{.1} 
\Text(88,52)[l]{{\tiny $k'$}} 
\Text(60,60)[l]{{\tiny $k'+im_\pi$}} 
\Text(60,30)[l]{{\tiny $k'-im_\pi$}} 
\Text(110,50)[l]{{\tiny $\Rea k$}} 
\Text(105,5)[l]{{\tiny $k+im_\pi$}} 
\SetWidth{2}
\Line(10,40)(100,40) 
\Line(100,40)(100,10)
\Line(100,10)(110,10)
\Line(110,10)(110,40)
\Line[arrow,arrowpos=1](110,40)(200,40)
\Text(100,-10)[l]{{\scriptsize c)}}
\end{axopicture}
&
\begin{axopicture}(200,100)(-20,0)
\LongArrow(10,0)(10,100) 
\LongArrow(0,40)(200,40) 
\Line[dash](90,45)(90,100) 
\Line[dash](90,15)(90,0) 
\Line[dash](105,15)(105,100) 
\Text(180,50)[l]{{\small ${\cal C}$}} 
\GCirc(90,30){1}{.1} 
\Text(88,35)[l]{{\tiny $k'$}} 
\Text(60,50)[l]{{\tiny $k'+im_\pi$}} 
\Text(60,15)[l]{{\tiny $k'-im_\pi$}} 
\Text(110,50)[l]{{\tiny $\Rea k$}} 
\Text(105,5)[l]{{\tiny $k+im_\pi$}} 
\SetWidth{2}
\Line(10,40)(100,40) 
\Line(100,40)(100,10)
\Line(100,10)(110,10)
\Line(110,10)(110,40)
\Line[arrow,arrowpos=1](110,40)(200,40)
\Text(100,-10)[l]{{\scriptsize d)}}
\end{axopicture}
\end{tabular}
\end{center}
\caption[pilf]{{\small Interplay between the added vertical contours and cuts associated
 with external variables for case ii) that corresponds to $|\Ima k|>m_\pi$ and $|\Ima k'|<m_\pi$. 
 For definiteness in the figure we assume that $\Rea k$ and $\Rea k'$ are positive.
The different possibilities for the signs of  $\Ima k$ and $\Ima k'$ are explicitly plotted.
}
\label{fig:261215.1}}
\end{figure}

\vskip 5pt
\noindent
{\bf  i)} Both $|\Ima k|$ and $|\Ima k'|$ are less than $m_\pi$, so that the 
integration contour is the original one. No dynamical cut crosses the integration 
contour and therefore the LS equation is regular in $k$. 
 In other terms, after partial-wave projecting the arguments of the  $v(p_i,p_{i+1})$ 
 in Eqs.~\eqref{swpvm} and \eqref{180923.1}, with $p_0=k$ and $p_{n+1}=k'$, 
the arguments in the logs have always positive real part and there is no discontinuity.

\vskip 5pt
\noindent
{\bf ii)} $|\Ima k|>m_\pi$ and $m_\pi>|\Ima k'|>0$: 
 The integration contour is the original one plus one vertical addition at $|\Rea k|$ according to the 
rules in Eq.~\eqref{261215.1} and \eqref{210315.1}, 
 which corresponds to the situation illustrated in Fig.~\ref{fig:261215.1} for $\Rea k>0$. 
 Then, the energy denominator does not vanish along the original integration contour, 
but it would vanish along the vertical addition if $\pm k'$ places along it 
(this is analyzed from the full paragraph after Eq.~\eqref{281215.1} until the end of the present case). 
In the following the $n$-times iteration of the potential $v(k,k')$ in the Neumann series 
is denoted by $t_n(k,k';\frac{{k'}^2}{m})$ and correspond to  
\begin{align}
\label{261215.2}
t_n(k,k';\frac{{k'}^2}{m})=\left(\frac{m}{2\pi^2}\right)^n \int_{{\cal C}_1}\frac{dp_1 p_1^2}{p_1^2-{k'}^2}
v(k,p_1)\int_{{\cal C}_2}\frac{dp_2 p_2^2}{p_2^2-{k'}^2}v(p_1,p_2)
\int_{{\cal C}_3}\ldots\int_{{\cal C}_n}\frac{dp_n p_n^2}{p_n^2-{k'}^2}v(p_{n-1},p_n)v(p_n,k')~,
\end{align}
where the different  integration contours ${\cal C}_i$ are fixed as discussed after Eq.~\eqref{210315.2}. 

 We first show that for $n=1$ there are no other cuts 
in the variable $k$ beyond those given in Eq.~\eqref{210315.3}. 
 The factor $v(k,p_1)$ induces the added vertical contour at $|\Rea k|$. 
 In the integral of $p_1$ the integration contour goes around the vertical addition
 but the variable $p_1$ has  cuts at $(\pm k')\pm i\sqrt{m_\pi^2+x^2}$ (the relevant ones indicated by dashed lines in Fig.~\ref{fig:261215.1}) because the next factor 
$v(p_1,k')$. When $k$ varies the added integration contour moves correspondingly
  so that only when it  overlaps with one of  these vertical cuts in $p_1$ at $|\Rea k'|$ 
 we have discontinuity in $k$. 
If the added contour at $|\Rea k|$ extends from 0 up to $|\Ima k|-m_\pi$, 
top panels in Fig.~\ref{fig:261215.1}, the condition for the referred
 overlapping implies that $\sgn(\Rea k') \Ima k'+ m_\pi<|\Ima k|- m_\pi$ (apart from the obvious one 
 $|\Rea k|=|\Rea k'|$). Notice that since 
$|\Ima k'|<m_\pi$ the cut  $|\Ima k'|-\sqrt{m_\pi^2+x^2}$ involves always negative values and does not overlap with the added positive vertical contour.    
 Then, $|\Ima k|>\sgn(\Rea k')\Ima k'+2m_\pi$ and this set of possible values for $k$
 belongs to the cuts indicated in Eq.~\eqref{210315.3}. 
If the added contour extends from $-|\Ima k|+m_\pi$ up to 0, we proceed analogously and 
the condition now is that $\sgn(\Rea k') \Ima k'-m_\pi>-|\Ima k|+m_\pi$, so 
that $|\Ima k|>-\sgn(\Rea k')\Ima k'+2m_\pi$, and again this set of values for $k$ belongs to the cuts in Eq.~\eqref{210315.3}. 
All the different possibilities are plotted in Fig.~\ref{fig:261215.1}, where the added contour at $|\Rea k|$ 
follows the rules in Eqs.~\eqref{261215.1} and \eqref{210315.1}.
 There is no discontinuity in $k$ stemming from the original integration contour because 
it does not intersect the cuts  $(\pm k')\pm i\sqrt{m_\pi^2+x^2}$.

For the $n$-times iterated amplitude $t_{n}(k,k';{k'}^2/{m})$ we write it as 
\begin{align}
\label{281215.1}
t_{n}(k,k';\frac{{k'}^2}{m})&=\frac{m}{2\pi^2} \int_{{\cal C}} \frac{dp_n p_n^2}{p_n^2-{k'}^2}
v(k,p_n)t_{n-1}(p_n,k';\frac{{k'}^2}{m})~,
\end{align}
and proceed by induction. Therefore, the integration contour is the one already 
fixed at the beginning of this case ii). Namely, it consists 
of the original extension plus the vertical addition at $|\Rea k|$ according to the standard rules in 
Eqs.~\eqref{261215.1} and \eqref{210315.1}. 
Then,  when $p_n$ goes around the vertical integration contour at $|\Rea k|$ 
there is a discontinuity in $k$ when this addition intersects the cuts in the next factor $t(p_n,k';{k'}^2/m)$,  
in the same way as discussed for $n=1$. 
As a result,  the cuts in $k$  are contained in  Eq.~\eqref{210315.3}. 

Another source of discontinuity occurs when the denominator $p^2-{k'}^2=(p-k')(p+k')$ 
vanishes along the vertical segment  $|\Rea k|\pm i\lambda (|\Ima k| -m_\pi)$,
with the sign  $\pm$ following the 
rules in Eq.~\eqref{210315.1}. 
 In such a case there is a discontinuity between $|\Rea k|$ being larger or smaller than $|\Rea k'|$. 
 As it is clear pictorially, e.g. from the left top  panel of Fig.~\ref{fig:261215.1},  
 we again have that the cut stemming from this discontinuity happens for $k$ 
 belonging to the set of values in Eq.~\eqref{210315.3}.

\begin{figure}
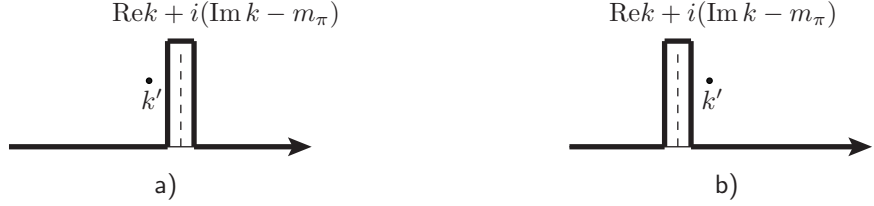

\begin{center}
\begin{tabular}{lr}
\begin{axopicture}(150,50)(20,-10)
\Text(50,18)[l]{{\small $k'$}}
\GCirc(53,25){1}{.1}
\Line(0,0)(110,0)
\Line[dash](65,0)(65,35)
\Text(55,-15)[l]{{\small a)}}
\Text(39,50)[l]{{\small $\Rea k+i(\Ima k-m_\pi)$}}
\SetWidth{2}
\Line(0,0)(60,0)
\Line(60,0)(60,40)
\Line(60,40)(70,40)
\Line(70,40)(70,0)
\Line[arrow,arrowpos=1](70,0)(110,0)
\end{axopicture}
&
\begin{axopicture}(150,50)(-30,-10)
\Text(50,18)[l]{{\small $k'$}}
\GCirc(53,25){1}{.1}
\Line(0,0)(110,0)
\Line[dash](41,0)(41,35)
\Text(55,-15)[l]{{\small b)}}
\Text(15,50)[l]{{\small $\Rea k+i(\Ima k-m_\pi)$}}
\SetWidth{2}
\Line(0,0)(36,0)
\Line(36,0)(36,40)
\Line(36,40)(46,40)
\Line(46,40)(46,0)
\Line[arrow,arrowpos=1](46,0)(110,0)
\end{axopicture}
\end{tabular}
\caption{{\small Diagrams for the calculation of the discontinuity when the added vertical contour at  
$\Rea k$ crosses $k'$. In this plot we are taking that $k$ and $k'$ are in the first quadrants of their 
complex planes.
\label{fig.180806.1}}}
\end{center}
\end{figure}

The explicit calculation of this discontinuity also exemplifies the techniques we are developing. 
 For definiteness assume that $k$ and $k'$ are both in the first quadrant of their complex planes. 
We also introduce a shorter notation for the real and imaginary parts of $k$ and $k'$ so that
\begin{align}
\label{180806.2}
k_r&=\Rea k~,~k_i=\Ima k_i~,\\
k'_r&=\Rea k'~,~k'_i=\Ima k'_i~.\nn
\end{align}
 When $k'$ is to the left of the added vertical contour, left panel in Fig.~\ref{fig.180806.1}, 
 $k_r=k'_r+\vep$ (as always we understand that $\vep\to 0^+$). 
 We write the integration variable as $p_1=k_r+i\nu_1\pm \delta$, $\nu_1\in [0,k_i-m_\pi]$, 
 with the plus sign for the vertical segment to the right of the dashed line and 
 the minus for the segment to the left of the same line and $\vep>\delta\to 0^+$ $(\delta/\vep\to 0)$. 
 In this form we have the following expression 
  for the contribution ${\cal I}_a$ stemming from the vertical contour in Fig.~\ref{fig.180806.1}a),
\begin{align}
\label{180806.3}
&{\cal I}_a=
\frac{m}{2\pi^2}\int_0^{k_i-m_\pi}d\nu_1\frac{(k_r+i\nu_1)^2\left[v(k,k_r-\delta+i\nu_1)
-v(k,k_r+\delta+i\nu_1)\right]t(k'_r+\vep+i\nu_1,k')}
{(\nu_1-k'_i-i\vep)(2k'_r+i(\nu_1+k'_i))}\nn\\
&=\frac{m}{2\pi^2}\dashint_0^{k_i-m_\pi}d\nu_1\frac{(k_r+i\nu_1)^2\left[v(k,k_r-\delta+i\nu_1)
-v(k,k_r+\delta+i\nu_1)\right]t(k'_r+\vep+i\nu_1,k')}
{(\nu_1-k'_i)(2k'_r+i(\nu_1+k'_i))}\nn\\
&+i\frac{mk'}{4\pi}\left[v(k,k'+\vep-\delta)-v(k,k'+\vep+\delta)\right]t(k'+\vep,k')~.
\end{align}

When $k$ is to the left of $k'$ we have to change $\vep\to -\vep$ in the first line of Eq.~\eqref{180806.3} so 
that we have for the contribution ${\cal I}_b$ corresponding to Fig.~\ref{fig.180806.1}b)
\begin{align}
\label{180806.3b}
{\cal I}_b&=\frac{m}{2\pi^2}\dashint_0^{k_i-m_\pi}d\nu_1\frac{(k_r+i\nu_1)^2\left[v(k,k_r-\delta+i\nu_1)
-v(k,k_r+\delta+i\nu_1)\right]t(k'_r-\vep+i\nu_1,k')}
{(\nu_1-k'_i)(2k'_r+i(\nu_1+k'_i))}\nn\\
&-i\frac{mk'}{4\pi}\left[v(k,k'-\vep-\delta)-v(k,k'-\vep+\delta)\right]t(k'-\vep,k')~.
\end{align}
Notice that the discontinuity $v(k,k'\pm\vep-\delta)-v(k,k'\pm\vep+\delta)$ is continuous in 
$\vep\to 0^+$, what matters is the difference between the values of the potential just 
to the left and right of the cut. The discontinuity is therefore
\begin{align}
\label{180806.4}
{\cal I}_a-{\cal I}_b=&
\frac{m}{2\pi^2}\dashint_0^{k_i-m_\pi}\!\!d\nu_1 \frac{(k_r+i\nu_1)^2\left[v(k,k_r-\delta+i\nu_1)
-v(k,k_r+\delta+i\nu_1)\right]}{(\nu_1-k'_i)(2k'_r+i(\nu_1+k'_i))}\nn\\
\times& \left[t(k'_r+\vep+i\nu_1,k')-t(k'_r-\vep+i\nu_1,k')\right]\nn\\
+& i\frac{mk'}{4\pi}\left[v(k+\delta,k')-v(k-\delta,k')\right]\left[t(k'+\vep,k')+t(k'-\vep,k')\right]~.
\end{align}
From this expression we can conclude explicitly  that $k$ should belong 
to the cut determined by the values in Eq.~\eqref{210315.3} (a conclusion already deduced from more general
 arguments). This is clear from the difference $v(k+\delta,k')-v(k-\delta,k')$ in the last term. 
One can also conclude the same from the non-vanishing of the discontinuity    
$t(k'_r+\vep+i\nu_1,k')-t(k'_r-\vep+i\nu_1,k')$ in the integrand of the first line because  
it requires that $k+i(\nu_1-k_i)=k'+i\sqrt{m_\pi^2+x^2}\to k=k'+i(k_i-\nu_1+\sqrt{m_\pi^2+x^2}$), with 
$k_i-\nu_1\geq m_\pi$. 

One can also easily conclude for this case with $|\Ima k'|<m_\pi$  that 
\begin{align}
\label{180806.5}
t(k'+\vep,k',k')-t(k'-\vep,k')=v(k'+\vep,k')-v(k'-\vep,k')~,
\end{align}
because the integration contour is not modified and then the integral term in the LS equation is 
regular in $k'$ as first argument. This result simplifies the last term of Eq.~\eqref{180806.4}.

\vskip5pt
\noindent
{\bf iii)} $|\Ima k|<m_\pi$ and $|\Ima k'|>m_\pi$: In order to follow an argument in close analogy 
with ii) we make use of Eq.~\eqref{180804.10} and write $t(k,k';{k'}^2/m)=t(k',k;{k'}^2/m)$. 
In this way we can apply the same argument as in ii) by just exchanging $k\leftrightarrow k'$.
 As a result,  for the $n$-times iterated amplitude $t_n(k,k';{k'}^2/m)$,
\begin{align}
\label{101216.1}
t_{n}(k',k;\frac{{k'}^2}{m})
&=\frac{m}{2\pi^2} \int_{{\cal C}} \frac{dp_n p_n^2}{p_n^2-{k'}^2}
v(k',p_n)t_{n-1}(p_n,k;\frac{{k'}^2}{m})~,
\end{align}
 there is a cut in $k$ when the vertical addition at $|\Rea k'|$ given by $p_n=|\Rea k'|\pm i(|\Ima k'|-m_\pi)$, 
 with the signs fixed as in Eq.~\eqref{210315.1} applied to $k'$,  overlaps with the 
cuts in $p_n$ located at $(\pm)k\pm i\sqrt{m_\pi^2+x^2}$. 
 When $k$ varies  these cuts move correspondingly so that  
 only when one of them overlaps with the added integration contour we have a discontinuity in $k$. 
This implies that the resulting DC extent is contained in the values of Eq.~\eqref{210315.3}.
 
  To show this last statement in more detail, we can just 
 follow the same arguments as in ii) after Eq.~\eqref{261215.2} with the roles of $k$ and $k'$ exchanged, 
as well as in Fig.~\ref{fig:261215.1}: 
 If the added vertical contour at $|\Rea k'|$ extends from 0 up to $|\Ima k'|-m_\pi$, 
 top panels in Fig.~\ref{fig:261215.1}, 
 the condition for the referred overlapping implies that $\sgn(\Rea k) \Ima k+ m_\pi<|\Ima k'|- m_\pi$ 
 (apart from the obvious one  $|\Rea k|=|\Rea k'|$). 
Notice that since $|\Ima k|<m_\pi$ the cut  $|\Ima k|-\sqrt{m_\pi^2+x^2}$ involves always negative values and 
does not overlap with the added positive vertical contour.    
 Then, $\sgn (\Rea k)\Ima k<|\Ima k'|-2m_\pi$ and  all this set of possible values for $k$ belongs to the cuts 
indicated in Eq.~\eqref{210315.3}. 
If the vertical addition extends from $-|\Ima k'|+m_\pi$ up to 0, we proceed analogously and 
the condition now is that $\sgn(\Rea k) \Ima k-m_\pi>-|\Ima k'|+m_\pi$ so 
that $\sgn (\Rea k)\Ima k>-|\Ima k'|+2m_\pi$ and again this set of values for $k$ belongs to the cuts in Eq.~\eqref{210315.3}. 
All the different possibilities are plotted in Fig.~\ref{fig:261215.1}, where the added contour at $|\Rea k'|$ 
follows the rules in Eqs.~\eqref{261215.1} and \eqref{210315.1} applied to $k'$.
 There is no discontinuity in $k$ stemming from the original integration contour because 
 it does not intersect the cuts at $(\pm k)\pm i\sqrt{m_\pi^2+x^2}$.

The only difference with respect to the case ii) is that now the added vertical contour at $|\Rea k'|$ does not 
intersect ever $k'$ because its vertical extension is smaller than $|\Ima k'|$ by an amount of $m_\pi$. 

\vskip 5pt

\begin{figure}[ht]
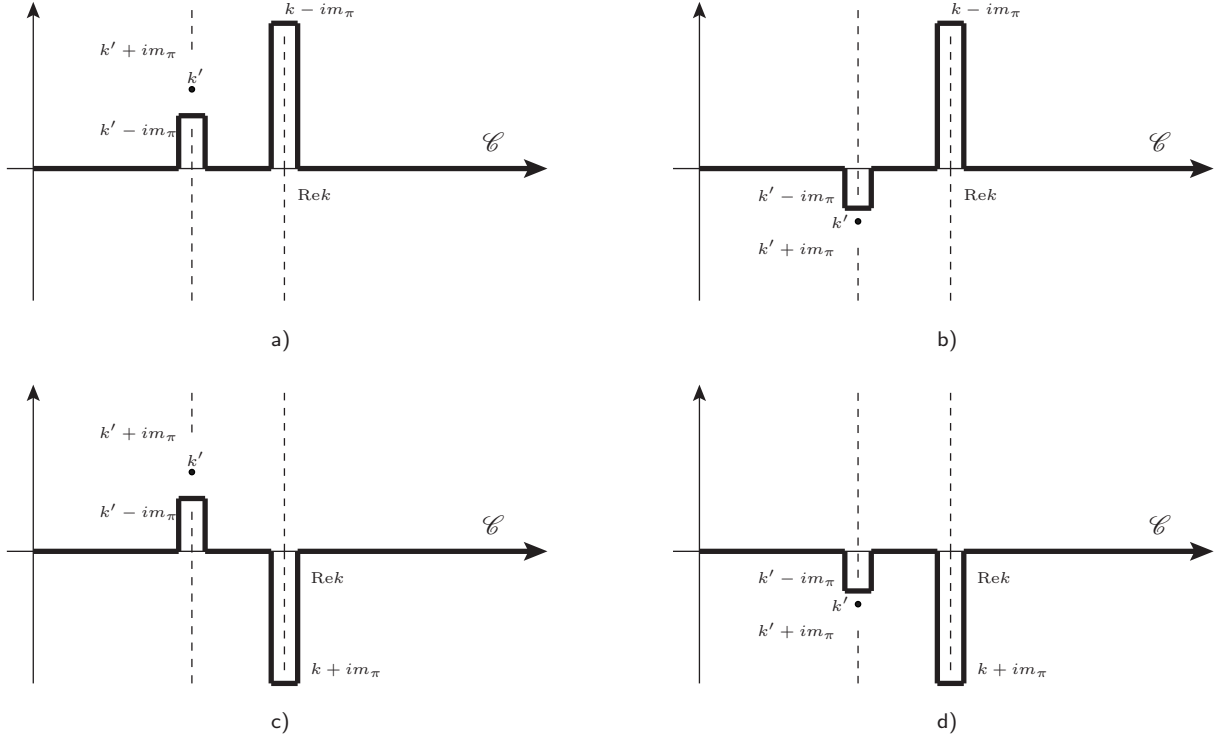

\begin{center}
\begin{tabular}{ll}
\begin{axopicture}(200,100)(20,0)
\LongArrow(10,-10)(10,100) 
\LongArrow(0,40)(200,40) 
\Line[dash](70,85)(70,100) 
\Line[dash](70,55)(70,-10) 
\Line[dash](105,40)(105,90) 
\Text(180,50)[l]{{\small ${\cal C}$}} 
\GCirc(70,70){1}{.1} 
\Text(68,75)[l]{{\tiny $k'$}} 
\Text(35,85)[l]{{\tiny $k'+im_\pi$}} 
\Text(35,55)[l]{{\tiny $k'-im_\pi$}} 
\Text(110,30)[l]{{\tiny $\Rea k$}} 
\Text(105,100)[l]{{\tiny $k-im_\pi$}}
\Line[dash](105,-10)(105,40) 
\SetWidth{2}
\Line(10,40)(65,40)
\Line(65,40)(65,60)
\Line(65,60)(75,60)
\Line(75,60)(75,40)
\Line(75,40)(100,40) 
\Line(100,40)(100,95)
\Line(100,95)(110,95)
\Line(110,95)(110,40)
\Line[arrow,arrowpos=1](110,40)(200,40)
\Text(100,-25)[l]{{\scriptsize a)}}
\end{axopicture}
&
\begin{axopicture}(200,100)(-20,0)
\LongArrow(10,-10)(10,100) 
\LongArrow(0,40)(200,40) 
\Line[dash](70,10)(70,-10) 
\Line[dash](70,30)(70,100) 
\Line[dash](105,40)(105,90) 
\Text(180,50)[l]{{\small ${\cal C}$}} 
\GCirc(70,20){1}{.1} 
\Text(60,20)[l]{{\tiny $k'$}} 
\Text(32,10)[l]{{\tiny $k'+im_\pi$}} 
\Text(32,30)[l]{{\tiny $k'-im_\pi$}} 
\Text(110,30)[l]{{\tiny $\Rea k$}} 
\Text(105,100)[l]{{\tiny $k-im_\pi$}}
\Line[dash](105,-10)(105,40) 
\SetWidth{2}
\Line(10,40)(65,40)
\Line(65,40)(65,25)
\Line(65,25)(75,25)
\Line(75,25)(75,40)
\Line(75,40)(100,40) 
\Line(100,40)(100,95)
\Line(100,95)(110,95)
\Line(110,95)(110,40)
\Line[arrow,arrowpos=1](110,40)(200,40)
\Text(100,-25)[l]{{\scriptsize b)}}
\end{axopicture}
\\
\\
\\
\\
\begin{axopicture}(200,100)(20,0)
\LongArrow(10,-10)(10,100) 
\LongArrow(0,40)(200,40) 
\Line[dash](70,85)(70,100) 
\Line[dash](70,55)(70,-10) 
\Line[dash](105,40)(105,-5) 
\Text(180,50)[l]{{\small ${\cal C}$}} 
\GCirc(70,70){1}{.1} 
\Text(68,75)[l]{{\tiny $k'$}} 
\Text(35,85)[l]{{\tiny $k'+im_\pi$}} 
\Text(35,55)[l]{{\tiny $k'-im_\pi$}} 
\Text(115,30)[l]{{\tiny $\Rea k$}} 
\Text(115,-5)[l]{{\tiny $k+im_\pi$}} 
\Line[dash](105,100)(105,40) 
\SetWidth{2}
\Line(10,40)(65,40)
\Line(65,40)(65,60)
\Line(65,60)(75,60)
\Line(75,60)(75,40)
\Line(75,40)(100,40) 
\Line(100,40)(100,-10)
\Line(100,-10)(110,-10)
\Line(110,-10)(110,40)
\Line[arrow,arrowpos=1](110,40)(200,40)
\Text(100,-25)[l]{{\scriptsize c)}}
\end{axopicture}
&
\begin{axopicture}(200,100)(-20,0)
\LongArrow(10,-10)(10,100) 
\LongArrow(0,40)(200,40) 
\Line[dash](70,10)(70,-10) 
\Line[dash](70,30)(70,100) 
\Line[dash](105,40)(105,-5) 
\Text(180,50)[l]{{\small ${\cal C}$}} 
\GCirc(70,20){1}{.1} 
\Text(60,20)[l]{{\tiny $k'$}} 
\Text(32,10)[l]{{\tiny $k'+im_\pi$}} 
\Text(32,30)[l]{{\tiny $k'-im_\pi$}} 
\Text(115,30)[l]{{\tiny $\Rea k$}} 
\Text(115,-5)[l]{{\tiny $k+im_\pi$}} 
 \Line[dash](105,100)(105,40) 
\SetWidth{2}
\Line(10,40)(65,40)
\Line(65,40)(65,25)
\Line(65,25)(75,25)
\Line(75,25)(75,40)
\Line(75,40)(100,40) 
\Line(100,40)(100,-10)
\Line(100,-10)(110,-10)
\Line(110,-10)(110,40)
\Line[arrow,arrowpos=1](110,40)(200,40)
\Text(100,-25)[l]{{\scriptsize d)}}
\end{axopicture}
\\
\\
\end{tabular}
\end{center}
\caption[pilf]{{\small Interplay between the added vertical contours and cuts associated
 with external variables for case iv) that corresponds to $|\Ima k|>m_\pi$ and $|\Ima k'|>m_\pi$. 
 In the figure it is assumed that $\Rea k$ and $\Rea k'$ are positive.
The different possibilities for the signs of the imaginary parts of $k$ and $k'$ are explicitly plotted.
}\label{fig:301215.1}}
\end{figure}

{\bf iv)} $|\Ima k|>m_\pi$ and $|\Ima k'|>m_\pi$: When calculating $t_n(k,k';{k'}^2/m)$  in Eq.~\eqref{281215.1}, 
 the factor \\$ v(k,p_n)t_{n-1}(p_n,k';{k'}^2/m)$  mixes both variables $k$ and $k'$ and   
 there are vertical contours added with bases at $|\Rea k|$ and $|\Rea k'|$, as shown in Fig.~\ref{fig:301215.1}. 
We can also follow similar steps as in case ii) and, then, when $p_n$ goes around the vertical integration contour at $|\Rea k|$ 
 it is necessary for having a discontinuity in $k$ that this addition intersects 
 the cuts in the next factor $v(p_n,k')$, located at $(\pm k')\pm i\sqrt{m_\pi^2+x^2}$. 
However, as a novelty here compared to ii), the intersection referred should involve only those cuts that are 
not avoided already by the deformation of the integration contour at $|\Rea k'|$.
 This is due to the fact that there is a cancellation between the discontinuities of the integrand when 
circumventing the vertical additions at $|\Rea k|$ and $|\Rea k'|$ and the discontinuities that stem 
from crossing them. For instance, for the top left panel in Fig.~\ref{fig:301215.1} when circumventing
the vertical segment at $|\Rea k|$ we have the discontinuity of $v(k,p)$, denoted by $[v(k,p)]$, times 
 $t(p,k')$ with $|\Rea p|$ to the right of $|\Rea k'|$, that we indicate by writing $t_R(p,k')$. 
 Within this compact notation this contribution is written as $[v(k,p)]t_R(p,k')$. 
Next, circumventing the vertical addition at $|\Rea k'|$ implies the discontinuity of the $T$ matrix, $[t(p,k')]$, 
times the potential function evaluated to the left of the 
vertical segment at $|\Rea k|$, $v_L(k,p)$. We then indicate this contribution as $v_L(k,p)[t(p,k')]$. 
Now, when crossing between $|\Rea k|$ and $|\Rea k'|$, so that 
now $|\Rea k'|>|\Rea k|$, the same contributions are $[v(k,p)]t_L(p,k')$ and $v_R(k,p)[t(p,k')]$. 
Therefore, the difference between these contributions are $[v(k,p)](t_R(p,k')-t_L(p,k')) + 
(v_L(k,p)-v_R(k,p))[t(p,k')]=0$ because $t_R(p,k')-t_L(p,k')=-[t(p,k')]$.

 This gives rise to an interesting analysis by requiring that the part of the vertical cut in $p_n$ at $|\Rea k'|$,
 which is not part of the added vertical segment to the integration contour,   
 overlaps with the added integration contour at $|\Rea k|$ 
 (of course when this happens $|\Rea k|=|\Rea k'|$). 
 By definiteness let us assume that $|\Ima k|>|\Ima k'|$ which corresponds 
to Fig.~\ref{fig:301215.1} (if this is not the case we would repeat the analysis exchanging 
$k\leftrightarrow k'$). 
If the contour at $|\Rea k|$ extends from 0 up to $|\Ima k|-m_\pi$ 
(top panels in Fig.~\ref{fig:301215.1}), then the requirement is 
\begin{align}
\sgn(\Rea k') \Ima k'+m_\pi<|\Ima k|-m_\pi~,
\label{301215.1}
\end{align}
so that
\begin{align}
 |\Ima k|>\sgn(\Rea k')\Ima k'+2 m_\pi~,
\label{301215.2}
\end{align}
and then this set of possible values for $k$ belongs to the cuts indicated in Eq.~\eqref{210315.3}. 
Similarly, for the 
two lower panels in Fig.~\ref{fig:301215.1}, 
that correspond to an added vertical cut at $|\Rea k|$ from $-|\Ima k|+m_\pi$ up to 0, 
the required condition is  
\begin{align}
\sgn(\Rea k') \Ima k' -m_\pi>-|\Ima k|+m_\pi~,
\label{301215.3}
\end{align}
which translates into 
\begin{align}
 |\Ima k|>-\sgn(\Rea k') \Ima k' + 2m_\pi~,
\label{301215.3a}
\end{align}
and the same conclusion follows.

Along the added vertical segment up to $\pm i (|\Ima k'|-m_\pi)$, with $|\Ima k'|>m_\pi$, the 
 energy denominator does not vanish. However, further discussion is needed along the added vertical 
contour at $|\Rea k|$ when $|\Ima k|>m_\pi$, because then it is not forbidden that $p_1^2-{k'}^2$ 
could vanish (which requires $|\Ima k|-m_\pi>|\Ima k'|$ and $|\Rea k|=|\Rea k'|$). 
This situation can be analyzed  as we did in the case ii) as a source of DC, where we 
concluded that its contribution implies a cut in $k$ contained in the set of values in Eq.~\eqref{210315.3}. 
The expression for the discontinuity due to this source is given by Eq.~\eqref{180806.4}. 

 Despite that now $|\Ima k'|>m_\pi$ the  Eq.~\eqref{180806.5} 
for the discontinuity of $t(k'\pm \vep,k')$ is still valid. 
To simplify the writing we also take here that both $k$ and $k'$ are in their first quadrant.
Circumventing the vertical segments by parallel lines displaced a distance $\delta$, $\vep>\delta\to 0^+$, to 
the left and right of the vertical segments at $\Rea k'$ and $\Rea k'\pm \vep$, we have 
\begin{align}
&t(k'+\vep,k';\frac{{k'}^2}{m})-t(k'-\vep,k';\frac{{k'}^2}{m})=v(k'+\vep,k)-v(k'-\vep,k)\nn\\
&+\frac{mi}{2\pi^2}\int_0^{\Ima k'-m_\pi}
\frac{d\nu_1 (\Rea k'+i\nu_1)^2}{(\Rea k'+i\nu_1)^2-{k'}^2}
\big\{
v(k'+\vep,\Rea k'+i\nu_1)\big[t(\Rea k'-\delta+i\nu_1,k';\frac{{k'}^2}{m})
-t(\Rea k'+\delta+i\nu_1,k';\frac{{k'}^2}{m})\big]\nn\\
&+\big[v(k'+\vep,\Rea k'+\vep-\delta+i\nu_1)-v(k'+\vep,\Rea k'+\vep+\delta+i\nu_1)\big]
t(\Rea k'+\vep+i\nu_1,k';\frac{{k'}^2}{m})
\big\}\nn\\
&-\frac{mi}{2\pi^2}\int_0^{\Ima k'-m_\pi}\frac{d\nu_1 (\Rea k'+i\nu_1)^2}{(\Rea k'+i\nu_1)^2-{k'}^2} 
\big\{
v(k'-\vep,\Rea k'+i\nu_1)\big[t(\Rea k'-\delta+i\nu_1,k';\frac{{k'}^2}{m})
-t(\Rea k'+\delta+i\nu_1,k';\frac{{k'}^2}{m})\big]\nn\\
&+\big[v(k'-\vep,\Rea k'-\vep-\delta+i\nu_1)-v(k'-\vep,\Rea k'-r+\delta+i\nu_1)\big]
t(\Rea k'-\vep+i\nu_1,k';\frac{{k'}^2}{m})
\big\}~,
\label{210315.10}
\end{align}
where we have taken into account that in the first and third integral terms $v(k'\pm \vep,\Rea k'+i\nu_1)$ is 
located to the right and left of the added contour at $\Rea k'$, respectively, and the same applies 
to $t(\Rea k'\pm \vep+i\nu_1,k';{k'}^2/m)$ at the second and fourth integral terms.
The two integrals in the previous equation 
can be grouped together by taking into account that the discontinuity of the function $v$
 is itself continuous in $\vep$, 
\begin{align}
\lim_{\vep\to 0^+}& \big[v(k'+\vep,\Rea k'+\vep-\delta+i\nu_1)-v(k'+\vep,\Rea k'+\vep+\delta+i\nu_1)\nn\\
=\lim_{\vep\to 0^+}& \big[v(k'-\vep,\Rea k'-\vep-\delta+i\nu_1)-v(k'-\vep,\Rea k'-\vep+\delta+i\nu_1)~.
\label{220315.1}
\end{align}
Then we are left with 
\begin{align}
&t(k'+\vep,k';\frac{{k'}^2}{m})-t(k'-\vep,k';\frac{{k'}^2}{m})=v(k'+\vep,k')-v(k'-\vep,k')\nn\\
&+\frac{mi}{2\pi^2}\int_0^{\Ima k'-m_\pi}
\frac{d\nu_1 \nu_1^2}{(\Rea k'+i\nu_1)-{k'}^2}
\big\{
\big[v(k'+\vep,\Rea k'+i\nu_1)-v(k'-\vep,\Rea k'+i\nu_1)\big]\nn\\
&\times 
\big[t(\Rea k'-\delta+i\nu_1,k';\frac{{k'}^2}{m})-t(\Rea k'+\delta+i\nu_1,k';\frac{{k'}^2}{m})\big]\nn\\
&+\big[v(k'+\vep,\Rea k'+\vep-\delta+i\nu_1)-v(k'+\vep,\Rea k'+\vep+\delta+i\nu_1)\big]\nn\\
&\times
\big[t(\Rea k'+\vep+i\nu_1,k';\frac{{k'}^2}{m})-t(\Rea k'-\vep+i\nu_1,k';\frac{{k'}^2}{m})\big]
\big\}~,
\label{220315.2}
\end{align}
and the terms in the integrand sum up to zero, so that
\begin{align}
t(k'+\vep,k';\frac{{k'}^2}{m})-t(k'-\vep,k';\frac{{k'}^2}{m})=v(k'+\vep,k')-v(k'-\vep,k')~.
\label{220315.3}
\end{align}
This result simplifies the last term of Eq.~\eqref{180806.4}.
Notice that this is not the LHC discontinuity of the on-shell $T$ matrix
but the dynamical cut induced in the first argument of the half-off-shell PWA.

 The dynamical cuts of $t(k,k';{k'}^2/m)$ as a function 
of $k'$ can be described analogously as done above for $k$ but exchanging 
$k\leftrightarrow k'$, given the reciprocity between these two variables in the arguments employed. 
Furthermore, note also that $t(k,k';{k'}^2/m)=t(k',k;{k'}^2/m)$,  Eq.~\eqref{180804.10}. 
 Of course, the energy is always ${k'}^2/m$ and varies 
with $k'$, but the energy denominator in the LS has played a limited role 
 in determining  the position of the dynamical cuts in 
the complex $k$ plane. These contributions to the cuts in $k$ arise when $k'$ intersects the vertical 
addition at $|\Rea k|$, as explained in detail for the case ii). 
 Reciprocally, when varying $k'$ the DCs in this variable from 
the energy denominator will stem in the same way. 

In summary,  the DCs of a half-off-shell PWAs in the $k$ and $k'$ variables are the same as those 
in the potential $v(k,k')$ and are given in the $k$ variable by Eq.~\eqref{210315.3},
 and its inverse for the $k'$ argument,
\begin{align}
\label{180806.6}
k'&=(\pm) k \pm i\sqrt{m_\pi^2+x^2}~.
\end{align}
Of course, in addition to the DCs, one also has the RHC 
in the variable $k'$ 
that happens for $k'$ real (${k'}^2>0$), as considered above and discussed in Eq.~\eqref{180804.15}.

As consistency remark of the developments in this section, notice that the DCs in the variable $k'$ for the 
half-off-shell PWA imply the same vertical deformations in the LS as we concluded making use of the Neumann 
series in Eq.~\eqref{210315.2}. Namely, from the integrand in the LS equation $v(k,p)t(p,k';{k'}^2/m)$ we should 
avoid the same cuts as those stemming from $v(k,p)v(p,k')$, and that give rise to circumventing the vertical additions 
defined by Eqs.~\eqref{261215.1}  and \eqref{210315.1} applied to both $k$ and $k'$. 

\section{Discontinuities of coupled PWAs along the DCs}
\def\theequation{\arabic{section}.\arabic{equation}}
\setcounter{equation}{0}   
\label{sec.071115.1}

In this section we consider  several partial waves coupled,  all of them made out of the same particles. 
The interactions are governed by a partial-wave projected  potential with an arbitrary spectral mass 
decomposition (which can be worked out from the ones of the different components in which the 
potential is decomposed making use of of rotational symmetry and spins of the interacting particles, cf. 
Appendix~\ref{app:170715.1}). As discussed after Eq.~\eqref{eq.sdv} the spectral 
function of the partial-wave projected potential might be also an entire function of the initial and final three-momenta, 
which goes beyond the purely local potential case.
 Because of rotational invariance 
all the partial waves coupled have the same total angular momentum $J$. 
 The orbital angular momenta associated to the channels $i$ and 
$j$ are indicated by $\ell_i$ and $\ell_j$, in this order, and because of parity invariance 
$(-1)^{\ell_i}=(-1)^{\ell_j}$, since one has the same intrinsic parities for the initial
 and final states.

The spectral representation for the potential is of the type in Eq.~\eqref{150116.6}   
but, since the polynomial part is an entire function, it does not contribute to the discontinuity of the 
potential. The cuts of the partial-wave projected 
potential  $v_{ij}(p,p')$ as a function of $k$ in its complex plane are
\begin{align}
p=(\pm)p'\pm i \sqrt{m_\pi^2+x^2}~,
\label{240116.2}
\end{align}
as follows from Sec.~\ref{sec:170116.1}, 
where the analytical properties of  $v(p,p')$ were studied.
The extent of the cuts in the $p'$ variable also corresponds to Eq.~\eqref{240116.2} 
 by exchanging $p$ and $p'$. 
 Let us recall that the contributions that stem from  $\mu>m_\pi$ in 
 the spectral-mass representation give rise to cuts in the partial-wave projected potential $v_{ij}(p,p')$ 
that are comprised in Eq.~\eqref{240116.2} as they extend along the values
\begin{align}
p=(\pm)p'\pm i \sqrt{\mu^2+x^2}~,~\mu\geq m_\pi~,
\label{250116.1}
\end{align}
as we discussed at the end of Sec.~\ref{sec:170116.1}.


 We are interested in the study of the half-off-shell partial-wave projected LS equation in coupled channels
  which reads
\begin{align}
t_{ij}(p_1,p_2;\frac{{p_2}^2}{m})&=v_{ij}(p_1,p_2)+\frac{m}{2\pi^2}\sum_n \int_{\cal C}\frac{dp_3 p_3^2}{p_3^2-p_2^2} 
v_{in}(p_1,p_3) t_{n j}(p_3,p_2;\frac{{p_2}^2}{m})~,
\label{240116.6}
\end{align}
where the sum in $n$ extends over the coupled PWAs.
 The integration contour ${\cal C}$ has 
to be fixed according to the external arguments $p_1$, $p_2$ by the analytical continuation 
of the original contour $p_3\in [0,\infty]$ for real $p_1$ and $p_2$, as discussed in Sec.~\ref{sec.180805.1}. 

  For the on-shell LS equation Eq.~\eqref{240116.6} reads
\begin{align}
t_{ij}(p,p;\frac{p^2}{m})&=v_{ij}(p,p)
+\frac{m}{2\pi^2}\sum_n \int_0^\infty \frac{dp_1\,p_1^2}{p_1^2-p^2} v_{in}(p,p_1)
t_{n j}(p_1,p;\frac{{p}^2}{m})~,
\label{240116.8}
\end{align}
with $p\in \mathbb{R}$. 
We next  extend the variable $p$ to its complex plane. We are 
interested in those values of $p$ corresponding to the DCs that result from 
Eq.~\eqref{240116.2}. When taking $p'=p$ in this equation  
the on-shell DCs originate from the solution of 
$p=-p\pm i\sqrt{m_\pi^2+x^2}$, $x\in \mathbb{R}$, which is 
\begin{align}
\label{180810.1}
p=\pm \frac{i}{2}\sqrt{m_\pi^2+x^2}~.
\end{align}
These values in the variable $p^2$ extends along $p^2\leq -m_\pi^2/4$ and gives rise to the LHC 
 for an on-shell PWA. 
We will demonstrate below that, because of parity invariance, an on-shell PWA only depends 
on $p^2$.

Our main objective in this section is to evaluate the discontinuity  when crossing 
the LHC of an on-shell PWA, 
so that we aim to calculate $t(ik+\vep,ik+\vep)-t(ik-\vep,ik-\vep)$ with $p=ik\pm \vep$ 
and $k\geq m_\pi/2$. Therefore, these values of $p$ are slightly displaced to the left and 
right from the cut,  
\begin{align}
p=i k\pm \varepsilon\rightarrow p^2=-k^2 \pm i\varepsilon~.
\label{240116.9}
\end{align}
 The values of $p$ in the previous equation generate the following cuts 
 in the integration variable $p_1$ of Eq.~\eqref{240116.8},
\begin{align}
p_1&= (\pm)(ik\pm \varepsilon) \pm i\sqrt{m_\pi^2+x^2}~,~x\in \mathbb{R}~,
\label{240116.10}
\end{align}
 The crossing with these DCs of the original 
integration contour in Eq.~\eqref{240116.8}, $p\in [0,+\infty]$,
 only happens when $k>m_\pi$.
  If this is the case, one has to deform the original integration contour by circumventing the cuts 
at $\vep$ with a vertical extension from 0 up to $\pm i (k-m_\pi)$, depending whether 
one has $\pm \vep$ in Eq.~\eqref{240116.10}, respectively. 
   Here one is applying Eq.~\eqref{210315.1} to $p$, whose real part is $\pm \vep$.

For clarification, let us stress that the contribution in the spectral representation of the potential 
with $\mu$ such that $m_\pi<\mu< k$, would give rise to the same type of vertical additions but with shorter extension 
of length $k- \mu$.  
These deformations are then incorporated in the vertical addition attached to the lowest 
value of $\mu$,  $\mu=m_\pi$.  
Needless to say that for $\mu>k$ the DCs do not cross the integration contour and 
there is no deformation induced.

\subsection{Relationships between PWAs with different arguments}
\label{sec:240116.1}

Now, we derive several relationships between PWAs that stem by transforming their arguments 
in different ways and that will be used in some of our derivations. 
These relations are well known for real values of the initial and final three-momenta, 
but we now extend them for complex values too. 
We offer a demonstration of these relations based on 
the contour deformation techniques introduced in this work.

$\bullet$ 
From the spectral decomposition of the potential it is clear the following property
\begin{align}
v_{ij}(p_1^*,p_2^*)&=v_{ij}(p_1,p_2)^*~.
\label{240116.4}
\end{align}
Let us see that this is also the case for the full PWA 
\begin{align}
t_{ij}(p_1^*,p_2^*;\frac{{p_2^*}^2}{m})&=t_{ij}(p_1,p_2;\frac{p_2^2}{m})^*~.
\label{240116.5}
\end{align}
 To obtain this result let us take into account that the deformation of the integration contour that is needed for calculating 
$t_{ij}(p_1^*,p_2^*;{p_2^*}^2/m)$, indicated by ${\cal C}^*$, is the complex conjugate of the one used for $t_{ij}(p_1,p_2;p_2^2/m)$, 
denoted by ${\cal C}$. 
This statement is clear because the DCs in $v_{ij}(p_1^*,p_2^*)$ occur at complex conjugate positions of those in $v_{ij}(p_1,p_2)$. 
 Then we have
\begin{align}
t_{ij}(p_1^*,p_2^*;\frac{{p_2^*}^2}{m})&=v_{ij}(p_1^*,p_2^*)+\frac{m}{2\pi^2}\sum_{n}\int_{{\cal C}^*} 
\frac{dp_3^* {p_3^*}^2}{{p_3^*}^2-{p_2^*}^2} v_{in}(p_1^*,p_3^*) 
t_{n j}(p_3^*,p_2^*;\frac{{p_2^*}^2}{m})\nn\\
&=\left\{v_{ij}(p_1,p_2)+\frac{m}{2\pi^2}\sum_{n}\int_{\cal C} \frac{dp_3 p_3^2}{p_3^2-p_2^2} 
v_{in}(p_1,p_3) t_{n j}(p_3^*,p_2^*;\frac{{p_2^*}^2}{m})^* \right\}^*~.
\label{240116.6a}
\end{align}
Hence, by taking the complex conjugate of the previous equation it is clear that
 both $t(p_1^*,p_2^*;{p_2^*}^2/m)^*$ and $t(p_1,p_2,p_2^2/m)$ satisfy the same LS equation in coupled channels, and 
as a result they can be taken the same.

Notice that the Schwarz reflection principle for functions of one complex variable 
 is a particular case of Eq.~\eqref{240116.5} when restricted to real $p_1$, so that  
\begin{align}
t_{ij}(p_1,p_2^*;\frac{{p_2^*}^2}{m})&=t_{ij}(p_1,p_2;\frac{p_2^2}{m})^*~~,~~p_1\in \mathbb{R}~,
\label{220116.2}
\end{align}
or when considering on-shell scattering $p_1=p_2=p$,
\begin{align}
t_{ij}(p^*,p^*;\frac{{p^*}^2}{m})&=t_{ij}(p,p;\frac{p^2}{m})^*~.
\label{220116.3}
\end{align}

$\bullet$ Now, for the partial-wave projected potential one has the well-known relations 
\begin{align}
\label{180811.1}
v_{ij}(-p_1,p_2)=(-1)^{\ell_i} v_{ij}(p_1,p_2)~,\nn\\
v_{ij}(p_1,-p_2)=(-1)^{\ell_j} v_{ij}(p_1,p_2)~,
\end{align}
that can be easily deduced by imposing parity invariance in the formulas of the 
footnote~\ref{foot.180607.1}.
 We want to show that these relations also hold for the PWAs. 
 For real $p_1$ and $p_2$ the half-off-shell LS equation is 
\begin{align}
t_{ij}(p_1,p_2;\frac{p_2^2}{m})&=v_{ij}(p_1,p_2)+\frac{m}{2\pi^2}\sum_n \int_0^\infty  
\frac{dp_3 p_3^2}{p_3^2-p_2^2}v_{in}(p_1,p_3)t_{n j}(p_3,p_2;\frac{p_2^2}{m})~.
\label{260116.1}
\end{align}
It is clear that if we exchange the sign of $p_1$ the rhs of the equation does change by a factor 
$(-1)^{\ell_i}$ and then 
\begin{align}
t_{ij}(-p_1,p_2;\frac{p_2^2}{m})=(-1)^{\ell_i} t_{ij}(p_1,p_2;\frac{p_2^2}{m})~.
\label{2260116.2}
\end{align}
 Had we changed the sign of  $p_2$ the resulting $T$-matrix element, $t_{ij}(p_1,-p_2;\frac{p_2^2}{m})$ 
 satisfies the same LS equation as 
$(-1)^{\ell_j} t_{ij}(p_1,p_2;\frac{p_2^2}{m})$, and since the solution is supposed to be unique, 
the $T$-matrix element is also the same. Explicitly,
\begin{align}
t_{ij}(p_1,-p_2;\frac{p_2^2}{m})&=v_{ij}(p_1,-p_2)+\frac{m}{2\pi^2}\sum_n \int_0^\infty  
\frac{dp_3 p_3^2}{p_3^2-p_2^2}v_{in}(p_1,p_3)t_{n j}(p_3,-p_2;\frac{p_2^2}{m})~,\nn\\
&=(-1)^{\ell_j}v_{ij}(p_1,p_2)+\frac{m}{2\pi^2}\sum_n \int_0^\infty  
\frac{dp_3 p_3^2}{p_3^2-p_2^2}v_{in}(p_1,p_3)t_{n j}(p_3,-p_2;\frac{p_2^2}{m})~,
\label{260116.3}
\end{align}
which is the same IE as the one satisfied by $(-1)^{\ell_j}t_{ij}(p_1,p_2;\frac{p_2^2}{m})$,
\begin{align}
(-1)^{\ell_j}t_{ij}(p_1,p_2;\frac{p_2^2}{m})&=
(-1)^{\ell_j} v_{ij}(p_1,p_2)
+\frac{m}{2\pi^2}\sum_n \int_0^\infty  
\frac{dp_3 p_3^2}{p_3^2-p_2^2}
v_{in}(p_1,p_3)(-1)^{\ell_j} t_{n j}(p_3,p_2;\frac{p_2^2}{m})~.
\label{260116.4}
\end{align}

Now we consider the generalization of Eq.~\eqref{2260116.2} to complex $p_1$ and $p_2$. The LS 
equation is in this case
\begin{align}
\label{260116.7} 
t_{ij}(p_1,p_2;\frac{p_2^2}{m})&=v_{ij}(p_1,p_2)+\frac{m}{2\pi^2}\sum_n 
\int_{\cal C} \frac{dp_3 p_3^2}{p_3^2-p_2^2} 
v_{in}(p_1,p_3) t_{n j}(p_3,p_2;\frac{{p_2}^2}{m})~,
\end{align}
where the integration contour is ${\cal C}$. 
 One important point to realize is that the same integration contour applies to  
 all  $t_{ij}(-p_1,p_2;\frac{p_2^2}{m})$, $t_{ij}(p_1,-p_2;\frac{p_2^2}{m})$ and 
$t_{ij}(p_1,p_2;\frac{p_2^2}{m})$, because of the 
 symmetry around the origin of the branch points in $p_3$ for $v(p_1,p_3)$ and $t(p_3,p_2;\frac{p_2^2}{m})$. 
 For instance, if  $p_1=q_r+i q_i$ with positive $q_r$ and $q_i>m_\pi$ the corresponding vertical addition at $q_r$ 
in the integration contour ${\cal C}$ extends from 0 up to $i(q_i-m_\pi)$. However, both the real and imaginary parts of $-p_1$ 
are negative and then, according to the rules in Eq.~\eqref{210315.1},  the vertical addition takes place 
at $-\Rea p_1=q_r$ and extends from 0 up to $i(-\Ima (-p_1)-m_\pi)=i(q_i-m_\pi)$, and the same integration contour results.
  Once this point is clear then one 
can proceed in completely analogous way as done above for real $p_1$ and $p_2$  to conclude that 
\begin{align}
\label{260116.6}
t(-p_1,p_2;\frac{p_2^2}{m})&=(-1)^{\ell_i} t(p_1,p_2;\frac{p_2^2}{m})~,\nn\\
t(p_1,-p_2;\frac{p_2^2}{m})&=(-1)^{\ell_j} t(p_1,p_2;\frac{p_2^2}{m})~,
\end{align}
also hold for complex arguments. Let us stress that because of parity conservation 
\begin{align}
(-1)^{\ell_i}=(-1)^{\ell_j}~.
\label{260116.5}
\end{align}
   As a corollary of Eq.~\eqref{260116.6}, we  also conclude that
 $t_{ij}(p_1,p_2;\frac{p_2^2}{m})/(p_1 p_2)^{\ell_i}$, $t_{ij}(p_1,p_2;\frac{p_2^2}{m})/(p_1 p_2)^{\ell_j}$ and \\ 
 $t_{ij}(p_1,p_2;\frac{p_2^2}{m})/(p_1^{\ell_i} p_2^{\ell_j})$ are  functions of the moduli squared of the CM 
three-momenta because these combinations are invariant under the change of sign of $p_1$ or $p_2$. In 
the same way we also conclude that the on-shell PWA $t_{ij}(p,p;p^2/m)$ is a function of $p^2$.

\subsection{Calculation of the on-shell discontinuity of PWAs}
\label{sec:270116.1}

We want to calculate 
\begin{align}
\label{180811.2}
\Delta_{ij}(p^2)=\Ima t_{ij}(ik+\ve,ik+\ve)~,
\end{align}
 with $p^2=-k^2+i\ve$, cf.~Eq.~\eqref{240116.9}, and $k\in \mathbb{R}^+$. 
 For that we take the discontinuity
\begin{align}
t_{ij}(ik+\ve,ik+\ve)-t_{ij}(ik-\ve,ik-\ve)=2i \Delta_{ij}(p^2)~.
\label{270116.1}
\end{align}
This equality follows by applying Eq.~\eqref{260116.6} and \eqref{240116.5}, 
in this order, to $t_{ij}(ik-\ve,ik-\ve)$ because
\begin{align}
t_{ij}(ik-\ve,ik-\ve)&=(-1)^{\ell_i+\ell_j} t_{ij}(-ik+\ve,-ik+\ve)=t_{ij}(ik+\ve,ik+\ve)^*~.
\label{270116.2}
\end{align}

\begin{figure}[ht]
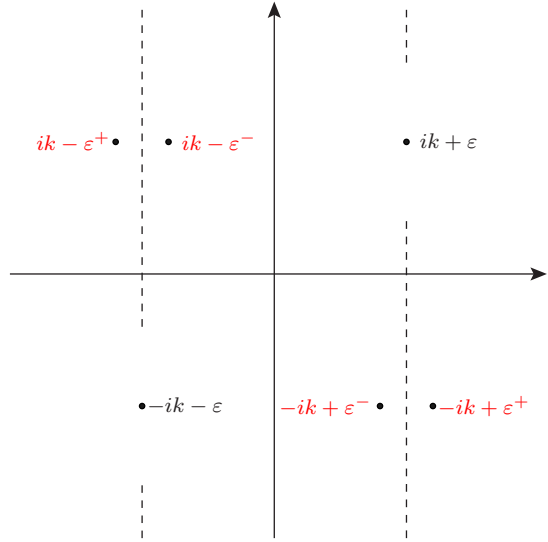

\begin{center}
\begin{axopicture}(200,200)
\LongArrow(100,0)(100,200) 
\LongArrow(0,100)(200,100) 
\GCirc(150,150){1}{.1} 
\GCirc(160,50){1}{.1} 
\GCirc(140,50){1}{.1} 
\GCirc(50,50){1}{.1} 
\GCirc(40,150){1}{.1} 
\GCirc(60,150){1}{.1} 
\Text(155,150)[l]{{\scriptsize $i k +\vep$}}
\Text(162,50)[l]{\textcolor{red}{{\scriptsize $-i k +\vep^+$}}}
\Text(102,50)[l]{\textcolor{red}{{\scriptsize $-i k +\vep^-$}}}
\Text(52,50)[l]{{\scriptsize $-i k-\vep$}}
\Text(65,150)[l]{\textcolor{red}{{\scriptsize $ik-\vep^-$}}}
\Text(10,150)[l]{\textcolor{red}{{\scriptsize $ik-\vep^+$}}}
\Line[dash](150,180)(150,200) 
\Line[dash](150,120)(150,0) 
\Line[dash](50,80)(50,200) 
\Line[dash](50,20)(50,0) 
\end{axopicture}
\end{center}
\caption[pilf]{{\small Location of the cuts of $t_{ij}(z_1,i k+\ve)$ 
in the variable $z_1$. This variable takes the values 
$-ik+\ve^{\mp}$  in the difference 
$t_{ij}(-ik+\ve^-,ik+\ve)-t_{ij}(-ik+\ve^+,ik+\ve)$. 
 Note that the figure is not drawn to scale because the gap between two vertical 
cuts at $\pm \ve$ is $2 m_\pi$, while the horizontal separation between vertical cuts is $2\ve\to 0^+$.}
\label{fig:270116.1}
}
\end{figure}

 Instead of taking the difference in Eq.~\eqref{270116.1}, 
\begin{align}
t_{ij}(ik+\ve,ik+\ve)-t_{ij}(ik-\ve,ik-\ve)~,
\label{270116.3}
\end{align}
 we now consider 
\begin{align}
t_{ij}(-i k+\ve^-,i k+\ve)-t_{ij}(-i k+\ve^+,i k+\ve)~,
\label{270116.4}
\end{align}
with $\ve^+>\ve>\ve^-$ and $\ve^\pm \to \ve$. 
To show the connection between both discontinuities let us exchange the sign 
in the first argument of Eq.~\eqref{270116.4}. 
By taking into account Eq.~\eqref{260116.6}  it results that 
 \begin{align}
t_{ij}(-ik+\ve^-,ik+\ve)-t_{ij}(-ik+\ve^+,ik+\ve)&=(-1)^{\ell_i} \big[ t_{ij}(ik-\ve^-,ik+\ve)-t_{ij}(ik-\ve^+,ik+\ve) \big]~.
\label{040615.1a}
\end{align}
The first arguments of the PWAs in the differences of the previous equation are shown by the  red  points 
in Fig.~\ref{fig:270116.1}.
 Now, inspection of Fig.~\ref{fig:270116.1} reveals that for $\ve\to 0^+$ the PWAs 
$t_{ij}(ik+\ve,ik+\ve)$ and $t_{ij}(ik-\ve^-,ik+\ve)$ can be continued analytically one to each other
 in that limit.\footnote{As remarked in the caption of Fig.~\ref{fig:270116.1}, 
this figure is not drawn to scale so that the vertical gap, in the middle of which one has the point $ik+\ve$, 
is  $2m_\pi$ wide while the horizontal distance between the vertical cuts is $2\ve\to 0^+$.\label{foot:270116.1}}
  A Taylor expansion of $t_{ij}(z_1,ik+\ve)$ can be performed at the point $z_1=ik+\ve$  within a circle of radius 
 $2\ve$ (because of the location of the nearest DC at $-ik-\ve\pm i\sqrt{m_\pi^2+x^2}$), 
 which incorporates the point $ik-\ve^-$ where $t_{ij}(ik-\ve^-,ik+\ve)$ is evaluated. In this way, we prove that 
$t_{ij}(ik-\ve^-,ik+\ve)- t_{ij}(ik+\ve,ik+\ve)={\cal O}(\ve)$. 
  We can also proceed similarly for $t_{ij}(ik-\ve,z_1)$  and, as a function of $z_1$,  
  take a Taylor expansion centered at $ik-\ve$ of radius 
$\ve+\ve'$, with $\ve^-<\ve'<\ve$,  that includes the point $t_{ij}(ik-\ve,ik+\ve^-)$. 
This Taylor series is possible because the location of the nearest DC at $-ik+\ve\pm i\sqrt{m_\pi^2+x^2}$. 
 The PWA $t_{ij}(ik-\ve,ik+\ve^-)$ is also equal to $t_{ij}(ik-\ve^+,ik+\ve)$ for $\ve^+\to 0$ because the second argument 
is located in the same relative position with respect to the cuts of $t_{ij}(ik-\ve^+,z_1)$. 
  As a result it follows that for $\ve \to 0^+$  
\begin{align}
\label{280116.1}
t_{ij}(ik+\ve,ik+\ve)-t_{ij}(ik-\ve,ik-\ve)&=
t_{ij}(ik-\ve^-,ik+\ve)-t_{ij}(ik-\ve^+,ik+\ve)\nn\\
&= (-1)^{\ell_i}\big[t_{ij}(-ik+\ve^-,ik+\ve)-t(-ik+\ve^+,ik+\ve)\big]~,
\end{align}
where in the last equality we have used Eq.~\eqref{040615.1a}. 

We now define the modified PWAs $\hht_{ij}(p_1,p_2;p_2^2/m)$ by
\begin{align}
\label{180811.3}
\hht_{ij}(p_1,p_2)&=p_1^{\ell_i+1}{p_2}^{\ell_j+1} t_{ij}(p_1,p_2)~,
\end{align}
because they are more convenient to properly treat the threshold behavior of PWAs. 
 In terms of their discontinuity we also introduce the function $\mff_{ij}(\nu)$ as
 \begin{align}
i \mff_{ij}(\nu)&=\hht_{ij}(i\nu+\ve^-,ik+\ve)-\hht_{ij}(i\nu+\ve^+,ik+\ve)~.
\label{280116.2}
\end{align}

Making use of these functions, we obtain from Eq.~\eqref{280116.1} the following expression for $\Delta_{ij}(p^2)$, 
\begin{align}
\label{280116.3}
\Delta_{ij}(p^2)&=(-1)^{\ell_i}i^{\ell_i-\ell_j}\frac{\mff_{ij}(-k)}{2k^{\ell_i+\ell_j+2}}~.
\end{align}
In the following we derive how to calculate $\mff_{ij}(\nu)$ for $\nu=-k$ in order to obtain $\Delta_{ij}(p^2)$ 
  for  $S$ and higher partial waves. To accomplish this aim we are going to determine an IE for 
  $\mff_{ij}(\nu)$ with $\nu$ in the interval $\nu\in [-k,k-m_\pi]$.

\begin{figure}[ht]
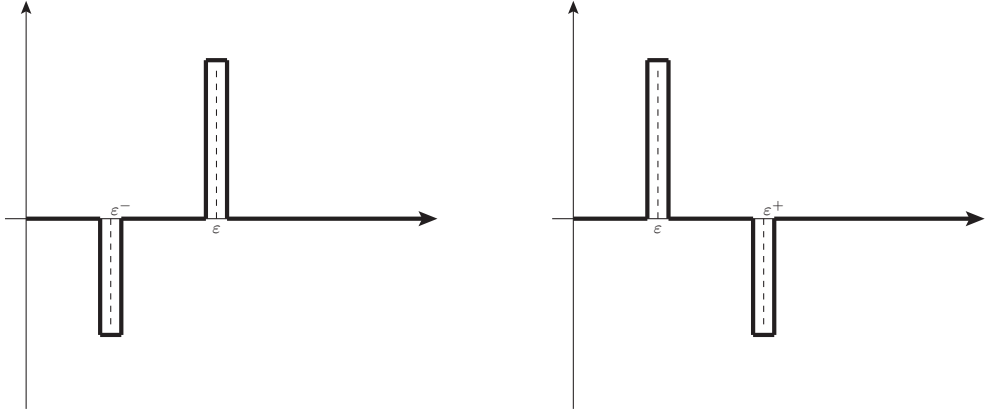

\begin{center}
\begin{tabular}{ll}
\scalebox{0.8}{
\begin{axopicture}(200,200)(-20,20)
\LongArrow(10,10)(10,200) 
\LongArrow(0,100)(200,100) 
\Line[dash](50,50)(50,100) 
\Line[dash](100,100)(100,170) 
\Text(50,105)[l]{{\scriptsize $\ve^-$}}
\Text(100,95){{\scriptsize $\ve$}}
\SetWidth{2}
\Line(10,100)(45,100)
\Line(45,100)(45,45)
\Line(45,45)(55,45)
\Line(55,45)(55,100)
\Line(55,100)(95,100)
\Line(95,100)(95,175)
\Line(95,175)(105,175)
\Line(105,175)(105,100)
\Line[arrow,arrowpos=1](105,100)(200,100)
\end{axopicture}
}
&
\scalebox{0.8}{
\begin{axopicture}(300,200)(-60,20)
\LongArrow(10,10)(10,200) 
\LongArrow(0,100)(200,100) 
\Line[dash](100,50)(100,100) 
\Line[dash](50,100)(50,170) 
\Text(100,105)[l]{{\scriptsize $\ve^+$}}
\Text(50,95){{\scriptsize $\ve$}}
\SetWidth{2}
\Line(10,100)(45,100)
\Line(45,100)(45,175)
\Line(45,175)(55,175)
\Line(55,175)(55,100)
\Line(55,100)(95,100)
\Line(95,100)(95,45)
\Line(95,45)(105,45)
\Line(105,45)(105,100)
\Line[arrow,arrowpos=1](105,100)(200,100)
\end{axopicture}
}
\end{tabular}
\end{center}
\caption{{\small Vertical contours used in Eq.~\eqref{280116.5} corresponding to the case 
$\nu<0$. The left panel applies to $t_{ij}(i\nu+\ve^-,ik+\ve)$ and 
 the right one to $t_{ij}(i\nu+\ve^+,ik+\ve)$.
 \label{fig:280116.2}}}
\end{figure}

We again start our derivations from the half-off-shell LS equation for PWAs in coupled channels,
\begin{align}
t_{ij}(p,p')&=
v_{ij}(p,p')
+\frac{m}{2\pi^2}\sum_n \int_{\cal C} \frac{dp_1\,p_1^2}{p_1^2-{k'}^2}v_{in}(p,p_1)t_{n j}(p_1,p')~,
\label{280116.4}
\end{align}
where the (deformed) integration contour ${\cal C}$ depends on the external variables $p$ and $p'$, 
as discussed in Sec.~\ref{sec.180805.1}. 

$\bullet$  Let us discuss first the case $\nu<0$. 
We take directly the difference $t(i\nu+\ve^-,ik+\ve)-t(i\nu+\ve^+,ik+\ve)$ and consider the 
added vertical contours shown in Fig.~\ref{fig:280116.2}, 
 which are the only ones  that give contribution to the difference. 
There is no contribution from the unperturbed original contour $p_1\in [0,\infty]$ because it does not 
intersect with any DC that has not been circumvented already, cf. Sec.~\ref{sec:dc}. 
 Then it results
\begin{align}
\label{280116.5}
&t_{ij}(i\nu+\ve^-,ik+\ve)-t_{ij}(i\nu+\ve^+,ik+\ve)= v_{ij}(i\nu+\ve^-,ik+\ve)-v_{ij}(i\nu+\ve^+,ik+\ve)\nn\\
&-\theta(-\nu-m_\pi)\frac{i m}{2\pi^2}\sum_n\int_0^{\nu+m_\pi}\frac{d\nu_1 \nu_1^2}{k^2-\nu_1^2}
 \big[v_{in}(i\nu+\ve^-,i\nu_1+\ve^--\delta)-v_{in}(i\nu+\ve^-,i\nu_1+\ve^-+\delta)\big]t_{n j}(i\nu_1+\ve^-,ik+\ve)\nn\\
&-\theta(k-m_\pi)\frac{i m}{2\pi^2}\sum_n \int_0^{k-m_\pi}\frac{d\nu_1 \nu_1^2}{k^2-\nu_1^2} 
v_{in}(i\nu+\ve^-,i\nu_1+\ve) \big[t_{n j}(i\nu_1+\ve-\delta,ik+\ve)-t_{n j}(i\nu_1+\ve+\delta,ik+\ve)\big]
\nn\\
&+\theta(-\nu-m_\pi)\frac{i m}{2\pi^2}\sum_n \int_0^{\nu+m_\pi}\frac{d\nu_1 \nu_1^2}{k^2-\nu_1^2}
 \big[v_{in}(i\nu+\ve^+,i\nu_1+\ve^+-\delta)-v_{in}(i\nu+\ve^+,i\nu_1+\ve^++\delta)\big]t_{n j}(i\nu_1+\ve^+,ik+\ve) \nn\\
&+\theta(k-m_\pi)\frac{i m}{2\pi^2}\sum_n \int_0^{k-m_\pi}\frac{d\nu_1 \nu_1^2}{k^2-\nu_1^2} 
v_{in}(i\nu+\ve^+,i\nu_1+\ve) \big[t_{n j}(i\nu_1+\ve-\delta,ik+\ve)-t_{n j}(i\nu_1+\ve+\delta,ik+\ve)\big]~,
\end{align}
with $\nu<0$. In the previous equation we have taken into account that $v_{in}(i\nu+\ve^\pm,i\nu_1)$ is continuous around the upwards vertical cut 
of extension $k-m_\pi$ and so it is $t_{n j}(i\nu_1,ik+\ve)$ around the downwards vertical cuts of extension $-\nu-m_\pi$ 
(with $|\nu|>m_\pi$). Combining the integrals up to $\nu+m_\pi$ and $k-m_\pi$ separately one can rewrite the 
previous equation as 
\begin{align}
\label{280116.6}
&t_{ij}(i\nu+\ve^-,ik+\ve)-t_{ij}(i\nu+\ve^+,ik+\ve)= v_{ij}(i\nu+\ve^-,ik+\ve)-v_{ij}(i\nu+\ve^+,ik+\ve)\nn\\
&+\theta(-\nu-m_\pi)\frac{i m}{2\pi^2}\sum_n \int_0^{\nu+m_\pi}\frac{d\nu_1 \nu_1^2}{k^2-\nu_1^2}
 \big[v_{in}(i\nu+\ve^-,i\nu_1+\ve^-+\delta)-v_{in}(i\nu+\ve^-,i\nu_1+\ve^--\delta)\big]\nn\\
\times & \big[t_{n j}(i\nu_1+\ve^-,ik+\ve)-t_{n j}(i\nu_1+\ve^+,ik+\ve)\big]\\
&-\theta(k-m_\pi)\frac{i m}{2\pi^2}\sum_n \int_0^{k-m_\pi}\frac{d\nu_1 \nu_1^2}{k^2-\nu_1^2} 
\big[v_{in}(i\nu+\ve^-,i\nu_1+\ve)-v_{in}(i\nu+\ve^+,i\nu_1+\ve) \big]\nn\\
&\times  \big[t_{n j}(i\nu_1+\ve-\delta,ik+\ve)-t_{n j}(i\nu_1+\ve+\delta,ik+\ve)\big]~.\nn
\end{align}
Exchanging the order of the integration limits in the first integral we can combine the two integrals 
in one as \begin{align}
\label{290116.1}
&t_{ij}(i\nu+\ve^-,ik+\ve)-t_{ij}(i\nu+\ve^+,ik+\ve)=
v_{ij}(i\nu+\ve^-,ik+\ve)-v_{ij}(i\nu+\ve^+,ik+\ve)\nn\\
&-\theta(k-m_\pi)\frac{m i}{2\pi^2}\sum_n \int_{(\nu+m_\pi)\theta(|\nu|-m_\pi)}^{k-m_\pi}\frac{d\nu_1\nu_1^2}{k^2-\nu_1^2}
\big[v_{in}(i\nu+\ve^-,i\nu_1+\ve)-v_{in}(i\nu+\ve^+,i\nu_1+\ve)\big]\nn\\
&\times  \big[t_{n j}(i\nu_1+\ve-\delta,ik+\ve)-t_{n j}(i\nu_1+\ve+\delta,ik+\ve)\big]~.
\end{align}

\begin{figure}[ht]
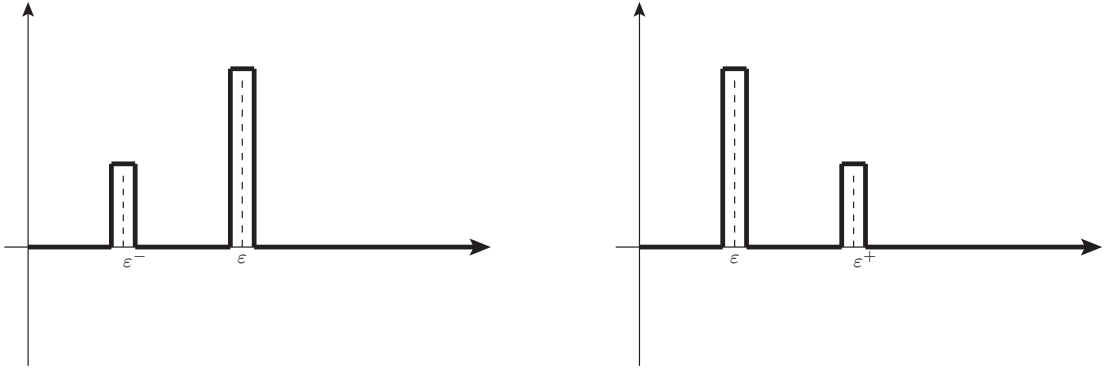

\begin{center}
\begin{tabular}{llll}
\scalebox{.9}{
\begin{axopicture}(200,200)(-20,30)
\LongArrow(10,50)(10,200) 
\LongArrow(0,100)(200,100) 
\Line[dash](50,100)(50,130) 
\Line[dash](100,100)(100,170) 
\Text(50,95)[l]{{\scriptsize $\ve^-$}}
\Text(100,95){{\scriptsize $\ve$}}
\SetWidth{2}
\Line(10,100)(45,100)
\Line(45,100)(45,135)
\Line(45,135)(55,135)
\Line(55,135)(55,100)
\Line(55,100)(95,100)
\Line(95,100)(95,175)
\Line(95,175)(105,175)
\Line(105,175)(105,100)
\Line[arrow,arrowpos=1](105,100)(200,100)
\end{axopicture}
}
& 
\scalebox{.9}{
\begin{axopicture}(300,200)(-60,30)
\LongArrow(10,50)(10,200) 
\LongArrow(0,100)(200,100) 
\Line[dash](100,100)(100,130) 
\Line[dash](50,100)(50,170) 
\Text(100,95)[l]{{\scriptsize $\ve^+$}}
\Text(50,95){{\scriptsize $\ve$}}
\SetWidth{2}
\Line(10,100)(45,100)
\Line(45,100)(45,175)
\Line(45,175)(55,175)
\Line(55,175)(55,100)
\Line(55,100)(95,100)
\Line(95,100)(95,135)
\Line(95,135)(105,135)
\Line(105,135)(105,100)
\Line[arrow,arrowpos=1](105,100)(200,100)
\end{axopicture}
}
\end{tabular}
\caption{{\small Vertical contours used in Eq.~\eqref{290116.2} corresponding to the case $\nu>0$.
 The left panel applies to $t_{ij}(i\nu+\ve^-,ik+\ve)$ and 
 the right one to $t_{ij}(i\nu+\ve^+,ik+\ve)$.
 \label{fig:290116.1}}}
\end{center}
\end{figure}

 For the case $\nu>0$ we have to consider the vertical contours shown in Fig.~\ref{fig:290116.1}, which 
 are present for $\nu$ and $k$ larger than $m_\pi$ and extend from 0 up to $\nu-m_\pi$ and 
 $k-m_\pi$, in order. These are the only ones that give contribution to the difference 
of partial waves under study. Then, following the same steps used to derive Eqs.~\eqref{280116.5} and \eqref{280116.6}, we have now
\begin{align}
\label{290116.2}
&t_{ij}(i\nu+\ve^-,ik+\ve)-t_{ij}(i\nu+\ve^+,ik+\ve)= v_{ij}(i\nu+\ve^-,ik+\ve)-v_{ij}(i\nu+\ve^+,ik+\ve)\nn\\
&+\theta(\nu-m_\pi)\frac{i m}{2\pi^2}\sum_n \int_0^{\nu-m_\pi}\frac{d\nu_1 \nu_1^2}{k^2-\nu_1^2}
 \big[v_{in}(i\nu+\ve^-,i\nu_1+\ve^-+\delta)-v_{in}(i\nu+\ve^-,i\nu_1+\ve^--\delta)\big]\nn\\
&\times  \big[t_{n j}(i\nu_1+\ve^-,ik+\ve)-t_{n j}(i\nu_1+\ve^+,ik+\ve)\big]\\
&-\theta(k-m_\pi)\frac{i m}{2\pi^2}\sum_n\int_0^{k-m_\pi}\frac{d\nu_1 \nu_1^2}{k^2-\nu_1^2} 
\big[v_{in}(i\nu+\ve^-,i\nu_1+\ve)-v_{in}(i\nu+\ve^+,i\nu_1+\ve) \big]\nn\\
&\times 
\big[t_{n j}(i\nu_1+\ve-\delta,ik+\ve)-t_{n j}(i\nu_1+\ve+\delta,ik+\ve)\big]~,~\nu>0~.\nn
\end{align}
Let us notice that now both integrals run through positive values of $\nu_1$ so that they subtract each other, because of the 
different sign in front of them,  and we have an analogous expression to Eq.~\eqref{290116.1},
\begin{align}
\label{290116.3}
&t_{ij}(i\nu+\ve^-,ik+\ve)-t_{ij}(i\nu+\ve^+,ik+\ve)=v_{ij}(i\nu+\ve^-,ik+\ve)-v_{ij}(i\nu+\ve^+,ik+\ve)\nn\\
&-\theta(k-m_\pi)\frac{m i}{2\pi^2}\sum_n \int_{(\nu-m_\pi)\theta(\nu-m_\pi)}^{k-m_\pi}\frac{d\nu_1\nu_1^2}{k^2-\nu_1^2}
\big[v_{in}(i\nu+\ve^-,i\nu_1+\ve)-v_{in}(i\nu+\ve^+,i\nu_1+\ve)\big] \nn\\
&\times  \big[t_{n j}(i\nu_1+\ve-\delta,ik+\ve)-t_{n j}(i\nu_1+\ve+\delta,ik+\ve)\big]~.
\end{align}

To shorten notation we indicate in the subsequent the lower limit of integration in Eqs.~\eqref{290116.1} and \eqref{290116.3} as  
\begin{align}
\label{290116.5}
\vartheta(\nu)=\sgn(\nu)(|\nu|-m_\pi)\theta(|\nu|-m_\pi)~.
\end{align}
We can then express Eqs.~\eqref{290116.1} and \eqref{290116.3} in the same manner as
\begin{align}
\label{290116.6}
&t_{ij}(i\nu+\ve^-,ik+\ve)-t_{ij}(i\nu+\ve^+,ik+\ve)=v_{ij}(i\nu+\ve^-,ik+\ve)-v_{ij}(i\nu+\ve^+,ik+\ve)\nn\\
&-\theta(k-m_\pi)\frac{m i}{2\pi^2}\sum_n \int_{\vartheta(\nu)}^{k-m_\pi}\frac{d\nu_1\nu_1^2}{k^2-\nu_1^2}
\big[v_{in}(i\nu+\ve^-,i\nu_1+\ve)-v_{in}(i\nu+\ve^+,i\nu_1+\ve)\big] \nn\\
&\times  \big[t_{n j}(i\nu_1+\ve-\delta,ik+\ve)-t_{n j}(i\nu_1+\ve+\delta,ik+\ve)\big]~.
\end{align}

 The previous equation allows us to conclude the continuity of 
$\Rea t_{ij}(i\nu+\ve\pm\delta,ik+\ve)$, $\nu\in [-k,k-m_\pi]$,  
 for $\delta\to 0$. This continuity is clear for $\Rea v_{ij}(i\nu+\ve^\pm,i k +\ve)$  and $\Rea v_{ij}(i\nu+\ve^\pm,i\nu_1+\ve)$ 
with $\ve^\pm\to \ve$, see e.g. Eq.~\eqref{180923.1}. As a result, the discontinuities in the potential
 in Eq.~\eqref{290116.6} are purely imaginary, which allows to rewrite this equation as
\begin{align}
\label{290116.7}
&t_{ij}(i\nu+\ve^-,ik+\ve)-t_{ij}(i\nu+\ve^+,ik+\ve)=i\big[\Ima v_{ij}(i\nu+\ve^-,ik+\ve)-\Ima v_{ij}(i\nu+\ve^+,ik+\ve)\big]\nn\\
&+\theta(k-m_\pi)\frac{m }{2\pi^2}\sum_n \int_{\vartheta(\nu)}^{k-m_\pi}\frac{d\nu_1\nu_1^2}{k^2-\nu_1^2}
\big[\Ima v_{in}(i\nu+\ve^-,i\nu_1+\ve)-\Ima v_{in}(i\nu+\ve^+,i\nu_1+\ve)\big]\nn\\
&\times  \big[t_{n j}(i\nu_1+\ve-\delta,ik+\ve)-t_{n j}(i\nu_1+\ve+\delta,ik+\ve)\big]~.
\end{align}
Because the independent term in this IE is purely imaginary while the kernel is purely real 
it is clear that the solution 
of this IE, 
that corresponds to the discontinuity of $t_{ij}(i\nu +\ve \pm \delta,ik+\ve)$ with $\delta\to 0$, 
is purely imaginary.\footnote{It is important to note that the use of this property of continuity for $\Rea \hhv_{ij}$
 and $\Rea \hht_{ij}$ is applied for finite $\ve$.  
 That is, the integrations are done with $\ve\neq 0$ but with $\ve^\pm\to \ve$ and  $\delta\to 0$.
\label{foot:290116.1}}  
 To hold this statement consider the resolvent of the kernel of Eq.~\eqref{290116.7} or its expansion 
 in a Neumann series. 

As a consequence of the continuity of $\Rea t_{ij}(i\nu + \ve \pm \delta,ik+\ve)$ for $\delta\to 0$ 
it follows that we can rewrite the definition of $\mff_{ij}(\nu)$, Eq.~\eqref{280116.2}, simply as
\begin{align}
\label{290116.8}
\mff_{ij}(\nu)&=\Ima \hht_{ij}(i\nu+\ve-\delta,ik+\ve)-\Ima \hht_{ij}(i\nu+\ve+\delta,ik+\ve)~.
\end{align}

 Taking now into account the property of continuity of $\Rea t_{ij}(i\nu_1+\ve\pm\delta,ik+\ve)$ for $\delta\to 0$, 
 the  IE in Eq.~\eqref{290116.7} becomes   
\begin{align}
\label{290116.9}
&\Ima t_{ij}(i\nu+\ve^-,ik+\ve)-\Ima t_{ij}(i\nu+\ve^+,ik+\ve)=\Ima v_{ij}(i\nu+\ve^-,ik+\ve)-\Ima v_{ij}(i\nu+\ve^+,ik+\ve)\nn\\
&+\theta(k-\nu-2m_\pi)\frac{m }{2\pi^2}\sum_n \int_{\vartheta(\nu)}^{k-m_\pi}\frac{d\nu_1\nu_1^2}{k^2-\nu_1^2}
\big[\Ima v_{in}(i\nu+\ve^-,i\nu_1+\ve)-\Ima v_{in}(i\nu+\ve^+,i\nu_1+\ve)\big]\nn\\
&\times  \big[\Ima t_{n j}(i\nu_1+\ve-\delta,ik+\ve)- \Ima t_{n j}(i\nu_1+\ve+\delta,ik+\ve)\big]~.
\end{align}

 From Eq.~\eqref{290116.9}  let us derive a new IE that allows one to treat in an explicit proper way 
   the infrared  behavior for $\ell_n\geq 1$, which stems from the fact that 
$\Ima v_{i n}(i\nu+\ve^\pm,i\nu_1+\ve)$ and $\Ima t_{n j}(i\nu_1+\ve\pm \delta,ik+\ve)$ diverge like 
$1/\nu_1^{\ell_n+1}$ for $\nu_1\to 0$, so that its product does like $1/\nu_1^{2(\ell_n+1)}$. 
  This behavior of the discontinuity at the origin  is explicit from the imaginary part of 
  Eq.~\eqref{swpvm} for the $S$-wave case. It is also clear from the integral of Eq.~\eqref{swpv}  
that a factor $t^{\ell_n}$ in the numerator of the integrand, e.g. from a Legendre polynomial $P_{\ell_n}(t)$, 
would imply an extra factor $1/\nu_1^{\ell_n}$ in $\Ima v_{i n}(i\nu+\ve^\pm,i\nu_1+\ve)$. 
That this divergent factor also appears in $\Ima t_{nj}(i\nu_1+\ve+\delta,ik+\ve)$ is  
clear from the Neumann series of the LS, cf. Eq.~\eqref{210315.2}. 
  
 To derive this new IE  we first rewrite Eq.~\eqref{290116.9} as 
\begin{align}
\label{290116.10}
&\Ima t_{ij}(i\nu+\ve^-,ik+\ve)-\Ima t_{ij}(i\nu+\ve^+,ik+\ve)=
\Ima v_{ij}(i\nu+\ve^-,ik+\ve)-\Ima v_{ij}(i\nu+\ve^+,ik+\ve)\nn\\
&-\frac{\theta(k-m_\pi)m }{2\pi^2} \sum_n \int_{\vartheta(\nu)}^{k-m_\pi}
\frac{d\nu_1\nu_1^2}{k^2-\nu_1^2}
\Rea \Big\{ \big[v_{in}(i\nu+\ve^-,i\nu_1+\ve)- v_{in}(i\nu+\ve^+,i\nu_1+\ve)\big] \nn\\
&\times \big[t_{nj}(i\nu_1+\ve-\delta,ik+\ve)-t_{nj}(i\nu_1+\ve+\delta,ik+\ve)\big]\Big\}~,
\end{align}
where we have used again the continuity  of $\Rea v_{in}(i\nu+\ve^\pm,i\nu_1+\ve)$ and 
$\Rea t_{nj}(i\nu_1+\ve\pm \delta,ik+\ve)$ for $\delta\to 0$ and $\ve^\pm\to \ve$,
 Eq.~\eqref{290116.7},  that allows us to neglect the contribution 
\begin{align}
\label{290116.11}
&-\frac{\theta(k-m_\pi)m }{2\pi^2} \sum_n \int_{\vartheta(\nu)}^{k-m_\pi}
 \frac{d\nu_1\nu_1^2}{k^2-\nu_1^2}
\big[\Rea v_{in}(i\nu+\ve^-,i\nu_1+\ve)- \Rea v_{in}(i\nu+\ve^+,i\nu_1+\ve)\big]\nn\\
&\times \big[\Rea t_{nj}(i\nu_1+\ve-\delta,ik+\ve)-\Rea t_{nj}(i\nu_1+\ve+\delta,ik+\ve)\big]
\underset{}\longrightarrow 0~.
\end{align}
We remark here again, as in the footnote \ref{foot:290116.1}, 
 that the continuity referred is in $\ve^\pm \to \ve$ and $\delta\to 0$. This is the situation in 
Eq.~\eqref{290116.11} because one first takes these limits while keeping  $\ve\neq 0$. 
In this way the integrals remain finite and there is then no  ambiguous limit of the type $0 \times \infty$, that could appear because 
of the infrared divergences if simultaneously with $\ve^\pm \to \ve$ and $\delta\to 0$ 
 one also took the limit $\ve\to 0$ in the integrations.

Next, we employ the functions $\hhv_{ij}(k,k')$ and $\hht_{ij}(k,k';z)$. The later was already defined 
in Eq.~\eqref{180811.3}, while for the former an analogous definition holds  
\begin{align}
\label{290116.12}
\hhv_{ij}(p,p')&=p^{\ell_i+1}{p'}^{\ell_j+1} v_{ij}(p,p')~,
\end{align}
In this way the divergent behavior of $\Ima v_{ij}(i\nu+\ve^\pm,i\nu_1+\ve)$ as $1/\nu^{\ell_i+1}$ 
 ($1/{\nu_1}^{\ell_j+1})$ for $\nu(\nu_1)\to 0$ is not longer present in $\hhv_{ij}(i\nu+\ve^\pm,i\nu_1+\ve)$. 
 The same can be said for $t_{ij}(i\nu_1+\ve+\pm \delta,ik+\ve)$ and $\hht_{ij}(i\nu+\ve^\pm,ik+\ve)$.

Then, we  extract $1/[(i\nu)^{\ell_i+1}(ik)^{\ell_j+1}]$ as a common factor in Eq.~\eqref{290116.10}
 because $i^{\ell_i+\ell_j+2}=(-1)^{(\ell_i+\ell_j)/2}=\pm 1$. 
To simplify the notation we also utilize  the function $\Delta \hhv_{ij}(\nu,\nu_1)$, defined as,
\begin{align}
\label{290116.13}
\Delta\hhv_{ij}(\nu,\nu_1)&=\Ima \hhv_{ij}(i\nu+\ve^-,i\nu_1+\ve)-\Ima \hhv_{ij}(i\nu+\ve^+,i\nu_1+\ve)~,
\end{align}
 and $\mff_{ij}(\nu_1)$,  Eq.~\eqref{290116.8}.
Expanding the real part in the integrand of  Eq.~\eqref{290116.10} we can rewrite it as
\begin{align}
\label{290116.14}
\mff_{ij}(\nu)&=\Delta\hhv_{ij}(\nu,k)\nn\\
&-\frac{\theta(k-m_\pi)m}{2\pi^2}\sum_n \int_{\vartheta(\nu)}^{k-m_\pi}
\frac{d\nu_1\nu_1^2}{k^2-\nu_1^2}\Rea \frac{1}{(i\nu_1+\ve)^{\on}}
\Rea \Big\{ \big[\hhv_{in}(i\nu+\ve^-,i\nu_1+\ve)- \hhv_{in}(i\nu+\ve^+,i\nu_1+\ve)\big] \nn\\
&\times \big[\hht_{nj}(i\nu_1+\ve-\delta,ik+\ve)-\hht_{nj}(i\nu_1+\ve+\delta,ik+\ve)\big]\Big\}\nn\\
&+\frac{\theta(k-m_\pi)m}{2\pi^2}\sum_n \int_{\vartheta(\nu)}^{k-m_\pi}\frac{d\nu_1\nu_1^2}{k^2-\nu_1^2}
\Ima \frac{1}{(i\nu_1+\ve)^{\on}}\Ima \Big\{ \big[\hhv_{in}(i\nu+\ve^-,i\nu_1+\ve)-\hhv_{in}(i\nu+\ve^+,i\nu_1+\ve)\big] \nn\\
&\times \big[\hht_{nj}(i\nu_1+\ve-\delta,ik+\ve)-\hht_{nj}(i\nu_1+\ve+\delta,ik+\ve)\big]\Big\}~,
\end{align}
with 
\begin{align}
\label{290116.15}
\on&=2\ell_n+2~.
\end{align}
 The last integral in Eq.~\eqref{290116.14} does not contribute because 
\begin{align}
\label{290116.16}
&\Ima \Big\{ \big[\hhv_{in}(i\nu+\ve^-,i\nu_1+\ve)- \hhv_{in}(i\nu+\ve^+,i\nu_1+\ve)\big]
 \big[\hht_{nj}(i\nu_1+\ve-\delta,ik+\ve)-\hht_{nj}(i\nu_1+\ve+\delta,ik+\ve)\big]\Big\}\nn\\
&=\Rea \big[\hhv_{in}(i\nu+\ve^-,i\nu_1+\ve)- \hhv_{in}(i\nu+\ve^+,i\nu_1+\ve)\big] 
\Ima \big[\hht_{nj}(i\nu_1+\ve-\delta,ik+\ve)-\hht_{nj}(i\nu_1+\ve+\delta,ik+\ve)\big]\nn\\
&+ \Ima \big[\hhv_{in}(i\nu+\ve^-,i\nu_1+\ve)- \hhv_{in}(i\nu+\ve^+,i\nu_1+\ve)\big] 
\Rea \big[\hht_{nj}(i\nu_1+\ve-\delta,ik+\ve)-\hht_{nj}(i\nu_1+\ve+\delta,ik+\ve)\big]
\end{align}
is zero in the limit $\ve^\pm\to \ve$ and $\delta\to 0$ because of the continuity of both 
$\Rea\hhv(i\nu+\ve^\pm,i\nu_1+\ve)$ and $\Rea \hht(i\nu_1+\ve\pm \delta,ik+\ve)$, in order.
 The same remark as already done twice above, in footnote \ref{foot:290116.1}  
 and just after Eq.~\eqref{290116.11}, is in order here.\footnote{Note also that 
\begin{align}
\label{290116.17}
\Ima \frac{1}{(i\nu_1+\ve)^{\on}} = \frac{1}{2i}\left(
\frac{1}{(i\nu_1+\ve)^{\on}} - \frac{1}{(-i\nu_1+\ve)^{\on}}
\right)
\end{align}
 is null for finite $\nu_1$ with vanishing $\ve$.}

 Then, Eq.~\eqref{290116.14} can be finally written as
\begin{align}
\label{290116.18a}
&\mff_{ij}(\nu)=\Delta\hhv_{ij}(\nu,k)
+\frac{\theta(k-m_\pi)m }{4\pi^2}\sum_n \int_{\vartheta(\nu)}^{k-m_\pi}
\frac{d\nu_1\nu_1^2}{k^2-\nu_1^2}
\left(\frac{1}{(i\nu_1+\ve)^{\on}}+\frac{1}{(i\nu_1-\ve)^{\on}}\right)
 \Delta\hhv_{in}(\nu,\nu_1) \mff_{nj}(\nu_1)~,
\end{align}
where we have used again the continuity  of $\Rea \hhv_{in}(i\nu+\ve,i\nu_1+\vep\pm \delta)$ and 
$\Rea \hht_{nj}(i\nu_1+\ve\pm \delta,ik+\ve)$  for $\delta\to 0$ to express
\begin{align}
\label{290116.18b}
&\Rea \Big\{ \big[\hhv_{in}(i\nu+\ve^-,i\nu_1+\ve)- \hhv_{in}(i\nu+\ve^+,i\nu_1+\ve)\big] 
 \big[\hht_{nj}(i\nu_1+\ve-\delta,ik+\ve)-\hht_{nj}(i\nu_1+\ve+\delta,ik+\ve)\big]\Big\}
\nn\\
&\to 
-\Ima \big[\hhv_{in}(i\nu+\ve^-,i\nu_1+\ve)- \hhv_{in}(i\nu+\ve^+,i\nu_1+\ve)\big] 
\Ima \big[\hht_{nj}(i\nu_1+\ve-\delta,ik+\ve)-\hht_{nj}(i\nu_1+\ve+\delta,ik+\ve)\big]~.
\end{align}
 
 Eq.~\eqref{290116.18a} can be furthered simplified by taking into account that 
 $\Delta \hhv_{ij}(\nu,\nu_1)$   stems from the imaginary part in the difference of log-terms
  in Eq.~\eqref{180923.1},  namely, 
\begin{align}
\label{290116.18}
i \Delta\hhv_{ij}(\nu,\nu_1)&= \frac{1}{\pi^2}\int_{m_\pi^2}^\infty d\mu^2 \eta_{ij}(\mu^2) \rho_{ij}(\nu^2,\nu_1^2;\mu^2)\\
&\times \Big( \log\big[-(\nu+\nu_1)^2+\mu^2+i(\nu+\nu_1)0^+\big]
-\log\big[-(\nu-\nu_1)^2+\mu^2-i(\nu-\nu_1)0^+\big]  \nn\\
 &
- \left\{\log\big[-(\nu+\nu_1)^2+\mu^2+i(\nu+\nu_1)0^+\big]-\log\big[-(\nu-\nu_1)^2+\mu^2+i(\nu-\nu_1)0^+\big]\right\}\Big)\nn\\
&=\frac{1}{\pi^2}\int_{m_\pi^2}^\infty d\mu^2 \eta_{ij}(\mu^2) \rho_{ij}(\nu^2,\nu_1^2;\mu^2)\nn\\
&\times \left(
-\log\big[-(\nu-\nu_1)^2+\mu^2-i(\nu-\nu_1)0^+\big]+\log\big[-(\nu-\nu_1)^2+\mu^2+i(\nu-\nu_1)0^+\big]\right)~,\nn
\end{align}
where the first two log terms  correspond to $\Ima \hhv_{ij}(i\nu+\ve^-,i\nu_1+\ve)$ 
and the last two (shown between curly brackets) to $\Ima\hhv_{ij}(i\nu+\ve^+,ik+\ve)$. 
 The integrand is proportional to  the function $\rho_{ij}(\nu^2,\nu_1^2;\mu^2)$, which 
 stems from the kinematical expressions for the 
partial-wave projections. It is a function of $\nu^2$ and $\nu_1^2$ because 
 $v_{ij}(-p_1,p_2)=(-1)^{\ell_i} v_{ij}(p_1,p_2)$ (an analogously for $p_2$), 
Eq.~\eqref{180811.1}, which determines  
that $\hhv_{ij}(-p_1,p_2)=-\hhv_{ij}(p_1,p_2)$ and $\hhv_{ij}(p_1,-p_2)=-\hhv_{ij}(p_1,p_2)$. 
This in turn implies that $\hhv_{ij}(p_1,p_2)$ is an odd function of its arguments. 
The discontinuity $\Delta\hhv_{ij}(\nu,\nu_1)$ stems entirely from the log factor in 
the partial wave projection, which changes sign when $p_1(p_2)$ changes sign while keeping 
$p_2(p_1)$ fixed. Therefore, the function multiplying this log factor must be even in its 
arguments. This prefactor function is what we define as $\rho_{ij}(\nu^2,\nu_1^2)/\pi$.
  For explicit expression one can consider Eq.~\eqref{swpvm}, or any other contribution 
to a partial wave projected potential.

It follows from Eq.~\eqref{290116.18} that in order to end with a non vanishing imaginary
 part it is required that $(\nu_1-\nu)^2>\mu^2$  or, in other terms, that 
\begin{align}
\label{290116.20}
\nu_1=\nu\pm \sqrt{\mu^2+x^2}~~,~x\in \mathbb{R}~. 
\end{align}
 Due to $\vartheta(\nu)$ in the integral of Eq.~\eqref{290116.18a} 
only the plus sign matters as $\nu_1>\nu-m_\pi>\nu-\mu$. Therefore,  
for the range of values of $\nu_1$ relevant in Eq.~\eqref{290116.18a}, Eq.~\eqref{290116.18} simplifies as
\begin{align}
\label{170525.1}
 \Delta\hhv_{ij}(\nu,\nu_1) &=
\frac{1}{\pi}\int_{m_\pi}^\infty 2\mu d\mu \eta_{ij}(\mu^2)(-2) \rho_{ij}(\nu^2,\nu_1^2;\mu^2) \theta(\nu_1-\nu-\mu) \\
&=-\frac{2}{\pi}\int_{m_\pi}^{\nu_1-\nu} 2\mu d\mu \eta_{ij}(\mu^2)\rho_{ij}(\nu^2,\nu_1^2;\mu^2)~.\nn
\end{align}

The fact that the integrand in Eq.~\eqref{290116.18a} is nonzero only for $\nu_1>\nu+m_\pi$ also implies 
that one can replace $\vartheta(\nu)$ by $\nu+m_\pi$ since 
\begin{align}
\label{290116.21}
\nu+m_\pi\geq \epsilon(\nu)(|\nu|-m_\pi)\theta(|\nu|-m_\pi)~.
\end{align}
Of course, it is then required that $k-m_\pi>\nu+m_\pi$, because otherwise $\Delta\hhv_{ij}(\nu,\nu_1)$ 
is zero as  $\nu_1$ extends up to $k-m_\pi$. This condition on $k$ and $\nu$ 
can be accounted for by the presence of $\theta(k-2m_\pi-\nu)$ 
in front of the integral in Eq.~\eqref{290116.18a}, which can then be recast as
\begin{align}
\label{290116.22}
\mathfrak{f}_{ij}(\nu)= \Delta\hhv_{ij}(\nu,k)+
\frac{\theta(k-2m_\pi-\nu)m}{4\pi^2}\sum_n \int_{m_\pi+\nu}^{k-m_\pi}\frac{d\nu_1\nu_1^2}{k^2-\nu_1^2}
\left\{\frac{1}{(i\nu_1+\ve)^{\on}}+\frac{1}{(i\nu_1-\ve)^{\on}}\right\}
 \Delta\hhv_{in}(\nu,\nu_1)\mathfrak{f}_{nj}(\nu_1)
\end{align}

It is convenient to express this equation explicitly in terms of the spectral 
representation of the partial-wave projected potential 
as follows from Eq.~\eqref{170525.1}. It results
\begin{empheq}[box=\widefbox]{align}
\label{170525.2}
\mathfrak{f}_{ij}(\nu)=& -\frac{2}{\pi}\int_{m_\pi}^{k-\nu} 2\mu d\mu \eta_{ij}(\mu^2)\rho_{ij}(\nu^2,k^2;\mu^2) \\
-&\frac{\theta(k-2m_\pi-\nu)m}{2\pi^3}\sum_n \int_{\nu+m_\pi}^{k-m_\pi}\frac{d\nu_1\nu_1^2}{k^2-\nu_1^2}
S_n(\nu_1)\mff_{nj}(\nu_1)\int_{m_\pi}^{\nu_1-\nu}2\mu d\mu \eta_{in}(\mu^2)\rho_{in}(\nu^2,\nu_1^2;\mu^2)~,\nn
\end{empheq}
where
\begin{align}
\label{170525.3}
S_{n}(\nu_1)&=\frac{1}{(i\nu_1+\ve)^{2\ell_n+2}}+\frac{1}{(i\nu_1-\ve)^{2\ell_n+2}}~.
\end{align}

In order to calculate $\Delta_{ij}(p^2)$ we need to obtain $f_{ij}(-k)$, Eq.~\eqref{280116.3}, so that it is 
enough with solving 
the IE in Eq.~\eqref{170525.2} for the symmetric interval of values $\nu\in [-k+m_\pi,k-m_\pi]$. 

We would like also to stress that the IE of Eq.~\eqref{170525.2} has finite limits of integration
 and its solution is completely fixed by the knowledge of the 
discontinuity of the potential, $\Delta\hhv_{ij}(\nu,\nu_1)$. 
No counterterms enter in the latter because these 
are contact interactions and do not originate any contribution to $\Delta\hhv_{ij}(\nu,\nu_1)$. 
Therefore, there is no ambiguity whatsoever in the calculation of $\mff_{ij}(\nu)$, it is a neat 
output from the knowledge of the potential at finite distances, 
contrarily to the LS equation.

Strictly speaking Eq.~\eqref{170525.2} is not an IE because $\nu_1\geq \nu+m_\pi$, so that 
$\mff_{ij}(\nu)$ does not appear as unknown in the integral on the rhs of the equality. 
It is a recursion relation that allows to obtain $\mff_{ij}(\nu)$ by integrating an expression 
involving the values of $\mff_{ij}(\nu_1)$ with $\nu_1\geq \nu+m_\pi$.

$\bullet$ Now, let us discuss how to handle the previous IE in an appropriate way ready to be used 
to cure numerically the problem of the infrared divergences 
for $\ell_n\geq 1$, for which $\nu_1^2 S_{n}(\nu_1)$ diverges 
like $1/\nu_1^{2\ell_n}$ for $\nu_1\to 0$.

 The problem with the infrared behavior of the integrand originates when the integration in Eq.~\eqref{170525.2} approaches or 
crosses the value $\nu=-m_\pi$, which starts happening for $\nu\leq -m_\pi+\Delta$, with $\Delta>0$ but small so that 
numerically one would get problems in the evaluation of the integral to solve the IE. 
The point is to rewrite the IE in such a way as to be able to evaluate algebraically those integrals for which the 
integration variable  crosses zero.
 One should stress that at $\nu=-m_\pi$ one would have at most a branch point singularity in the 
discontinuity represented by $\mff_{ij}(\nu)$, without any divergence in the PWAs.

 A key point in our manipulations that follow is to make use of the 
symmetry under the exchange $\nu_1 \leftrightarrow -\nu_1$ in the function $\rho_{in}(\nu^2,\nu_1^2;\mu^2)$. 
  First, we symmetrize partially 
 the integral in Eq.~\eqref{170525.2} for $\nu+m_\pi<0$, which can then be expressed also as
\begin{align}
\label{170526.1}
&\theta(k-2m_\pi-\nu)\int_{\nu+m_\pi}^{-\nu-m_\pi}\frac{d\nu_1 \nu_1^2}{k^2-\nu_1^2}S_{n}(\nu_1)
\int_{m_\pi}^{\nu_1-\nu}2\mu d\mu \eta_{in}(\mu^2)\rho_{in}(\nu^2,\nu_1^2;\mu^2)\mff_{nj}(\nu_1)=
\left\{ \begin{array}{l}\nu_1\to -\nu_1\end{array}\right\}\\
=&\theta(k-2m_\pi-\nu)\int_{\nu+m_\pi}^{-\nu-m_\pi}\frac{d\nu_1 \nu_1^2}{k^2-\nu_1^2}S_{n}(\nu_1)
\int_{m_\pi}^{-\nu_1-\nu}2\mu d\mu \eta_{in}(\mu^2)\rho_{in}(\nu^2,\nu_1^2;\mu^2)\mff_{nj}(-\nu_1)~.\nn
\end{align}
 Taking this result into account we rewrite Eq.~\eqref{170525.2} as
\begin{align}
\label{170526.2}
\mathfrak{f}_{ij}(\nu)& = -\frac{2}{\pi}\int_{m_\pi}^{k-\nu} 2\mu d\mu \eta_{ij}(\mu^2)\rho_{ij}(\nu^2,k^2;\mu^2) 
-\frac{\theta(k-2m_\pi-\nu)m}{4\pi^3}\sum_n \int_{\nu+m_\pi}^{-\nu-m_\pi}\frac{d\nu_1\nu_1^2}{k^2-\nu_1^2}
S_{n}(\nu_1)\\
\times &\Big[\mff_{nj}(\nu_1)\int_{m_\pi}^{\nu_1-\nu}2\mu d\mu \eta_{in}(\mu^2)\rho_{in}(\nu^2,\nu_1^2;\mu^2)
+ \mff_{nj}(-\nu_1)\int_{m_\pi}^{-\nu_1-\nu}2\mu d\mu \eta_{in}(\mu^2)\rho_{in}(\nu^2,\nu_1^2;\mu^2)
\Big]\nn\\
-&\frac{\theta(k-2m_\pi-\nu)m}{2\pi^3}\sum_n\int_{-\nu-m_\pi}^{k-m_\pi}\frac{d\nu_1\nu_1^2}{k^2-\nu_1^2}
 S_{n}(\nu_1)  \mff_{nj}(\nu_1)\int_{m_\pi}^{\nu_1-\nu}2\mu d\mu \eta_{in}(\mu^2)\rho_{in}(\nu^2,\nu_1^2;\mu^2)~,\nn
\end{align}
 Notice  that $k-m_\pi\geq -\nu-m_\pi$ since $\nu\geq -k$. 
The first integral in $\nu_1$ can be restricted to positive values of this variable  because the integrand is even and then 
we have for $\nu>-m_\pi$,
\begin{empheq}[box=\widefbox]{align}
\label{170526.3}
\mathfrak{f}_{ij}(\nu)& = -\frac{2}{\pi}\int_{m_\pi}^{k-\nu} 2\mu d\mu \eta_{ij}(\mu^2)\rho_{ij}(\nu^2,k^2;\mu^2) 
-\frac{\theta(k-2m_\pi-\nu)m}{2\pi^3}\sum_n \int_{0}^{-\nu-m_\pi}\frac{d\nu_1\nu_1^2}{k^2-\nu_1^2}
S_{n}(\nu_1)\\
\times &\Big[\mff_{nj}(\nu_1)\int_{m_\pi}^{\nu_1-\nu}2\mu d\mu \eta_{in}(\mu^2)\rho_{in}(\nu^2,\nu_1^2;\mu^2)
+ \mff_{nj}(-\nu_1)\int_{m_\pi}^{-\nu_1-\nu}2\mu d\mu \eta_{in}(\mu^2)\rho_{in}(\nu^2,\nu_1^2;\mu^2)
\Big]\nn\\
-&\frac{\theta(k-2m_\pi-\nu)m}{2\pi^3}\sum_n \int_{-\nu-m_\pi}^{k-m_\pi}\frac{d\nu_1\nu_1^2}{k^2-\nu_1^2}
 S_{n}(\nu_1)\mff_{nj}(\nu_1)\int_{m_\pi}^{\nu_1-\nu}2\mu d\mu \eta_{in}(\mu^2)\rho_{in}(\nu^2,\nu_1^2;\mu^2)~,\nn
\end{empheq}

 However, Eq.~\eqref{170526.3} is still  troublesome for its numerical implementation  for $\ell_n\geq 1$ because 
  the variable $\nu_1$ in the integrals can be 
arbitrarily close to zero when $\nu\to -m_\pi$. 
We then proceed to express the first iterated form of Eq.~\eqref{170525.2}. 

\begin{figure}
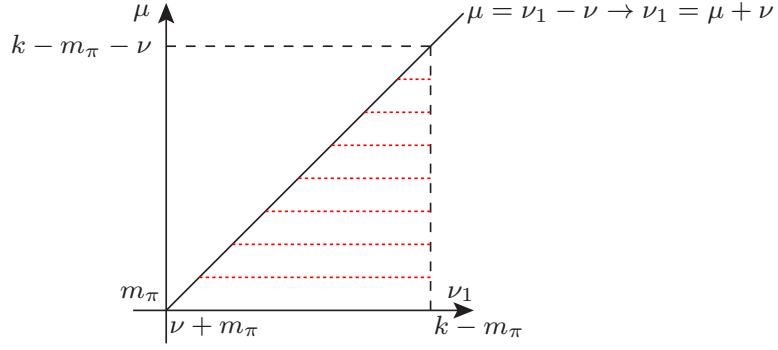

\begin{center}
\scalebox{1.25}{
\begin{axopicture}(100,100)
\LongArrow(10,0)(10,100) 
\Text(0,100)[l]{\scriptsize{$\mu$}}
\LongArrow(0,10)(100,10) 
\Text(95,15)[l]{\scriptsize{$\nu_1$}}
\Line(10,10)(100,100) 
\Line[dash](10,90)(90,90)
\Line[dash](90,90)(90,10)
\Line[color=red,dash,dsize=1](20,20)(90,20)
\Line[color=red,dash,dsize=1](30,30)(90,30)
\Line[color=red,dash,dsize=1](40,40)(90,40)
\Line[color=red,dash,dsize=1](50,50)(90,50)
\Line[color=red,dash,dsize=1](60,60)(90,60)
\Line[color=red,dash,dsize=1](70,70)(90,70)
\Line[color=red,dash,dsize=1](80,80)(90,80)
\Text(11,5)[l]{\scriptsize{$\nu+m_\pi$}}
\Text(91,5)[l]{\scriptsize{$k-m_\pi$}}
\Text(-4,15)[l]{\scriptsize{$m_\pi$}}
\Text(-37,90)[l]{\scriptsize{$k-m_\pi-\nu$}}
\Text(101,100)[l]{\scriptsize{$\mu=\nu_1-\nu\to \nu_1=\mu+\nu$}}
\end{axopicture}
}
\caption{{\small Integration region for the double integral of $\nu_1$, $\mu$ in Eq.~\eqref{170525.2}. For
 $\mu\in[m_\pi,k-m_\pi-\nu]$ then $\nu_1\in[\mu+\nu,k-m_\pi]$.}
 \label{fig.170526.1}}
\end{center}
\end{figure}

In order to end with integrals which can be calculated algebraically 
 when $\nu_1$ crosses zero, we first exchange the order of the integrals in the last term 
of Eq.~\eqref{170525.2}. This is straightforward, particularly taking into account 
the integration region explicitly shown in Fig.~\ref{fig.170526.1}. Then we have the following 
equivalent form of Eq.~\eqref{170525.2}
\begin{align}
\label{170526.4}
\mathfrak{f}_{ij}(\nu)=& -\frac{2}{\pi}\int_{m_\pi}^{k-\nu} 2\mu d\mu \eta_{ij}(\mu^2)\rho_{ij}(\nu^2,k^2;\mu^2) \\
-&\frac{\theta(k-2m_\pi-\nu)m}{2\pi^3}\sum_n \int_{m_\pi}^{k-m_\pi-\nu} 2\mu d\mu \eta_{in}(\mu^2)
\int_{\mu+\nu}^{k-m_\pi}\frac{d\nu_1\nu_1^2}{k^2-\nu_1^2}
S_{n}(\nu_1)\rho_{in}(\nu^2,\nu_1^2;\mu^2)\mff_{nj}(\nu_1)~.\nn
\end{align}

The contribution of only the first iteration of the previous equation reads
\begin{align}
\label{170526.5}
\frac{\theta(k-2m_\pi-\nu)m}{\pi^4}\sum_n \int_{m_\pi}^{k-m_\pi-\nu}2\mu d\mu \eta_{in}(\mu^2)
&\int_{\mu+\nu}^{k-m_\pi}
\frac{d\nu_1 \nu_1^2}{k^2-\nu_1^2}S_{n}(\nu_1)\rho_{in}(\nu^2,\nu_1^2;\mu^2)\nn\\
\times&\int_{m_\pi}^{k-\nu_1}2\mu'd\mu'\eta_{nj}({\mu'}^2)\rho_{nj}(\nu_1^2,k^2;{\mu'}^2)~.
\end{align}

\begin{figure}
\begin{center}
\scalebox{1.25}{
\begin{axopicture}(100,100)
\LongArrow(10,0)(10,100) 
\Text(0,100)[l]{\scriptsize{$\mu'$}}
\LongArrow(0,10)(100,10) 
\Text(95,15)[l]{\scriptsize{$\nu_1$}}
\Line(90,10)(10,90) 
\Line[color=red,dash,dsize=1](10,20)(80,20)
\Line[color=red,dash,dsize=1](10,30)(70,30)
\Line[color=red,dash,dsize=1](10,40)(60,40)
\Line[color=red,dash,dsize=1](10,50)(50,50)
\Line[color=red,dash,dsize=1](10,60)(40,60)
\Line[color=red,dash,dsize=1](10,70)(30,70)
\Line[color=red,dash,dsize=1](10,80)(20,80)
\Text(11,5)[l]{\scriptsize{$\mu+\nu$}}
\Text(91,5)[l]{\scriptsize{$k-m_\pi$}}
\Text(-4,15)[l]{\scriptsize{$m_\pi$}}
\Text(-27,90)[l]{\scriptsize{$k-\mu-\nu$}}
\Text(37,70)[l]{\scriptsize{$\mu'=-\nu_1+k\to \nu_1=-\mu'+k$}}
\end{axopicture}
}
\caption{{\small Integration region for the double integral of $\nu_1$, $\mu'$ in Eq.~\eqref{170526.5}. For
 $\mu'\in[m_\pi,k-\mu-\nu]$ then $\nu_1\in[\mu+\nu,k-\mu']$.}
 \label{fig.170526.2}}
\end{center}
\end{figure}

We have to change again the order of the last two integrals in the variables $\nu_1$ and $\mu'$. This is clear from the integration region 
given in Fig.~\ref{fig.170526.2}, and we have the equivalent expression for Eq.~\eqref{170526.5}
\begin{align} 
\label{170526.6}
\frac{\theta(k-2m_\pi-\nu)m}{\pi^4}\sum_n &\int_{m_\pi}^{k-m_\pi-\nu}2\mu d\mu \eta_{in}(\mu^2)
\int_{m_\pi}^{k-\mu-\nu}2\mu'd\mu'\eta_{nj}({\mu'}^2)\nn\\
\times &\int_{\mu+\nu}^{k-\mu'}
\frac{d\nu_1 \nu_1^2}{k^2-\nu_1^2}S_{n}(\nu_1)\rho_{in}(\nu^2,\nu_1^2;\mu^2)\rho_{nj}(\nu_1^2,k^2;{\mu'}^2)~.
\end{align}
This expression is adequate to perform the last integral algebraically for $\ell_n\geq 1$.\footnote{If this is too hard, an option 
is then to perform the expansion in $\nu_1^2$ around $\nu_1=0$ of 
$\sum_n \rho_{in}(\nu^2,\nu_1^2;\mu^2)\rho_{nj}(\nu_1^2,k^2;{\mu'}^2)$ up to order $\nu_1^{2\ell_n-2}$, $\ell_n\geq 1$, that we indicate by 
$\sum_m a_m \nu_1^{2m}$. 
One then subtracts and adds this to the previous term. The difference 
$\sum_n \rho_{in}(\nu^2,\nu_1^2;\mu^2)\rho_{nj}(\nu_1^2,k^2;{\mu'}^2)-\sum_m a_m \nu_1^{2m}$ can be integrated 
numerically, while the integration involving only the expansion can be done algebraically. This kind of trick is always at 
our disposal in any other case with similar integrals.}

Now, we consider the term that comprises all the higher order iterations that result from Eq.~\eqref{170526.4}, it reads
\begin{align}
\label{170526.7}
&
\theta(k-2m_\pi-\nu) \left(\frac{m}{2\pi^3}\right)^2
\sum_{n_1,n_2} \int_{m_\pi}^{k-m_\pi-\nu} 2\mu d\mu \eta_{in_1}(\mu^2)
\int_{\mu+\nu}^{k-m_\pi}\frac{d\nu_1\nu_1^2}{k^2-\nu_1^2}
S_{n_1}(\nu_1)\rho_{in_1}(\nu^2,\nu_1^2;\mu^2)\\
\times&
\theta(k-2m_\pi-\nu_1)\int_{m_\pi}^{k-m_\pi-\nu_1}2\mu' d\mu' \eta_{n_1 n_2}({\mu'}^2)
\int_{\mu'+\nu_1}^{k-m_\pi}\frac{d\nu_2 \nu_2^2}{k^2-\nu_2^2}S_{n_2}(\nu_2)\rho_{n_1n_2}(\nu_1^2,\nu_2^2;{\mu'}^2)\mff_{n_2j}(\nu_2)~.\nn
\end{align}
An important point  is to take into account the step function $\theta(k-2m_\pi-\nu_1)$ present in the second line 
of Eq.~\eqref{170526.7}. 
This implies that $\nu_1<k-2m_\pi$, which 
also limits 
the range of the integration in the variable $\mu$, so that $\mu<k-2m_\pi-\nu$ to guarantee that $\mu+\nu<k-2m_\pi$. 
 Then, one also must require that $k-2m_\pi-\nu>m_\pi$. 
 As a result, Eq.~\eqref{170526.7} becomes
\begin{align}
\label{170526.8}
&\theta(k-3m_\pi-\nu)\left(\frac{m}{2\pi^3}\right)^2\sum_{n_1,n_2} \int_{m_\pi}^{k-2m_\pi-\nu} 2\mu d\mu \eta_{in_1}(\mu^2)
\int_{\mu+\nu}^{k-2m_\pi}\frac{d\nu_1\nu_1^2}{k^2-\nu_1^2}
S_{n_1}(\nu_1)\rho_{in_1}(\nu^2,\nu_1^2;\mu^2)\\
\times&
\int_{m_\pi}^{k-m_\pi-\nu_1}2\mu' d\mu' \eta_{n_1 n_2}({\mu'}^2)
\int_{\mu'+\nu_1}^{k-m_\pi}\frac{d\nu_2 \nu_2^2}{k^2-\nu_2^2}S_{n_2}(\nu_2)\rho_{n_1n_2}(\nu_1^2,\nu_2^2;{\mu'}^2)\mff_{n_2j}(\nu_2)~.\nn
\end{align}

\begin{figure}
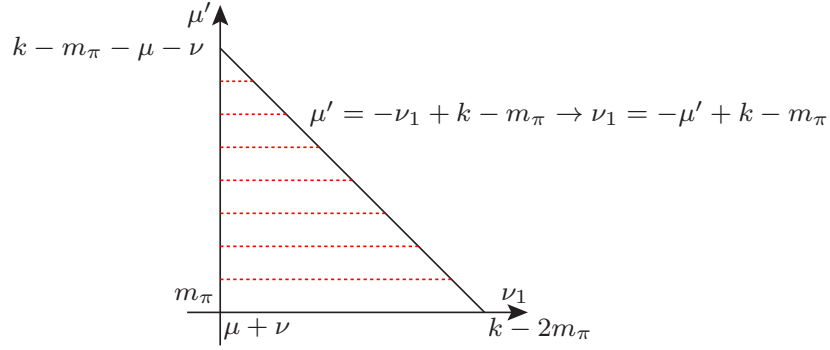

\begin{center}
\scalebox{1.25}{
\begin{axopicture}(100,100)
\LongArrow(10,0)(10,100) 
\Text(0,100)[l]{\scriptsize{$\mu'$}}
\LongArrow(0,10)(100,10) 
\Text(95,15)[l]{\scriptsize{$\nu_1$}}
\Line(90,10)(10,90) 
\Line[color=red,dash,dsize=1](10,20)(80,20)
\Line[color=red,dash,dsize=1](10,30)(70,30)
\Line[color=red,dash,dsize=1](10,40)(60,40)
\Line[color=red,dash,dsize=1](10,50)(50,50)
\Line[color=red,dash,dsize=1](10,60)(40,60)
\Line[color=red,dash,dsize=1](10,70)(30,70)
\Line[color=red,dash,dsize=1](10,80)(20,80)
\Text(11,5)[l]{\scriptsize{$\mu+\nu$}}
\Text(91,5)[l]{\scriptsize{$k-2 m_\pi$}}
\Text(-4,15)[l]{\scriptsize{$m_\pi$}}
\Text(-53,90)[l]{\scriptsize{$k-m_\pi-\mu-\nu$}}
\Text(37,70)[l]{\scriptsize{$\mu'=-\nu_1+k-m_\pi\to \nu_1=-\mu'+k-m_\pi$}}
\end{axopicture}
}
\caption{{\small Integration region for the double integral of $\nu_1$, $\mu'$ in Eq.~\eqref{170526.8}. For
  $\mu'\in[m_\pi,k-m_\pi-\mu-\nu]$ then $\nu_1\in[\mu+\nu,k-m_\pi-\mu']$.}
 \label{fig.170526.3}}
\end{center}
\end{figure}

Let us first exchange the order of the integration in the variables $\nu_1$ and $\mu'$ in Eq.~\eqref{170526.8}, 
which is analogous to the one already done 
to obtain Eq.~\eqref{170526.6} with $k-m_\pi$ instead of $k$ in the limits of integration. Nonetheless, 
attending to the region of integration shown in Fig.~\ref{fig.170526.3} the result is also clear. One obtains
\begin{align}
\label{170526.9}
&\theta(k-3m_\pi-\nu)\left(\frac{m}{2\pi^3}\right)^2\sum_{n_1,n_2} \int_{m_\pi}^{k-2m_\pi-\nu} 2\mu d\mu \eta_{i n_1}(\mu^2)
\int_{m_\pi}^{k-m_\pi-\mu-\nu}2\mu' d\mu' \eta_{n_1 n_2}({\mu'}^2)\\
\times & 
\int_{\mu+\nu}^{k-m_\pi-\mu'}\frac{d\nu_1\nu_1^2}{k^2-\nu_1^2}
S_{n_1}(\nu_1)\rho_{in_1}(\nu^2,\nu_1^2;\mu^2)
\int_{\mu'+\nu_1}^{k-m_\pi}\frac{d\nu_2 \nu_2^2}{k^2-\nu_2^2}S_{n_2}(\nu_2)\rho_{n_1n_2}(\nu_1^2,\nu_2^2;{\mu'}^2)\mff_{n_2j}(\nu_2)~.\nn
\end{align}

\begin{figure}
\begin{center}
\scalebox{1.25}{
\begin{axopicture}(100,100)
\LongArrow(10,0)(10,100) 
\Text(-2,100)[l]{\scriptsize{$\nu_2$}}
\LongArrow(0,10)(100,10) 
\Text(95,15)[l]{\scriptsize{$\nu_1$}}
\Line(10,10)(100,100) 
\Line[dash](10,90)(90,90)
\Line[dash](90,90)(90,10)
\Line[color=red,dash,dsize=1](10,20)(20,20)
\Line[color=red,dash,dsize=1](10,30)(30,30)
\Line[color=red,dash,dsize=1](10,40)(40,40)
\Line[color=red,dash,dsize=1](10,50)(50,50)
\Line[color=red,dash,dsize=1](10,60)(60,60)
\Line[color=red,dash,dsize=1](10,70)(70,70)
\Line[color=red,dash,dsize=1](10,80)(80,80)
\Text(11,5)[l]{\scriptsize{$\mu+\nu$}}
\Text(91,5)[l]{\scriptsize{$k-m_\pi-\mu'$}}
\Text(-27,15)[l]{\scriptsize{$\nu+\mu+\mu'$}}
\Text(-22,90)[l]{\scriptsize{$k-m_\pi$}}
\Text(101,100)[l]{\scriptsize{$\nu_2=\nu_1+\mu'\to \nu_1=\nu_2-\mu'$}}
\end{axopicture}
}
\caption{{\small Integration region for the double integral of $\nu_1$, $\nu_2$ in Eq.~\eqref{170526.9}. For
 $\nu_2\in[\mu+\mu'+\nu,k-m_\pi]$ then $\nu_1\in[\mu+\nu,\nu_2-\mu']$.}
 \label{fig.170526.4}}
\end{center}
\end{figure}

Next, we change the order of integration in the variables $\nu_1$ and $\nu_2$, which is also straightforward by considering the 
integration region shown in Fig.~\ref{fig.170526.4}. In terms of this change Eq.~\eqref{170526.9} becomes
\begin{align}
\label{170526.10}
&\frac{\theta(k-3m_\pi-\nu)m^2}{(2\pi^3)^2}\sum_{n_1,n_2}\int_{m_\pi}^{k-2m_\pi-\nu}2\mu d\mu \eta_{in_1}(\mu^2)\int_{m_\pi}^{k-m_\pi-\mu-\nu} 
2\mu' d\mu' \eta_{n_1n_2}({\mu'}^2)\int_{\mu+\mu'+\nu}^{k-m_\pi} \frac{d\nu_2 \nu_2^2}{k^2-\nu_2^2}S_{n_2}(\nu_2) \mff_{n_2 j}(\nu_2)\\
\times & 
\int_{\mu+\nu}^{\nu_2-\mu'} \frac{d\nu_1 \nu_1^2}{k^2-\nu_1^2}S_{n_1}(\nu_1)\rho_{in_1}(\nu^2,\nu_1^2;\mu^2)\rho_{n_1n_2}(\nu_1^2,\nu_2^2;{\mu'}^2)~.
\nn
\end{align}
Putting together the intermediate expressions obtained in Eqs.~\eqref{170526.6} and \eqref{170526.10},
 we have for the first-iterated form of Eq.~\eqref{170526.4}
\begin{empheq}[box=\widefbox]{align}
\label{170526.11}
\mff_{ij}(\nu)&= -\frac{2}{\pi}\int_{m_\pi}^{k-\nu} 2\mu d\mu \eta_{ij}(\mu^2)\rho_{ij}(\nu^2,k^2;\mu^2)
+\frac{\theta(k-2m_\pi-\nu)m}{\pi^4}\sum_n \int_{m_\pi}^{k-m_\pi-\nu}2\mu d\mu \eta_{in}(\mu^2)\\
\times & \int_{m_\pi}^{k-\mu-\nu}2\mu'd\mu'\eta_{nj}({\mu'}^2) \int_{\mu+\nu}^{k-\mu'}
\frac{d\nu_1 \nu_1^2}{k^2-\nu_1^2}S_{n}(\nu_1)\rho_{in}(\nu^2,\nu_1^2;\mu^2)\rho_{nj}(\nu_1^2,k^2;{\mu'}^2)\nn\\
+&\theta(k-3m_\pi-\nu)\left(\frac{m}{2\pi}\right)^2
\sum_{n_1,n_2}\int_{m_\pi}^{k-2m_\pi-\nu}2\mu d\mu \eta_{in_1}(\mu^2)\int_{m_\pi}^{k-m_\pi-\mu-\nu} 
2\mu' d\mu' \eta_{n_1n_2}({\mu'}^2)\nn\\
\times & 
\int_{\mu+\mu'+\nu}^{k-m_\pi} \frac{d\nu_2 \nu_2^2}{k^2-\nu_2^2}S_{n_2}(\nu_2) \mff_{n_2 j}(\nu_2)
\int_{\mu+\nu}^{\nu_2-\mu'} \frac{d\nu_1 \nu_1^2}{k^2-\nu_1^2}S_{n_1}(\nu_1)\rho_{in_1}(\nu^2,\nu_1^2;\mu^2)\rho_{n_1n_2}(\nu_1^2,\nu_2^2;{\mu'}^2)~.
\nn
\end{empheq}

Then, we have at our disposal several expressions that are used in different intervals of $\nu$ so as  
to avoid the infrared singularities in $S_{n}$ for $\ell_n>0$
\begin{align}
\label{170526.12}
\mathfrak{f}_{ij}(\nu)&=\left\{
\begin{array}{ll}
\text{Eq.~\eqref{170525.2} or \eqref{170526.4}} & -m_\pi/2<\nu~,\\
\text{Eq.~\eqref{170526.11}} &  -3m_\pi/2<\nu<-m_\pi/2~,\\
\text{Eq.~\eqref{170526.3}} &  \nu<-3m_\pi/2~,
\end{array}
\right.
\end{align}

 $\bullet$  In order to apply numerically Eq.~\eqref{170526.3} for  $-3m_\pi/2>\nu$ 
one could  add and subtract the first $\ell_n$ terms in the Taylor 
series in powers of  $\nu_1^2$ of the even function between brackets in this equation, namely, 
\begin{align}
\label{170526.13}
h(\nu_1^2)&=\mff_{nj}(\nu_1)\int_{m_\pi}^{\nu_1-\nu}2\mu d\mu \eta_{in}(\mu^2)\rho_{in}(\nu^2,\nu_1^2;\mu^2)
+ \mff_{nj}(-\nu_1)\int_{m_\pi}^{-\nu_1-\nu}2\mu d\mu \eta_{in}(\mu^2)\rho_{in}(\nu^2,\nu_1^2;\mu^2)~.
\end{align}
Let us denote by $Q_{\ell_n}(\nu_1^2)$ the $(\ell_n-1)$-degree polynomial in the variable $\nu_1^2$ 
resulting from the first $\ell_n$ terms 
in the Taylor expansion of the $h(\nu_1^2)$ around 0, namely, 
\begin{align}
\label{170601.1}
Q_{\ell_n}(\nu_1^2)=&
\sum_{i=0}^{\ell_n-1}\frac{1}{i !}
 \frac{\partial^i h(0)}{\partial {\nu_1^2}^i} \nu_1^{2i}~,
\end{align}
with the understanding that $Q_{\ell_n}(\nu_1^2)=0$ for $\ell_n=0$. Due to the complicated expression of $h(\nu_1^2)$ 
it seems that the most appropriate strategy is to evaluate numerically the different order derivatives of $h(\nu_1^2)$ at $\nu_1=0$ 
 needed in Eq.~\eqref{170601.1} by 
making use of the already known values of $\mathfrak{f}(\nu_1)$ for $\nu_1\geq m_\pi$. In this way, the integral 
\begin{align}
\label{170601.2}
\int_{0}^{-\nu-m_\pi}\frac{d\nu_1\nu_1^2}{k^2-\nu_1^2} S_{n}(\nu_1)  Q_{\ell_n}(\nu_1^2)
\end{align}
can be done algebraically because the polynomial $Q_{\ell_n}(\nu_1^2)$ 
only involves even powers of $\nu_1$.  
Furthermore, as  the  difference $h(\nu_1^2)-Q_{\ell_n}(\nu_1^2)$ times the extra factor of
 $\nu_1^2$ from the measure vanishes as $\nu_1^{2l+2}$ for $\nu_1\to 0$, one has the cancellation of the  divergence in 
 $S_{\ell_n}(\nu_1)\underset{\nu_1\to 0}\longrightarrow \nu_1^{-2\ell_n-2}$.  Therefore,  the integral 
\begin{align}
\label{170601.2b}
\int_{0}^{-\nu-m_\pi}\frac{d\nu_1\nu_1^2}{k^2-\nu_1^2} S_{n}(\nu_1) \big[h(\nu_1^2)-Q_{\ell_n}(\nu_1^2)\big]
\end{align}
can be done numerically.

$\bullet$ With the final version of the IE used for calculating $\mff_{ij}(\nu)$, according Eq.~\eqref{170526.12} and 
the previous remark on how to treat numerically with Eq.~\eqref{170526.3}, 
let us explain in more detail the absence of divergences that might stem from the divergent behavior of 
$S_{n}(\nu_1)\nu_1^2$ for $\nu_1\to 0$ as $1/\nu_1^{\bar{n}-2}$. 
For the values $\nu>-m_\pi/2$ this statement is clear because the integration variable $\nu_1$ 
in Eq.~\eqref{170525.2}  never approaches to zero since the lower limit of integration $\nu+m_\pi>m_\pi/2$.

In the case of the partially symmetric form of the IE for $-3m_\pi/2<\nu$, Eq.~\eqref{170526.3}, we already 
mentioned above that the integration of Eq.~\eqref{170601.2} exists and can be done algebraically
 because the integrand is a function of $\nu_1^2$, by construction all the functions in the integral 
 are even in $\nu_1$. 
 Therefore, the primitive of the 
integral can be taken to be zero at the lower limit of integration ($\nu_1=0$), while the upper one is always 
larger than $m_\pi/2$. Namely, we are referring to the fact that 
 \begin{align}
 \label{180820.1}
& \int_0^{-\nu-m_\pi}d\nu_1\left(\frac{1}{(\nu_1+i\vep)^{2m}}+\frac{1}{(\nu_1-i\vep)^{2m}}\right)=
 -\frac{1}{2m-1}\left(\frac{1}{(\nu_1+i\ve)^{2m-1}}+\frac{1}{(\nu_1-i\ve)^{2m-1}}\right)_{\nu_1=-\nu-m_\pi}\\
& +\frac{1}{2m-1}\left(\frac{1}{(\nu_1+i\ve)^{2m-1}}+\frac{1}{(\nu_1-i\ve)^{2m-1}}\right)_{\nu_1=0}
=\frac{2}{2m-1}\frac{1}{(\nu+m_\pi)^{2m-1}}~,
\nn
 \end{align}
 for $m\geq 1$. 
Giving more details in this respect, let us prove that the primitive of any integral of the form
\begin{align}
\label{180124.1b}
\int \frac{d\nu_1\,\nu_1^{2m}}{k^2-\nu_1^2}\big[\frac{1}{(\nu_1+i\ve)^{2n}}+\frac{1}{(\nu_1-i\ve)^{2n}}\big]
={\cal F}_m(\nu)~,
\end{align}
can be chosen such that ${\cal F}_m(0)=0$, without any additive integration constant, which does not matter since 
we subtract this primitive between two integration limits.\footnote{This is nothing but an 
exemplification of the fact that the primitive of an even function is an odd function, 
if we do not include any additive integration constant. 
As a result the primitive function vanishes in the origin.}

For $m=n$ the result is clear
\begin{align}
\label{180124.2}
\int\frac{d\nu_1}{k^2-\nu_1^2}=\frac{1}{k}{\rm{arctanh}}\frac{\nu_1}{k}~.
\end{align}
For $m>n$ the result follows because we can always reduce in steps of one the power of $\nu_1^2$ in the numerator, e.g. for 
$m=1$ we have
\begin{align}
\label{180124.3}
\int\frac{d\nu_1 \nu_1^2}{k^2-\nu_1^2}=
\int\frac{d\nu_1 (\nu_1^2-k^2+k^2)}{k^2-\nu_1^2}=-\nu_1+k\,{\rm{arctanh}}\frac{\nu_1}{k}~.
\end{align}
In more general terms for $m>n$ the numerator is a polynomial $Q(\nu_1^2)$ of degree $m-n$ in $\nu_1^2$. Thus,
\begin{align}
\label{180124.4}
\int\frac{d\nu_1 Q(\nu_1^2)}{k^2-\nu_1^2}=\int\frac{d\nu_1 \big[Q(\nu_1^2)-Q(k^2)+Q(k^2)\big]}{k^2-\nu_1^2}
=-\int d\nu_1 Q'(\nu_1)^2+Q(k^2)\int\frac{d\nu_1}{k^2-\nu_1^2}~,
\end{align}
with $Q'(\nu_1^2)=dQ(k^2)/d\nu_1^2+\ldots$. 
The first integral in the last term is zero at $\nu_1=0$ because it is the integration 
of a polynomial of degree $m-n-1$ in $\nu_1^2$ (no additive subtraction constant is included).

Let us now consider $m<n$. We reach the same conclusion by applying the following trick, $p=n-m$,
\begin{align}
\label{180124.5}
\int\frac{d\nu_1}{k^2-\nu_1^2}\big[\frac{1}{(\nu_1+i\ve)^{2p}}+\frac{1}{(\nu_1-i\ve)^{2p}}\big]
=\frac{1}{k^2}\int\frac{d\nu_1(k^2-\nu_1^2+\nu_1^2)}{k^2-\nu_1^2}
\big[\frac{1}{(\nu_1+i\ve)^{2p}}+\frac{1}{(\nu_1-i\ve)^{2p}}\big]\\
=\frac{1}{k^2}\int d\nu_1 \big[\frac{1}{(\nu_1+i\ve)^{2p}}+\frac{1}{(\nu_1-i\ve)^{2p}}\big]
+\frac{1}{k^2}\int\frac{d\nu_1}{k^2-\nu_1^2}\big[\frac{1}{(\nu_1+i\ve)^{2(p-1)}}+\frac{1}{(\nu_1-i\ve)^{2(p-1)}}\big]~.\nn
\end{align}
The integral before the last one is
\begin{align}
\label{180124.6}
\frac{1}{k^2}\int d\nu_1 \big[\frac{1}{(\nu_1+i\ve)^{2p}}+\frac{1}{(\nu_1-i\ve)^{2p}}\big]
=-\frac{1}{k^2(2p-1)}\big[\frac{1}{(\nu_1+i\ve)^{2p-1}}+\frac{1}{(\nu_1-i\ve)^{2p-1}}\big]
\end{align}
which is zero for $\nu_1=0$. Finally, 
the last term in Eq.~\eqref{180124.5} is of the same type as the starting point and can be treated in the same way
 giving rise again to a finite amount of terms of the form in Eq.~\eqref{180124.6}.

Regarding Eq.~\eqref{170526.11}, applied when $-3m_\pi/2<\nu<-m_\pi/2$, the integrals for which the integration 
variable $\nu_1$ can cross zero correspond to
\begin{align}
\label{180820.2}
\int_{\mu+\nu}^{\nu_2-\mu'} \frac{d\nu_1 \nu_1^2}{k^2-\nu_1^2}
S_{n_1}(\nu_1)\rho_{in_1}(\nu^2,\nu_1^2;\mu^2)\rho_{n_1n_2}(\nu_1^2,\nu_2^2;{\mu'}^2)~.
\end{align}
When this integration is particularized to $\nu_2=k$ we obtain the integral at the end of the second term in 
the rhs of Eq.~\eqref{170526.11}.  For $\mu\in [m_\pi,3m_\pi/2]$ the lower limit of integration 
might cross zero and the upper limit of integration is in the range 
$\nu_2-\mu'\in [\mu+\nu,k-2m_\pi]$. The key point for the absence of divergences in the previous integral 
is the appearance of the product of the two $\rho's$, 
$\rho_{in_1}(\nu^2,\nu_1^2;\mu^2)\rho_{n_1n_2}(\nu_1^2,\nu_2^2;{\mu'}^2)$. 
This makes that when calculating the primitive of this integral and taking its difference in the 
corresponding limits of integrations, the latter have a finite limit when any of them (or both simultaneously) 
tends to zero.

To discuss this point, let us first consider a $P$-wave projection of a central potential for which
 we pick up in the numerator  only a factor 
of $\cos\theta$, so that from the integral 
\begin{align}
\label{180819.3}
\int_{-1}^{+1}\frac{zdz}{z-\xi}=2+\xi\int_{-1}^{+1}\frac{dz}{z-\xi}
\end{align}
the only contribution to the discontinuity of the potential arises from the last term which is proportional 
to $\xi(\nu^2,\nu_1^2,\mu^2)$. 
Therefore, $\rho(\nu^2,\nu_1^2;\mu^2)$ for a $P$ wave is proportional to 
\begin{align}
\label{180820.3}
2\nu\nu_1\xi(\nu^2,\nu_1^2,\mu^2)&=\nu^2-\mu^2+\nu_1^2~.
\end{align}
 As a result, from  
 $\rho_{in_1}(\nu^2,\nu_1^2;\mu^2)$ in the numerator of Eq.~\eqref{180820.2}
  we pick up either a factor $\nu^2-\mu^2$ or a factor  $\nu_1^2$. 
Both of them times $ S_2(\nu_1)$  does not give rise to any divergent contribution when $\nu\to 0$. 
For the $\nu_1^2$ factor this is clear because 
 $\nu_1^2 S_2(\nu_1)$ is finite, while for the former $(\nu^2-\mu^2)=(\nu-\mu)(\nu+\mu)$ and 
 cancels the divergence that would stem from the integration of $S_2(\nu_1)$ around $\nu_1=0$ 
 when $\nu+\mu\to 0$ or $\nu+\mu<0$. For a projection of a non-necessarily central potential
 there could be  other contribution not proportional to $\xi$, but it would be 
 proportional to $(\nu\nu_1)^2$ after extracting $(pp')^{\ell+1}$ as common factor in calculating $\hhv$.
 Then we would have a factor $\nu_1^2$ that removes the divergence in the integrand.
 There could also be involved higher powers of the rhs of Eq.~\eqref{180820.3} but the argument
holds straightforwardly.
 A similar analysis can be done when considering the upper limit $\nu_2-m_\pi$ 
making use of the function $\rho(\nu_2^2,\nu_1^2;{\mu'}^2)$.  

 For higher angular momentum,  a generic function 
 $\rho_{in}(\nu^2,\nu_1^2;\mu^2)$ is not just proportional  to a factor $(\nu_1\xi)^{\ell_n}$, 
 because several terms with different powers of $\cos\theta$ enter in the 
partial-wave projection formula. For instance, the Legendre polynomials $P_\ell(\cos\theta)$ with $\ell\geq 2$ 
involve terms  with different powers of $\cos\theta$. Indeed from a partial-wave projection formula
\begin{align}
\label{180820.4}
\int_{-1}^{+1}\frac{dz P_\ell(z)}{z-\xi}
\end{align} 
the contribution to the discontinuity, following the argument driving to Eq.~\eqref{210615.6}, 
is proportional to $P_\ell(\xi)$. In these cases the absence of divergences stem from the product of the 
two functions $\rho$ in Eq.~\eqref{180820.2}, $\rho_{in_1}(\nu^2,\nu_1^2;\mu^2)\rho_{n_1n_2}(\nu_1^2,\nu_2^2;{\mu'}^2)$. 
We cannot give here a direct demonstration that there are not divergences in the calculation of 
$\mff_{ij}(\nu)$ despite the diverging behavior of $S_{n}(\nu_1)$ when $\nu_1\to 0$. 
However, we can give strong arguments in favor of this conclusion. 
First, we recall the  general remark stated above about the fact that the discontinuity of the PWA $\hht$ 
involved in the definition of $\mff_{ij}(\nu)$, Eq.~\eqref{280116.2}, is for general arguments 
finite and not divergent. 
In this respect, notice that  Eq.~\eqref{180820.2} for $\nu=-k$,  and 
$\nu_2=k$ is the same that, after the two integrals in $\mu$ and $\mu'$ with arbitrary spectral functions have been done 
in the second term of Eq.~\eqref{170526.11},
 gives the once iterated contribution for $f_{ij}(-k)$, and therefore to $\Delta_{ij}(-k^2)$ in this 
 approximation (which is exact for $k<3m_\pi/2$). Since the integrands of the integrals 
 in $\mu$ and $\mu'$ are arbitrary  this implies that Eq.~\eqref{180820.2} with 
 $\nu=-k$ and $\nu_2=k$ must be finite. 
Of course, this strongly supports the correctness
 of our statement concerning the finiteness of the integration in Eq.~\eqref{180820.2}. 
 Furthermore, we discuss in detail in Sec.~\ref{sec.180819.2} the  calculation of $\Delta(p^2)$ for the 
$^1D_2$ PWA.  The different contributions when evaluating Eq.~\eqref{180820.2} are given 
 and we show how a $1/(\nu+m_\pi)+1/(\nu_2-m_\pi)$ divergent term appearing in partial contributions 
 finally cancel in the total result. Similarly happens for the $^3D_1$ wave, which is also worked out 
in Sec.~\ref{sec.180829.1b}.

$\bullet$ Once we know $\mff_{ij}(\nu)$, $\nu\in[-k+m_\pi,k-m_\pi]$, we can calculate $\mff_{ij}(-k)$ 
by applying the symmetric form of Eq.~\eqref{170526.3} to $\nu=-k$. 
 For this value the integration interval is symmetric and the last integral in this equation vanishes because 
the lower and upper limits of integration coincide. 
We then use the following expression for calculating $\mff_{ij}(-k)$, 
\begin{align}
\label{180821.9}
\mathfrak{f}_{ij}(-k)& = -\frac{2}{\pi}\int_{m_\pi}^{2k} 2\mu d\mu \eta_{ij}(\mu^2)\rho_{ij}(\nu^2,k^2;\mu^2) 
-\frac{\theta(k-m_\pi)m}{2\pi^3}\sum_n \int_{0}^{k-m_\pi}\frac{d\nu_1\nu_1^2}{k^2-\nu_1^2}
S_{n}(\nu_1)\\
\times &\Big[\mff_{nj}(\nu_1)\int_{m_\pi}^{k+\nu_1}2\mu d\mu \eta_{in}(\mu^2)\rho_{in}(k^2,\nu_1^2;\mu^2)
+ \mff_{nj}(-\nu_1)\int_{m_\pi}^{k-\nu_1}2\mu d\mu \eta_{in}(\mu^2)\rho_{in}(k^2,\nu_1^2;\mu^2)\Big]~.\nn
\end{align}
Regarding the issue of the divergences of $\nu_1^2 S_{n}(\nu_1)$ for $\nu_1\to 0$ and $\ell\geq 2$, 
the most of the discussion held in connection with this issue for Eq.~\eqref{170526.3} holds here. 
The difference now is that $k$ can be near $m_\pi$, while in Eq.~\eqref{170526.3} the upper limit of integration 
is always larger than $m_\pi/2$ because $\nu<-3m_\pi/2$.  
However, if $k<m_\pi$ the integral does not contribute because of the Heaviside function, and then $\mff(-k)$ is 
given by the independent term. 
 On the other hand, if $m_\pi<k<3m_\pi/2$ the extra contribution is just the 
once-iterated one, which is given by substituting in the integration of Eq.~\eqref{180821.9} 
the independent term for $\mff_{ij}(\pm\nu)$. This is because for $k<3m_\pi/2$ the Heaviside function 
$k-2m_\pi-\nu>0$ in Eq.~\eqref{170525.2} is not fulfilled for $\nu\in[-k+m_\pi,k-m_\pi]$. 
Therefore, $\mff_{ij}(\nu)$ is calculated with machine precision and the integration in Eq.~\eqref{180821.9} 
for calculating $\mff(-k)$ can be done safely numerically, unless $k$ is almost coincident with $m_\pi$ from above. 
Nonetheless, even if this is the case we can calculate $\mff(-k)$ algebraically, 
 getting rid of any possible numerical problem for $k\to m_\pi+\ve$. Indeed, let us recall that 
 the algebraic calculation of the first-iterated contribution to $\mff_{ij}(\nu)$ is needed 
to implement Eq.~\eqref{170526.11}.

$\bullet$ The fact that $\Delta_{ij}(p^2)=\Delta_{ji}(p^2)$, which follows from time-reversal symmetry and the analytic continuation 
in $p$ of the on-shell PWA $t_{ij}(p,p)$, is equivalent to the statement that $\mff_{ij}(-k)=\mff_{ji}(-k)$, 
cf. Eq.~\eqref{280116.3}. However, it does not imply that $f_{ij}(\nu)=f_{ji}(\nu)$ and 
we have to solve separately $\mff_{ij}(\nu)$ and $\mff_{ji}(\nu)$. 
 Let us also show that $f_{ij}(-k)=f_{ji}(-k)$,
\begin{align}
\label{300116.10}
i\mff_{ij}(-k)&=
\hht_{ij}(-ik+\ve^-,ik+\ve)-\hht_{ij}(-ik+\ve^+,ik+\ve)
=\hht_{ji}(ik+\ve,-ik+\ve^-)-\hht_{ji}(ik+\ve,-ik+\ve^+)\nn\\
&=\hht_{ji}(-ik+\ve,ik+\ve^-)^*-\hht_{ji}(-ik+\ve,ik+\ve^+)^*
\end{align}
The discontinuity only affects the imaginary part of $t_{ij}$ so that we can remove the complex 
conjugation in the last equation and instead write a global minus sign, 
\begin{align}
\label{300116.11}
i\mff_{ij}(-k)&=-\big[\hht_{ji}(-ik+\ve,ik+\ve^-)-\hht_{ji}(-ik+\ve,ik+\ve^+) \big]
=\hht_{ji}(-ik+\ve,ik+\ve^+)-\hht_{ji}(-ik+\ve,ik+\ve^-)\nn\\
&=i\mff_{ji}(-k)~.
\end{align}

For the $2\times 2$ coupled-channel case the 
linear system of IEs of Eq.~\eqref{290116.22} explicitly decouples in two systems of two IEs each (because every IE only involves 
the first subscript $i$ while $j$ is kept fixed),
\begin{align}
\label{300116.8}
\mff_{11}(\nu)&=\Delta \hhv_{11}(\nu,k)+\theta(k-2m_\pi-\nu)\frac{m}{4\pi^2}\int_{m_\pi+\nu}^{k-m_\pi}\frac{d\nu_1\nu_1^2}{k^2-\nu_1^2}
\left[\Delta\hhv_{11}(\nu,\nu_1)S_1(\nu_1) \mff_{11}(\nu_1)+\Delta\hhv_{12}S_2(\nu_1)\mff_{21}(\nu_1)
\right]~,\nn\\
\mff_{21}(\nu)&=\Delta \hhv_{21}(\nu,k)+\theta(k-2m_\pi-\nu)\frac{m}{4\pi^2}\int_{m_\pi+\nu}^{k-m_\pi}\frac{d\nu_1\nu_1^2}{k^2-\nu_1^2}
\left[\Delta\hhv_{21}(\nu,\nu_1)S_1(\nu_1) \mff_{11}(\nu_1)+\Delta\hhv_{22}S_2(\nu_1)\mff_{21}(\nu_1)
\right]~,
\end{align}
and 
\begin{align}
\label{300116.9}
\mff_{22}(\nu)&=\Delta \hhv_{22}(\nu,k)+\theta(k-2m_\pi-\nu)\frac{m}{4\pi^2}\int_{m_\pi+\nu}^{k-m_\pi}\frac{d\nu_1\nu_1^2}{k^2-\nu_1^2}
\left[\Delta\hhv_{21}(\nu,\nu_1)S_1(\nu_1) \mff_{12}(\nu_1)+\Delta\hhv_{22}S_2(\nu_1)\mff_{22}(\nu_1)
\right]~,\nn\\
\mff_{12}(\nu)&=\Delta \hhv_{12}(\nu,k)+\theta(k-2m_\pi-\nu)\frac{m}{4\pi^2}\int_{m_\pi+\nu}^{k-m_\pi}\frac{d\nu_1\nu_1^2}{k^2-\nu_1^2}
\left[\Delta\hhv_{11}(\nu,\nu_1)S_1(\nu_1) \mff_{12}(\nu_1)+\Delta\hhv_{12}S_2(\nu_1)\mff_{22}(\nu_1)
\right]~.
\end{align}
For the general case of $n$ coupled partial waves we would have $n$ subsystems of coupled IE, each containing $n$ equations.
 
In subsequent sections we are going to calculate the LHC discontinuity $\Delta_{ij}(p^2)$ for potentials 
with a spectral function arising from the exchange of only the lightest quanta, so that 
$\eta_{ij}(\mu^2)=\pi\delta(\mu^2-m_\pi^2)$. For later use we rewrite Eqs.~\eqref{170525.2}, 
\eqref{170526.3} and \eqref{170526.11} for such simpler case, in order:
\begin{align}
\label{180814.1}
&\mathfrak{f}_{ij}(\nu)= -2\rho_{ij}(\nu^2,k^2)-
\frac{\theta(k-2m_\pi-\nu)m}{2\pi^2}\sum_n\int_{m_\pi+\nu}^{k-m_\pi}\frac{d\nu_1\nu_1^2}{k^2-\nu_1^2}
S_{n}(\nu_1)
\rho_{in}(\nu^2,\nu_1^2)\mathfrak{f}_{nj}(\nu_1)~,\\
\label{180814.2}
&\mathfrak{f}_{ij}(\nu)= -2\rho_{ij}(\nu^2,k^2)-
\frac{\theta(k-2m_\pi-\nu)m}{2\pi^2}\sum_n\int_{-m_\pi-\nu}^{k-m_\pi}\frac{d\nu_1\nu_1^2}{k^2-\nu_1^2}
S_{n}(\nu_1)
\rho_{in}(\nu^2,\nu_1^2)\mathfrak{f}_{nj}(\nu_1)\\
&-\frac{\theta(k-2m_\pi-\nu)m}{2\pi^2}\sum_n \int_0^{-\nu-m_\pi}\frac{d\nu_1\nu_1^2}{k^2-\nu_1^2}
S_{n}(\nu_1)
\rho_{in}(\nu^2,\nu_1^2)
\big[\mathfrak{f}_{nj}(\nu_1)+\mathfrak{f}_{nj}(-\nu_1)\big]~,\nn\\
\label{180814.3}
&\mathfrak{f}_{ij}(\nu) = -2 \rho_{ij}(\nu^2,\nu_1^2)+\theta(k-2m_\pi-\nu)\frac{m}{\pi^2}
\sum_n\int_{m_\pi+\nu}^{k-m_\pi} 
\frac{d\nu_1  \nu_1^2}{k^2-\nu_1^2}S_{n}(\nu_1)\rho_{in}(\nu^2,\nu_1^2)\rho_{nj}(\nu_1^2,k^2)\\
&+\theta(k-3m_\pi-\nu)\left(\frac{m}{2\pi^2}\right)^2 
\sum_{n_1,n_2}\int_{2m_\pi+\nu}^{k-m_\pi}\frac{d\nu_2 \nu_2^2}{k^2-\nu_2^2}S_{n_2}(\nu_2)\mathfrak{f}_{n_2 j}(\nu_2)
 \int_{m_\pi+\nu}^{\nu_2-m_\pi}\frac{d\nu_1 \nu_1^2}{k^2-\nu_1^2}S_{n_1}(\nu_1) 
\rho_{in_1}(\nu^2,\nu_1^2) \rho_{n_1n_2}(\nu_1^2,\nu_2^2)~.\nn
\end{align}

For the case of regular potentials  A. Martin derived long time ago \cite{martin1961.19.1257} an equation  
 for evaluating the discontinuity along the LHC for the uncoupled $S$-wave scattering. 
This equation is analogous to the restriction of Eq.~\eqref{170525.2}, in conjunction with Eq.~\eqref{280116.3}, to such case. 
 His method is entirely different to ours and is based on the solution of the Schr\"odinger 
equation in configuration space, solving for the wave functions which fulfill standard 
boundary conditions at the origin. 
Notice how our method based on the analytical extrapolation of the LS equation allows one 
to avoid the complication on the appropriate boundary condition for a wave function when one 
has singular potentials, a problem not treated by Martin as noticed by himself in 
Refs.~\cite{martin1961.20.390,martin1961.21.158}.
  In a previous paper \cite{martin1959.14.403}, and always for regular interactions,  
Martin  developed a method to calculate the on-shell $S$ matrix 
for uncoupled $S$ wave in terms of the above mentioned equation analogous 
to Eq.~\eqref{170525.2}. However, the extension of the results of Martin 
 for the uncoupled $S$-wave case to others is much more involved. 
 In this regard, new equations are derived in Ref.~\cite{martin1960.15.99} 
 to calculate the on-shell $S$ matrix for uncoupled higher partial waves that are  
 much more cumbersome than for the $S$-wave case. For the coupled case 
 Ref.~\cite{martin1961.20.390} elaborated on the scattering of two spin 1/2 particles 
 with non-central forces present. A compilation of these results by Martin {\it et al.} 
 can be found in Ref.~\cite{martin1961.21.158}.  
In contrast, it is worth stressing that our main equation \eqref{170525.2} 
offers a uniform framework to evaluate $\Delta_{ij}(p^2)$ 
 for regular 
 and singular potentials, $S$- and higher partial waves as well as for 
 coupled and uncoupled scattering. 
 We also offered a previous derivation \cite{entem.170930.1},  by iterating directly  the LS equation,
  of an equation analogous to Eq.~\eqref{180814.1} for 
the calculation of the discontinuity along the LHC of  an 
uncoupled $S$-wave PWA for a pure Yukawa potential.

\section{Explicit calculation of $\Delta_{ij}(p^2)$ along the LHC for some regular and singular potentials}
\label{sec.180814.1}
\def\theequation{\arabic{section}.\arabic{equation}}
\setcounter{equation}{0}   

We consider in this section the calculation of $\Delta_{ij}(p^2)$ for some potentials so as to 
illustrate the use of the master Eq.~\eqref{170525.2} [and its appropriate reshuffling, cf.~Eq.~\eqref{170526.12}]. 
In particular, it will be evident the qualitative differences in the behavior of $\Delta_{ij}(p^2)$ for 
regular potentials, on the one hand, and for singular potentials, on the other hand. Among the latter ones  
the attractive and repulsive cases will also have $\Delta_{ij}(p^2)$ with different qualitative behaviors. 
 We will also exemplify the use of Eq.~\eqref{170526.12} by considering PWAs up to and including $D$ waves. 
 All the examples that follow refer to $NN$ scattering and we use the spectroscopical notation $^S \ell_J$  
 to denote the $NN$ PWAs in the $\ell SJ$ partial-wave  decomposition.

\begin{figure}[H]
\begin{center}
\scalebox{0.8}{
\begin{tabular}{lll}
\includegraphics[width=.4\textwidth]{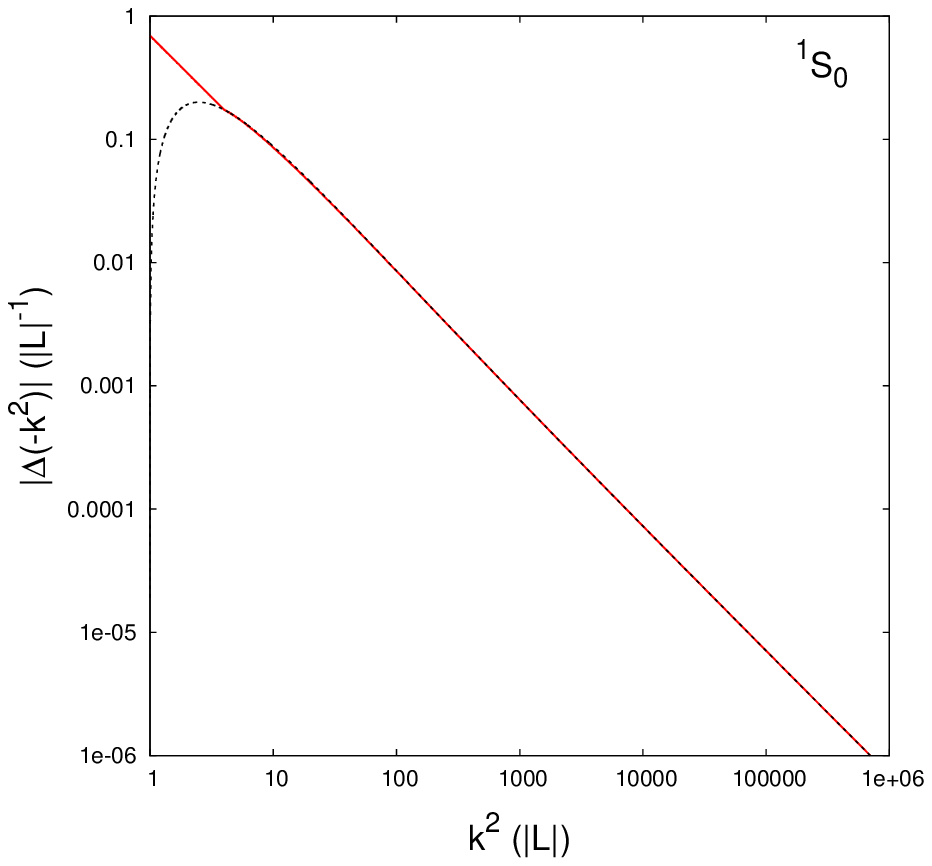} &
\includegraphics[width=.4\textwidth]{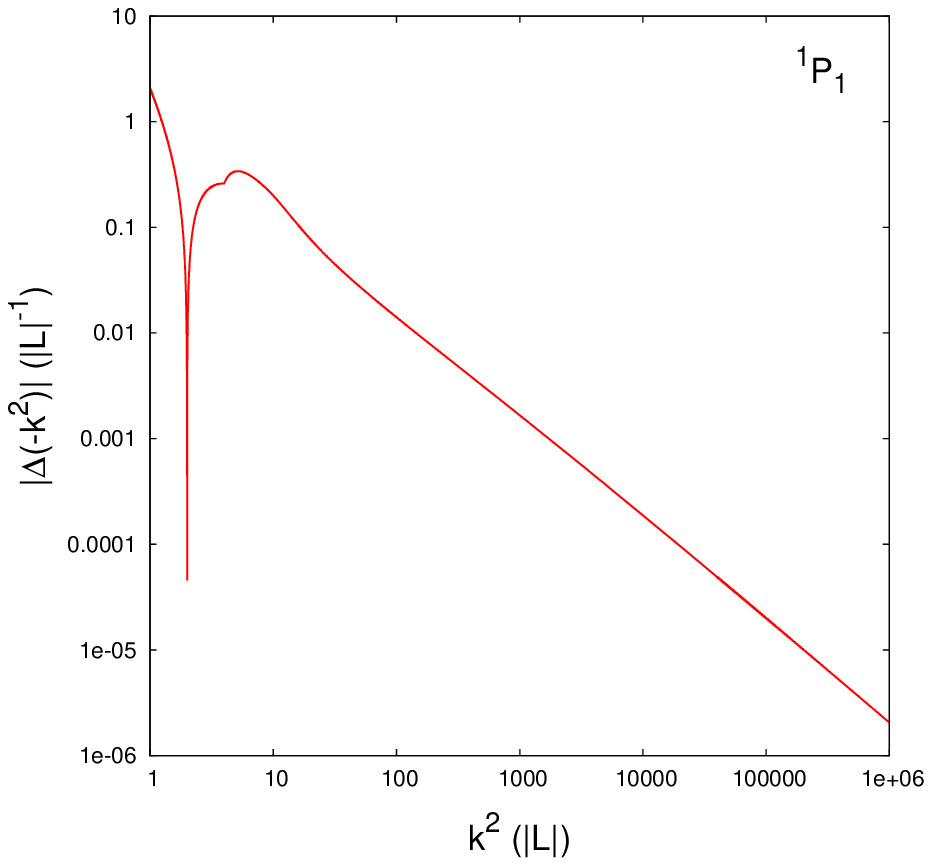} \\  
&\includegraphics[width=.4\textwidth]{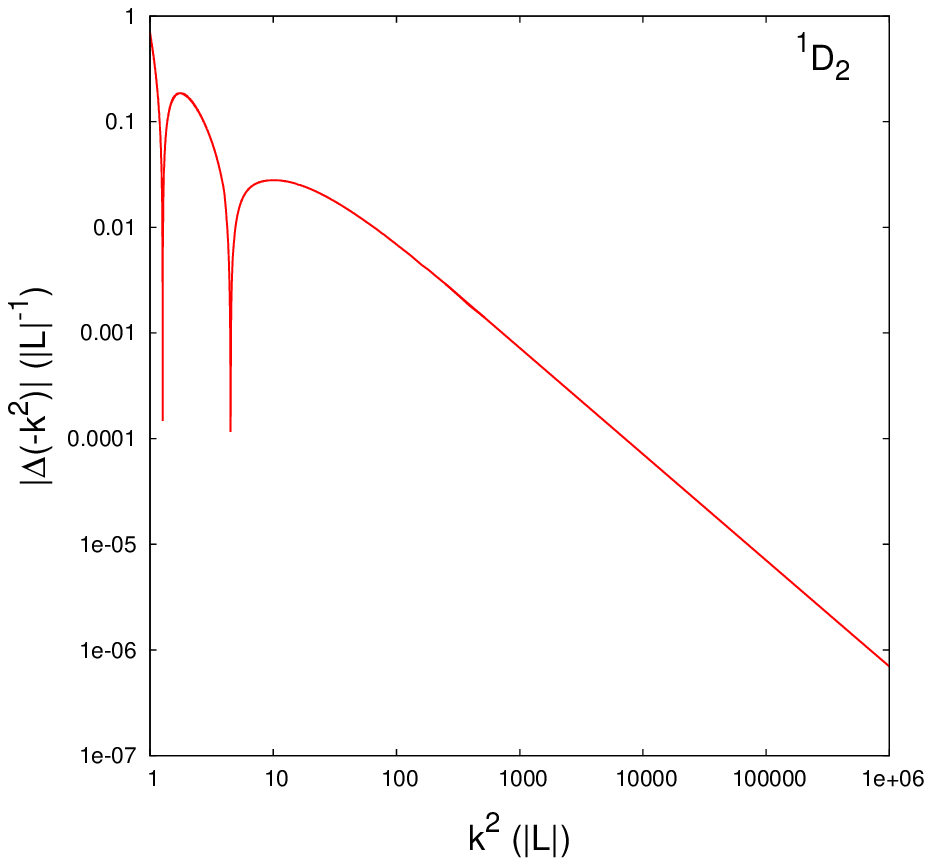} 
\end{tabular}
}
\end{center}
\caption{{\small Log-log plots of the discontinuities $\Delta(-k^2)$ for the partial 
waves $^1S_0$ (top left), $^1P_1$ (top right) and $^1D_2$ (bottom right panel) 
by the red solid lines. 
The black dotted line for the $^1S_0$ PWA represents the 
 approximate algebraic solution of Eq.~\eqref{170715.3}.  
The results for these panels correspond to regular potentials  obtained from the partial-wave projection of the OPE $NN$ potential.}
\label{fig.180817.1}
}
\end{figure} 
\subsection{Regular potentials}
\label{sec.180814.2}

We study the $NN$ uncoupled  partial waves $^1S_0$, $^1P_1$ and $^1D_2$ driven by  the OPE potential. 
 Since all these waves are singlet ($S=0$) the resulting partial-wave projected OPE potential of Eq.~\eqref{180505.2} is 
 regular, corresponding to a Yukawa one. 
 A characteristic feature of all these examples is that the discontinuity $\Delta(-k^2)$ vanishes for 
 $k\to \infty$ as $1/k^2$.  Apart from its explicit calculation, this also follows from the discussion at the end of 
Sec.~\ref{sec.180814.3b} because for these potentials $\mff(-k)$ tends to its Born-term contribution in such limit.

  We indicate the total isospin of the two nucleons by $I$,  
it can be 0 or 1, and it is such that $(-1)^{\ell+S+I}=-1$, because of the Pauli exclusion principle. 
Thus, $I=1$ for the waves $^1S_0$ and $^1D_2$, and $I=0$ for $^1P_1$. 
In terms of the total isospin $I$ and spin $S$  we have for the scalar products $\mathbf{i}_1\cdot \mathbf{i}_2=I-3/4$ and 
$\mathbf{s}_1\cdot\mathbf{s}_2=S-3/4$. 
After this brief preamble,
the partial-wave projected potentials in configuration space resulting from Eq.~\eqref{180505.2} for the referred partial 
waves, being the only difference among them the value of $I$, can be expressed as
\begin{align}
V(r)&=-\frac{4I-3}{\pi}\left(\frac{g_A m_\pi}{4f_\pi}\right)^2 \frac{e^{-m_\pi r}}{r}~.
\end{align}
Therefore, this is an attractive Yukawa potential for $^1S_0$, $^1D_2$ ($I=1$) and repulsive for $^1P_1$ ($I=0$).

Except for illustrative purposes, we are going to use the potentials in momentum space. 
The appropriate formulas for the partial-wave projection of a spin 1/2-spin 1/2 off-shell 
scattering amplitude in the $\ell SJ$ basis can be found in the Appendix \ref{app:170715.1}. 
For the OPE case in order to use these formulas the only non-vanishing invariant function 
is $W_T$ \cite{kaiser.080516.1},
\begin{align}
\label{180815.1}
W_T&=\frac{g_A^2}{4f_\pi^2(\vq^2+m_\pi^2)}~.
\end{align}

\subsubsection{$\Delta(p^2)$ for the $^1S_0$ PWA}
\label{sec.180814.3b}

By straightforward calculation the $^1S_0$ potential in momentum space is
\begin{align}
\label{180815.2a}
v(p,p')&=\frac{g}{p{p'}}\Theta(p,p')~,\\
g&=\frac{g_A^2 m_\pi^2}{8 f_\pi^2}~.\nn
\end{align}
Apart of the constant $g$ we have introduced the function $\Theta(p,p')$ 
to represent in a shorter way the difference of logs 
\begin{align}
\label{180815.2}
\Theta(p,p')&=\log\left[(p+p')^2+m_\pi^2\right]-\log\left[(p-p')+m_\pi^2\right]~,
\end{align}
that typically appears in the partial wave projections of the potential, 
cf. Eq.~\eqref{swpvm}. 
 From Eq.~\eqref{180815.2a} we deduce that the function $\rho(\nu^2,\nu_1^2)$ is simply a constant
\begin{align}
\label{180815.3}
\rho(\nu^2,\nu_1^2)&=\pi g~.
\end{align}
The IE of Eq.~\eqref{180814.1} for the calculation of $\Delta(p^2)$ in this case is
\begin{align}
\label{180815.4}
\mff(\nu)&=-\pi g+\theta(k-2m_\pi-\nu)\lambda \int_{m_\pi+\nu}^{k-m_\pi}
\frac{d\nu_1}{k^2-\nu_1^2}\mff(\nu_1)~,\\
\lambda&=\frac{m g}{2\pi}~.\nn
\end{align}
This IE is solved by recursion. 
First, we have that $\mff(\nu)=-\pi g$ for $\nu\in[k-2m_\pi,k-m_\pi]$. 
With these values we can then integrate Eq.~\eqref{180815.4} and obtain $\mff(\nu)$ for $\nu\in [k-3m_\pi,k-2m_\pi]$. 
This process is iterated numerically so that we can calculate $\mff(\nu)$ for any $\nu\in[-k+m_\pi,k-m_\pi]$. 
The solution obtained is shown by the red solid line in the top left panel of Fig.~\ref{fig.180817.1}.\footnote{In the following 
 we use the values $g_A=1.26$, $f_\pi=92.4$~MeV, $m=939$~MeV and $m_\pi=138$~MeV.} 
 We already discussed with detail in Ref.~\cite{entem.170930.1} the comparison of the full solution with 
the results obtained by iterating a finite number of times Eq.~\eqref{180815.4}. If this equation is iterated 
$n$ times then we obtain the exact solution for $\nu\in[k-(2+n)m_\pi,k-m_\pi]$. Of course, for $n$ large enough we cover 
the whole range of values desired, namely for $n>2k/m_\pi-3$. 

We now explain the numerical method that we have followed for solving the IE of  Eq.~\eqref{180815.4} recursively. 
 We take a symmetric partition of $\nu_1\in[-k+m_\pi,k-m_\pi]$ around the origin. 
 It is numerically more convenient to take the partition for positive values of $\nu_1$ and 
then to reflect it for negative ones. This is so because of the symmetrization process 
needed for higher partial waves in the calculation of $\mff_{ij}(\nu)$, cf. Eq.~\eqref{170526.3}.
For positive values of $\nu$ we perform the following change of variables 
and define the dimensionless variable $x$ as
\begin{align}
\label{180818.1}
x=\frac{m_\pi}{k-\nu_1}~,\\
\nu_1=k-\frac{m_\pi}{x}~,\nn
\end{align}
so that $x\in[m_\pi/k,1]$. 
 Then, we take a Gau{\ss}ian partition of $N$ points in this interval of $x$ and 
 denote by $x_i$ the points along this partition in decreasing order, $x_{i+1}<x_i$, and 
 by $\omega_i$ the Gau{\ss}ian weights. 
 For negative values of $\nu$, we reflect symmetrically around the origin 
 and  we have for $N+1\leq i\leq 2N$ the 
 reflection $\nu_i=-\nu_{2N+1-i}=-k-m_\pi/x_i$, with $x_i=-x_{2N+1-i}$,
  and $\omega_i=\omega_{2N+1-i}$. 

A loop in $i$ runs from 1 to $2N$, so that for $i$ small enough  
 the condition $k-2m_\pi-\nu_i<0$ is fulfilled and
  then $\mff(\nu_i)$ is given by the independent term 
\begin{align}
\mff(\nu_i)&=-\pi g~.
\end{align}
Once this condition is not satisfied, it follows that this is also the case  
for larger values of $i$, or smaller values of $\nu_i$ (it is necessary that $k>3m_\pi/2$).
 Let us then indicate by $i_0$ the largest value of $i$ for which $\nu_{i_0}>\nu_i+m_\pi$. 
 Then Eq.~\eqref{180815.4} becomes
\begin{align}
\label{180818.2}
\mff(\nu_i)&=-\pi g+\lambda \sum_{j=1}^{i_0}\frac{\omega_j^2}{x_j^2} 
\frac{\mff(\nu_j)}{k^2-\nu_j^2}~.
\end{align}
This numerical implementation is used in all the IEs involved in Eq.~\eqref{170526.12} when they 
are solved by recursion.

An algebraic solution of Eq.~\eqref{180815.4} can be obtained in the limit $k\gg m_\pi$ by 
transforming it into a differential equation (DE). For that we differentiate this equation 
with respect to $\nu$ and obtain
\begin{align}
\label{140715.3}
\frac{\mff'(\nu)}{\mff(\nu+m_\pi)}&=-\lambda 
\frac{\theta(k-2m_\pi-\nu)}{k^2-(m_\pi+\nu)^2}
\end{align}
Notice that the term with the derivative of the Heaviside function $\theta(k-2m_\pi-\nu)$ 
from Eq.~\eqref{180815.4}, that 
results in a Dirac-delta function, does not contribute because its coefficient, an integral of zero extent,
 is zero.

The DE of Eq.~\eqref{140715.3} is not easy to solve due to the shift in $m_\pi$ in the argument of 
the denominator in the left-hand side. We are then tempted to neglect 
this shift because we are assuming that 
$m_\pi$ is very small compared with $k$. However, the variable $|\nu|$ could take values around $|k|$,
 in which case the factor $1/(k^2-(m_\pi+\nu)^2)$ on the rhs  
 becomes huge $\propto 1/k m_\pi$. 
 That is, one is also sensitive to $k-|\nu|$ which could be also very small. 
 Then, we  first assume that we can transform the DE of Eq.~\eqref{140715.3} into the simpler form
\begin{align}
\label{140715.4}
\frac{\mff'(\nu)}{\mff(\nu)}&=-\lambda 
\frac{\theta(k-2m_\pi-\nu)}{k^2-(m_\pi+\nu)^2}~,
\end{align}
which keeps the possible enhanced behavior of the rhs, as in the original DE. 
The solution of Eq.~\eqref{140715.4} is straightforward and is given by
\begin{align}
\label{140715.5}
\mff(\nu)&=\mff(k-2m_\pi)\exp\left[-\frac{\lambda}{k}{\rm{arctanh}}\left(\frac{\nu+m_\pi}{k}\right)
\right]\theta(k-2m_\pi-\nu)~.
\end{align}
Now, $\mff(\nu)=-\pi g$ for $\nu\geq k-2m_\pi$, as follows  from  Eq.~\eqref{180815.4}. 
 This allows us to fix the integration constant in Eq.~\eqref{140715.5} and hence, it results
\begin{align}
\label{140715.6}
\mff(\nu)&=-\pi g 
\exp\left[-\frac{\lambda}{k}{\rm {arctanh}}\left(\frac{\nu+m_\pi}{k}\right)\right]
\exp\left[\frac{\lambda}{k}{\rm {arctanh}}\left(\frac{k-m_\pi}{k}\right)\right]
\theta(k-2m_\pi-\nu)-\pi g \theta(\nu-k+2m_\pi)~\nn\\
&=-\pi g \left(\frac{k+\nu+m_\pi}{k-\nu-m_\pi}\frac{m_\pi}{2k-m_\pi}\right)^{-\lambda/2k}\theta(k-2m_\pi-\nu)
-\pi g \theta(\nu-k+2m_\pi)~.
\end{align}
We can check this solution by reinserting it back in the original DE Eq.~\eqref{140715.3}:
\begin{align}
\label{160715.6}
\frac{\mff'(\nu)}{\mff(\nu+m_\pi)}+\frac{\lambda}{k^2-(\nu+m_\pi)^2}=
\frac{\lambda}{k^2-(\nu+m_\pi)^2}\left\{
1-\exp\left[-\frac{\lambda}{k}{\rm arctanh}\frac{\nu+m_\pi}{k}\right]
\exp\left[\frac{\lambda}{k}{\rm arctanh}\frac{\nu+2m_\pi}{k}\right]
\right\}~.
\end{align}
If our solution Eq.~\eqref{140715.6} were exact the rhs of Eq.~\eqref{160715.6}, that we call error function, 
 would be zero. 
It is not so but for $k>>m_\pi$, $|\nu|$ the error function is 
\begin{align}
\label{170715.1}
-\frac{\lambda^2 m_\pi}{k^4}+\frac{1}{k}\times {\cal O}(\ve^5)~,
\end{align}
with $\ve=\nu/k$ or $m_\pi/k$. 
For $|\nu|\sim k$ the error function is 
\begin{align}
\label{170715.2}
-\frac{\lambda^2}{4 m_\pi k^2}\times \log {\rm const}~,
\end{align} 
where the constant depends on the value of $\nu$. Of course, for $k-m_\pi>\nu>k-2m_\pi$ the 
solution of Eq.~\eqref{160715.6} is exact  because of the 
 Heaviside function in Eq.~\eqref{180815.4}.

The discontinuity $\Delta(-k^2)$ along the LHC in the case $k\gg m_\pi$
 can be expressed directly in terms of $\mff(-k)$, cf. Eq.~\eqref{280116.3}. 
 From  Eq.~\eqref{140715.6} the following approximate algebraic expression follows,
\begin{align}
\label{170715.3}
\Delta(-k^2)&=\frac{\mff(-k)}{2k^2}
\xrightarrow[k\to \infty]{}
-\frac{\pi g}{2 k^2}\left(\frac{m_\pi}{2k-m_\pi}\right)^{-\lambda/k}~.
\end{align}
This equation implies that for $k\to \infty$ the function $\Delta(-k^2)$ vanishes as $k^{-2}$. 
For $\nu=-k$ the factor $\log {\rm const}=\log 2$ in  Eq.~\eqref{170715.2} and the error function 
rapidly goes to zero with increasing $k$.
 The  outcome from Eq.~\eqref{170715.3} is represented by the black dotted line in the top left panel of 
Fig.~\ref{fig.180817.1}. We see that the full solution (solid line) and the approximated one  (dotted line) 
 quickly overlap each others with increasing $k$. Reference \cite{entem.170930.1} considered also much bigger values 
of $g_A$ for illustrative purposes, since then the non-perturbative effects in the calculation of $\Delta(p^2)$ are 
more prominent. In such circumstances, the asymptotic behavior in which the approximate solution of Eq.~\eqref{170715.3} 
becomes accurate starts for much larger values of $k$.

We also observe from Eq.~\eqref{170715.3} that the function $\Delta(-k^2)$ for $k\to \infty$ corresponds to using 
the independent term for $\mff(-k)$ in Eq.~\eqref{280116.3}. 
 Indeed, this is the expected behavior when the OPE function $\rho(\nu^2,\nu_1^2;m_\pi^2)$ is a 
 polynomial in $\nu$ and $\nu_1$ of degree $m$, as it is the case 
in the examples that we solve explicitly in this work. 
To show it let us scale by a factor $\tau$ the variables $k$, $\nu$ and $\nu_1$. 
 Therefore, the $n$-times iterated partial contribution 
to $\mff(-k)$ from Eq.~\eqref{170525.2} scales as 
\begin{align}
\label{180825.1}
\tau^{(n+1)m-(2\ell+1)n}=\tau^{(m-2\ell-1)n+m}~.
\end{align} 
The amount $2\ell+1$ keeps track of the scaling power in $\nu_1$ of 
$d\nu_1 \nu_1^2 S_{n}/(k^2-\nu_1^2)$ in the integrand of Eq.~\eqref{170525.2}. 
Compared to the Born-term contribution (the one stemming from the independent term),
 that scales as $\tau^{m}$, we have the relative scaling of Eq.~\eqref{180825.1} as 
 $\tau^{(m-2\ell-1)n}$, with a negative exponent for $m<2\ell+1$. 
If this is the case, every iteration of this equation has a larger power in $1/k$  
 by at least $2\ell+1-m>0$.  
The pion mass is not scaled because we take the limit $k\gg m_\pi$. 
As a matter of fact, when a power $m_\pi^{2p}$ happens in a function $\rho$ 
it drives to a smaller scaling and gives rise to subleading contributions in the limit under 
consideration. 
On the other hand, the function $1/(k^2-\nu_i^2)$, when integrated in $\nu_i$, could give rise to 
enhancement in the near region close to the limits of integrations for $\nu_i$ approaching $\pm (k-n' m_\pi)$. 
However, the enhancement is only of logarithmic type as it is clear by writing 
$1/(k^2-\nu_i^2)=1/[(k-\nu_i)(k+\nu_i)]$, so that one factor tends in modulus to $1/k$ 
and the other gives the mentioned logarithmic enhancement once integrated around this region. 
The former factor sets the order of magnitude (modulo logarithms) expected from the scaling in $\tau$. 
 Thus, the resulting $\Delta(-k^2)$ is dominated for $k\gg m_\pi$ by the independent term contribution, since 
 any further iteration is suppressed by extra powers of $1/k$. 
The onset of this asymptotic behavior depends on the strength of the interaction. 
E.g. in the explicit algebraic expression of Eq.~\eqref{170715.3} this is controlled by the parameter $g$. 

 
For the case of a general  spectral function, one considers  the weighted function $\rho(\nu,\nu_1)$ defined by
\begin{align}
\label{180825.2}
\rho(\nu,\nu_1)&=
\frac{1}{\pi}\int_{m_\pi}^{\nu_1-\nu} 2\mu d\mu \eta(\mu^2)\rho(\nu^2,\nu_1^2;\mu^2)~,~\nu_1\geq \nu+m_\pi~,
\end{align}
as it is clear from Eq.~\eqref{170525.2}. 
We should determine whether this function has a scaling behavior in its arguments for $|\nu_1|\sim|\nu|\sim k\gg m_\pi$ 
as $\nu^m$ and $\nu_1^m$, times possible logarithmic factors. 
If this is the case we could apply a similar scaling argument 
as the one driving to Eq.~\eqref{180825.1} and expect the dominant behavior of the Born-term contribution in the 
asymptotic region $k\to\infty$ for $m<2\ell+1$.
 This implies that $\Delta(-k^2)$ for $k\to \infty$ vanishes faster than $1/k$ because $\mff(-k)$ tends to its Born-term 
contribution divided by $k^{2\ell+2}$, cf. Eq.~\eqref{280116.3}, and the former diverges less rapid than $k^{2\ell+1}$. 
 For coupled partial waves we would add the subscripts $i$ and $j$ in the functions $\rho$ and $\eta$
 present in both sides of Eq. \eqref{180825.2}. The coupled case is treated below in detail when also discussing singular
 interactions in Sec.~\ref{sec.180829.1}.
 
Indeed, the outcome of the previous discussion is to be expected because 
of the following interesting connection with a result 
of Ref.~\cite{oller.180722.2}. 
As it is derived in this reference  the solution to any $N/D$ IE exists 
 when $|\Delta(-k^2)|$ vanishes faster than  $1/k$ for $k\to \infty$.
By making use of the exact $N/D$ method, 
 developed in Sec.~\ref{sec.180819.1}, we can always obtain from the $\Delta(p^2)$ calculated  
 the standard solution of the LS equation for a regular potential.  

\subsubsection{$\Delta(p^2)$ for the $^1P_1$ PWA}
\label{sec.180814.3}

 The OPE partial-wave projected potential in the $^1P_1$ wave is
\begin{align}
\label{240915.1}
v(p,p')&=\frac{3g}{pp'}-\frac{3g}{4(pp')^2}(p^2+{p'}^2+m_\pi^2) \Theta(p,p')~.
\end{align}
Only the second term on the rhs of the previous equation contributes to the discontinuity of the potential. 
 From this equation we deduce $\rho(\nu^2,\nu_1^2)$ for the $^1P_1$ PWA,\footnote{For brevity 
we do not write the full $\rho(\nu^2,\nu_1^2;m_\pi^2)$ an omit the last argument whenever there is no ambiguity in this respect.} 
\begin{align}
\label{180819.1}
\rho(\nu^2,\nu_1^2)&=\frac{3g \pi}{4}(\nu^2+\nu_1^2-m_\pi^2)~.
\end{align}

 The IE we have to solve is the particularization of Eq.~\eqref{170526.12} to the present case with $\ell=1$. 
Let us write down the IEs that result here for the different intervals of  $\nu$.  
In the following we take the pion mass as the unit of energy ($m_\pi=1$), unless it is explicitly written. 
This simplifies the equations to some extent since $m_\pi$ appear profusely. 
For $\nu>-1/2$ we have the standard form of the IE,
\begin{align}
\label{240915.4}
\mff(\nu)&=-\frac{3\pi g}{2}(k^2+\nu^2-1)+\theta(k-2-\nu)\frac{3g m}{8\pi}\int_{1+\nu}^{k-1}d\nu_1
\frac{\nu^2+\nu_1^2-1}{k^2-\nu_1^2} S_2(\nu_1) \mff(\nu_1)~,
\end{align}
where the function $S_2(\nu_1)$ is given in Eq.~\eqref{170525.3} with $2\ell_n+2\to 2$. 
We tried to transform this IE into a DE but because of the $\nu^2$ dependence in the integrand one needs to 
consider at least a third order DE, which makes that a simple algebraic solution cannot be 
found even in the limit 
$k\gg m_\pi$ as done in Sec.~\ref{sec.180814.3b} for the $^1S_0$ wave. 
 In the range $\nu>k-2$, which is the only one  in the interval of interest for $\nu\in [-k+1,k-1]$ 
if $k<3/2$, we have that $\mff(\nu)$ is given by the independent term $-3\pi g(k^2+\nu^2-1)/2$.

 Now we present the partially symmetrized form of the IE around $\nu_1=0$ that is used for 
 $\nu_1<-3/2$, Eq.~\eqref{180814.2}. 
\begin{align}
\label{270915.1}
\mff(\nu)&=-\frac{3\pi g}{2}(k^2+\nu^2-1)+\theta(k-2-\nu)\frac{3g m}{8\pi}\int_{0}^{-1-\nu}d\nu_1
\frac{\nu^2+\nu_1^2-1}{k^2-\nu_1^2} S_2(\nu_1) (\mff(\nu_1)+\mff(-\nu_1))\nn\\
&+\theta(k-2-\nu)\frac{3g m}{8\pi}\int_{-1-\nu}^{k-1}d\nu_1\frac{\nu^2+\nu_1^2-1}{k^2-\nu_1^2} S_2(\nu_1) \mff(\nu_1)~.
\end{align}

In order to perform the numerical integration around $\nu_1=0$ we apply the procedure 
explain after Eq.~\eqref{170526.12}, so that we add and subtract $2\mff(0)$  to 
the symmetric combination $\mff(\nu_1)+\mff(-\nu_1)$. Then, Eq.~\eqref{270915.1} becomes
\begin{align}
\label{270915.3}
\mff(\nu)&=-\frac{3\pi g}{2}(k^2+\nu^2-1)+\theta(k-2-\nu)\frac{3g m}{8\pi}\int_{0}^{-1-\nu}d\nu_1
\frac{\nu^2+\nu_1^2-1}{k^2-\nu_1^2} S_2(\nu_1) (\mff(\nu_1)+\mff(-\nu_1)-2\mff(0)) \nn\\
&+\theta(k-2-\nu)\frac{3g m}{4\pi}\mff(0)\int_{0}^{-1-\nu}d\nu_1
\frac{\nu^2+\nu_1^2-1}{k^2-\nu_1^2} S_2(\nu_1) \\
&+\theta(k-2-\nu)\frac{3g m}{8\pi}\int_{-1-\nu}^{k-1}d\nu_1\frac{\nu^2+\nu_1^2-1}{k^2-\nu_1^2} S_2(\nu_1) \mff(\nu_1)~.\nn
\end{align}
In this way the integrand in the first integral is finite for $\nu_1\to 0$ and poses no problems for its 
numerical evaluation, while the  integral  multiplied by $\mff(0)$ can be done algebraically with the result
\begin{align}
\label{270915.4}
\int_{0}^{-1-\nu}d\nu_1\frac{\nu^2+\nu_1^2-1}{k^2-\nu_1^2} S_2(\nu_1)
=\frac{1}{k^3}\left(
2k - 2k \nu + (k^2+\nu^2-1)\log\frac{k+1+\nu}{k-1-\nu}
\right)~.
\end{align}
Inserting back this expression into Eq.~\eqref{270915.3} we obtain the final expression of the IE for 
$-3/2<\nu$:
\begin{align}
\label{270915.5}
\mff(\nu)&=-\frac{3\pi g}{2}(k^2+\nu^2-1)+\theta(k-2-\nu)\frac{3g m}{8\pi}\int_{0}^{-1-\nu}d\nu_1
\frac{\nu^2+\nu_1^2-1}{k^2-\nu_1^2} S_2(\nu_1) (\mff(\nu_1)+\mff(-\nu_1)-2\mff(0))\nn\\
&-\theta(k-2-\nu)\frac{3g m \mff(0)}{4\pi k^3}\left(2k - 2k \nu + (k^2+\nu^2-1)\log\frac{k+1+\nu}{k-1-\nu}\right)
\nn\\
&+\theta(k-2-\nu)\frac{3g m}{8\pi}\int_{-1-\nu}^{k-1}d\nu_1\frac{\nu^2+\nu_1^2-1}{k^2-\nu_1^2} S_2(\nu_1) \mff(\nu_1)~.
\end{align}
The calculation of $\mff(0)$ is straightforward, since it only involves values of $\nu_1\geq 1$ for which there 
is no infrared divergences in the integrand, and it can be calculated directly making use of Eq.~\eqref{240915.4}
\begin{align}
\label{270915.6}
\mff(0)&=-\frac{3\pi g}{2}(k^2-1)+\theta(k-2)\frac{3g m}{8\pi}\int_{1}^{k-1}d\nu_1
\frac{\nu_1^2-1}{k^2-\nu_1^2} S_2(\nu_1) \mff(\nu_1)~.
\end{align}

 We now consider Eq.~\eqref{180814.3} for the interval of values $-3/2<\nu<-1/2\,$.  
 For pedagogical reasons we exemplify the derivations driving to this equation 
 explicitly here for the $^1P_1$ PWA. Then,  Eq.~\eqref{240915.4}, 
  after its first iteration, can be written as
\begin{align}
\label{270915.7}
\mff(\nu)&=-\frac{3\pi g}{2}(k^2+\nu^2-1)
-\theta(k-2-\nu)\left(\frac{3g}{2}\right)^2\frac{m}{4}\int_{\nu+1}^{k-1}
\frac{d\nu_1(\nu^2+\nu_1^2-1)(k^2+\nu_1^2-1)}{k^2-\nu_1^2}S_2(\nu_1) \\
&+\theta(k-2-\nu)\left(\frac{3gm}{8\pi}\right)^2\int_{\nu+1}^{k-1}
\frac{d\nu_1(\nu^2+\nu_1^2-1)}{k^2-\nu_1^2}S_2(\nu_1)\theta(k-2-\nu_1)\int_{\nu_1+1}^{k-1}
\frac{d\nu_2(\nu_1^2+\nu_2^2-1)}{k^2-\nu_2^2}S_2(\nu_2)\mff(\nu_2)~.\nn
\end{align}

The order of the integration variables $\nu_1\leftrightarrow \nu_2$  
in the last integral is exchanged, attending to the integration region shown in
  Fig.~\ref{fig.170526.4} (here $\mu=\mu'=m_\pi$). Let us notice also that 
in this integral one has to fulfill $k-2>\nu_1$ and $\nu_1>\nu+1$ which in turn implies that 
$k-3>\nu$.    It then results
\begin{align}
\label{270915.8}
\mff(\nu)&=-\frac{3\pi g}{2}(k^2+\nu^2-1)
+\theta(k-2-\nu)\left(\frac{3g}{2}\right)^2\frac{m}{4}\int_{\nu+1}^{k-1}
\frac{d\nu_1(\nu^2+\nu_1^2-1)(k^2+\nu_1^2-1)}{k^2-\nu_1^2}S_2(\nu_1)\nn\\
&+\theta(k-3-\nu)\left(\frac{3gm}{8\pi}\right)^2\int_{\nu+2}^{k-1}\frac{d\nu_2S_2(\nu_2)}{k^2-\nu_2^2}\mff(\nu_2) 
\int_{\nu+1}^{\nu_2-1}\frac{d\nu_1(\nu^2+\nu_1^2-1)(\nu_1^2+\nu_2^2-1)S_2(\nu_1)}{k^2-\nu_1^2}~.
\end{align}
The last integral can be done algebraically with the result
\begin{align}
\label{270915.9}
{\cal F}(\nu_2)&=\int_{\nu+1}^{\nu_2-1}\frac{d\nu_1(\nu^2+\nu_1^2-1)(\nu_1^2+\nu_2^2-1)S_2(\nu_1)}{k^2-\nu_1^2}
=-\frac{2}{k^3}\left[
k(2+\nu-\nu_2)(k^2-(\nu-1)(\nu_2+1))\right.\nn\\
&-\left.(k^2+\nu^2-1)(k^2+\nu_2^2-1)
\left(
{\rm arctanh}\frac{\nu+1}{k}-{\rm arctanh}\frac{\nu_2-1}{k}
\right)
\right]~.
\end{align}
The particularization of this expression for $\nu_2=k$ corresponds to the first integral in the rhs of 
Eq.~\eqref{270915.8}. In terms of the function ${\cal F}(\nu_2)$ we can rewrite Eq.~\eqref{270915.8}, 
used for $-3/2<\nu<-1/2$, as
\begin{align}
\label{270915.10}
\mff(\nu)&=-\frac{3\pi g}{2}(k^2+\nu^2-1)
+\theta(k-2-\nu)\left(\frac{3g}{2}\right)^2\frac{m}{4}{\cal F}(k)
-\theta(k-3-\nu)\left(\frac{3gm}{8\pi}\right)^2\int_{\nu+2}^{k-1}\frac{d\nu_2S_2(\nu_2)}{k^2-\nu_2^2}{\cal F}(\nu_2) 
\mff(\nu_2) ~.
\end{align}

 Once $\mff(\nu)$ is obtained we calculate $\Delta(p^2)$ according to  Eq.~\eqref{280116.3}, 
\begin{align}
\label{270915.11}
\Delta(-k^2)&=-\frac{\mff(-k)}{2 k^4}~.
\end{align}

We show in the top right panel of Fig.~\ref{fig.180817.1} a log-log plot with the resulting 
$\Delta(-k^2)$ for the $^1P_1$ PWA, where its  modulus as a function of $k^2$ is shown in appropriate units of $|L|^{-1}$. 
 We explicitly see the vanishing of $|\Delta(-k^2)|$ for $k\to \infty$ with the same slope as for the $^1 S_0$, 
 so that it decreases again as $k^{-2}$. We have also tested this conclusion by performing a fit to $\Delta(-k^2)$ 
 in this asymptotic  region. This is the expected behavior following the discussion 
 at the end of Sec.~\ref{sec.180814.3b}, because from Eq.~\eqref{180819.1} $m=2$, which is less than 
$2\ell+1=3$. Therefore, we should have asymptotically the  progressive on-set of an increasingly
 dominant Born-term contribution. It results that the full $\Delta(-k^2)$ for the $^1P_1$ is rather 
 similar to the Born contribution.

\subsubsection{$\Delta(p^2)$ for the $^1D_2$ PWA}
\label{sec.180819.2}

 The $^1D_2$ partial-wave potential in momentum space reads
\begin{align}
\label{180819.4}
v(p,p')&=\frac{g}{16(pp')^3}\left\{-12 p{p'}(p^2+{p'}^2+m_\pi^2)+
\left(3(p^2+{p'}^2+m_\pi^2)^2-4p^2{p'}^2\right) \Theta(p,p') \right\}~.
\end{align}
Since this is a singlet PWA, the partial-wave projection formula for OPE is proportional to
 Eq.~\eqref{180820.4} with $\ell=2$ and divided by $2pp'$. 
  From Eq.~\eqref{180819.4} we deduce that
\begin{align}
\label{180820.5}
\rho(\nu^2,\nu_1^2)&=\frac{\pi g}{16}\left[
3\left(\nu^2+\nu_1^2-m_\pi^2\right)^2-4\nu^2\nu_1^2
\right]~.
\end{align}
Notice that this function is just $\pi g P_2(\xi)(\nu\nu_1)^2/8$, as it was also discussed in connection 
with Eq.~\eqref{180820.4}.

For the different ranges of $\nu$ the resulting IEs from Eqs.~\eqref{180814.1}-\eqref{180814.3} are
\begin{align}
\label{180821.1}
&-1/2<\nu \nn\\
\mff(\nu)&=- 2\rho(\nu^2,k^2)
+\theta(k-2-\nu)\frac{m}{\pi^2}\int_{\nu+1}^{k-1}\frac{d\nu_1}{(k^2-\nu_1^2)\nu_1^4}
\rho(\nu^2,\nu_1^2) \mff(\nu_1)~,\\
\label{180821.2}
&-3/2<\nu<-1/2\nn\\
\mff(\nu)&=-2\rho(\nu^2,k^2)-\theta(k-2-\nu)\frac{m}{\pi^2}\int_{\nu+1}^{k-1}\frac{d\nu_1}{k^2-\nu_1^2}
S_4(\nu_1)\rho(\nu^2,\nu_1^2)\rho(k^2,\nu_1^2)\\
&+\theta(k-3-\nu)\frac{m^2}{2\pi^4}\int_{\nu+2}^{k-1}\frac{d\nu_2}{(k^2-\nu_2^2)\nu_2^4}\mff(\nu_2)
\int_{\nu+1}^{\nu_2-1}\frac{d\nu_1}{k^2-\nu_1^2}S_4(\nu_1)\rho(\nu^2,\nu_1^2)\rho(\nu_2^2,\nu_1^2)~,\nn\\
\label{180821.3}
&\nu<-3/2 \nn\\
\mff(\nu)&=-2\rho(\nu^2,k^2)+\theta(k-2-\nu)\frac{m}{\pi^2}\int_{-\nu-1}^{k-m_\pi}\frac{d\nu_1}{(k^2-\nu_1^2)\nu_1^4}
 \rho(\nu^2,\nu_1^2)\mff(\nu_1)\\
&+\theta(k-2-\nu)\frac{m}{2\pi^2}\int_0^{-\nu-1}\frac{d\nu_1}{k^2-\nu_1^2}S_4(\nu_1)\rho(\nu^2,\nu_1^2)
\left[\mff(\nu_1)+\mff(-\nu_1)-2\mff(0)-\mff''(0)\nu_1^2\right]\nn\\
&+\theta(k-2-\nu)\left[\mff(0)+\frac{1}{2}\mff''(0)\nu_1^2\right]
\frac{m}{\pi^2}\int_0^{-\nu-1}\frac{d\nu_1}{k^2-\nu_1^2}S_4(\nu_1)\rho(\nu^2,\nu_1^2)~.\nn
\end{align}
 We indicate the derivative of a function with respect to $\nu$ by a prime. The function $S_4(\nu_1)$ is given by 
Eq.~\eqref{170525.3} with the substitution $2\ell_n+2\to 4$.

As it was explained in Eq.~\eqref{180124.1b} the last integral in Eq.~\eqref{180821.3} is finite. 
We then concentrate on the evaluation of the integral
\begin{align}
\label{180821.4}
\int_{\nu+1}^{\nu_2-1}\frac{d\nu_1}{k^2-\nu_1^2}S_4(\nu_1)\rho(\nu^2,\nu_1^2)\rho(\nu_2^2,\nu_1^2)~,
\end{align} 
which is the analogous one to that discussed around Eq.~\eqref{180820.2}. 
 We are going to consider the different contributions to Eq.~\eqref{180821.4} by performing an expansion 
in powers of $\nu_1^2$ of the product of the two functions $\rho$. Then, we have terms with 
$\nu^{2i}$ and $i=0,\dots,4$ that we write as 
\begin{align}
\label{180821.5}
\rho(\nu^2,\nu_1^2)\rho(\nu_2^2,\nu_1^2)=\sum_{i=0}^4 c_i \nu^{2i}~.
\end{align}
  One picks up divergent integrals out of these contributions for the integrations  
\begin{align}
\label{180823.1}
\int_{\nu+1}^{\nu_2-1}\frac{d\nu_1}{k^2-\nu_1^2}S_4(\nu_1)\nu_1^{2i}
\end{align}
and $i=0,1$, with leading divergences as $1/(\nu+1)^{3-2i}+1/(\nu_2-1)^{3-2i}$. 
However, the divergences $1/(\nu+1)^3$ for $\nu\to -1$ do not actually appear because the coefficient
 of the independent term ($i=0$) contains the factor $(\nu^2-1)^2=(\nu-1)^2(\nu+1)^2$, and then we are left with only 
the divergences of first degree in $1/(\nu-1)$. The latter finally cancels because of finite 
arrangements between the coefficients $c_0$ and $c_2$ and the actual calculation of the integrals 
with $i=0,1$. Indeed we have for the integrals a power expansion of $\nu$ around $\nu=-1$ such that
\begin{align}
\label{180821.6}
c_0 \int_{\nu+1}^{\nu_2-1}\frac{d\nu_1}{k^2-\nu_1^2}S_4(\nu_1)=\frac{24(1-\nu_2^2)^2}{k^2(\nu+1)}+\ldots\\
c_2 \int_{\nu+1}^{\nu_2-1}\frac{d\nu_1}{k^2-\nu_1^2}S_4(\nu_1)\nu_1^2=-\frac{24(1-\nu_2^2)^2}{k^2(\nu+1)}+\ldots\nn
\end{align}
where the ellipsis indicate finite contributions. We see explicitly the cancellation between the divergent 
contributions for $\nu\to -1$. 
We could also repeat a similar analysis to conclude that the  only remaining first degree divergences
  as $1/(\nu_2-1)$ for $\nu_2\to 1$ from the integrals in Eq.~\eqref{180823.1} with $i=0,2$ finally cancel.
 
Making use of Eq.~\eqref{180821.1} we have the following expressions for $\mff(0)$ and $\mff''(0)$  in terms of 
$\mff(\nu)$ with $\nu\geq 1$,
\begin{align}
\label{180821.7}
\mff(0)&=-2\rho(0,k^2)+\theta(k-2)\frac{m}{\pi^2}\int_1^{k-1}\frac{d\nu_1}{(k^2-\nu_1^2)\nu_1^4}
\rho(0,\nu_1^2)\mff(\nu_1)~,\\
\mff''(0)&=-2\rho''(0,k)+\theta(k-2)\frac{m}{\pi^2}\int_1^{k-1}\frac{d\nu_1}{(k^2-\nu_1^2)\nu_1^4}
\rho''(0,\nu_1^2)\mff(\nu_1)~,\nn
\end{align}
where a prime denotes a derivative with respect to $\nu$. 
 In obtaining these expressions simplifications arise by noticing that 
\begin{align}
\label{180821.8}
\rho(0,1)=0~,\\
\rho'(\nu^2,(1+\nu)^2)|_{\nu=0}&=0~.\nn
\end{align}
 It is also simpler to calculate $\mff''(0)$ algebraically by evaluating the derivatives 
 of the symmetric combination $\mff(\nu)+\mff(-\nu)$ and then to particularize at the origin.

Once $\mff(\nu)$ is determined in the range of values of interest of $\nu\in[-k+1,k-1]$, we then 
apply Eq.~\eqref{180821.9} (or Eq.~\eqref{180821.3} with $\nu=-k$) 
to calculate $\mff(-k)$, and then Eq.~\eqref{280116.3} to calculate $\Delta(-k^2)$. 
The modulus of the latter is shown in 
the log-log plot of the  bottom right panel of Fig.~\ref{fig.180817.1}. 
 Let us mention that in a log-log plot a change of sign of $\Delta(-k^2)$ originates the abrupt behavior that 
 can be observed twice in this plot. 
 It is also clear from the figure that asymptotically $\Delta(-k^2)$ also decreases as $1/k^2$ for $k\to \infty$, 
 since one has the same slope as for the $^1S_0$ curve (this is also checked numerically).
 Again, this is the expected behavior because  the discussion 
 at the end of Sec.~\ref{sec.180814.3b} is applicable here, as $m=4$ from Eq.~\eqref{180820.5} is less than 
$2\ell+1=5$. Thus, the  progressive on-set of a dominant Born-term contribution to $\Delta(-k^2)$ should happen 
asymptotically for $k\gg m_\pi$. Indeed, $\Delta(-k^2)$ for the $^1D_2$ is very similar to the Born contribution in 
the whole range shown in Fig.~\ref{fig.180817.1}.

\subsection{Attractive singular  potentials}
\label{sec.180823.1}

We study in this section two examples of attractive singular potentials. 
We discussed the $NN$ OPE potential for the $^3P_0$ partial wave.
Then, we move to consider higher order contributions to the $NN$ potential up to and including next-to-leading order (NLO) and 
next-to-next-to-leading order (NNLO) contributions in its calculation within $\chi$PT \cite{kaiser.080516.1}, and apply 
it to the $^1S_0$ PWA. 
The functions $\Delta(p^2)$ for the NLO and NNLO $^1S_0$ potentials were already calculated in 
Ref.~\cite{entem.170930.1}, though the actual method for its calculation was not given in this reference. 
We fill now this gap and discuss its calculation in detail. Furthermore, this case also exemplifies 
the method to calculate the discontinuity along the LHC for a potential with an involved spectral function.

\begin{figure}[H]
\begin{center}
\scalebox{0.8}{
\begin{tabular}{lll}
\includegraphics[width=.4\textwidth]{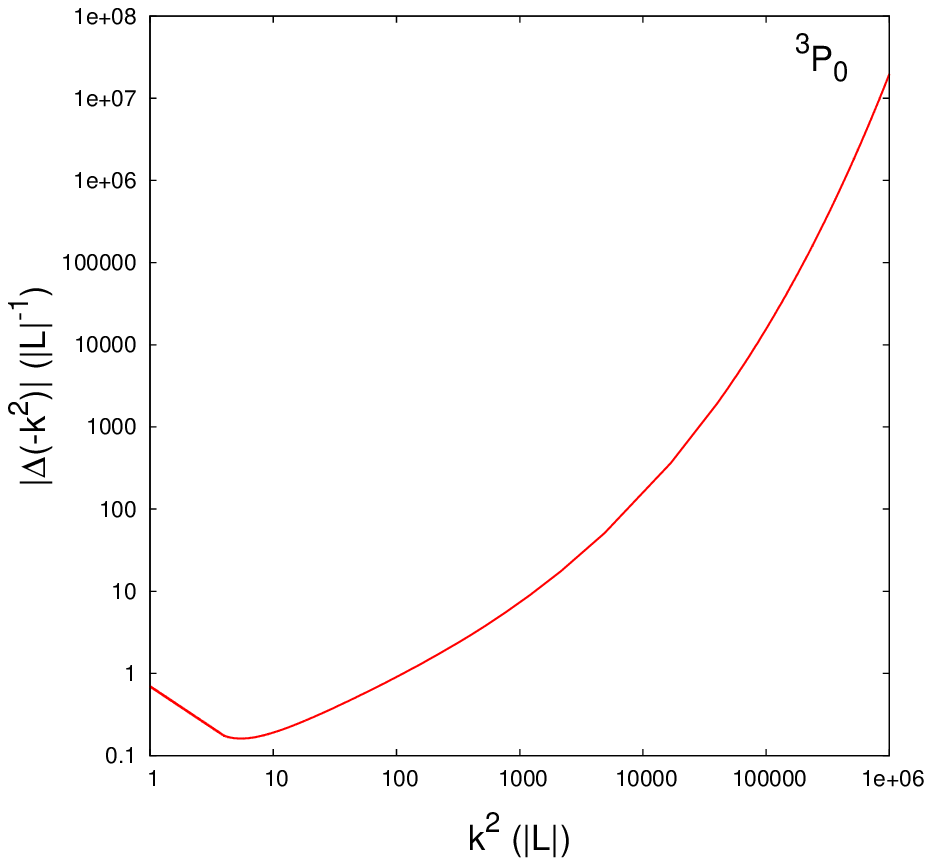} &
\includegraphics[width=.4\textwidth]{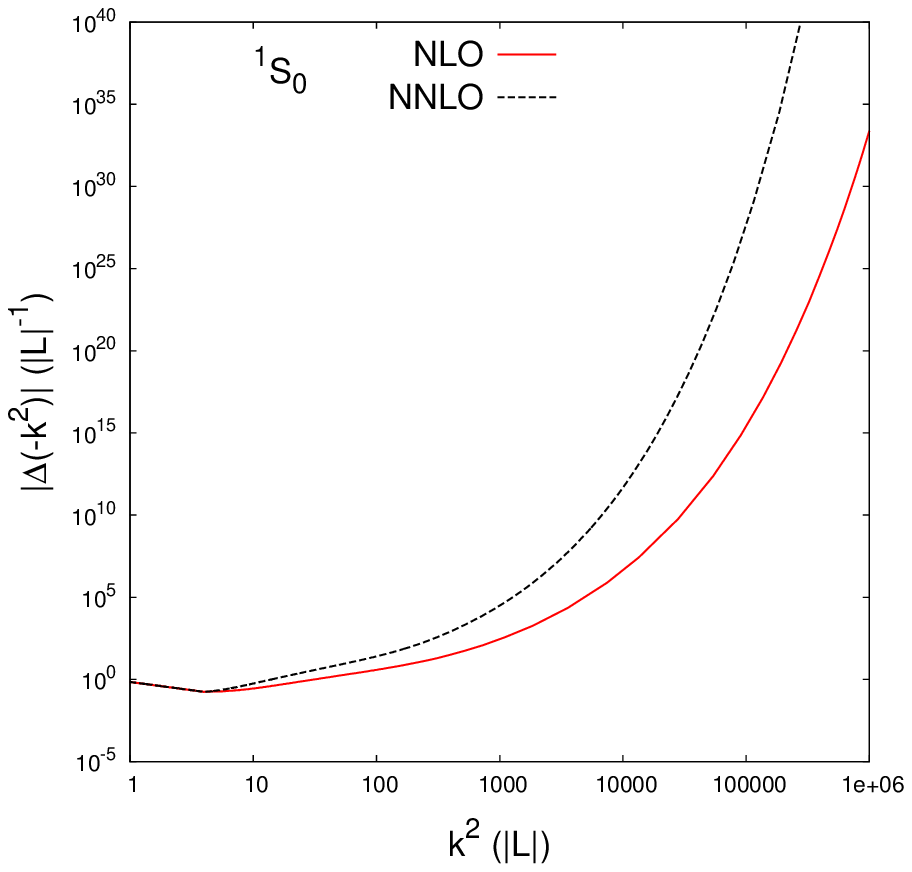} \\ 
\includegraphics[width=.4\textwidth]{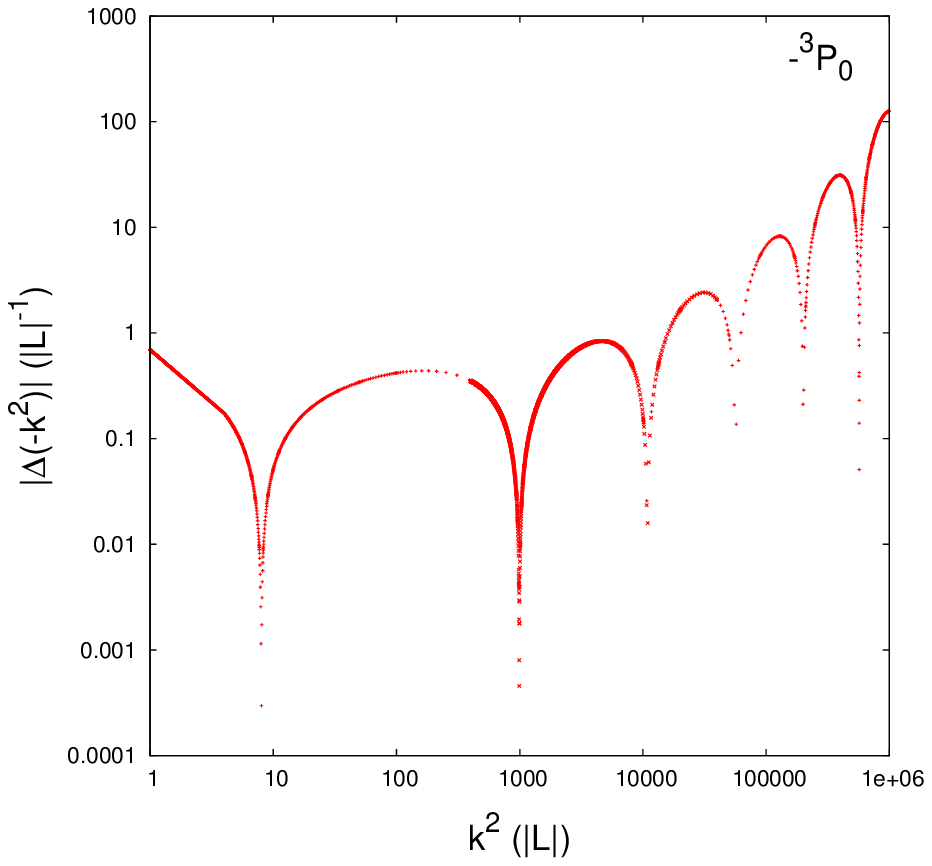} & 
\includegraphics[width=.4\textwidth]{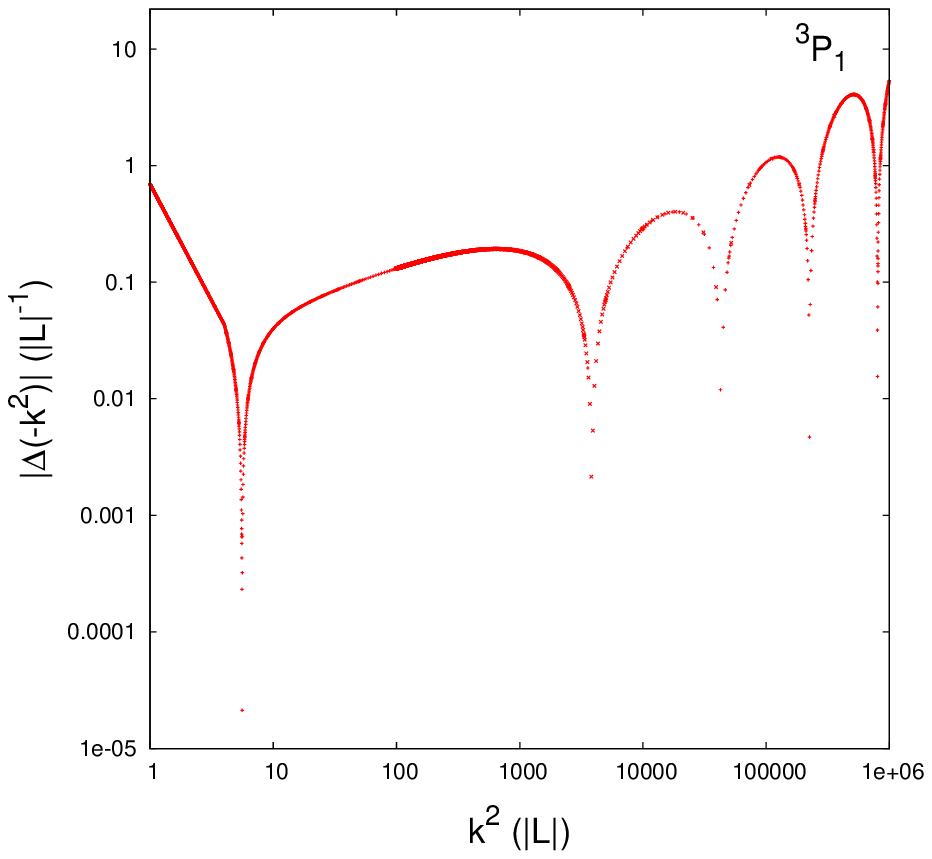} \\ 
\end{tabular}
}
\end{center}
\caption{{\small Log-log plot of the discontinuity $\Delta(-k^2)$ for the PWAs
 $^3P_0$ (top left), $^1S_0$ (top right), $-^3P_0$ (bottom left) and $^3P_1$ (bottom right panel). 
The first two waves correspond to attractive singular interactions and the last two to repulsive 
singular ones. 
The PWA denoted as $-^3P_0$ refers to the $^3P_0$ PWA but calculating 
 with a potential equal to minus the OPE potential for this PWA.
The OPE $NN$ potential is used in the $^3P_0$, $-^3P_0$ and $^3P_1$ PWAs. 
For the $^1S_0$ case we employ the NLO (red solid line) and NNLO (black dashed line)
  $\chi$PT $NN$ potentials, which also involve two-pion-exchange contributions \cite{kaiser.080516.1}.}
\label{fig.180829.1}
}
\end{figure}

For these two examples we obtain that $|\Delta(-k^2)|$ grows faster than any power of $k^2$ for $k^2\to \infty$. 
This is clearly observed in a log-log plot, where the slope of the curve continuously grows with $k^2$, as we show 
 in the top panels of Fig.~\ref{fig.180829.1} (and all the panels in Fig.~\ref{fig.180130.1}). 
 
 Indeed, we expect that this growing divergent behavior of $\Delta(-k^2)$ for $k^2\to \infty$ holds for any 
singular attractive potential for which $\rho(\nu^2,\nu_1^2;m_\pi^2)$ stems from OPE and it is a polynomial 
in $\nu$ and $\nu_1$ of degree $m$, with $m>2\ell+1$. 
This expectation is based on the discussion at the end of Sec.~\ref{sec.180814.2} driving to Eq.~\eqref{180825.1}, 
where we concluded that an $n$-times iterated contribution to $\Delta(-k^2)$ scales with 
 a power $k^{(m-2\ell-1)n}$ relative to the Born-term contribution. 
 As a result, the scaling power in $k$ of $\Delta(-k^2)$ for $k\to \infty$ grows with $k$, 
  since the number of iterations needed to solve the recursive IE of Eq.~\eqref{170525.2} increases in this variable. 


As also remarked at the end of Sec.~\ref{sec.180814.3b}, for a potential with 
 a general  spectral function, one applies the previous argument to the weighted function  
 $\rho(\nu,\nu_1)$ defined by
\begin{align}
\label{180825.2b}
\rho(\nu,\nu_1)&=
\frac{1}{\pi}\int_{m_\pi}^{\nu_1-\nu} 2\mu d\mu \eta(\mu^2)\rho(\nu^2,\nu_1^2;\mu^2)~,~\nu_1\geq \nu+m_\pi~,
\end{align}
 by establishing whether it scales in $\nu$, $\nu_1$ for  
$|\nu|\sim |\nu_1|\sim k\gg m_\pi$ as $\nu^{m}$, $\nu_1^{m}$ times logarithmic factors, in order. 
If this is the case and $m$ is larger than $2\ell+1$,  
 then one expects having this exceedingly strong divergent behavior for  $\Delta(-k^2)$ 
 along the LHC.  

When $\Delta_{ij}(-k^2)$ diverges faster than any power of $k$ for $k\to \infty$ and  keeps the same sign, 
it follows  the important conclusion  
that a DR for the PWA $t_{ij}(p,p)$  is not possible in this case because an infinite number 
of subtractions would be needed.\footnote{Nonetheless, it is still possible to use the $N/D$ method as we 
develop in Sec.~\ref{sec.180819.1}.} Let us recall that because of unitarity the PWA along the physical axis 
decreases as $1/p$ for $p\to \infty$, cf. Eq.~\eqref{130116.3}. Thus, the contribution from the RHC in the DR of a 
PWA cannot compensate the exceedingly strong divergence of $\Delta_{ij}(-k^2)$ for $k^2\to \infty$.

\subsubsection{$\Delta(p^2)$ for the $^3P_0$ PWA}
\label{sec.180827.1}

At the level of OPE the $^3P_0$ potential in configuration space is from Eq.~\eqref{180505.2}\footnote{Reference 
\cite{kolck.180505.1} provides the matrix elements of the operator $\mathbf{i}_1\cdot \mathbf{i}_2 S_{12}$ for spin triplet channels with 
total angular momentum $J$.}
\begin{align}
\label{180828.1}
V(r)&=-\frac{g}{2\pi}\frac{e^{-m_\pi r}}{r}\left[1+\frac{4}{m_\pi r}+\frac{4}{(m_\pi r )^2}\right]~,
\end{align}
which is an attractive singular potential. In momentum space we have the potential function $v(p,p')$ given by
\begin{align}
\label{180828.2}
v(p,p')&=\frac{g}{2pp' m_\pi^2}\left[
2(p^2+{p'}^2)-\frac{(p^2-{p'}^2)^2+m_\pi^2(p^2+{p'}^2)}{2pp'}\Theta(p,p')
\right]~.
\end{align}
From this equation we read the following expression for the function $\rho(\nu^2,\nu_1^2)$, 
\begin{align}
\label{180828.3}
\rho(\nu^2,\nu_1^2)&=-\frac{\pi g}{4 m_\pi ^2}\left[
\left(\nu^2-\nu_1^2\right)^2-m_\pi^2(\nu^2+\nu_1^2)
\right]~.
\end{align}
This function is a polynomial in $\nu$ and $\nu_1$ of fourth degree (larger than $2\ell+1$). Therefore, 
according to the analysis above at the beginning of Sec.~\ref{sec.180823.1}, we conclude from 
this other mean that indeed this is a singular potential. 
 The on-shell discontinuity of this potential along the LHC ($\nu=\nu_1=k$) 
is $\rho(k^2,k^2)/k^4=\pi g /2k^2$.

The IEs that we have to solve depending on the values of $\nu$ are:
\begin{align}
\label{180828.4}
&-1/2<\nu \nn\\
\mff(\nu)&=- 2\rho(\nu^2,k^2)
-\theta(k-2-\nu)\frac{m}{\pi^2}\int_{\nu+1}^{k-1}\frac{d\nu_1}{(k^2-\nu_1^2)\nu_1^2}
\rho(\nu^2,\nu_1^2) \mff(\nu_1)~,\\
\label{180828.5}
&-3/2<\nu<-1/2\nn\\
\mff(\nu)&=-2\rho(\nu^2,k^2)-\theta(k-2-\nu)\frac{m}{\pi^2}\int_{\nu+1}^{k-1}\frac{d\nu_1}{k^2-\nu_1^2}
S_2(\nu_1)\rho(\nu^2,\nu_1^2)\rho(k^2,\nu_1^2)\\
&-\theta(k-3-\nu)\frac{m^2}{2\pi^4}\int_{\nu+2}^{k-1}\frac{d\nu_2}{(k^2-\nu_2^2)\nu_2^2}\mff(\nu_2)
\int_{\nu+1}^{\nu_2-1}\frac{d\nu_1}{k^2-\nu_1^2}S_2(\nu_1)\rho(\nu^2,\nu_1^2)\rho(\nu_2^2,\nu_1^2)~,\nn\\
\label{180828.6}
&\nu<-3/2 \nn\\
\mff(\nu)&=-2\rho(\nu^2,k^2)+\theta(k-2-\nu)\frac{m}{\pi^2}\int_{-\nu-1}^{k-m_\pi}\frac{d\nu_1}{(k^2-\nu_1^2)\nu_1^2}
 \rho(\nu^2,\nu_1^2)\mff(\nu_1)\\
&+\theta(k-2-\nu)\frac{m}{2\pi^2}\int_0^{-\nu-1}\frac{d\nu_1}{k^2-\nu_1^2}S_2(\nu_1)\rho(\nu^2,\nu_1^2)
\left[\mff(\nu_1)+\mff(-\nu_1)-2\mff(0)\right]\nn\\
&+\theta(k-2-\nu) \mff(0)\frac{m}{\pi^2}\int_0^{-\nu-1}\frac{d\nu_1}{(k^2-\nu_1^2)}S_2(\nu_1)\rho(\nu^2,\nu_1^2)~.\nn
\end{align}
We also use Eq.~\eqref{180828.4} for $\nu=0$ for calculating $\mff(0)$, which is needed in Eq.~\eqref{180828.6}. 
The integrals 
\begin{align}
\label{180901.1}
I(\nu)&=\int_0^{-\nu-1}\frac{d\nu_1 \nu_1^2}{k^2-\nu_1^2}S_2(\nu_1)\rho(\nu^2,\nu_1^2)~,\\
J(\nu,\nu_2)&=\int_{\nu+1}^{\nu_2-1}\frac{d\nu_1}{k^2-\nu_1^2}S_2(\nu_1)\rho(\nu^2,\nu_1^2)\rho(\nu_2^2,\nu_1^2)~,\nn
\end{align}
appearing in Eqs.~\eqref{180828.5} and \eqref{180828.6}, respectively, are done algebraically. 
 These IEs are analogous to those deduced for the  $^1P_1$ PWA in Sec.~\ref{sec.180814.3}. 
 However, we gather them here all together and in a way that is valid for any uncoupled $P$ wave.

From the solution of these IEs we obtain $\Delta(p^2)$ along the LHC, and this function is 
represented by the log-log plot in the top left panel of Fig.~\ref{fig.180829.1}. 
 We observe clearly the referred increasing degree of divergence of $|\Delta(-k^2)|$ with $k$ as $k\to \infty$. 
 
\subsubsection{$\Delta(p^2)$ for the $^1S_0$ PWA from the  NLO and NNLO chiral potentials}
\label{sec.180901.1}

In this section we give an example of calculating the discontinuity along the LHC for a potential 
involving a continuous spectral function. We consider the $^1S_0$ $NN$ PWA and the $NN$ potential calculated at NLO and 
NNLO in baryon $\chi$PT from Ref.~\cite{kaiser.080516.1}. 
Since at this level the $NN$ potential involves also two-pion exchange contributions the resulting 
spectral function  is continuous for $\mu\geq 2m_\pi$. The phase shifts obtained from the resulting $\Delta(p^2)$ 
by applying the $N/D$ method were already discussed in Ref.~\cite{entem.170930.1}, but the calculation of 
$\Delta(p^2)$ for this case was not treated in detail, and we discuss it here.  

Reference~\cite{kaiser.080516.1} expresses the NLO $NN$ potential in terms of the function 
$L(q)$ given by
 \begin{align}
   \label{180903.1}
L(q)&=\frac{\omega}{q}\log\frac{\omega+q}{2m_\pi}~,~\omega=\sqrt{4m_\pi^2+q^2}~,
 \end{align}
and $L(-q)=L(q)$.  As a function of $q^2$ it is convenient to introduce the variable  $x=q^2/4m_\pi^2$ and write the function
$L(x)$ as 
 \begin{align}
   \label{170714.1}
   L(x)&=\frac{\sqrt{1+x}}{\sqrt{x}}\log\left(\sqrt{x}+\sqrt{1+x}\right)~.
 \end{align}
This function has only LHC for $x<-1$, as a required for a 
$NN$ potential. It is  the correct analytical 
extrapolation in the complex $q^2$ plane of $L(q)$ in Eq.~\eqref{180903.1},
it coincides with the original one  in an open domain around the physical region, namely for
$q^2>0$.
The function $\log z$ present in Eq.~\eqref{170714.1} should be interpreted with ${\rm arg}z\in[-\pi,\pi]$.
Regarding the square roots, one can define them either having the cut 
along the negative or positive real numbers.
For definiteness we use the standard convention so that $\sqrt{z}$ is calculated
with  ${\rm arg}z\in [-\pi,\pi]$. 
For simplicity in the writing, we keep the same symbol for the function $L$
despite that now its argument is $x$ instead of $q$. 
The discontinuity of $L(x)$ is given by
\begin{align}
  \label{170714.2}
  \lim_{\ep\to 0^+}\left\{L(x+i\ep)-L(x-i\ep)\right\}&=
  2i\Ima L(x)=i\pi\frac{\sqrt{-1-x}}{\sqrt{-x}}~,~x<-1~.
\end{align} 
As a check of this result we write down the following once-subtracted DR for $L(x)$
\begin{align}
  \label{170714.3}
  L(x)&=L(0)+\frac{x}{\pi}\int_{-\infty}^{-1} dy \frac{\Ima L(y+i\ep)}{y(y-x)}\\
  &=1+\frac{x}{2}\int_{-\infty}^{-1}dy\frac{\sqrt{-1-y}}{\sqrt{-y}}\frac{1}{y(y-x)}~,
\end{align}
where we have used that $L(0)=1$, as can be easily obtained by taking the limit 
$x\to 0$ in either Eq.~\eqref{180903.1} or Eq.~\eqref{170714.1}. 
The equality between the previous DR and the algebraic result 
in Eq.~\eqref{170714.1}  is very well fulfilled numerically in the complex $q^2$ plane. 
In its numerical implementation we have performed the change of
integration variable $t=-1/y$ and have the simpler and more convenient expression
\begin{align}
  \label{170714.4}
  L(x)&=1+\frac{x}{2}\int_0^1 dt \frac{\sqrt{1-t}}{1-t x}~.
  \end{align}

Reference~\cite{kaiser.080516.1} gives all the NLO contributions in terms of the functions $W_C(q^2)$,
$V_T(q^2)=-V_S(q^2)/q^2$. Although its explicit expressions can be found in this reference, we reproduce here 
the parts of these functions that have LHC:
\begin{align}
\label{180903.1b}
W_C(q^2)&=\frac{1}{384 \pi^2 f_\pi^4}\left[
4m_\pi^2(5g_A^4-4g_A^2-1)+q^2(23 g_A^4-10 g_A^2-1)+\frac{48g_A^4m_\pi^4}{4m_\pi^2+q^2}
\right]L(x)+\ldots~,\\
V_T(q^2)&=\frac{3g_A^4}{64\pi^2 f_\pi^4}L(x)+\ldots~.\nn
\end{align}
The ellipsis indicate polynomial terms in $q^2$ without LHC. 
These polynomials are of first and zeroth degree for $W_C$ and $V_T$, in order.

We follow the procedure explained in Appendix~\ref{app:170715.1} to determine the partial-wave projected
$NN$ potential.
The resulting formula for the NLO contribution of the $^1S_0$ potential is
\begin{align}
    \label{170714.10b}
    v(p,p')&=\frac{1}{2}\int_{-1}^{+1}dx\left\{W_C(q^2)+2q^2 V_T(q^2)\right\}~,\\
    q^2&=p^2+{p'}^2-2p{p'}x~.\nn
  \end{align}
In order to proceed forward for calculating $\Delta \hhv(\nu,\nu_1)$ we need the spectral decomposition of the integrand in the
previous equation, which can be calculated from  the imaginary parts of the
potential functions $W_C(q^2)$ and $V_T(q^2)$ at $q^2+i\epsilon$ for $q^2<-4m_\pi^2$ or $x<-1$. 
The latter can be expressed in terms of the imaginary part of the function $L(x)$.
 The most delicate structure for this purpose is
\begin{align}
  \label{170714.11}
\frac{L(x)}{x+1}~,
\end{align}
appearing  in $W_C(q^2)$, Eq.~\eqref{180903.1b}.  
It is important to stress for the following discussion that $L(-1)=0$, as it is clear
from Eq.~\eqref{170714.1}. In this way the contribution from the imaginary part of $1/(x+1)$
for $x+i\ep$ and $x\leq -1$, which is $-\pi\delta(x+1)$, finally vanishes when used
within the integral of the spectral representation of a potential because it is multiplied by $L(-1)=0$.
 The corresponding expression for the spectral functions of $W_C$ 
  and $V_T$ can be worked by employing the relation of Eq.~\eqref{140116.9}, resulting 
\begin{align}
\eta_C(\mu^2)&=-\frac{1}{768 \pi f_\pi^4}\left[
4m_\pi^2(5g_A^4-4g_A^2-1)-\mu^2(23 g_A^4-10 g_A^2-1)+\frac{48g_A^4m_\pi^4}{4m_\pi^2-\mu^2}
\right]\frac{\sqrt{\mu^2-4m_\pi^2}}{\mu} \theta(\mu-2m_\pi)~,\\
\eta_T(\mu^2)&=-\frac{3g_A^4}{128\pi f_\pi^4}\frac{\sqrt{\mu^2-4m_\pi^2}}{\mu}\theta(\mu-2m_\pi)~,\nn
\end{align}
 respectively.
 
Since $\eta_C(\mu^2)$ diverges like $\mu^2$ for $\mu\to \infty$ and the regular terms in $W_C(q^2)$ are polynomial 
of first degree in $q^2$, it follows that this function obeys a twice-subtracted DR. 
In turn, $\eta_T(\mu^2)$ tends to constant for $\mu\to \infty$ and the regular part of $V_C(q^2)$ is a zeroth degree polynomial. 
Thus, $V_T(q^2)$ can be expressed as a once-subtracted DR. This allows us to check the 
calculated spectral functions by writing down the corresponding DRs for $W_C(q^2)$ and $V_T(q^2)$, as in Eq.~\eqref{150116.2}, 
that perfectly agree with their algebraic expressions.

Additionally, we are interested in
$\Delta \hat{v}(\nu,\nu_1)$, which requires integrating the spectral decomposition for the
potential up to $\nu_1-\nu$, cf.~Eq.~\eqref{170525.1}. Thus, 
\begin{align}
  \label{170714.11b}
  \Delta \hhv(\nu,\nu_1)&=pp'\Delta v(\nu,\nu_1)
  =\frac{pp'}{2\pi}\int_0^{\nu_1-\nu}2\mu d\mu\left[\eta_C(\mu^2)-2\mu^2\eta_T(\mu^2)\right]
  \Delta \int_{-1}^{+1}\frac{dx}{\mu^2+q^2}\\
  &=-\frac{1}{2}\int_0^{\nu_1-\nu}2\mu d\mu \left\{\eta_C(\mu^2)-2\mu^2\eta_T(\mu^2)\right\}~.\nn
\end{align}
Notice that for the calculation of the discontinuity we have replaced the factor $q^2$ in front
of $V_T(q^2)$ in Eq.~\eqref{170714.10b} by $-\mu^2$ because the difference $q^2+\mu^2$
cancels the denominator and no discontinuity would result then. 
Comparing with Eq.~\eqref{170525.1} we have that 
\begin{align}
\rho(\nu^2,\nu_1^2;\mu^2)=\pi/4~.
\end{align} 

We have checked that Eq.~\eqref{170714.11b} reproduces the perturbative NLO contribution to $\Delta(p^2)$,
already used in Ref.~\cite{oller.180722.2}. The expression is 
\begin{align}
  \label{170714.12}
  \Delta(p^2)&=\frac{\Delta \hhv(-k,k)}{2k^2}
  =\frac{1}{4p^2}\int_0^{-4p^2}d\mu^2 \left\{\eta_C(\mu^2)-2\mu^2\eta_T(\mu^2)\right\}~.
  \end{align}

For the present case one can perform algebraically the integration in Eq.~\eqref{170714.11b}
and obtain a close expression for   $\Delta\hhv(\nu,\nu_1)$, and the result is (let us recall that $\nu_1\geq \nu+m_\pi$)
\begin{align}
\Delta\hhv(\nu,\nu_1)&=\frac{1}{3072\pi f_\pi^4}\left[
24 m_\pi^4 (1+2g_A^2+5g_A^4)\log\frac{\sqrt{y}+\sqrt{y-4m_\pi^2}}{2m_\pi}\right.\\
&\left.+\sqrt{y(y-4m_\pi^2)}\left\{(-59 g_A^4+10 g_A^2+1)y+2 m_\pi^2(79g_A^4-26 g_A^2-5)\right\}
\right]~,\nn\\
   y&=(\nu_1-\nu)^2~.\nn
\end{align}
In terms of it we can then apply the standard procedure to calculate $\mff(\nu)$ 
by solving the recursive IE of Eq.~\eqref{290116.22}.
 We show in the top right panel of Fig.~\ref{fig.180829.1} a log-log of the function $|\Delta(-k^2)|$ 
calculated. It is clear from this figure that the $NN$ potential at NLO in baryon $\chi$PT for
the $^1S_0$ PWA is an attractive singular potential, with $|\Delta(-k^2)|$ growing faster than any 
power of $k$ for $k\to \infty$. This behavior is expected from the expression giving $\Delta \hhv(\nu,\nu_1)$  
in Eq.\eqref{170714.11b}, from where we can infer the function $\rho(\nu,\nu_1)$ as
 $\rho(\nu,\nu_1)=-\pi\Delta\hhv(\nu,\nu_1)/2$. 
This function scales as $(\nu_1-\nu)^m$, $m=4$, in the limit $|\nu_1|\sim |\nu|\sim k\gg m_\pi$. 
Therefore, this exceedingly growing 
behavior in $k$ of $|\Delta(k)|$ follows by applying the argument derived from Eq.~\eqref{180825.2b}, as $m=4>2\ell+1=1$. 

The NNLO case is very similar. Now the relevant function is
\begin{eqnarray}
	A(q) &=& \frac{1}{2q} \arctan \frac{q}{2m_\pi} = \frac{1}{2q} \frac{i}{2}
	\left[ \log\left( 1-i\frac{q}{2m_\pi} \right)-\log\left( 1+i\frac{q}{2m_\pi} \right) \right]~,
\end{eqnarray}
and $A(-q)=A(q)$. In terms of the variable $x$ this function reads
\begin{eqnarray}
	A(x) &=& \frac{1}{4m_\pi \sqrt{x}} \frac{i}{2}
	\left[ \log\left( 1-i\sqrt{x} \right)-\log\left( 1+i\sqrt{x} \right) \right]~.
\end{eqnarray}
The discontinuity of $A(x)$ is given by
\begin{align}
  \lim_{\ep\to 0^+}\left\{A(x+i\ep)-A(x-i\ep)\right\}&=
  2i\Ima A(x)=-i\pi\frac{1}{4m_\pi} \frac{1}{\sqrt{-x}}~,~x<-1~.
\end{align} 
Now we can perform the non-subtracted DR
\begin{align}
  A(x)&=\frac{1}{\pi}\int_{-\infty}^{-1} dy \frac{\Ima A(y+i\ep)}{(y-x)}\\
  &=-\frac{1}{8m_\pi}\int_{-\infty}^{-1}dy\frac{1}{\sqrt{-y}(y-x)}~,
\end{align}
which recovers the original function $A(x)$.

Again the contributions at NNLO are given in Reference~\cite{kaiser.080516.1}, which read
\begin{eqnarray}
  V_C &=& \frac{3g_A^2}{16\pi f_\pi^4} \bigg\{
    \bigg[ 2m_\pi^2(2c_1-c_3) - q^2 \bigg(c_3+\frac{3g_A^2}{16M_N}\bigg) \bigg] (2m_\pi^2+q^2)A(q) 
    - \frac{g_A^2 m_\pi^5}{16 M_N (4m_\pi^2+q^2)} 
    \bigg\}
    +\ldots
\nn  \\
    W_C &=& \frac{g_A^2}{128\pi M_Nf_\pi^4} \bigg\{
      \bigg[ 4m_\pi^2+2q^2-g_A^2(4m_\pi^2+3q^2) \bigg] (2m_\pi^2+q^2) A(q) 
    - \frac{3g_A^2 m_\pi^5}{4m_\pi^2+q^2} 
    \bigg\}
    +\ldots
    \\
    V_T &=& -\frac{9g_A^4}{512\pi M_N f_\pi^4} (2m_\pi^2+q^2) A(q) 
    +\ldots
\nonumber    \\
    W_T &=& \frac{g_A^2}{32\pi f_\pi^4} A(q) \bigg[ \bigg(c_4+\frac{1}{4M_N} \bigg) (4m_\pi^2+q^2) -\frac{g_A^2}{8M_N} (10m_\pi^2+3q^2)
    \bigg]
    +\ldots
\nonumber    \\
    V_{SO} &=& \frac{3g_A^4}{64\pi M_N f_\pi^4} (2m_\pi^2+q^2) A(q)
    +\ldots
\nonumber    \\
    W_{SO} &=& \frac{g_A^2(1-g_A^2)}{64\pi M_N f_\pi^4}  (4m_\pi^2+q^2) A(q)
    +\ldots
\nonumber    
\end{eqnarray}
where the $c_i$ are the $\pi N$ LEC's.

The $^1S_0$ partial wave projected $NN$ potential is now given by
\begin{align}
    v(p,p')&=\frac{1}{2}\int_{-1}^{+1}dx\left\{V_C(q^2)+W_C(q^2)
    +2q^2 (V_T(q^2)+W_T(q^2))
    \right\}~.
  \end{align}
The spectral functions in accordance with the relation of Eq.~\eqref{140116.9} are then
\begin{eqnarray}
\eta_C(\mu^2)  
&=& \frac{3g_A^2}{64 f_\pi^4} \bigg\{
    -\frac{g_A^2 m_\pi^5}{4 M_N} \delta(\mu^2-4m_\pi^2)
    \\ &&
    + \bigg[ 2m_\pi^2(2c_1-c_3) + \mu^2 \bigg(c_3+\frac{3g_A^2}{16M_N}\bigg) \bigg] 
    \frac{2m_\pi^2-\mu^2}{\mu} \theta\left(\mu-2m_\pi\right)\bigg\}
\nonumber    \\
\eta_{WC}(\mu^2)  
&=& \frac{g_A^2}{512 M_Nf_\pi^4} \bigg\{
        - 12g_A^2 m_\pi^5 \delta(\mu^2-4m_\pi^2)
      \nonumber \\ &&
    + \bigg[ 4m_\pi^2-2\mu^2-g_A^2(4m_\pi^2-3\mu^2) \bigg] 
    \frac{2m_\pi^2-\mu^2}{\mu} \theta\left(\mu-2m_\pi\right)\bigg\}
\nonumber    \\
\eta_T(\mu^2)  
&=& -\frac{9g_A^4}{2048 M_N f_\pi^4} \frac{2m_\pi^2-\mu^2}{\mu}
\theta\left(\mu-2m_\pi\right)
\nonumber    \\
\eta_{WT}(\mu^2)  
&=& \frac{g_A^2}{128 f_\pi^4} \frac{1}{\mu} 
\bigg[ \bigg(c_4+\frac{1}{4M_N} \bigg) (4m_\pi^2-\mu^2) -\frac{g_A^2}{8M_N} (10m_\pi^2-3\mu^2) \bigg]
\theta\left(\mu-2m_\pi\right)
\nonumber    
\end{eqnarray}
Now for central terms we recover the original potentials with a twice-subtracted DR relation, whereas for the
tensor term only once-subtracted DR is needed.

Finally we get $\Delta \hat v(\nu,\nu_1)$ that we obtain using
\begin{align}
  \Delta \hhv(\nu,\nu_1)&=pp'\Delta v(\nu,\nu_1)
  \\
  &=\frac{pp'}{2\pi}\int_0^{\nu_1-\nu}2\mu d\mu\left[\eta_C(\mu^2)+\eta_{WC}(\mu^2)-2\mu^2
	  (\eta_T(\mu^2)+\eta_{WT}(\mu^2)) 
  \right]
  \Delta \int_{-1}^{+1}\frac{dx}{\mu^2+q^2} \nn \\
  &=-\frac{1}{2}\int_0^{\nu_1-\nu}2\mu d\mu \left\{\eta_C(\mu^2)+\eta_{WC}(\mu^2)
  -2\mu^2
  (\eta_T(\mu^2)+\eta_{WT}(\mu^2))
  \right\}~.\nn
\end{align}
and reads
\begin{eqnarray}
  \Delta \hat v(\nu,\nu_1) &=&
  \frac{g_A^2}{1920 f_\pi^4}
  \bigg\{
          \bigg( 480 c_1 - 336 c_3 - 128 c_4 - \frac{4}{M_N} - \frac{25g_A^2}{4M_N}\bigg)  m_\pi^5
   \\ &&
   + \bigg(-720 c_1 + 360 c_3          - \frac{30}{M_N} + \frac{30g_A^2}{M_N}\bigg)  m_\pi^4\sqrt{y}
   \nonumber \\ &&
   + \bigg( 120 c_1 - 120 c_3 + 40 c_4 + \frac{20}{M_N} - \frac{95g_A^2}{2M_N}\bigg)  m_\pi^2 y^{3/2}
   \nonumber \\ &&
    +       \bigg(            18 c_3 -  6 c_4 - \frac{3}{M_N} + \frac{45g_A^2}{4M_N}\bigg)  y^{5/2}
  \bigg\}
   \nonumber 
\end{eqnarray}
With $\Delta \hat v(\nu,\nu_1)$ we use the same procedure to calculate $\mff(\nu)$. 
We show in the top right panel of Fig.~\ref{fig.180829.1} a log-log of the function $|\Delta(-k^2)|$
where the singular attractive behavior is clear and we see that the NNLO contribution is more singular 
than the NLO. This is expected since now $\rho(\nu,\nu_1)$ scales as $(\nu_1-\nu)^m$ with
$m=5>2l+1=1$.

\subsection{Repulsive singular potentials}
\label{sec.180829.1}

We study in this section the function $\Delta(p^2)$ again for the $^3P_0$ but using a potential that is 
minus the one derived from OPE and used in Sec.~\ref{sec.180827.1}. 
To shorten the notation we denote this 
case as the $-^3P_0$ PWA. Next, we consider the PWA $^3P_1$. A characteristic feature of $\Delta(-k^2)$ for these 
potentials is that the envelope of $|\Delta(-k^2)|$ grows faster than any power for $k\to \infty$, but 
at the same $\Delta(-k^2)$ oscillates and changes sign.

We can understand the origin of his behavior  by 
applying the discussion at the beginning of Sec.~\ref{sec.180823.1}. 
From this analysis we conclude again that $|\Delta(-k^2)|$ diverges faster than any power of $k^2$ in the limit $k\to \infty$,
 for $\rho(\nu^2,\nu_1^2;m_\pi^2)$ having a degree $m>2\ell+1$. 
 Now, we particularize Eq.~\eqref{180814.1} for the uncoupled case, extract the coefficient $\lambda$ of the asymptotic 
 behavior [$\lambda$ is the coefficient multiplying  the higher power 
of $\nu^2$ (or $\nu_1^2$) in $\rho(\nu^2,\nu_1^2;m_\pi^2)$], and rewrite Eq.~\eqref{180814.1} divided by $-\lambda$ as
\begin{align}
\label{180831.1}
\frac{\mathfrak{f}(\nu)}{-\lambda}
&=  2\frac{\rho(\nu^2,k^2)}{\lambda} \\
&+(-1)^{\ell}\lambda \frac{\theta(k-2m_\pi-\nu)m}{2\pi^2}
\int_{m_\pi+\nu}^{k-m_\pi}\frac{d\nu_1\nu_1^2}{k^2-\nu_1^2}
\left\{\frac{1}{(\nu_1+i\ve)^{2\ell+2}}+\frac{1}{(\nu_1-i\ve)^{2\ell+2}}\right\}
\frac{\rho(\nu^2,\nu_1^2)}{\lambda}\frac{\mathfrak{f}(\nu_1)}{-\lambda}~.\nn
 \end{align}
It is clear from this equation that if $(-1)^\ell \lambda>0$ we have a situation of continuous growing 
of the degree of divergence of $\Delta(-k^2)$ for $k\to \infty$; this is the case of an attractive singular potential. 
On the contrary, for $(-1)^\ell\lambda <0$ we have an  oscillatory situation since the sign in front of the 
$n_{\rm th}$ iteration of the IE flips compared to the previous and next orders of iteration 
[the sign is indeed $(-1)^n$]. 
We then have the situation corresponding to a repulsive singular potential. 
This discussion can be extended to the case of a potential given by a spectral function, 
as done at the end of Sec.~\ref{sec.180814.3b}, 
by introducing the weighted spectral function $\rho(\nu,\nu_1)$, Eq.~\eqref{180825.2},
 which is assumed to scale in its arguments faster than $2\ell+1$. 

The extension of this result to coupled partial waves can be performed after a diagonalization
in the asymptotic region $|\nu|\sim |\nu_1|\sim k\ggg m_\pi$ of
\begin{align}
  \label{181028.1}
\Ima t_{ij}(i\nu+\ve^-,ik+\ve)-\Ima t_{ij}(i\nu+\ve^+,ik+\ve)&=  \frac{\mff_{ij}(\nu)}{(i\nu)^{\ell_i+1}(ik)^{\ell_j+1}}~,\\
\Ima v_{ij}(i\nu+\ve^-,i\nu_1+\ve)-\Ima v_{ij}(i\nu+\ve^+,i\nu_1+\ve)&=
\frac{\Delta \hhv_{ij}(\nu,\nu_1)}{(i\nu)^{\ell_i+1}(i\nu_1)^{\ell_j+1}}~,\nn
\end{align}
 by a real orthogonal matrix $M$,
\begin{align}
  \label{181028.2}
  \Delta^{d}(\nu,\nu_1)&=M \Delta \hhv(\nu,\nu_1) M^T~,
\end{align}
This diagonalization
process would imply the diagonalization of the IE of Eq.~\eqref{290116.9} in the aforementioned asymptotic region as 
\begin{align}
  \label{181028.3}
\Delta^{d}(\nu)&=\Delta {\cal v}^{d}(\nu,k)
+\theta(k-\nu-2m_\pi)\frac{m }{2\pi^2} \int_{\nu+m_\pi}^{k-m_\pi}\frac{d\nu_1\nu_1^2}{k^2-\nu_1^2}
\Delta^{d}(\nu,\nu_1) \Delta^{d}(\nu_1)~,
\end{align}
with
\begin{align}
  \Delta^{d}(\nu)&=M\left[\Ima t(i\nu+\ve^-,ik+\ve)-\Ima t(i\nu+\ve^+,ik+\ve)\right]M^T~.
\end{align}
We would then apply the same considerations discussed for the uncoupled case to every of the resulting eigenfunctions comprising the
entries of the diagonal matrix $\Delta {\cal t}^{d}(\nu)$. Namely, we should first consider the scaling factor $m$ for
$\Delta^{d}_i(\nu,\nu_1)$; if this function times $k$ has a positive scaling factor then it behaves as a singular interaction,
otherwise it does as a regular one. In the former case we should then determine the asymptotic coefficient $\lambda$ and by dividing
Eq.~\eqref{181028.3} by this factor we have
\begin{align}
  \label{181028.4}
\frac{\Delta {\cal t}^{d}(\nu)}{\lambda}&=
\frac{\Delta {\cal v}^{d}(\nu,k)}{\lambda}
+
\lambda \theta(k-\nu-2m_\pi)\frac{m}{2\pi^2}
\int_{\nu+m_\pi}^{k-m_\pi}\frac{d\nu_1\nu_1^2}{k^2-\nu_1^2}
\frac{\Delta{\cal v}^{d}(\nu,\nu_1)}{\lambda}
\frac{\Delta{\cal t}^{d}(\nu_1)}{\lambda}~.
\end{align}
Therefore, if $\lambda$ is positive we have an attractive singular eigen-interaction, while for negative $\lambda$
the eigen-interaction is a repulsive singular one.  From this criterion we might come back to the uncoupled case by
taking into account Eqs.~\eqref{181028.2} and Eq.~\eqref{170525.1}.

\subsubsection{$\Delta(p^2)$ for the $-^3P_0$ and $^3 P_1$ PWAs}
\label{sec.180830.1}

In this section we discuss the calculation of $\Delta(p^2)$ for the $P$-waves, 
$-^3P_0$ and $^3P_1$. We apply the same IEs as in Eqs.~\eqref{180828.4}-\eqref{180828.6}, the only change 
for every case is in the function $\rho(\nu^2,\nu_1^2)$. 

For the $-^3P_0$ PWA the function $\rho$ is the same as in Eq.~\eqref{180828.3} but multiplied by a minus sign. 
The resulting $|\Delta(-k^2)|$ along the LHC is shown in a log-log plot in the bottom left panel of Fig.~\ref{fig.180829.1}. 
We can see clearly the frequent changes of sign in $\Delta(-k^2)$ for $k\gg m_\pi$. 

In the case of the $^3P_1$ PWA the potential in configuration space is
\begin{align}
\label{180831.4}
V(r)&=\frac{g}{2\pi }\frac{e^{-m_\pi r }}{ r}\left[1+\frac{1}{m_\pi r}+\frac{1}{(m_\pi r)^2}\right]~.
\end{align}
It clearly corresponds to a repulsive singular potential. 
 In momentum space the potential is given by 
\begin{align}
\label{180831.2}
v(p,p')&=-\frac{g}{8(pp')^2m_\pi^2}\left[
4pp'\left(p^2+{p'}^2-m_\pi^2\right)
+\left(p^2-{p'}^2+m_\pi^2\right)\left({p'}^2-{p}^2+m_\pi^2\right)\Theta(p,p')
\right]~.
\end{align}
From where the function $\rho(\nu^2,\nu_1^2)$ reads
\begin{align}
\label{180831.3}
\rho(\nu^2,\nu_1^2)&=\frac{g \pi}{8 m_\pi^2}((\nu^2-\nu_1^2)^2-m_\pi^4)~.
\end{align}
The resulting $\Delta(-k^2)$ is shown in the bottom right panel of Fig.~\ref{fig.180829.1}. 
The qualitative behavior is the same as for the $-^3P_0$ PWA, though the change of signs are somewhat less frequent. 
This is due to the fact that, as we deduce by inspection of Eq.~\eqref{180831.3} and minus Eq.~\eqref{180828.3}, 
 the $\rho$ function for the $-^3P_0$ PWA is a factor of two larger than the $\rho$ function for the $^3P_1$ PWA 
 in the limit $\nu^2$ and $\nu_1^2$ much larger than $m_\pi^2$. 
In this limit they scale as $\nu^4$ times a positive coefficient $\lambda$ (so that $(-1)^\ell \lambda<0$), and the discussion 
at the beginning of Sec.~\ref{sec.180829.1}, on the expected oscillatorily diverging behavior of $\Delta(-k^2)$ for $k\gg m_\pi$, is fulfilled.

\subsection{Coupled channels: $\Delta_{ij}(p^2)$ for the $^3S_1-{^3D_1}$ PWAs}
\label{sec.180829.1b}

We proceed to calculate the discontinuities $\Delta_{ij}(p^2)$ for the coupled waves with $S=1$ and $J=1$, corresponding 
to the PWAs $^3S_1$ ($i=1$) and $^3D_1$ ($i=2$), and taking the OPE $NN$ potential. 
The expression for the transition amplitudes between the two different PWAs in configuration space can be worked out  
directly from Eq.~\eqref{180505.2}:
 \begin{align}
\label{180907.1}
V_{11}(r)&=-\frac{m_\pi^2\hg}{2\pi}Y(r)~,\\
V_{12}(r)&=-\frac{m_\pi^2\hg}{2\pi}\left[2\sqrt{2}T(r)+Y(r)\right]~,\nn\\
V_{22}(r)&=\frac{m_\pi^2\hg}{2\pi}\left[2 T(r)-Y(r)\right]~.\nn
\end{align}
In the limit $r\to 0$ the potentials $V_{ij}(r)$ are dominated by the singular behavior of $T(r)$. 
The corresponding eigenvalues of $V_{ij}(r)$ in the limit $r\to 0$ tend to 
\begin{align}
\label{180907.2}
\lambda_1 &=-\frac{3\hg}{\pi   r^3}+{\cal O}(r^{-1})~,\\ 
\lambda_2 &=+\frac{6\hg}{\pi   r^3}+{\cal O}(r^{-1})~.\nn
\end{align}
Therefore, we have one eigenchannel with an attractive singular potential and another with a 
 repulsive singular one. 

For the calculation of the potential in momentum space we apply the projections formulas in 
Appendix \ref{app:170715.1} and the expression for $W_T$ in Eq.~\eqref{180815.1}.  
 It results:
\begin{align}
  \label{180122.1}
\xi&=\frac{{p'}^2+p^2+m_\pi^2}{2 p p'}~,\\
\hg&=\frac{g}{m_\pi^2}~,\nn\\
  \label{180122.2}
v_{11}(p',p)&=-2\hg+\frac{\hg m_\pi^2 \Theta(p',p) }{ 2p p'}~,\\
v_{12}(p',p)&=\sqrt{8} \hg  \left(\Theta(p',p)  \left(\xi-\frac{p}{2 p'}-\frac{\left(3 \xi ^2-1\right) p'}{4 p} \right)
+\frac{3 \xi  p'}{2 p}-2\right)~,\nn\\
v_{21}(p',p)&=\sqrt{8} \hg  \left(\Theta(p',p)  \left(\xi-\frac{p'}{2 p}-\frac{\left(3 \xi ^2-1\right) p}{4 p'} \right)
+\frac{3 \xi  p}{2 p'}-2\right)~,\nn\\
v_{22}(p',p)&=2\hg  \left(1-\frac{3 \xi  \left(p^2+{p'}^2\right)}{4 p p'}+\Theta(p',p)  \left(\frac{\left(3 \xi ^2-1\right)
  \left(p^2+{p'}^2\right)}{8 p p'}-\frac{\xi }{2}\right)\right)~.\nn
\end{align}
Regarding the discontinuities we present them for the intervals of the arguments needed to calculate
$\Delta_{ij}(p^2)$, $p^2=-k^2$, according to Eqs.~\eqref{300116.8} and \eqref{300116.9}: 
\begin{align}
\label{180122.3}
 & k>m_\pi~,~\nu\in[-k,k-2m_\pi]~,~\nu_1\in [\nu+m_\pi,k-m_\pi]~,\nn\\
  \Delta \hat{v}_{11}(\nu,\nu_1)&=-\pi   \hg m_\pi^2~,\\
  \Delta \hat{v}_{12}(\nu,\nu_1)&=\frac{\pi  \hg}{2 \sqrt{2}} 
\left(3 \left(\nu^2-\nu_1^2\right)^2+3 m_\pi^4+2 m_\pi^2 \left(\nu_1 ^2-3 \nu^2\right)\right)~,\nn\\
  \Delta \hat{v}_{21}(\nu,\nu_1)&=\frac{\pi \hg}{2 \sqrt{2}}
 \left(3 \left(\nu ^2-\nu_1^2\right)^2+3 m_\pi^4+2 m_\pi^2 \left(\nu ^2-3 \nu_1^2\right)\right)~,\nn\\
  \Delta \hat{v}_{22}(\nu,\nu_1)&=\frac{\pi \hg}{8} \left(3 \left(\nu ^2-\nu_1^2\right)^2 \left(\nu ^2+\nu_1^2\right)+3 m_\pi^4 \left(\nu ^2+\nu_1^2\right)-2
   m_\pi^2 \left(3 (\nu ^4+ \nu_1^4)+2 \nu ^2 \nu_1^2\right)\right)~.\nn
\end{align}
We also need $\Delta \hat{v}_{ij}(\nu,k)$ for $\nu\in [-k,k-m_\pi]$. Since $k>\nu+m_\pi$ for $k>m_\pi/2$ (the starting
point of the LHC)
we can particularize the expressions in Eq.~\eqref{180122.3} with $\nu_1=k$ to obtain $\Delta \hat{v}_{ij}(\nu,k)$.

We use the formalism developed in Sec.~\ref{sec:270116.1} to calculate $\Delta_{ij}(p^2)$ for the 
$^3S_1-^3D_1$ system, with $\ell_1=0$ and $\ell_2=2$. 
 We then proceeded to solve the IEs of Eq.~\eqref{170526.12}, according to the value of $\nu$.
The IEs couple $\mff_{ij}(\nu)$ with different $i$ but with the same $j$.
Therefore with end with two separated coupled IEs,
one for the set $\{\mff_{11}(\nu),\mff_{21}(\nu)\}$ and another one for $\{\mff_{12}(\nu),\mff_{12}(\nu)\}$,
 cf. Eqs.~\eqref{300116.8} and \eqref{300116.8}. 
For the $^3S_1-^3D_1$ coupled PWAs these IEs can be rewritten as
\begin{align}
\label{180123.2}
\mff_{ij}(\nu)&=\Delta\hhv_{ij}(\nu,k)
-\theta(k-2m_\pi-\nu)\frac{m}{2\pi^2}\int_{m_\pi+\nu}^{k-m_\pi}
\frac{d\nu_1}{k^2-\nu_1^2}\Delta\hhv_{i1}(\nu,\nu_1)\mff_{1j}(\nu_1)\nn\\
&-\theta(k-2m_\pi-\nu)\frac{m}{4\pi^2}\int_{m_\pi+\nu}^{k-m_\pi}
\frac{d\nu_1}{k^2-\nu_1^2}
\left[\frac{1}{(\nu_1+i\ve)^4}+\frac{1}{(\nu_1-i\ve)^4}\right]\Delta\hhv_{i2}(\nu,\nu_1)\mff_{2j}(\nu_1)~.
\end{align}
 
In order to avoid the infrared singularity for $\nu_1\to 0 $ in the last term of the previous equation
 we need, as explained after Eq.~\eqref{170526.12}, the Taylor expansion  around $\nu=0$ of the symmetric 
combination $\mff_{2i}(\nu)+\mff_{2i}(-\nu)$ up to ${\cal O}(\nu^2)$. 
The latter can be obtained by employing Eq.~\eqref{180123.2} which, for $\nu>-m_\pi/2$, simplifies further as 
\begin{align}
\label{180123.3}
\mff_{ij}(\nu)=\Delta\hhv_{ij}(\nu,k)
&-\theta(k-2m_\pi-\nu)\frac{m}{2\pi^2}\int_{m_\pi+\nu}^{k-m_\pi}
\frac{d\nu_1}{k^2-\nu_1^2}\Delta\hhv_{i1}(\nu,\nu_1)\mff_{1j}(\nu_1)\\
&-\theta(k-2m_\pi-\nu)\frac{m}{2\pi^2}\int_{m_\pi+\nu}^{k-m_\pi}
\frac{d\nu_1}{(k^2-\nu_1^2)\nu_1^4}
\Delta\hhv_{i2}(\nu,\nu_1)\mff_{2j}(\nu_1)~.\nn
\end{align}

From here we directly get 
\begin{align}
\label{180123.4}
\mff_{ij}(0)=\Delta\hhv_{ij}(0,k)
&-\theta(k-2m_\pi)\frac{m}{2\pi^2}\int_{m_\pi}^{k-m_\pi}
\frac{d\nu_1}{k^2-\nu_1^2}\Delta\hhv_{i1}(0,\nu_1)\mff_{1j}(\nu_1)\\
&-\theta(k-2m_\pi)\frac{m}{2\pi^2}\int_{m_\pi}^{k-m_\pi}
\frac{d\nu_1}{(k^2-\nu_1^2)\nu_1^4}\Delta\hhv_{i2}(0,\nu_1)\mff_{2j}(\nu_1)~.\nn
\end{align}
For the following steps to obtain the derivatives of $f_{2j}(\nu)$ at $\nu=0$ 
it is important to keep in mind that  $\Delta \hhv_{ij}(\nu,\nu_1)$ 
depends only quadratically in the arguments, as explicitly shown in Eq.~\eqref{180122.3}. 
 Another result that we are going to use is that
\begin{align}
\label{180123.5}
\Delta \hhv_{ij}(0,m_\pi)=0 ~\text{for}~\ell_i\neq 0~.
\end{align}
The reason is because $ 2 p' p \xi=0$ for $p'=0$ and $p=im_\pi$, cf.~Eq.~\eqref{180122.1} with $p'=i\nu$ and $p=i\nu_1$. 
As a result all the terms in $\Delta \hhv_{ij}$ vanish when 
we multiply by $\nu^{\ell_i+1} \nu_1^{\ell_j+1}$ in order to pass from $v_{ij}$ to $\Delta \hhv_{ij}$
 (this is why it is needed that $\ell_i\neq 0$, without any constraint in $\ell_j$, because if $\ell_i=0$  we 
do not multiply by ${p'}^{\ell _i}$, that is zero for $\nu=0$.) 

The first derivative of $\mff_{2j}(\nu)$ from \eqref{180123.3} is
\begin{align}
\label{180123.6}
&\mff_{2j}'(\nu)
=
\Delta \hhv_{2j}'(\nu,k)
+\theta(k-2m_\pi-\nu)\frac{m}{2\pi^2}\frac{\Delta\hhv_{21}(\nu,m_\pi+\nu)\mff_{1j}(m_\pi+\nu)}{k^2-(m_\pi+\nu)^2} \\
&+\theta(k-2m_\pi-\nu)\frac{m}{2\pi^2}\frac{\Delta\hhv_{22}(\nu,m_\pi+\nu)\mff_{2j}(m_\pi+\nu)}{(k^2-(m_\pi+\nu)^2)(m_\pi+\nu)^4}
-\theta(k-2m_\pi-\nu)\frac{m}{2\pi^2}\int_{m_\pi+\nu}^{k-m_\pi}\frac{d\nu_1}{k^2-\nu_1^2}\Delta\hhv'_{21}(\nu,\nu_1)\mff_{1j}(\nu_1)\nn\\
&-\theta(k-2m_\pi-\nu)\frac{m}{2\pi^2}\int_{m_\pi+\nu}^{k-m_\pi}\frac{d\nu_1}{(k^2-\nu_1^2)\nu_1^4}\Delta\hhv'_{22}(\nu,\nu_1)\mff_{2j}(\nu_1)~.\nn
\end{align} 
The terms involving the derivative of the Heaviside function $\theta(k-2m_\pi- \nu)$ in Eq.~\eqref{180123.3} 
do not contribute because they give rise to a Dirac delta function, $- \delta(k-2m_\pi-\nu)$, times an 
integral evaluated at $\nu=k-2m_\pi$. The latter vanishes because its integration interval shrinks to zero 

We now proceed to evaluate the second derivative of Eq.~\eqref{180123.3} at $\nu=0$. For this particular value many 
terms in the derivative of Eq.~\eqref{180123.6} do vanish. 
 The derivatives of the Heaviside functions in Eq.~\eqref{180123.6} do not give contribution. 
Those multiplied by an integral do not contribute because of the reason already explained in the previous paragraph. 
Regarding the others without integrals they also vanish because they are multiplied by $\Delta\hhv_{2i}(0,m_\pi)=0$.  
 
The second an third terms in the rhs of Eq.~\eqref{180123.6} do not give contribution to this derivative because 
the terms that do not involve the derivative of $\Delta\hhv_{2i}(\nu,m_\pi\pm\nu)$ 
 are then multiplied by $\Delta\hhv_{2i}(0,m_\pi)=0$. 
Those terms involving  $\Delta \hhv'_{2i}(\nu,m_\pi\pm \nu)|_{\nu=0}$ also vanish. 
The reason is the following
\begin{align}
\label{180123.7}
\left.\frac{\partial \Delta\hhv_{2i}(\nu,m_\pi + \nu)}{\partial \nu}\right|_{\nu=0}=
\left.\frac{\partial \Delta\hhv_{2i}(\nu',m_\pi)}{\partial \nu'}\right|_{\nu'=0} 
+ \left.\frac{\partial \Delta\hhv_{2i}(0,\nu')}{\partial \nu'}\right|_{\nu'=m_\pi}~.
\end{align}
The first term on the rhs of the previous equation is zero because 
$\Delta \hhv_{ij}(\nu,m_\pi)$ for OPE is a polynomial in $\nu^2$ and $\Delta \hhv_{2j}(0,m_\pi)=0$,
 so that $\Delta \hhv_{2i}(\nu,m_\pi)\propto \nu^2$. 
The last term is also zero because for calculating $\Delta\hhv_{2j}(p',p)$ we have to multiply by ${p'}^{3}p^{\ell_j+1}$. 
Therefore, only the term with the highest power of $\xi$ survives,  in our case $\xi^2$,  
but then $\partial ({p'}^2+p^2+m_\pi^2)^2/\partial p=2 p ({p'}^2+p^2+m_\pi^2)$ is zero for  $p'=0$ and $p=i m_\pi$. 
Of course, Eqs.~\eqref{180123.5} and \eqref{180123.7} can be explicitly checked from the algebraic expression given in 
Eq.~\eqref{180122.3}.

We are then left with the following expression for $\mff''_{2i}(0)$,
\begin{align}
\label{180123.8}
\mff''_{2j}(0)&=\Delta\hhv''_{2j}(0,k)
-\theta(k-2m_\pi)\frac{m}{2\pi^2}\int_{m_\pi}^{k-m_\pi}\frac{d\nu_1}{k^2-\nu_1^2}\Delta\hhv''_{21}(0,\nu_1)\mff_{1j}(\nu_1)\\
&-\theta(k-2m_\pi)\frac{m}{2\pi^2}\int_{m_\pi}^{k-m_\pi}\frac{d\nu_1}{(k^2-\nu_1^2)\nu_1^4}\Delta\hhv''_{22}(0,\nu_1)\mff_{2j}(\nu_1)~.\nn
\end{align}
This second derivative of $\mff_{2j}(\nu)$ at $\nu=0$ could be also evaluated numerically
 from Eq.~\eqref{180123.3} and we have 
checked that both results agree within numerical precision. We also discuss below another method applicable 
for $k>3m_\pi/2$ by fitting 
$\mff_{2j}(\nu)+\mff_{2j}(-\nu)$ in the region $\nu\in[-m_\pi/2,m_\pi/2]$ with a polynomial of fourth degree in $\nu$, 
cf. Eq.~\eqref{180124.1}.

We put together the IEs to get $\mff_{ij}(\nu)$ in the OPE case distinguishing the three ranges of values of $\nu$, 
particularizing Eq.~\eqref{170526.12} to this case:
\begin{align}
\label{180123.9}
&-m_\pi/2<\nu \nn\\
\mff_{ij}(\nu)&=\Delta\hhv_{ij}(\nu,k)
-\theta(k-2m_\pi-\nu)\frac{m}{2\pi^2}\int_{m_\pi+\nu}^{k-m_\pi}\frac{d\nu_1}{k^2-\nu_1^2}\Delta\hhv_{i1}(\nu,\nu_1)\mff_{1j}(\nu_1)\\
&-\theta(k-2m_\pi-\nu)\frac{m}{2\pi^2}\int_{m_\pi+\nu}^{k-m_\pi}\frac{d\nu_1}{(k^2-\nu_1^2)\nu_1^4}
\Delta\hhv_{i2}(\nu,\nu_1)\mff_{2j}(\nu_1)~. \nn\\
\label{180123.11}
&-3m_\pi/2<\nu<-m_\pi/2\nn\\
\mff_{ij}(\nu)&=\Delta\hhv_{ij}(\nu,k)
-\theta(k-2m_\pi-\nu)\frac{m}{2\pi^2}\int_{m_\pi+\nu}^{k-m_\pi}\frac{d\nu_1}{k^2-\nu_1^2}\Delta\hhv_{i1}(\nu,\nu_1)\mff_{1j}(\nu_1) \\
&-\theta(k-2m_\pi-\nu)\frac{m}{2\pi^2}\int_{m_\pi+\nu}^{k-m_\pi}\frac{d\nu_2}{(k^2-\nu_2^2)\nu_2^4}
\Delta\hhv_{i2}(\nu,\nu_2)\Delta\hhv_{2j}(\nu_2,k)\nn\\
&+\theta(k-3m_\pi-\nu)\left(\frac{m^2}{2\pi^2}\right)^2\int_{2m_\pi+\nu}^{k-m_\pi}\frac{d\nu_1}{(k^2-\nu_1^2)\nu_1^4}
\mff_{2j}(\nu_1)\int_{m_\pi+\nu}^{\nu_1-m_\pi}\frac{d\nu_2}{(k^2-\nu_2^2)\nu_2^4}\Delta\hhv_{i2}(\nu,\nu_2)\Delta\hhv_{22}(\nu_2,\nu_1)\nn\\
&+\theta(k-3m_\pi-\nu)\left(\frac{m^2}{2\pi^2}\right)^2\int_{2m_\pi+\nu}^{k-m_\pi}\frac{d\nu_1}{k^2-\nu_1^2}
\mff_{1j}(\nu_1)\int_{m_\pi+\nu}^{\nu_1-m_\pi}\frac{d\nu_2}{(k^2-\nu_2^2)\nu_2^4}\Delta\hhv_{i2}(\nu,\nu_2)\Delta\hhv_{21}(\nu_2,\nu_1)~.\nn\\
\label{180123.10}
&\nu<-3m_\pi/2\nn\\
\mff_{ij}(\nu)&=\Delta\hhv_{ij}(\nu,k)\\
&-\theta(k-2m_\pi-\nu)\frac{m}{2\pi^2}\int_{m_\pi+\nu}^{k-m_\pi}\frac{d\nu_1}{k^2-\nu_1^2}\Delta\hhv_{i1}\mff_{1j}(\nu_1)
-\theta(k-2m_\pi-\nu)\frac{m}{2\pi^2}\int_{-(m_\pi+\nu)}^{k-m_\pi}\frac{d\nu_1}{(k^2-\nu_1^2)\nu_1^4}\Delta\hhv_{i2}(\nu,\nu_1)\mff_{2j}(\nu_1)
\nn\\
&
-\theta(k-2m_\pi-\nu)\frac{m}{2\pi^2}\int_{0}^{-m_\pi-\nu}\frac{d\nu_1}{(k^2-\nu_1^2)\nu_1^4}\Delta\hhv_{i2}(\nu,\nu_1)
\big[ \mff_{2j}(\nu_1) + \mff_{2j}(-\nu_1)-2 \mff_{2j}(0) - \nu_1^2 \mff_{2j}''(0) \big]\nn\\
&-\theta(k-2m_\pi-\nu)\frac{m}{4\pi^2}\big[2\mff_{2j}(0)+\nu_1^2\mff_{2j}''(0)\big]
\int_0^{-m_\pi-\nu}\frac{d\nu_1}{k^2-\nu_1^2}\left[\frac{1}{(\nu_1+i\ve)^4}+\frac{1}{(\nu_1-i\ve)^4}\right]
\Delta\hhv_{i2}(\nu,\nu_1)~,\nn
\end{align}
The last integral in the previous equation is done algebraically with $\Delta\hhv_{i2}(\nu,\nu_1)$  given in Eq.~\eqref{180122.3}, 
 and it is finite because of Eq.~\eqref{180124.1b}. 
Similarly  the integrations in the variable $\nu_2$ in Eq.~\eqref{180123.11} are done algebraically and they 
are also finite because of cancellations of the type already discussed in the example of the $^1D_2$ PWA in 
Sec.~\ref{sec.180819.2} regarding Eq.~\eqref{180821.4}. General arguments on why these cancellations should occur 
were given after Eq.~\eqref{180820.2}. 

The Taylor expansion of $\mff_{21}(\nu)$ and of $\mff_{22}(\nu)$ 
for $\nu$ around 0 is numerically demanding and it  gives rise to numerical instabilities for too small values of 
$\nu$. Namely, for $\nu\to 0$ one observes that the subtracted combination 
\begin{align}
  \label{180123.1}
\frac{f_{2j}(\nu)+f_{2j}(-\nu)-2 f_{2j}(0)-f''_{2j}(0) \nu^2}{\nu^4}~,
\end{align}
that appears in Eq.~\eqref{180123.10}, 
 is first stable and reach a constant value but, continuing towards smaller values of $\nu$, at some point
 this limit turns out to be numerically unstable.  
In order to improve the efficiency of the numerical solution we  present another method.

 A typically more accurate strategy consists of making a polynomial fit of $\mff_{2i}(\nu)+\mff_{2i}(-\nu)$ 
in the region around  $\nu=0$, namely 
for $\nu\in[-m_\pi/2,m_\pi/2]$, when used in Eq.~\eqref{180123.10} for $k>5m_\pi/2$. 
As discussed above we need at least to work out the second derivative at the origin 
of $\mff_{2i}(\nu)$, so that we use a fourth degree polynomial to fit the previous symmetric combination:
\begin{align}
 \label{180124.1}
Q_i(\nu^2)&=\alpha_i+\beta_i \nu^2+\gamma_i \nu^4~,\nn\\
\chi^2&=\sum_j \left(\mff_{2i}(\nu_j)+\mff_{2i}(-\nu_j)-Q_i(\nu_j) \right)^2~.
\end{align}
The minimization process implies the following three equations
\begin{align}
  \label{180124.2b}
\frac{\partial \chi^2}{\partial \alpha_i}&=-2\sum_j\left(\mff_{2i}(\nu_i)+\mff_{2i}(-\nu_j)-\alpha_i-\beta_i\nu_j^2-\gamma_i \nu_j^4\right)=0~,\\
\frac{\partial \chi^2}{\partial \beta_i}&=-2\sum_j\left(\mff_{2i}(\nu_i)+\mff_{2i}(-\nu_j)-\alpha_i-\beta_i\nu_j^2-\gamma_i \nu_j^4\right)\nu_j^2=0~,\nn\\
\frac{\partial \chi^2}{\partial \gamma_i}&=-2\sum_j\left(\mff_{2i}(\nu_i)+\mff_{2i}(-\nu_j)-\alpha_i-\beta_i\nu_j^2-\gamma_i \nu_j^4\right)\nu_j^4=0~.\nn
\end{align}
These equations can be expressed in a simpler form that also makes clear their generalization to even higher-degree polynomials
\begin{align}
  \label{180124.3b}
\sum_j \big[ \mff_{2i}(\nu_j)+\mff_{2i}(-\nu_j) \big]&= \alpha_i \sum_j +\beta_i \sum_i \nu_j^2+\gamma_i\sum_j \nu_j^4~,\\
\sum_j \nu_j^2\big[ \mff_{2i}(\nu_j)+\mff_{2i}(-\nu_j) \big]&=\alpha_i\sum_j \nu_j^2+\beta_i \sum_j \nu_j^4+\gamma_i\sum_j \nu_j^6~,\nn\\
\sum_j \nu_j^4\big[ \mff_{2i}(\nu_j)+\mff_{2i}(-\nu_j) \big]&=\alpha_i\sum_j \nu_j^4+\beta_i \sum_j \nu_j^6+\gamma_i\sum_j \nu_j^8~.\nn
\end{align}
Its algebraic solution is implemented in the Fortran codes  to evaluate
 $\Delta_{ij}(p^2)$, so that we use $Q_i(\nu_1)$ for $-m_\pi/2<\nu_1<m_\pi/2$, instead 
of $\mff_{2i}(\nu_1)$ in the integrands of Eq.~\eqref{180123.10} for $\nu<-3m_\pi/2$. 
In connection with this, we also employ $\alpha_i/2$ and $\beta_i$ instead of $\mff_{2i}(0)$ and 
$\mff_{2i}''(0)$, respectively, so that the subtracted combination is 
$Q_i(\nu_1)-2 \mff_{2i}(0) - \nu_1^2 \mff_{2i}''(0)=\gamma_i \nu_i^4$.

\begin{figure}[H]
\begin{center}
\scalebox{0.8}{
\begin{tabular}{cc}
\includegraphics[width=.4\textwidth]{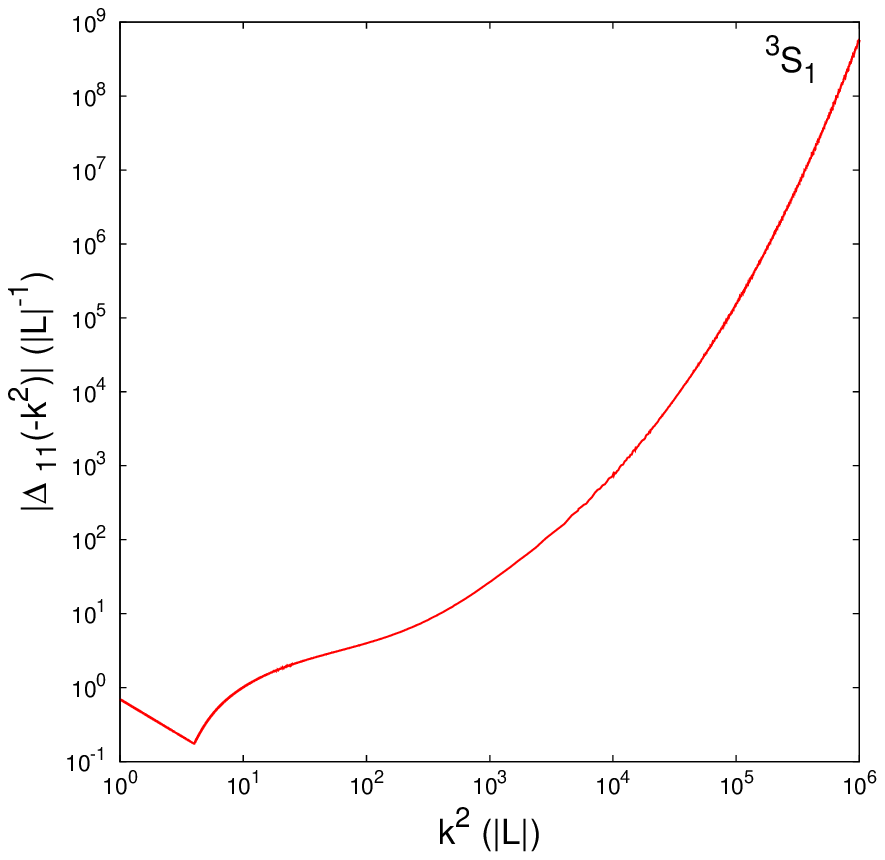} & 
\includegraphics[width=.4\textwidth]{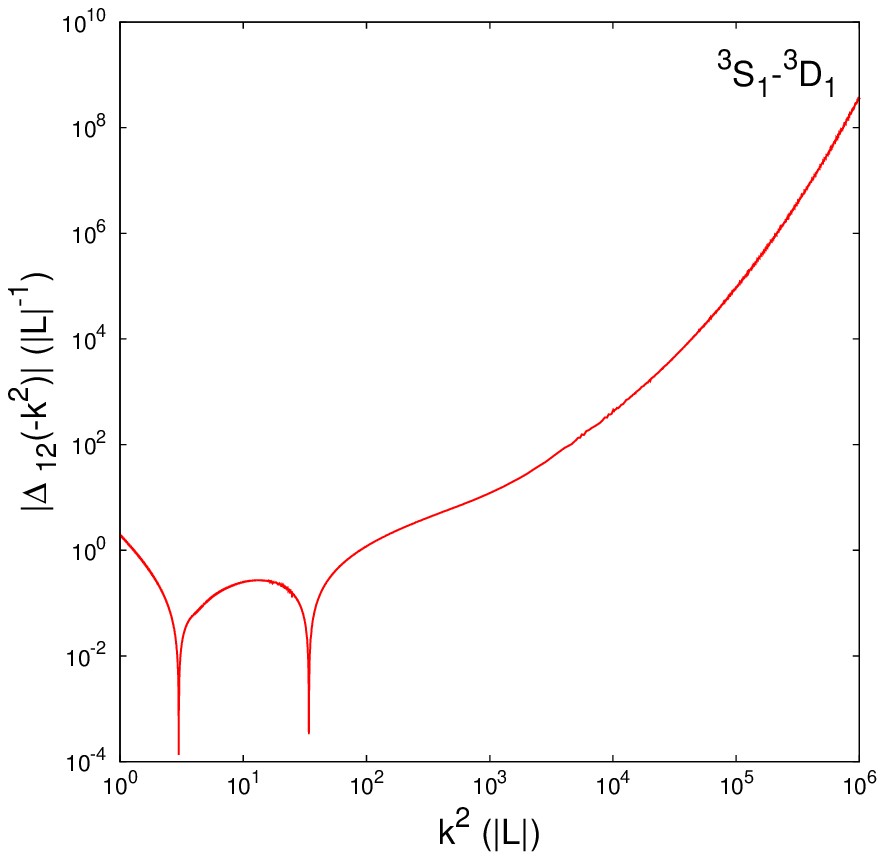}\\  
& \includegraphics[width=.4\textwidth]{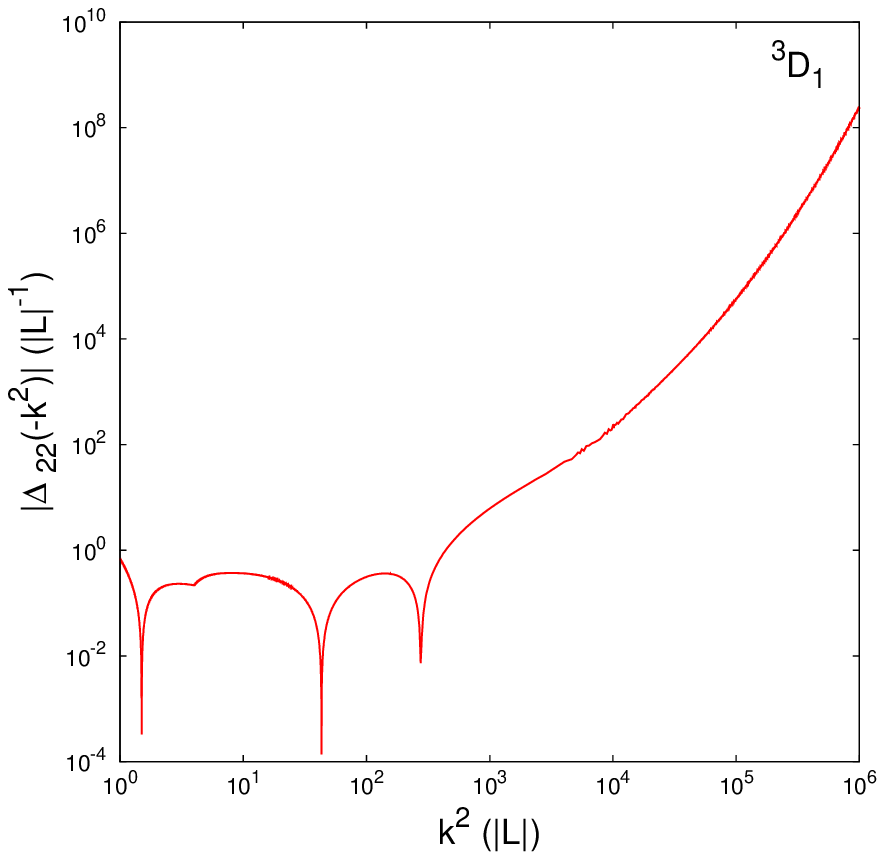}  
\\  
\end{tabular}}
\end{center}
\caption[pilf]{
From top to bottom and left to right, $\Delta_{11}(-k^2)$, $\Delta_{12,21}(-k^2)$ and $\Delta_{22}(-k^2)$ for the $^3S_1-{^3D_1}$ 
coupled PWAs. 
It is fulfilled numerically that $\Delta_{12}(-k^2)=\Delta_{21}(-k^2)$. 
\label{fig.180130.1}
}
\end{figure}

We show in Fig.~\ref{fig.180130.1} a log-log plot of the  $|\Delta_{ij}(-k^2)|$ calculated. 
It is clear that this is a scattering driven by a singular potential since $|\Delta_{ij}(-k^2)|$ 
grows faster than any power of $ k$ for $k\gg m_\pi$. 
In this limit, they behave as if they were driven from an attractive singular potential since 
no oscillation behavior is observed. 
We also have that $\Delta_{12}(p^2)=\Delta_{21}(p^2)$, a general consequence of 
the formalism, as derived in Eq.~\eqref{300116.11}.

\section{The exact $N/D$ method}
\label{sec.180819.1}
\def\theequation{\arabic{section}.\arabic{equation}}
\setcounter{equation}{0}   

We define the exact $N/D$ method as the $N/D$ method that employs the exact discontinuity 
$\Delta_{ij}(p^2)$ along the LHC. The latter is obtained unambiguously by solving 
the recursive IE of Eq.~\eqref{170526.12}. 
A characteristic feature of the exact $N/D$ method is that it drives to solutions that 
coincide with those of the LS equation, Eq.~\eqref{180804.5}. 
We already exemplified the exact $N/D$ method for the $^1S_0$ PWA in Ref.~\cite{entem.170930.1} and we now 
extend this analysis by  considering higher angular-momentum PWAs,  coupled waves and 
providing more details. 
We do not show the phase shifts that stem from the solution of the LS equation 
because they agree within numerical precision with those from the exact $N/D$ method and any difference will 
not be visible in the scale of the figures. A very detailed comparison in this respect for the $^1S_0$ PWA can be found 
in Ref.~\cite{entem.170930.1}. Therefore, the main point of the discussions that follow is to show 
which are the $N/D$ equations that drive to the same solution as the LS equation for regular and 
singular potentials.

The $N/D$ method was introduced in Ref.~\cite{chew.180624.1} by Chew and Mandesltam to study $\pi\pi$ scattering. 
It takes advantage of the fact that an on-shell PWA $t_{ij}(p,p;E)\equiv t(E)$, $p=\sqrt{m E}$,
 is a meromorphic function in the cut complex $E$ plane in the physical Riemann  sheet 
of $p(E)$ (${\rm arg} E\in[0,2\pi]$). The cuts are the RHC, Eq.~\eqref{180804.16}, and the LHC, Eq.~\eqref{180810.1}. 
The on-shell unitarity relation of Eq.~\eqref{180804.16} can be expressed more concisely  in matrix notation. 
For that let us introduce the matrix ${\mathfrak t}(E)$ whose matrix elements are the PWAs $t_{ij}(E)$ and then Eq.~\eqref{180804.16} becomes 
\begin{align}
\label{180909.1}
\Ima \hft (E+i\ve)&=\theta(E)\frac{m p}{4\pi}\hft (E+i\ve)\hft (E+i\ve)^*~.
\end{align}
Let us recall here that because of the Schwarz reflection principle, Eq.~\eqref{220116.3}, $\hft (E^*)=\hft (E)^*$. 
If one also takes into account time-reversal symmetry, Eq.~\eqref{180804.10}, 
the matrix $\hft (E)$ is  symmetric and then 
\begin{align}
\label{180909.2}
\hft^\dagger(E)=\hft (E)^*=\hft (E^*)~.
\end{align}
By writing in Eq.~\eqref{180909.1} that
 $2i\,\Ima \hft (E+i\ve)=\hft (E+i\ve)-\hft (E-i\ve)$ and multiplying it to the left and right by 
$\hft^{-1}(E+i\ve)$ and ${\hft^{-1}(E+i\ve)}^*$ (which we assume that exists at $E$ by now), 
Eq.~\eqref{180909.1} can be rewritten as
\begin{align}
\label{180909.3}
\Ima \hft^{-1}(E+i\ve)&=-\frac{m p}{4\pi} \theta(E)~,
\end{align}
with the rhs multiplied by the identity matrix in the space of coupled PWAs.

Let us consider first the case of an uncoupled PWA. 
The $N/D$ method writes it as the quotient of two functions, each of them having only one of the two cuts. 
Namely, 
\begin{align}
\label{180909.4}
t(E)&=\frac{n(p^2)}{d(p^2)}~,
\end{align}
where  the function $d(p^2)$ only has the RHC while the LHC is the only one in $n(p^2)$. 
This splitting is achieved by construction. It follows from  Eq.~\eqref{180909.3} that
\begin{align}
\label{180909.5}
\Ima d(p^2+i\ve)&=-\theta(p^2)\frac{m p}{4\pi}n(p^2)~.
\end{align}
In turn, from Eq.~\eqref{180909.4} we can write the $\Ima n(p^2+i\ve)$ in terms of $\Delta(p^2)=\Ima t(p^2+i\ve)$ for 
$p^2<L$,\footnote{For $NN$ scattering $L=-m_\pi^2/4$.}
Eq.~\eqref{180811.2}, as
\begin{align}
\label{180909.6}
\Ima n(p^2)&=\theta(L-p^2)d(p^2)\Delta(p^2)~.
\end{align}
Since the functions $t(E)$, $n(p^2)$ and $d(p^2)$ satisfy the Schwarz reflection principle\footnote{The functions 
$d(p^2)$ and $n(p^2)$ are real for negative and positive $p^2$, respectively.} their discontinuities along the real energy 
axis are given by $2i$ their imaginary parts. Making use of them, we can write down DRs for these functions that we 
denote by $\ND_{ij}$, where the subscripts mean $i$ subtractions in $n(p^2)$ and $j$ in  $d(p^2)$. 
The general form of these DRs are given in Ref.~\cite{oller.180722.2} to which we refer for further details. We concentrate in 
this work on those DRs that are going to be of use here in connection with the PWAs whose discontinuities along the LHC 
were calculated in Sec.~\ref{sec.180814.1}. 
The extension of the $N/D$ method to coupled waves will be treated in Sec.~\ref{sec.180909.4} when discussing the ${^3S_1}-{^3D_1}$ system.

The relation between the $t$ matrix and phase shifts can be derived from Eq.~\eqref{130116.1} and reads 
\begin{align}
\label{180909.12}
S_{ij}(E)&=\delta_{ij}+i\frac{mp}{2\pi}t_{ij}(E)~.
\end{align}
For the diagonal matrix elements it reads
\begin{align}
\label{180909.13}
S_{ii}(E)&=\eta_i e^{2i \delta_i(E)}~,
\end{align}
where $0\leq \eta_i\leq 1$ is the inelasticity parameter and $\delta_i$ the phase shift for the $i_{{\rm th}}$ PWA.

\subsection{Regular potentials}
\label{sec.180909.2}

The simplest DRs are of the type $\ND_{01}$, which do not involve any free parameter.
Despite that there is a subtraction in the function $d(p^2)$, it can be fixed at will because 
we can always divide the function $n$ and $d$ by a common constant while leaving $t(E)$ invariant. 
We choose such that $d(0)=1$, and the DRs that result are
\begin{align}
\label{180909.7}
n(p^2)&=\frac{1}{\pi}\int_{-\infty}^L dk^2\frac{\Delta(k^2)}{k^2-p^2}~,\\
d(p^2)&=1-\frac{mp^2}{4\pi^2}\int_0^\infty dk^2\frac{k}{k^2-p^2}n(k^2)~.\nn
\end{align}
From them the standard procedure is to work out an IE to obtain $d(p^2)$ along the LHC by substituting $n(k^2)$
 in terms of its DR as given in Eq.~\eqref{180909.7}. The resulting IE is
\begin{align}
\label{180909.8}
d(p^2)&=1-\frac{m p}{4\pi^2}\int_{-\infty}^L dk^2\frac{\Delta(k^2)d(k^2)}{(i\sqrt{-k^2}+p)\sqrt{-k^2}}~.
\end{align}

We should recall that $p=\sqrt{p^2}$ defined in the first RS, with ${\rm arg}\,p^2\in[0,2\pi]$, so that 
$\Ima p> 0$. On the contrary, in the second RS we have  ${\rm arg}\,p^2\in [2\pi,4\pi]$ and then $\Ima p< 0$. 
 In deriving this equation we have performed algebraically the integration along the  RHC, $k^2\in[0,\infty]$, 
\begin{align}
\label{180909.9}
g(p^2,q^2)&=\frac{1}{\pi}\int_0^\infty dk^2\frac{mk/4\pi}{(k^2-p^2)(k^2-q^2)}=\frac{im/4\pi}{p+k}~.
\end{align}
The $\ND_{01}$ IE provides the solution that agrees with that of the LS equation when applied to a regular potential in standard 
non-relativistic QM. 
This is so because it does not involve any free parameter and shares the same analytical properties, so that the solution 
is fixed.

\begin{figure}
\begin{center}
\scalebox{0.8}{
\begin{tabular}{cc}
\includegraphics[width=.4\textwidth]{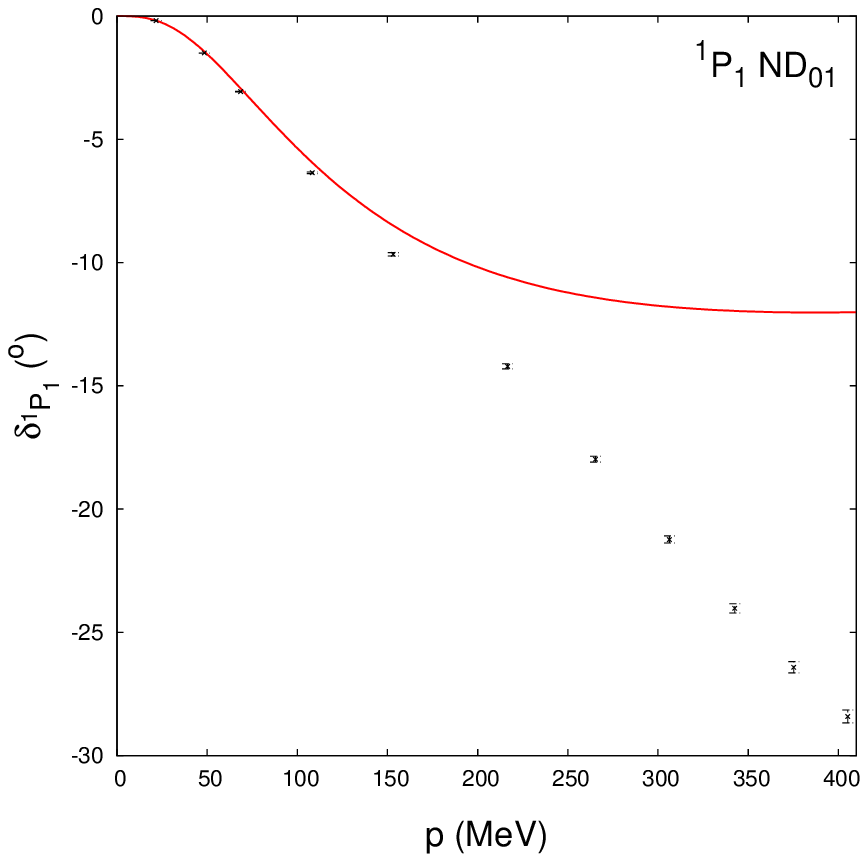} & \includegraphics[width=.4\textwidth]{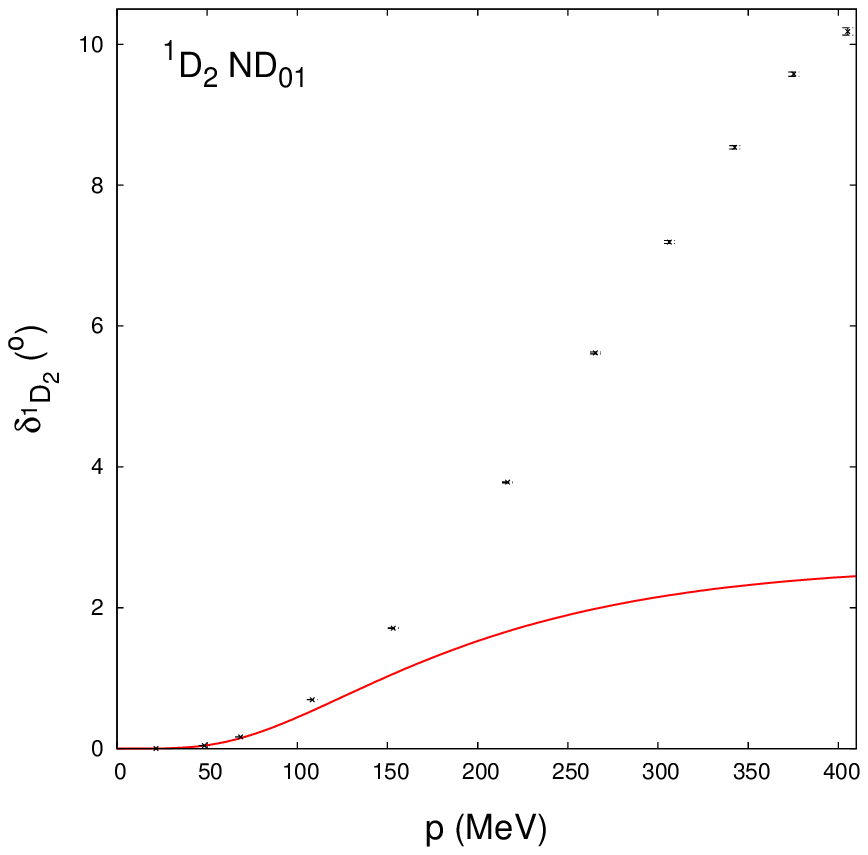}\\ 
\includegraphics[width=.4\textwidth]{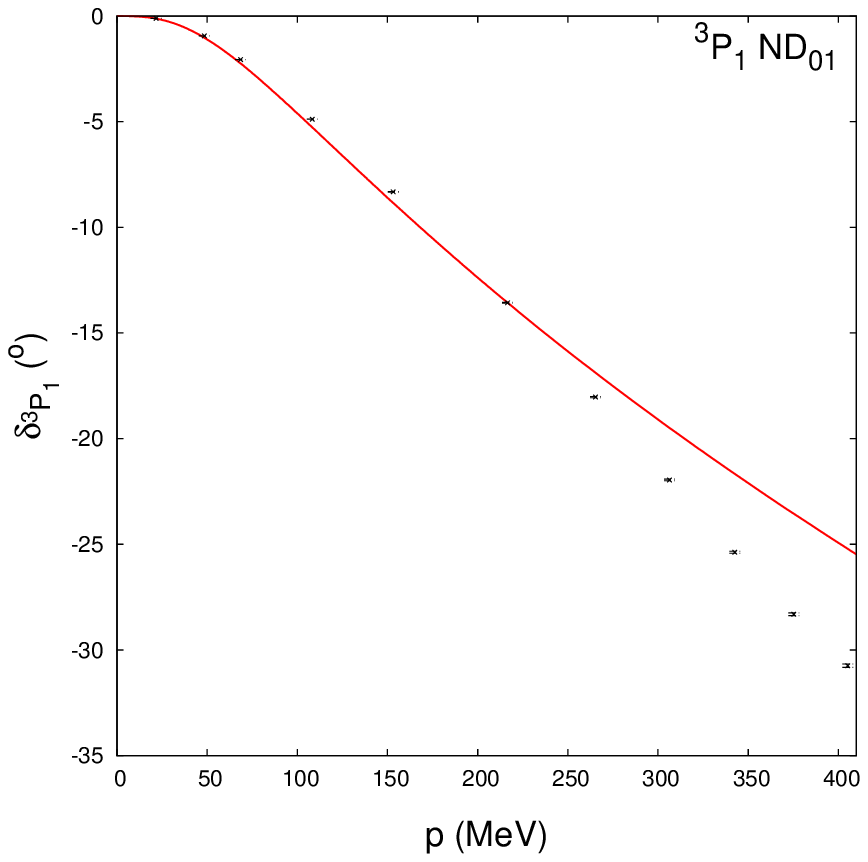} & \includegraphics[width=.4\textwidth]{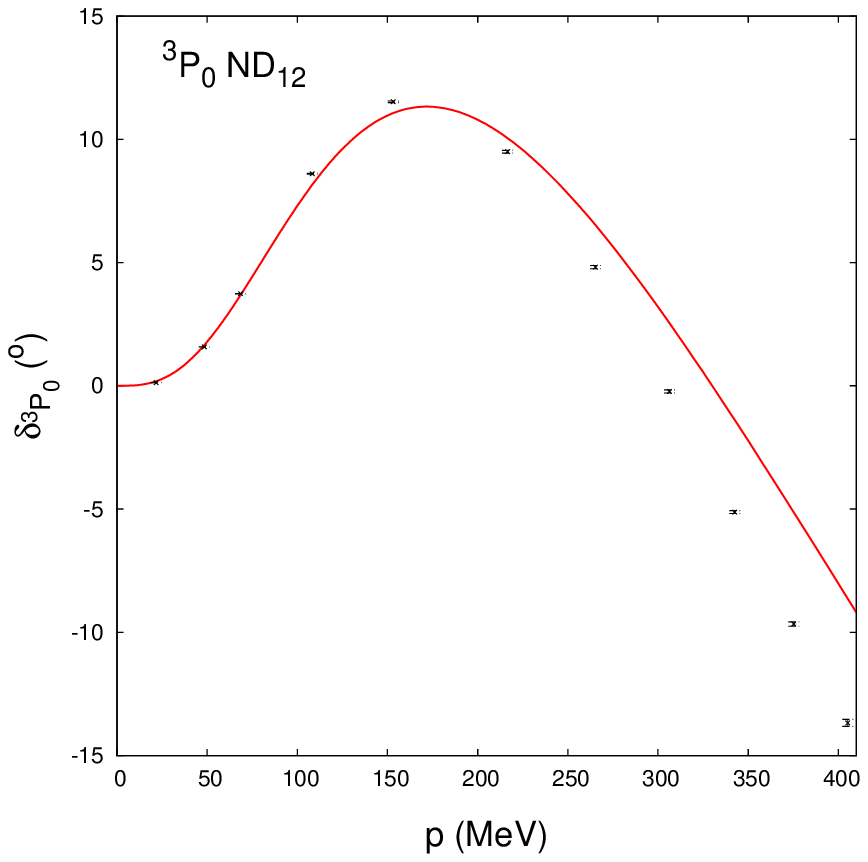}  
\end{tabular}}
\end{center}
\caption[pilf]{Phase shifts for different uncoupled PWAs obtained from the OPE $NN$ potential.
Top panels: Regular Yukawa-type potentials for the $^1P_1$ (left) and $^1D_2$ (right) PWAs. 
Bottom left panel: Repulsive singular potential applied to the $^3P_1$. 
In all these panels the  $\ND_{01}$ IE and the LS equation provide the same solution. 
Bottom right panel: Attractive singular potential for the $^3P_0$ PWA. The $\ND_{12}$ IE 
 and the LS equation with scattering volume fixed provide the same 
solution  (the former is equivalent to an $\ND_{01}$  IE with a CDD-pole-at-threshold, see the text). 
The points with errorbars correspond to the Granada $np$ phase shifts analysis \cite{navarro.180910.1}.
\label{fig.180910.1}
}
\end{figure}

We now apply the ${\rm {ND}}_{01}$ IE, Eq.~\eqref{180909.8}, to calculate the phase shifts of the 
singlet PWAs $^1P_1$ and $^1D_2$ whose $\Delta(p^2)$  was obtained in Secs.~\ref{sec.180814.3} and \ref{sec.180819.2}
 when taking the OPE $NN$ potential. We do not study the $^1S_0$ PWA since it was considered already in Ref.~\cite{entem.170930.1}. 
 The corresponding LS equation given in Eq.~\eqref{180804.5} reads for an uncoupled PWA,
\begin{align}
\label{180909.11}
t(p,p)&=v(p,p)+\frac{m}{2\pi^2}\int_0^\infty k^2dk\frac{v(p,k)t(k,p)}{k^2-p^2-i0^+}~,
\end{align} 
for physical $p\geq 0$.
The potentials in momentum space are given in Eqs.~\eqref{240915.1} and \eqref{180819.4} for the $^1P_1$ and $^1D_2$ PWAs, respectively.
 Notice that in order to calculate the on-shell $T$ matrix we need first to work out the half-off-shell $T$ matrix $t(k,p)$ for 
real and positive $k$. A brief account of the process for calculating it numerically is outlined in the 
Appendix \ref{app.180909.1}.

Now, we discuss succinctly how we have proceeded to solve the IEs for the $N/D$ method with the simplest example of the 
$\ND_{01}$, Eq.~\eqref{180909.8}. 
We make the change of variable 
\begin{align}
\label{270915.19}
x&=\frac{1}{k^2}~,
\end{align}
for having a finite range of integration $x\in [1/L,0]$ for $k^2\in[-\infty,L]$. 
A partition is taken with $N$ Gauss points, $x_i$, $i=1,\ldots,N$ ($k^2_i=1/x_i$), and weights $\omega_i$ corresponding to the 
 Gauss method for integration. We then have the linear system of $N$ equations
\begin{align}
\label{270915.21}
\sum_{j=1}^N \left(
\delta_{ij}
+\frac{m w_j}{4\pi^2 x_j^2}\frac{\Delta(k_j^2)\sqrt{k^2_i/k_j^2}}{\sqrt{-k_j^2}+\sqrt{-k_i^2}}
\right)d(k^2_j)=1~.
\end{align}
This is solved by inverting numerically the $N\times N$ matrix. 
Once $d(p^2)$ is known along the LHC we can apply Eq.~\eqref{180909.8} to obtain $d(p^2)$ for any complex value of $p^2$. 
In the same way, the function $n(p^2)$ can be calculated by applying Eq.~\eqref{180909.7} 
and then $t(p,p)=n(p^2)/d(p^2)$ can be determined.

The resulting phase shifts for the $^1P_1$ and $^1D_2$ PWAs are plotted in the top left and right panels of Fig.~\ref{fig.180910.1}, 
in order. We also include the phase shifts from the $np$ phase-shift analysis of the Granada group \cite{navarro.180910.1}. 
As indicated above, these results coincide with the ones obtained by employing a LS equation with the same potentials.

The same phase shifts for $P$- and higher PWAs are also obtained if we used  an $\ND_{11}$ IE,  
because the extra subtraction in the $n(p^2)$ can be fixed by the requirement that $n(0)=0$, as it is the case for 
any $\ell\geq 1$. In the case of an $S$ wave the subtraction constant in $n(p^2)$ could be fixed in terms of the 
scattering length $a$, which in our notation is $a_s=-m t(0,0)/4\pi$. 
The $\ND_{11}$ DRs are 
\begin{align}
  \label{170714.1b}
    n(p^2)&=-\frac{4\pi a_s}{m}+\frac{p^2}{\pi}\int_{-\infty}^L dk^2 \frac{\Delta(k^2)d(k^2)}{k^2(k^2-p^2)}~,\\
  d(p^2)&=1+ia_s p +\frac{im p^2}{4\pi^2}\int_{-\infty}^L dk^2\frac{\Delta(k^2)d(k^2)}{k^2(i\sqrt{-k^2}+p)}~.\nn
  \end{align}
The $\ND_{11}$ IE along the LHC reads then:
\begin{align}
  \label{170703.1}
  d(p^2)&=1-a_s\sqrt{-p^2}+\frac{m p^2}{4\pi^2}\int_{-\infty}^L dk^2\frac{\Delta(k^2)D(k^2)}{k^2(\sqrt{-k^2}+\sqrt{-p^2})}~.
\end{align}

\subsection{Repulsive and attractive singular potentials}
\label{sec.180909.3}

In the case of singular potentials we first analyze the case of the repulsive singular OPE $NN$ potential in the 
$^3P_1$ PWA. We obtain the $T$ matrix by applying the $\ND_{01}$ IE 
of Eq.~\eqref{180909.8}, without any free parameter, as in the regular-potential case (again the same results are obtained if one used the 
$\ND_{11}$ IE as explained above). 
We then obtain the same solution as by applying the LS equation in momentum space with an infinite cutoff;  
the latter method was discussed in Sec.~\ref{sec.180910.1} with respect to 
 the method in configuration space of Arriola {\it et al} \cite{arriola.180502.1}, 
 and the coincident results of Ref.~\cite{entem.180512.1} in momentum space. 
 The resulting phase shifts are given in the bottom left panel of Fig.~\ref{fig.180910.1}.

Next, we move to the case of attractive singular  potentials. 
We already analyzed this case in length along the Introduction,  and an  
extra parameter is needed to fix the (unitary) solution  from the Schr\"odinger or LS equations 
when the radial cutoff $r_0\to 0$ (or the momentum cutoff is taken to infinity)  \cite{case.180502.1,arriola.180502.1,kolck.180504.1}.  
Within the exact $N/D$ method this is always the situation that we have encountered when applying it to 
 attractive singular potentials, so that one free parameter has be introduced  at least. 

We already analyzed 
in Ref.~\cite{entem.170930.1} the phase shifts of the $^1S_0$ PWA obtained by employing the NLO and 
NNLO $NN$ chiral potentials, and we refer to that reference for further details on that analysis. 
These potentials are of the attractive singular type 
and the scattering amplitudes in every case were worked out in Ref.~\cite{entem.170930.1}. 
They perfectly agree with the ones from the 
LS equation with an extra counterterm added to the potential so as to fix the scattering length and renormalize  
the solutions, cf. Eqs.~(11) and (12) of Ref.~\cite{entem.170930.1}. 

Now we consider the $^3P_0$ PWA with the $NN$ OPE potential. The function $\Delta(p^2)$ was evaluated in Sec.~\eqref{sec.180827.1} and 
we now study the application of the exact $N/D$ method to obtain the  PWA that reproduces the experimental 
value of the scattering volume $a_V$. The latter is the free parameter that we also include to renormalize the solution of the 
LS equation with the same potential when taking the cutoff to infinity. In order to fit $a_V$ we need at least to consider 
the  $\ND_{12}$ DRs. 
The DR for the function $n(p^2)$ is the same as in Eq.~\eqref{170714.1b} but with $a_s=0$ because it is a $P$ wave, namely,
\begin{align}
\label{180910.1}
 n(p^2)&=\frac{p^2}{\pi}\int_{-\infty}^L dk^2 \frac{\Delta(k^2)d(k^2)}{k^2(k^2-p^2)}~.
\end{align}
The twice-subtracted DR for $d(p^2)$ reads
\begin{align}
\label{180910.2}
d(p^2)&=1+\frac{p^2}{\gamma}-\frac{(p^2)^2 m}{4\pi^2}\int_0^\infty dk^2\frac{k n(k^2)}{(k^2)^2(k^2-p^2)}~.
\end{align}
The integrand along the RHC is finite because from Eq.~\eqref{180910.1}  $n(k^2)\to 0$ for $k\to 0$ 
as $k^2$.\footnote{We do not include  more subtractions in the function $n(p^2)$ than in $d(p^2)$ because the 
integration along the RHC in the DR for the previous function would be infrared divergent,  after the 
exchange of the order of integration between the RHC and LHC integrations have been done \cite{oller.180722.2}.} 
After substituting in the previous the DR of Eq.~\eqref{180910.1} for $n(k^2)$ we have
\begin{align}
\label{180910.3}
d(p^2)&=1+\frac{p^2}{\gamma}
+\frac{(p^2)^2 m }{4 \pi^3}\int_{-\infty}^L dk^2\frac{\Delta(k^2)D(k^2)}{k^2}\int_0^\infty dq^2\frac{q}{q^2(q^2-p^2)(q^2-k^2)}~.
\end{align}
The integration along the RHC can be done algebraically and the final DR for $d(p^2)$ is
\begin{align}
\label{180910.4}
d(p^2)&=1+\frac{p^2}{\gamma}
-\frac{p^2 p m }{4 \pi^2}\int_{-\infty}^L dk^2\frac{\Delta(k^2)d(k^2)}{k^2\sqrt{-k^2}(i\sqrt{-k^2}+p)}~.
\end{align}
The problem with the direct use of the $\ND_{12}$ DRs is that  an implicit equation for $\gamma$ is obtained 
 in order to impose a given limit for $p^2\to 0$  of $t(E)/p^2$. 
 From Eqs.~\eqref{180910.1} and \eqref{180910.4} it follows that
\begin{align}
\label{180910.5}
a_V=-\lim_{p^2\to 0}\frac{m t(E)}{4\pi p^2}=
-\frac{m}{4\pi^2}\int_{-\infty}^L dk^2 \frac{\Delta(k^2)d(k^2)}{(k^2)^2}~,
\end{align}
and $\gamma$ enters implicitly through the dependence of the previous integral in $d(k^2)$. 

We can derive an explicit equation for $\gamma$ by multiplying simultaneously $n(p^2)$ and $d(p^2)$ by 
$\gamma/p^2$. This operator does not modify the $t(E)$ neither the cut structure of these functions. 
However, it introduces a pole in $d(p^2)$ at threshold, $p^2=0$. The resulting DRs can be easily worked out from the 
original ones in Eqs.~\eqref{180910.1} and \eqref{180910.4} by performing the transformations $n(p^2)\to n(p^2)\gamma/p^2$ and 
$d(p^2)\to d(p^2)\gamma/p^2$. They read,
\begin{align}
\label{180910.6}
 n(p^2)&= \frac{1}{\pi}\int_{-\infty}^L dk^2 \frac{\Delta(k^2)d(k^2)}{k^2-p^2}~,\\
\label{180910.7}
d(p^2)&=1+\frac{\gamma}{p^2}
-\frac{ m p}{4 \pi^2 }\int_{-\infty}^L dk^2\frac{\Delta(k^2)d(k^2)}{\sqrt{-k^2}(i\sqrt{-k^2}+p)}~.
\end{align}
Notice that this form of the $\ND_{12}$ IE is the same as the $\ND_{11}$ of Eq.~\eqref{180909.8}, plus a pole 
at threshold in the function $d(p^2)$. This pole is also referred as a CDD pole \cite{Castillejo.171020.1}.
  
Now we can derive an explicit equation for $\gamma$ in terms of $a_V$, it reads
\begin{align}
\label{180911.1}
\gamma&=-\frac{m}{a_V4\pi^2}\int_{-\infty}^L dk^2\frac{\Delta(k^2)d(k^2)}{k^2}~,
\end{align}
which is then substituted in Eq.~\eqref{180910.7}. As a result we have a DR for $d(p^2)$ in terms of the 
taken value of $a_V$,
\begin{align}
\label{180911.2}
d(p^2)&=1-\frac{m}{4 \pi^2}\int_{-\infty}^L dk^2 \Delta(k^2)d(k^2) \left[
\frac{p}{\sqrt{-k^2}(i\sqrt{-k^2}+p)}+\frac{1}{a_V p^2 k^2}
\right]~.
\end{align}
For $p^2\in [-\infty,L]$ the previous DR provides us with an IE whose solution allows to know 
the on-shell $T$ matrix $t(E)$ in the complex $p^2$ plane. The resulting phase shifts are given in the 
bottom right panel of Fig.~\ref{fig.180910.1} for a value $a_V=-0.89~m_\pi^{-3}$ \cite{oller.180722.2}.

\subsection{Coupled PWAs $^3S_1-{^3P_1}$ with an attractive singular eigenchannel}
\label{sec.180909.4}

As discussed in Sec.~\ref{sec.180910.1}, following the conclusions of Ref.~\cite{arriola.180502.1} that makes use 
of the Schr\"odinger equation in configuration space taking the radial cutoff $r_0\to 0$,  
there are as many free parameter as there are attractive eigenvalues of the potential in 
the limit $r\to 0$. 
 For the OPE $NN$ interaction in the $^3S_1-{^3D_1}$ system we established in 
Eq.~\eqref{180907.2} that there is one attractive eigenvalue. 
Therefore, the LS equation is renormalized in terms 
of one free parameter that we choose to be the input value for the $^3S_1$ scattering length. 
This is imposed in the LS equation by adding a contact interaction to $V_{11}(r)$ in Eq.~\eqref{180907.1}. 
In momentum space this gives rise to an $S$-wave constant counterterm $v_0$ added to $v_{11}(p,p')$, Eq.~\eqref{180122.2}. 
We discuss in Appendix \ref{app.180911.2} the LS equation used for the $^3S_1-{^3D_1}$ coupled PWAs, and move on 
to the $N/D$ IE that reproduces the LS solution.

\subsubsection{The simplest formalism for the $N/D$ method with coupled channels}
\label{subsec.180130.1}

We now extend the discussion at the beginning of Sec.~\ref{sec.180819.1} and consider the $N/D$ method in 
coupled channels. The simplest version of it consists of using a matrix notation, already introduced above, 
 which basically drives by itself to the resulting formulas,  as originally done in Ref.~\cite{bjorken.180203.2}. 
 We then have the $n\times n$ matrices  $\hft(E)$, $\hat{n}(p^2)=||n_{ij}(p^2)||$ and $\hat{d}(p^2)=||d_{ij}(p^2)||$, 
and write  the $T$ matrix as  
\begin{align}
\label{180130.1}
\hft   &=\hat{d}^{-1}\hat{n}~,\\
\hat{d}&=\hat{n}\,\hft^{-1}~,\nn\\
\hat{n}&=\hat{d}\,\hft~.\nn
\end{align} 
Unitarity implies that:
\begin{align}
\label{180130.2}
\Ima \hat{d}(p^2)&=\hat{n}(p^2) \Ima \hft(p^2)^{-1}=-\hat{n}(p^2) \hat{\rho}(p^2)~~,~~p>0~,
\end{align}
where $\hat{\rho}(p^2)$ is a diagonal matrix 
 whose matrix elements are  $mp/4\pi$.

The LHC discontinuity $\Delta(p^2)$ implies that:
\begin{align}
\label{180130.3}
\Ima \hat{n}(p^2)=\hat{d}(p^2)\Ima \hft(p^2)=\hat{d}(p^2)\hat{\Delta}(p^2)~~,~~p^2<L~,
\end{align}
where $\hat{\Delta}(p^2)$ is the matrix $||\Delta_{ij}(p^2)||$.

Since we want to impose a given value for the $^3S_1$ scattering length $a_s$, we consider a once-subtracted DR 
for $\hat{n}(p^2)$, as well as the standard once-subtracted DR for $\hat{d}(p^2)$. It results:
\begin{align}
\hat{n}(p^2)&=\hat{n}_0+\frac{p^2}{\pi}\int_{-\infty}^L dk^2\frac{\hat{d}(k^2)\hat{\Delta}(k^2)}{k^2(k^2-p^2)}~,\nn\\
\label{180130.4}
\hat{d}(p^2)&=\hat{d}_0-\frac{p^2}{\pi}\int_0^\infty dk^2\frac{\hat{n}(k^2)\hat{\rho}(k^2)}{k^2(k^2-p^2)}~,
\end{align}
where $\hat{n}_0$ and $\hat{d}_0$ are constant matrices,  $\hat{d}_0=\hat{d}(0)$ and $\hat{n}_0=\hat{n}(0)$. 
 Next, we multiply simultaneously the original matrices $\hat{d}$ and $\hat{n}$ by $\hat{d}_0^{-1}$, 
an operation that leaves invariant $\hft(p^2)$ according to Eq.~\eqref{180130.1}~. 
In this way, we can always take that $\hat{d}(0)=I$ in Eq.~\eqref{180130.5} and rewrite the DRs for 
$\hat{n}(p^2)$ and $\hat{d}(p^2)$ as
\begin{align}
\label{180130.5}
\hat{n}(p^2)&=\hat{n}_0+\frac{p^2}{\pi}\int_{-\infty}^L dk^2\frac{\hat{d}(k^2)\hat{\Delta}(k^2)}{k^2(k^2-p^2)}~,\\
\label{180130.5b}
\hat{d}(p^2)&=I-\frac{p^2}{\pi}\int_0^\infty dk^2\frac{\hat{n}(k^2)\hat{\rho}(k^2)}{k^2(k^2-p^2)}~.
\end{align}
Replacing $\hat{n}(k^2)$ in the last equation as given by its DR one has
\begin{align}
\label{180130.6}
\hat{d}(p^2)&=I
-\hat{n}_0\frac{p^2}{\pi}\int_0^\infty dk^2\frac{\hat{\rho}(k^2)}{k^2(k^2-p^2)}
+\frac{p^2}{\pi^2}\int_{-\infty}^L dk^2\frac{\hat{d}(k^2)\hat{\Delta}(k^2)}{k^2}
\int_0^\infty dq^2\frac{\hat{\rho}(q^2)}{(q^2-p^2)(q^2-k^2)}~.
\end{align}
Recalling Eq.~\eqref{180909.9} for $g(p^2,k^2)$, we can express Eq.~\eqref{180130.6} as:
\begin{align}
\label{180130.7}
\hat{d}(p^2)&=I-\frac{im p }{4\pi}\hat{n}_0+\frac{im p^2}{4\pi^2}\int_{-\infty}^L dk^2
\frac{\hat{d}(k^2)\hat{\Delta}(k^2)}{k^2(i\sqrt{-k^2}+p)}~.
\end{align}
This representation provides us with the following IE for $\hat{d}(p^2)$ with $p^2$ along the LHC,
\begin{align}
\label{180130.8}
\hat{d}(p^2)&=I+\frac{m\sqrt{-p^2}}{4\pi}\hat{n}_0+\frac{m p^2}{4\pi^2}\int_{-\infty}^L dk^2
\frac{\hat{d}(k^2)\hat{\Delta}(k^2)}{k^2(\sqrt{-p^2}+\sqrt{-k^2})}~,~p^2<L~.
\end{align}
Notice that in the IE for $d_{ij}(p^2)$ 
the row index $i$ only plays a parametric role and it is held fixed. 

The imposition of the scattering length $a_s$ is easily achieved by taking
\begin{align}
\label{180130.9}
\hat{n}_0&=\left(
\begin{array}{ll}
t_0 & 0\\
0 & 0
\end{array}
\right)~,
\end{align}
where we have used that $t_{11}(0)=t_0$ and $t_{ij}(0)=0$ for $i+j>2$. 
The former reads in terms of $a_s$,
\begin{align}
\label{180129.7}
t_0&=-\frac{4\pi a_s}{m}~.
\end{align}

\begin{figure}[H]
\begin{center}
\scalebox{0.8}{
\begin{tabular}{cc}
\includegraphics[width=.4\textwidth]{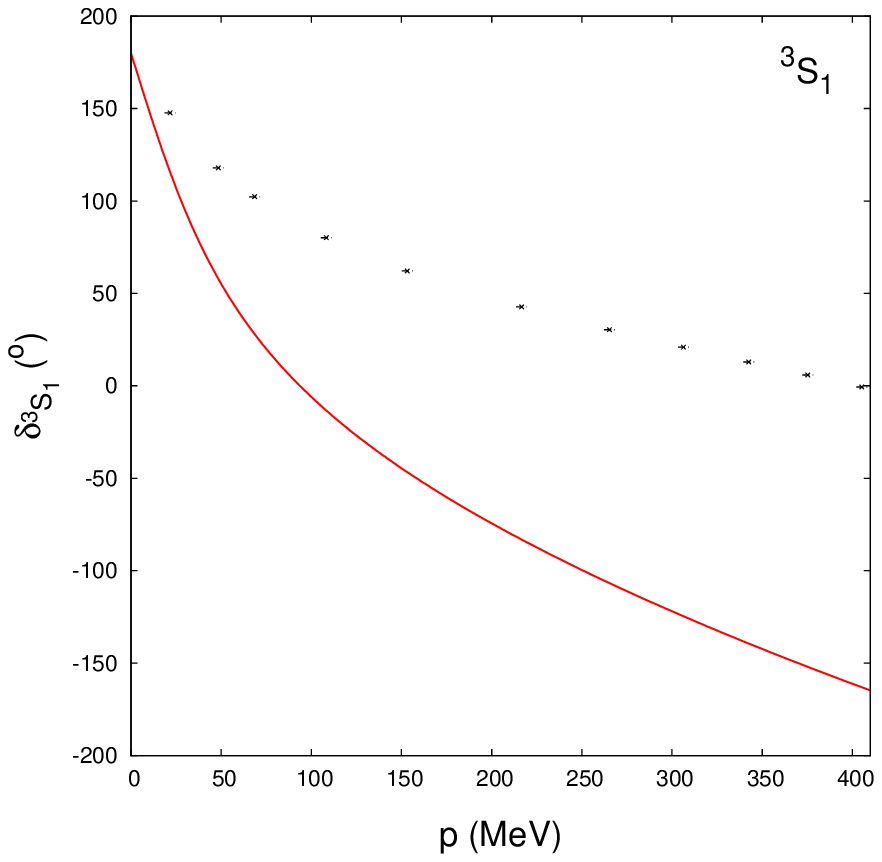} & \includegraphics[width=.4\textwidth]{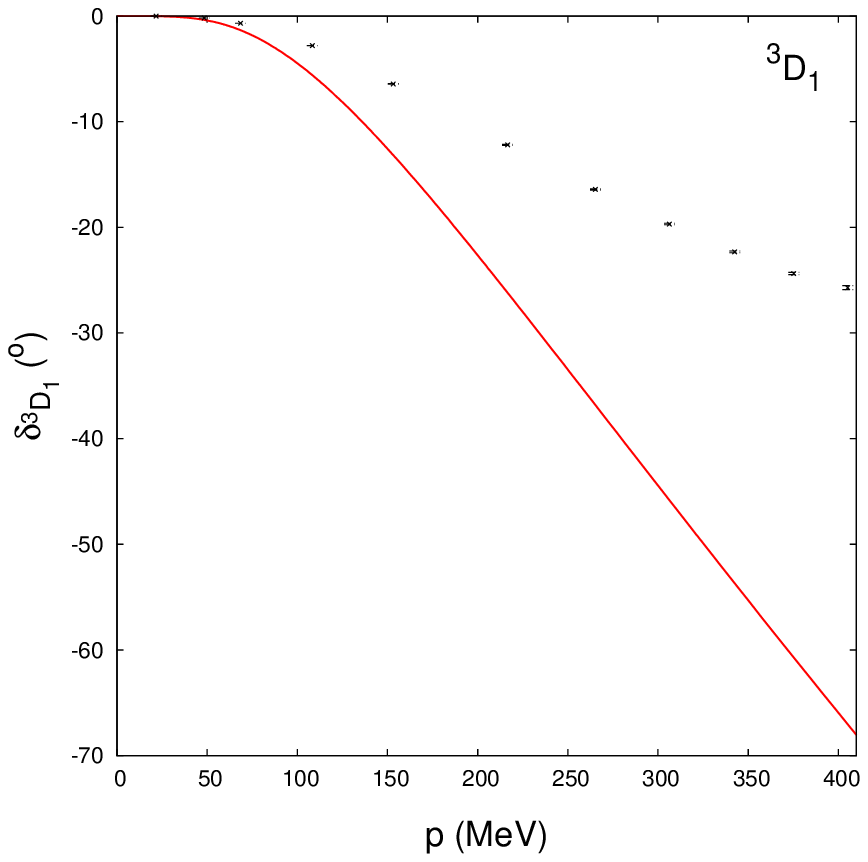}\\ 
 & \includegraphics[width=.4\textwidth]{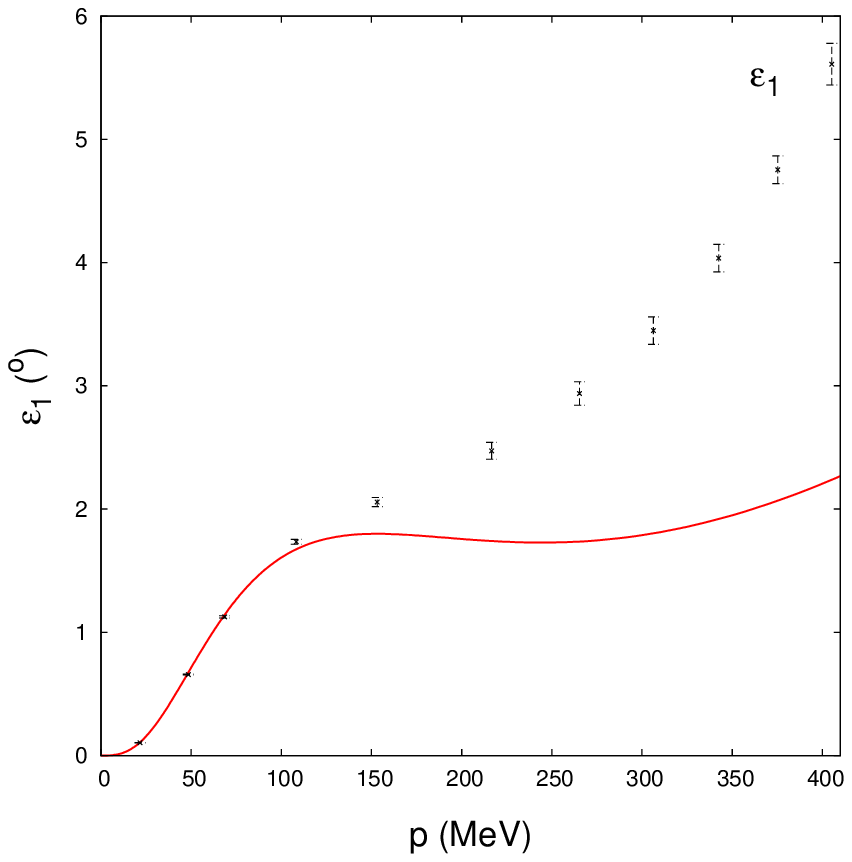}  
\end{tabular}}
\end{center}
\caption[pilf]{
From left to right and top to bottom, we plot the phase shifts 
for the $^3S_1$, $^3D_1$ waves and the mixing angle  
$\varepsilon_1$  as a function of $p$ in units of MeV. 
The dots with errorbars are the phase analysis of Ref.~\cite{navarro.180910.1}.  
\label{fig.180131.3}
}
\end{figure}

Regarding the numerical implementation of Eq.~\eqref{180130.8}, we first take 
 a partition along the LHC, $\{ p^2_\alpha, \alpha=1,\ldots,N\}$. 
Given the two-coupled-wave structure of the IEs we double the LHC partition, 
so that a primed index runs from 1 to $2N$. 
We define the vector $||\widetilde{d}_{i \alpha'}||$ (the index $i$ is fixed) as 
\begin{align}
\label{180130.10}
\widetilde{d}_{i \alpha'}&=\left\{
\begin{array}{ll}
d_{i 1}(p^2_{\alpha'}) & 1\leq \alpha'\leq N~,\\
&\\
d_{i 2}(p^2_{\alpha'}) & N+1\leq \alpha'\leq 2N~,
\end{array}\right.
\end{align}
and also the $2N\times 2N$ matrix $||\widetilde{\Delta}_{\alpha'\beta'}||$ as
\begin{align}
\label{180130.11}
\widetilde{\Delta}_{\alpha'\beta'}&=\left(
\begin{array}{l|ll}
 & 1\leq \beta'\leq N & N+1\leq \beta'\leq 2N \\
\hline
1\leq \alpha' \leq   N& \Delta_{11}(p^2_{j'}) & \Delta_{21}(p^2_{j'})\\ 
N+1\leq \alpha'\leq 2N& \Delta_{12}(p^2_{j'}) & \Delta_{22}(p^2_{j'})\\ 
\end{array}
\right)~.
\end{align}
We denote by $[\alpha']$ the integer part of $\alpha'/N$ plus 1 (1 if $1\leq \alpha'\leq N$ and 
2 for $N+1\leq \alpha'\leq 2N$).
In this notation Eq.~\eqref{180130.8} becomes 
\begin{align}
\label{180130.8a}
\widetilde{d}_{i \alpha'}&=\delta_{i [\alpha']} \left( 1 + \delta_{i 1}\frac{m \sqrt{-p^2_{\alpha'}}}{4\pi}t_0 \right)
+\frac{m p^2_{\alpha'}}{4\pi^2}\sum_{\beta'=1}^{2N} 
\frac{\omega_{\beta'}  \widetilde{\Delta}_{\alpha'\beta'}}{p^2_{\beta'}(\sqrt{-p^2_{\alpha'}}+\sqrt{-p^2_{\beta'}})}
\widetilde{d}_{i \beta'}~.
\end{align}
The matrix to be inverted numerically in the previous equation is also independent of $i$. 
The difference between $i=1$ and 2 stems from the independent term.

The results obtained for the phase shifts and mixing angle  
 are shown in Fig.~\ref{fig.180131.3} by the red solid lines. 
The mixing angle $\varepsilon_1$ and  the elasticity parameter, Eq.~\eqref{180909.13}, are related by 
 \begin{align}
\label{180917.1}
\eta_1=\sin 2\varepsilon~.
\end{align}
In the top left panel we plot the $^3S_1$ phase shift, in the top right one the $^3D_1$ phase shifts and 
the mixing angle is depicted in the bottom right panel.  
We also evaluate the effective range $r_s$ that follows from our calculation with the value $r_s=1.62$~fm,
 while the resulting Deuteron binding energy is $-2.11$~MeV. 
Within our normalization for an uncoupled PWA the effective-range expansion reads
\begin{align}
\label{180917.2}
\frac{4\pi}{m}\frac{d}{n}+i p&=-\frac{1}{a_s}+\frac{1}{2}r_s p^2+\sum_{m=1}^\infty v_{2m}p^{2m}~.
\end{align}
For the case of coupled PWAs one could apply the previous equation to the 
scattering eigenchannels  that diagonalize the $T$ matrix in the physical region, though 
the mixing process does not affect the calculation of $r_s$. Thus, in order to evaluate the latter, one can still use 
Eq.~\eqref{180917.2} expressed in terms of  $n_{11}(p^2)$ and $d_{11}(p^2)$, so that  
 one would have $n_{11}/d_{11}$ on the lhs. 
We refer to the Appendix of Ref.~\cite{oller.180724.1} for a detailed account 
of how to calculate the effective range and higher-order shape parameters in the $N/D$ method.
This method is by far the most accurate one  for evaluating numerically higher-order shape parameters, e.g. 
 Ref.~\cite{oller.180724.1} could evaluate  $v_{10}$ with a numerical accuracy of $1\%$.

\section{Conclusions}
 \label{sec.171001.1}

 A main result of this investigation is the derivation of a recursive 
non-linear integral equation (IE) that allows one to evaluate the exact discontinuity 
of an on-shell partial-wave amplitude along the left-hand cut in two-body non-relativistic scattering. 
This IE can be applied to regular and singular potentials, and its solution is completely fixed in
 terms of the discontinuities of the partial-wave projection of the potential. 
These discontinuities stem from the finite-range of the potential and the whole process is free of
regularization.

The derivation of this  IE is based on the analytical extrapolation  to complex values of three-momenta 
of the partial-wave projected half-off-shell 
Lippmann-Schwinger (LS) equation. This is by itself an interesting 
outcome of our study since, apart from the application worked out here,
 it would also allow e.g. to study the bound-, virtual- and 
resonance-state  pole positions and their residues that stem from the LS equation used.
As a result  the couplings of these states to the continuum could be calculated by evaluating the
 residues of the on-shell partial-wave amplitudes (which would involve complex momenta). 
The residues of the half-off-shell $T$ matrix would provide  the coupling functions with off-shell three-momentum, 
that are needed when the scattering-pole acts as intermediate state \cite{oller.180506.1}.

The exact $N/D$ method arises by  employing the exact discontinuity of partial-wave amplitudes along the left-hand cut 
within the $N/D$ method. We have given some examples in which
 the simplest dispersion relations that can be worked out from the 
exact $N/D$ method agree with the solutions of the Lippmann-Schwinger equation  
for regular and singular potentials. For the case of a singular attractive potential in the limit $r\to 0$, 
an extra parameter, apart from the finite-distance potential given, is needed in order to end with a renormalized scattering amplitude
 from the LS equation with infinite cutoff \cite{case.180502.1,arriola.180502.1}. 
Within the exact $N/D$ method this manifests 
in the necessity to include an extra subtraction constant to the $N/D$ dispersion relation without any free parameter. 
The latter reproduces the solutions obtained with the LS equation by sending the cutoff to infinity 
for regular and  repulsive singular potentials.

As we showed in Ref.~\cite{entem.170930.1} one could also obtain extra solutions from the exact $N/D$ 
 method involving more subtraction constants. 
These solutions could reproduce much better low-energy $NN$ phase shifts (e.g. they could accommodate a positive 
value for an $S$-wave effective range, like the $^1S_0$ one). 
This clearly shows the important role that the exact $N/D$ method is called for to play in phenomenology.

\subsection*{Acknowledgments}
JAO would like to acknowledge informative discussions with Serdar Elhatisari on QDT. 
This work is supported in part by the MINECO (Spain) and EU grants FPA2016-77313-P and FPA2016-77177-C2-2-P.

\medskip

\appendix

\section{Basis for the tensor operators in the $NN$ potential: \cite{epelbaum.101015.1}
  EGM's \& \cite{kaiser.080516.1} KBW's forms}
\label{app:170715.1}
\def\theequation{\Alph{section}.\arabic{equation}}
\setcounter{equation}{0}   

Ref.~\cite{epelbaum.101015.1} gives the $NN$ potential as
\begin{align}
  \label{170714.5}
  V&=V_C^E
  +V_\sigma^E \vsi_1\cdot \vsi_2
  +\frac{i}{2}V_{SL}^E(\vsi_1+\vsi_2)\cdot (\vk\times \vq)
  +V_{\sigma L}^E \vsi_1\cdot (\vq\times \vk) \vsi_2\cdot (\vq\times \vk)
  +V_{\sigma q}^E \vsi_1\cdot \vq \vsi_2\cdot \vq
  +V_{\sigma k}^E \vsi_1\cdot \vk \vsi_2\cdot \vk~,
  \end{align}
where  $\vq=\vp'-\vp$, $\vk=(\vp'+\vp)/2$ and the initial and final three-momenta are 
$\vp$ and $\vp'$, respectively. The functions $V_C(p,p',z)$, \ldots, $V_{\sigma k}(p,p',z)$ may depend on $p$, $p'$ and the
 cosine of the scattering angle  $\theta$. They could also depend on isospin through the isospin invariant operator 
$\mathbf{i}_1\cdot \mathbf{i}_2$.

However, in Eq.~\eqref{170714.5} there are three tensor operators involved, while two are just enough.
The idea is to use an orthogonal triad made from the vectors $\vq$ and $\vk$ as
\begin{align}
  \label{170714.6}
  \mathbf{u}_1&=\frac{\vq}{|\vq|}~\\
  \mathbf{u}_2&=\frac{\vq\times \vk}{|\vq\times \vk|}~,\nn\\
  \mathbf{u}_3&=\frac{\vq\times(\vq\times \vk)}{|\vq\times(\vq\times \vk)|}~.\nn
\end{align}
Then we have the identity,
\begin{align}
  \label{170714.7}
  \vsi_1\cdot \vsi_2&=\vsi_1\cdot \mathbf{u}_1 \vsi_2\cdot \mathbf{u}_1
  +\vsi_1\cdot \mathbf{u}_2 \vsi_2\cdot \mathbf{u}_2
  +\vsi_1\cdot \mathbf{u}_3 \vsi_2\cdot \mathbf{u}_3~,
\end{align}
which allows to express one of the tensor structures in terms of the other two.
In practice we remove the structure proportional to $V_{\sigma k}^E$ in favor
of $V_{\sigma q}^E$ and $V_{\sigma L}^E$. 
Notice that this result does not depend on whether the scattering is on-the-energy
shell or not. Related to this, one has the relation
\begin{align}
  \label{170714.9}
  \vq\times(\vq\times\vk)&=-\vp({\vp'}^2-\vp\cdot\vp')-\vp'({\vp}^2-\vp\cdot\vp')~.
  \end{align}
Therefore, for on-the-energy-shell scattering $\vq\times(\vq\times\vk)$ is just
proportional to $\vk$ but, in general, this is not the case and one should deliver the
argument taking the triad in Eq.~\eqref{170714.6}.

In this way, we can use  the decomposition of the $NN$ potential
introduced in Ref.~\cite{kaiser.080516.1}, that only involves two tensor structures
and five structures in total,
\begin{align}
  \label{170714.8}
  V&=V_C^K+V_S^K \vsi_1\cdot \vsi_2
  +V_{SO}^Ki(\vsi_1+\vsi_2)\cdot (\vq\times\vp)
  +V_{T}^K\vsi_1\cdot \vq \vsi_2\cdot \vq
  +V_{Q}^K\vsi_1\cdot(\vq\times \vp)\vsi_2\cdot(\vq\times\vp)~,
\end{align}
and proceed normally with the partial-wave projection, cf. footnote~\ref{foot.180607.1}. 

Comparing with the structures in Eq.~\eqref{170714.5} (once the term proportional to $V_{\sigma k}$ is
removed), we have a one to one correspondence between them, with the $V_i$ related as
\begin{align}
  \label{170714.10}
  V_C^E&=V_C^K~,\\
  V_\sigma^E&=V_S^K~,\nn\\
  V_{\sigma q}^E&=V_T^K~,\nn\\
  V_{\sigma L}^E&=V_Q^K~,\nn\\
V_{SL}^E&=-2V_{SO}^K~.\nn
\end{align}

For the partial-wave projections we will use the five structures given
in Ref.~\cite{kaiser.080516.1} but within the formulas for the more general
off-the-energy shell case  given in Ref.~\cite{epelbaum.101015.1} (divided by a factor $4\pi$ to match our normalization), 
taking into account the relations given in Eq.~\eqref{170714.10}. For completeness we reproduce the expressions that we use here, 
where the quantum numbers $\ell S J$ are referred in the final/initial states in this same order:

\noindent
a) $S=0$, $\ell=J$:
\begin{align}
\label{180814.4}
\langle J 0 J|t|J 0 J\rangle&=\frac{1}{2}\int_{-1}^{+1}dz\left[V_C-3V_S-q^2 V_T+p^2{p'}^2(z^2-1)V_Q\right]P_J(z)~,\\
q^2&=p^2+{p'}^2-2pp' z~.\nn
\end{align}

\noindent
b) $S=1$, $\ell=J$:
\begin{align}
\label{180814.5}
\langle J 1 J|t|J 1 J\rangle&=\frac{1}{2}\int_{-1}^{+1}dz\left\{
\left[V_C+V_S-4p{p'}z V_{SO}-p^2{p'}^2(1+3z^2)V_Q+4k^2 V_T\right]P_J(z)\right.\nn\\
&\left.+2p{p'}\left[V_{SO}+p{p'}zV_Q-V_T\right]\left(P_{J-1}(z)+P_{J+1}(z)\right)
\right\}~,\\
k^2&=\frac{1}{4}\left(p^2+{p'}^2+2p{p'}z\right)~.\nn
\end{align}

\noindent
c) $S=1$, $\ell=J\pm 1$:
\begin{align}
\label{180814.6}
\langle J\pm1\,1 J|t|J\pm 1\,1J\rangle&=\frac{1}{2}\int_{-1}^{+1}dz\left\{
2p{p'}\Big[V_{SO}\pm\frac{1}{2J+1}\left(V_T-p{p'}z V_Q\right)\Big]P_J(z)
+\Big[V_C+V_S-2p{p'}zV_{SO} \right. \nn\\
&\left. +p^2{p'}^2(1-z^2)V_Q\pm\frac{1}{2J+1}\left(2p^2{p'}^2 V_Q-(p^2+{p'}^2)V_T \right)
\Big]P_{J\pm 1}(z)
\right\}~.
\end{align}

\noindent
d) $S=1$, $\ell'=J-1$, $\ell=J+1$:
\begin{align}
\label{180814.7}
\langle J-1\,1 J|t|J+1\,1 J\rangle&=\frac{\sqrt{J+1}}{\sqrt{J}(2J+1)}\int_{-1}^{+1}dz\left\{
\left[V_T\left(2p{p'} J - p^2z(2J+1)\right)+p^2{p'}^2z V_Q\right]P_J(z)\right.\\
&\left.+\left[V_T\left(p^2(J+1)-{p'}^2J\right)-p^2{p'}^2V_Q\right]P_{J+1}(z)
\right\}~.
\end{align}
 Because of time-reversal symmetry, Eq.~\eqref{180804.10}, this matrix element is equal to 
$\langle J+1\,1 J |t| J-1\,1 J\rangle$ by simultaneously exchanging $p\leftrightarrow {p'}$.

We follow the notation of taking into account the isospin dependence of the different functions 
$V_i^K(p,p',z)$ by rewritten them as $V_i+(4I-3)W_i$, as done in Ref.~\cite{kaiser.080516.1}. 
Here the subscript $i$ refers to $C$, $S$, $T$, $SO$ or $Q$.

\section{Numerical procedure to solve the LS equation}
\label{app.180909.1}
\def\theequation{\Alph{section}.\arabic{equation}}
\setcounter{equation}{0}   

We want to obtain the on-shell $T$ matrix $t(p,p)$ for physical three-momentum $p$. 
We need to solve first the IE for the half-off-shell $T$ matrix $t(k,p)$, which we write 
 in a convenient way by resolving explicitly the pole in the 
denominator for $E=p^2/m+i\ve$, with a slightly positive imaginary part in $p^2$. In this way we have
\begin{align}
\label{200215.10}
t(k,p)&=v(k,p)+\frac{m}{2\pi^2}\dashint_0^\infty \frac{dq q^2}{q^2-{p}^2}v(k,q)
t(q,p)+i\frac{m p}{4\pi} v(k,p)t(p,p)~,
\end{align}
where the Cauchy principal value of an integral is indicated by the symbol $\dashint$. 
We can get rid of the latter by subtracting to the numerator of the integrand its value at the residue, 
because
\begin{align}
\dashint_0^\infty \frac{dq}{q^2-p^2}=0~,~p^2>0
\label{200215.10a}
\end{align}
In this way, Eq.~\eqref{200215.10} can be written as
\begin{align}
t(k,p)&=v(k,p)+\frac{m}{2\pi^2}\int_0^\infty \frac{dq }{q^2-p^2}\big[
{q}^2v(k,q)t(q,p)-p^2 v(k,p)t(p,p)
\big]\nn\\
+&i\frac{m}{4\pi}p v(k,p)t(p,p)~,
\label{200215.12}
\end{align}
which is more suitable for numerical manipulations. 
Indeed, we can discretize  Eq.~\eqref{200215.12} but, 
in order to end with a square matrix, we have also to include an 
extra point corresponding to the on-shell PWA $t(p,p)$ that also appears as an unknown in this equation. 
As a result we end with a system of $N+1$ equations,
 being $N$ the number of points in the partition, with each point indicated by $p_i$, $i=1,\ldots,N$, 
 and fulfilling the restriction that $p\neq p_i$ for all $i$ involved: 
\begin{align}
\label{200215.13}
t(p_i,p)&=v(p_i,p)
+\frac{m}{2\pi^2}
\sum_{j=1}^N \frac{\omega_j}{p_j^2-p^2} \big[ p_j^2
v(p_i,p_j) t(p_j,p)
-{p}^2  v(p_i,p) t(p,p) \big] \\
&+i\frac{m p}{4\pi} v(p_i,p) t(p,p)~,\nn\\
t(p,p)&=v(p,p)+\frac{m}{2\pi^2}\sum_{j=1}^N\frac{\omega_j}{p_j^2-{p}^2}
\big[p_j^2 v(p,p_j) t(p_j,p)-{p}^2v(p,p)t(p,p) \big]\nn\\
&+i\frac{m p}{4\pi} v(p,p)t(p,p)~.\nn
\end{align}
Here, $\omega_i$ refers to the weight of each point in the partition corresponding to the numerical method 
employed for integration. To have a finite interval of integration we use the change of variable
\begin{align}
\label{270915.12}
p/m_\pi&=\frac{1}{x}- 1~,\nn\\
x&=\frac{1}{1+p/m_\pi}~,
\end{align}
 In this way, 
\begin{align}
\label{270915.13}
x\in[0,1]~ {\text{for}}~p\in[0,\infty]~. 
\end{align}
 The linear system of $N+1$ that results then is ($m_\pi=1$)
\begin{align}
\label{270915.14}
&\sum_{j=1}^N\left\{
\delta_{ij}-\frac{m}{2\pi^2}\frac{\omega_j}{x_j^2}\frac{p_j^2}{p_j^2-p^2}v(p_i,p_j)\right\}t(p_j,p)
+v(p_i,p) \sigma_p t(p,p)=v(p_i,p)~,\nn\\
&\left[1+\sigma_p v(p,p)
\right]t(p,p)
-\frac{m}{2\pi^2}\sum_{j=1}^N\frac{\omega_j}{x_j^2}\frac{p_j^2}{p_j^2-p^2}v(k,p_j)t(p_j,p)
=v(p,p)~,
\end{align}
with
\begin{align}
\label{270915.15}
\sigma_p&=\frac{m p^2}{2\pi^2}\sum_{j=1}^N \frac{\omega_j}{x_j^2}\frac{1}{p_j^2-p^2}-\frac{i m p}{4\pi}~.
\end{align}

\section{LS equation for the $^3S_1-{^3D_1}$ and  $^3P_0$ systems}
\label{app.180911.1}
\def\theequation{\Alph{section}.\arabic{equation}}
\setcounter{equation}{0}   

We derive in this Appendix the LS equations used to calculate the half-off-shell $T$-matrix 
elements  with the threshold behavior fixed for the $^3S_1-{^3D_1}$ and $^3P_0$ PWAs.
For the former the $S$-wave $^3S_1$ scattering length, $a_s$, is an input and 
the $P$-wave scattering volume, $a_V$, is so for the latter. 

\subsection{The LS equation for the $^3S_1-{^3D_1}$ coupled PWAs}
\label{app.180911.2}
 We develop first the  LS equation in the coupled $^3S_1-{^3D_1}$ PWAs  with the restriction of reproducing 
a given value of the $^3S_1$ scattering length, $a_s$. 
  We employ the same trick as in Eq.~\eqref{200215.12} to remove the principal value when considering the IE 
for obtaining numerically the solution of the LS equation. As there, we consider  the half-off-shell $T$-matrix element 
$t_{ij}(k,p;\frac{{p}^2}{m}+i\ve)=t_{ij}(k,p)$, 
where $i,\,j=1,2$ refer to the coupled partial waves (1 for  $^3S_1$ and 2 for  $^3D_1$).  
 In this way, we write for the coupled case
\begin{align}
t_{ij}(k,p)&=v_{ij}(k,p)
+\frac{m}{2\pi^2}\sum_{\gamma=1}^2\int_0^\infty \frac{dq }{q^2-p^2}\big[
q^2v_{i\gamma}(k,q)t_{\gamma j}(q,p)-p^2 v_{i\gamma}(k,p)t_{\gamma j}(p,p)
\big]\nn\\
+&i\frac{m p}{4\pi}\sum_{\gamma=1}^2 v_{i\gamma}(k,p)t_{\gamma j}(p,p)~.
\label{180118.1}
\end{align}

Now, let us discuss how to impose a given value of the scattering length $a_s$. 
For that we  add a counterterm $v_0$ to $v_{11}(p,k)$ and define 
\begin{align}
\label{180129.4}
v_{11}(k,p)&=v_0+\hat{v}_{11}(k,p)\\
v_{ij}(k,p)&= \hat{v}_{ij}(k,p)~\text{ , } i+j>2 ~,\nn 
\end{align}
where $\hat{v}_{ij}(k,p)$ is the original $v_{ij}(k,p)$  defined in Eq.~\eqref{180122.2}. 
Let us note that 
\begin{align}
\label{180129.5}
\hat{v}_{ij}(0,0)&=\lim_{p,k\to 0}\hat{v}_{ij}(k,p)=0~,
\end{align}
as follows from Eq.~\eqref{180122.2} by direct evaluation (for $i+j>2$ this is clear because of the required near-threshold 
behavior of $v_{ij}(p,p)$ that vanishes as $p^{\ell_i+\ell_j}$ for $p\to 0$). 

In order to fix $v_0$ we particularize Eq.~\eqref{180118.1} for $p=0$ and $i=j=1$. It reads
\begin{align}
\label{180129.6}
t_0&=\lim_{p\to 0}t_{11}(p,p)=
v_{0}+\frac{m}{2\pi^2}\sum_{\gamma=1}^{2}\int_0^\infty dq \,v_{1\gamma}(0,q)t_{\gamma 1}(q,0)~,
\end{align}
with $t_0=-4\pi a_s/m$, cf. Eq.~\eqref{180129.7}. 
Taking into account the dependence of $v_{1\gamma}(0,p)$ 
on the rhs of the previous equation on $v_0$, we then have the following equation for the latter
\begin{align}
\label{180916.1}
v_0\left(1+\frac{m}{2\pi^2}\int_0^\infty dq\, t_{11}(q,0)\right)&=
t_0-\frac{m}{2\pi^2}\sum_{\gamma=1}^2\int_0^\infty dq\, \hat{v}_{1\gamma}(0,q)t_{\gamma 1}(q,0)~.
\end{align}
In order to evaluate $v_0$ from the previous equation we need first to calculate the half-off-shell $T$ matrix elements 
$t_{i 1}(k,0)$ for $0\leq k<\infty$. 
We can calculate them from Eq.~\eqref{180118.1} which, after taking into account the result in Eq.~\eqref{180916.1}, can be 
written directly in terms of $t_0$ as 
\begin{align}
\label{180916.2}
t_{i 1}(k,0)&=t_0 \delta_{i 1}+\hat{v}_{i 1}(k,0)
+\frac{m}{2\pi^2}\sum_{\gamma=1}^2\int_0^\infty dq \left[\hat{v}_{i\gamma}(k,q)-\delta_{i 1}\hat{v}_{i\gamma}(0,q)\right]t_{\gamma 1}(q,0)~.
\end{align}
Once this IE is solved we can then calculate $v_0$ from Eq.~\eqref{180916.1} and then  Eq.~\eqref{180118.1}
is given in terms of known quantities from the OPE $NN$ potential and the scattering length $a_s$. 
It is numerically convenient to introduce as an intermediate step some sort of cutoff $\Lambda$ on $v_{ij}(k,p)$, and then 
taking the limit $\Lambda\to \infty$.

\subsection{The LS equation for the $^3P_0$ PWA}
\label{app.180911.3}

The threshold behavior of the on-shell $T$ matrix is restricted so as to impose a given value of 
$a_V$, defined in Eq.~\eqref{180910.5}. This is achieved by adding a contact interaction to the 
$^3P_0$ partial-wave projected potential of Eq.~\eqref{180828.2}, which now is denoted by $\hat{v}(k,p)$. 
The potential reads 
\begin{align}
\label{180916.7}
v(k,p)=C k p +\hat{v}(k,p)~.
\end{align}
We discuss how $C$ can be fixed in terms of $a_V$.
 The threshold behavior  of the on-shell $T$ matrix is
\begin{align}
\label{180916.8}
t(p,p)\xrightarrow[p\to 0]{} -\frac{4\pi a_V p^2}{m}+{\cal O}(p^3)~.
\end{align}
It is also clear that the threshold behavior for the potential [this can be also explicitly worked out from Eq.~\eqref{180828.2}] is 
such that
\begin{align}
\label{180916.9}
v(k,p)\xrightarrow[p\to 0]{} p f(k)+{\cal O}(p^3)~,
\end{align}
with $f(k)$ a function of $k$.  
 From the last equation and the LS equation  it is also clear that $t(k,p)$ is proportional to $p$ for $p\to 0$. 

We make use of Eqs.~\eqref{180916.8} and \eqref{180916.9} to simplify 
Eq.~\eqref{200215.13} in the limit $p\to 0$ for the half-off-shell and on-shell $T$-matrix elements as
\begin{align}
\label{180916.10}
t(k,p)&=v(k,p)+\frac{m}{2\pi^2}\int_0^\infty dq v(k,q)t(q,p)+{\cal O}(p^4)~,\\
\label{180916.11}
t(p,p)&=v(p,p)+\frac{m}{2\pi^2}\int_0^\infty dq v(p,q)t(q,p)+{\cal O}(p^5)~.
\end{align}
From the last relation we can express $C$ in terms of $a_V$. 
In accordance with the  limit behaviors of $\hat{v}(k,p)$ and $t(k,p)$ for $p\to 0$ 
 we introduce the limit functions
\begin{align}
\label{180916.12}
\widetilde{t}(q)=\lim_{p\to 0}\frac{t(q,p)}{p}~,\\
\widetilde{v}(q)=\lim_{p\to 0}\frac{\hat{v}(q,p)}{p}~.
\end{align}
In terms of them, we have  from Eq.~\eqref{180916.11} the following expression for $C$,
\begin{align}
\label{180916.13}
C\left(1+\frac{m}{2\pi^2}\int_0^\infty dq q \widetilde{t}(q)\right)&=
-\frac{4\pi a_V}{m}-\frac{4g}{m_\pi^2}-\frac{m}{2\pi^2}\int_0^\infty dq \widetilde{v}(q)\widetilde{t}(q)~,
\end{align}
where $\lim_{k\to 0}\hat{v}(k,k)/k^2=4g/m_\pi^2$.  
For the calculation of $C$ from Eq.~\eqref{180916.13} we still need 
to work out $\widetilde{t}(q)$. For that we divide  Eq.~\eqref{180916.10} by 
$p$, take the limit $p\to 0$, and replace Eq.~\eqref{180916.13} into it. It results that
 \begin{align}
\label{180916.14}
\widetilde{t}(k)&=\widetilde{v}(k)-k\left(\frac{4\pi a_V}{m}+\frac{4g}{m_\pi^2}\right)
+\frac{m}{2\pi^2}\int_0^\infty dq\left[\hat{v}(k,q)-k \widetilde{v}(q)\right]
\widetilde{t}(q)~.
\end{align}
Once this IE is solve, the potential in Eq.~\eqref{180916.7} is known 
and the solution of the LS equation, Eq.~\eqref{200215.13}, can be worked out.

The procedure discussed here and in the previous Appendix \ref{app.180911.2}
is similar to the solution of the LS equation using
subtractive renormalization \cite{subtractive.181027.1}. 


\addcontentsline{toc}{section}{References}

\end{document}